\documentclass{article}

\usepackage{graphicx}
\usepackage{amsmath}
\usepackage{amsfonts}
\usepackage{amssymb}
\usepackage{dsfont}
\usepackage{wrapfig}
\usepackage{pdflscape}
\usepackage{hhline}

\numberwithin{equation}{section}

\def\be{\begin{eqnarray}}
\def\ee{\end{eqnarray}}
\def\nn{\nonumber}

\hoffset= - 1.0in         

\title{\textbf{Cabling procedure \\ for the colored HOMFLY polynomials} \vspace{.5cm}}
\author{\textbf{A.Anokhina}\footnote{ {\small \textit{MIPT} and \textit{ITEP, Moscow, Russia}};
anokhina@itep.ru}, \ \textbf{An.Morozov}\thanks{{\small \textit{Moscow State University} and \textit{ITEP, Moscow, Russia}};
Andrey.Morozov@itep.ru}\date{ }}

\begin{document}

\maketitle

\vspace{-6.5cm}

\begin{center}
\hfill ITEP/TH-23/13\\
\end{center}

\vspace{5.5cm}

\begin{abstract}
In the present paper, we discuss the cabling procedure for the colored HOMFLY polynomial. We describe how it can be used and how one can find all the quantities such as projectors and $\mathcal{R}$-matrices, which are needed in this procedure. The constructed matrix expressions for the projectors and $\mathcal{R}$-matrices in the fundamental representation allow one in principle to find the HOMFLY polynomial in any representation for any knot. Our computational algorithms allowed us to do this for the knots and links with $|Q|m\le 12$ where $m$ is the braid width of the knot or link the number of strands in a braid representation of the knot or link and $|Q|$ is the size of the representation.

We also discuss the cabling procedure from the group theory standpoint, deriving the expressions for the fundamental $\mathcal{R}$-matrices and illuminating several conjectures proposed in previous papers.
\end{abstract}

\tableofcontents

\bigskip

\section{Introduction}

Some quantum effects are known to be non-perturbative phenomena. Some of them arise already in quantum mechanics, e.g., the Aharonov-Bohm effect~\cite{ES},\cite{AB}, and there are much more of them in the gauge field theory~\cite{Col},\cite{Rub}. The non-perturbative phenomena now are attracting more and more attention. One of the main reasons for this is that there is a huge amount of exactly solvable problems where, unlike the realistic QFT, the answer itself is well defined and thus available to a rigorous analysis. On the other hand, there are crucial observable phenomena including confinement in QCD~\cite{2dQED}, which probably could be understood while studying non-perturbative effects. The studies of these effects gave birth to several new types of theories such as Seiberg-Witten theory~\cite{SW1},\cite{SW2} and conformal field theory~\cite{CFT} where non-perturbative effects play crucial role.

Quite important class of such non-perturbative theories are the so-called \emph{topological field theories}~\cite{Atiyah}. They are a very special class of QFT-s, where the observables, e.g., amplitudes, are unaffected by small perturbations of for example the coupling constant, and in this sense are topological invariants. The advances in the theories of this type led to the study of various topological objects either familiar ones or newly discovered. The most direct way to adopt QFT for the study of topological objects is to construct a QFT with an action that is in a sense a total derivative so that the corresponding partition function
\begin{equation}
\mathcal{Z}=\int\limits_{\mathcal{M}^{\prime}}[D\mathcal{A}] e^{\frac{i}{\hbar}\int\limits_{\mathcal{M}}\mathcal{L}[\mathcal{A}]}
\label{partfunc}
\end{equation}
remains constant under smooth deformations of the manifold $\mathcal{M}$~\cite{SemTSh},\cite{DH}. Such partition function can be quite sophisticated for a complicated manifold with nontrivial topological properties (e.g., see~\cite{DHex}). The other way is to consider a very simple manifold, e.g., a sphere, but to insert an additional structure into the integral~(\ref{partfunc}), i.e., try to study the averages of some observables. For a gauge theory the usual candidate is the \emph{Wilson average}~\cite{QFT},
\begin{equation}
\left<W^{\mathcal{K}}\right>=\cfrac{1}{\mathcal{Z}}\int\limits_{\mathcal{M}^{\prime}}[D\mathcal{A}]\mathrm{Tr}\ \mathrm{Pexp}\left(\oint\limits_{\mathcal{K}}\mathcal{A}dx \right) e^{\frac{i}{\hbar}\int\limits_{\mathcal{M}}\mathcal{L}[\mathcal{A}]}
\label{Wilson}
\end{equation}
Wilson average is an average of the Wilson loop, which is the path exponent of the integral of the connection $\mathcal{A}$ over the closed contour $\mathcal{K}$. Then even for a topologically trivial manifold a nontrivial embedding of the contour $\mathcal{K}\rightarrow\mathcal{M}$ can be considered. This embedding can be characterized by the corresponding Wilson loop average. The simplest theory, in which the observables of this type can be studied, is the $3$-dimensional topological gauge theory called the $3D$ Chern-Simons theory~\cite{CS}. The Wilson averages then correspond to knots, i.e., embeddings $S_1\in S_3$. This theory has a cubic Lagrangian,
\begin{equation}
\mathcal{L}_{CS}=\frac{k}{4\pi} \left(\mathcal{A}\wedge d\mathcal{A}-\frac{2}{3}\mathcal{A}\wedge\mathcal{A}\wedge\mathcal{A}\right).
\label{CS}
\end{equation}
The Wilson loop average $\left<W^{\mathcal{K}}_Q\right>_{CS(N,k)}$ then is defined for the knot $\mathcal{K}$ in the Chern-Simons theory with the coupling constant $k$ and the Wilson loop carrying the representation $Q$ of the gauge group $SU(N)$. In the case of link, i.e., of several knots intertwined with each other, an arbitrary representation $Q_i$ can be put at each component of the link.

In 1989, E.Witten suggested~\cite{Witt} that the Wilson averages of the Chern-Simons theory on the $S^3$ manifold with the $SU(2)$ gauge group are equal to the colored Jones polynomials known from the mathematical knot theory. This fact can be generalized to the gauge group $SU(N)$ with arbitrary $N$: the Jones polynomials should be replaced then by the HOMFLY polynomials~\cite{KauffTB}.

HOMFLY polynomials $H^{\mathcal{K}}(A,q)$ for the case of fundamental representation of $SU(N)$ in mathematical knot theory can be defined via the skein relations~\cite{HOMFLY},\cite{PrzTrac}. The skein relations are the set of equations on the HOMFLY polynomials of the given knot $\mathcal{K}$ and of the knots $\mathcal{K}^{\prime}$ and $\mathcal{K}^{\prime\prime}$ obtained by inverting and resolving any crossing (see Fig.\ref{eq:skeinpic}).

\begin{figure}
\begin{center}
\begin{picture}(180,50)(0,-10)
\put(-25,15){\makebox(0,0)[cc]{$A$}}
\put(-5,-5){\makebox(0,0){$\mathcal{K}$}}
\put(-15,0){\vector(2,3){20}}
\put(5,0){\line(-2,3){8}}
\put(-7,18){\vector(-2,3){8}}
\put(25,15){\makebox(0,0)[cc]{$- \ A^{-1}$}}
\put(50,-5){\makebox(0,0){$\mathcal{K}^{\prime}$}}
\put(60,0){\vector(-2,3){20}}
\put(40,0){\line(2,3){8}}
\put(52,18){\vector(2,3){8}}
\put(100,15){\makebox(0,0)[cc]{=\ \ $\Big(q-q^{-1}\Big)$}}
\put(150,-5){\makebox(0,0){$\mathcal{K}^{\prime\prime}$}}
\qbezier(140,0)(150,15)(140,30)
\put(142,27){\vector(-2,3){3}}
\qbezier(160,0)(150,15)(160,30)
\put(158,27){\vector(2,3){3}}
\end{picture}
\end{center}
\label{eq:skeinpic}
\caption{Skein relations in the topological framing (see Sec.~\ref{framing}).}
\end{figure}

\noindent The skein relations and the fundamental HOMFLY polynomial for the unknot provides the full definition
of all fundamental HOMFLY polynomials, i.e., any knot can be reduced to the unknot using the skein relations and
the answer does not depend on the order, in which the skein relations for different crossings
are applied\footnote{In principle, the skein relations exist also for colored polynomials (see further)
but they are much more complicated and cannot be used to reduce any knot to the unknot. See~\cite{MM} for details.}
\cite{HOMFLY}. The proof of topological invariance of the answer, i.e., that it does not change with smooth deformations
of the knot is provided, e.g., in~\cite{HOMFLY},\cite{PrzTrac}. It is also known that the HOMFLY polynomial is a Laurent
polynomial in $A$ and $q$ up to the factor $(q-q^{-1})$ for knots ~\cite{HOMFLY},\cite{PrzTrac} and diverges as  $(q-q^{-1})^{-n}$ for an $n$-component
link in the fundamental representation,~\cite{linksing}. Thus, the skein relations give a constructive though
quite sophisticated definition of the HOMFLY polynomials. Anyway, with the use of this method a huge amount of data for
different knots was obtained~\cite{katlas}.

According to the ideas of E.Witten~\cite{Witt} the HOMFLY polynomial is equal to the Chern-Simons Wilson average:
\begin{equation}
\left<W^{\mathcal{K}}\right>_{CS(N,k)}=H^{\mathcal{K}}\left(A=q^N,\ q=e^{\frac{2\pi i}{k+N}}\right).
\label{eq:WilHOM}
\end{equation}
The numerous studies of the connections between the Chern-Simons theory and the knot theory ~\cite{CSknots:MR}-\cite{Mar2},\cite{Inds1}-\cite{Inds4} as well as those of the  Chern-Simons theory~\cite{CS-CFT:MS}-\cite{CS-CFT:AGS} and of the knot theory~\cite{Schwarz}-\cite{Kauff},\cite{Inds5}-\cite{Morton4} themselves were performed recently.

The approach relating the knot invariants to the Chern-Simons theory~\cite{Witt},\cite{CSknots:MR}-\cite{Mar2},\cite{Inds1}-\cite{Inds4} as well as more mathematical approaches based on the Hecke algebras~\cite{Schwarz}-\cite{Kauff},\cite{Morton1}-\cite{Morton4} involves the construction, in which a vector space $V$ is associated with each connection component of a link. More precisely this $V$ is a space of a Lie group representation. Defined in a such way, a knot invariant depends (for a given knot and for a given group) on the discrete variable that marks the choice of the representation. This variable is called the color of the knot invariant~\cite{CSknots:KM}. Considering various representations of the group $SU(N)$, one can define the so called colored HOMFLY polynomials that generalize the HOMFLY polynomials defined earlier by skein relations~(\ref{eq:skeinpic}), which in fact correspond to the fundamental representation. Reshetikhin-Turaev approach~\cite{Tur}-\cite{GuadMarMin} used, e.g., in the series of publications~\cite{MMM1}-\cite{MMM3} provides a possible way to define colored HOMFLY basing on this idea.

\

In the present paper, we use the method to evaluate the HOMFLY polynomials based on the Reshetikhin-Turaev formalism~\cite{Tur}-\cite{GuadMarMin}. According to this formalism the HOMFLY polynomials can be described as a specially weighted trace of the product of the $\mathcal{R}$-matrices:
\begin{equation}
H^{\mathcal{K}}_{T_1\otimes T_2\ldots}=\mathrm{Tr}_{T_1\otimes T_2\ldots} \prod\limits_{\alpha}\tilde{\mathcal{R}}_{\alpha},
\end{equation}
where $\alpha$ enumerates all the crossings in the braid. The knot is obtained by closing this braid. It can be shown~\cite{Pras} that each knot can be represented as a closure of a braid.  A knot can have several different braid representations, even with different number of strands. E.g., the simplest braid representation of the trefoil knot is:
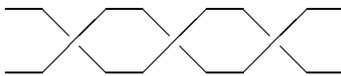
\begin{center}
\begin{picture}(130,30)(0,0)
\put(0,0){\line(1,0){14}}
\put(0,24){\line(1,0){14}}
\put(14,0){\line(1,1){24}}
\put(14,24){\line(1,-1){10}}
\put(28,10){\line(1,-1){10}}
\put(38,0){\line(1,0){14}}
\put(38,24){\line(1,0){14}}
\put(52,0){\line(1,1){24}}
\put(52,24){\line(1,-1){10}}
\put(66,10){\line(1,-1){10}}
\put(76,0){\line(1,0){14}}
\put(76,24){\line(1,0){14}}
\put(90,0){\line(1,1){24}}
\put(90,24){\line(1,-1){10}}
\put(104,10){\line(1,-1){10}}
\put(114,0){\line(1,0){14}}
\put(114,24){\line(1,0){14}}
\end{picture}
\end{center}
In principle, the Reshetikhin-Turaev formalism does not require a braid representation of the knot, but we will describe all the methods using this representation of a knot for simplicity.

One of the crucial ideas of this approach is that all the vectors of the same irreducible representation are eigenvectors of the $\mathcal{R}$ and, moreover, have the same eigenvalue:
\begin{equation}
\tilde{\mathcal{R}}=\sum_{Q}\lambda_Q \,I_{d_Q}\!\!\otimes \mathcal{R},
\end{equation}
where $d_Q$ is the dimension of the representation $Q$.
Since the $\mathcal{R}$-matrix acts as a constant  on any irreducible representation it can be said that the $\mathcal{R}$-matrix acts in the space of intertwining
 operators.
This can be understood as if the $\mathcal{R}$-matrix acts on the highest weight vector $\xi_i$ of the representation $Q_i$ such that
$\mathcal{R}_{ij}=[\mathcal{R}\xi_{i}]_{\xi_j}$.
In this sense,  it can be said that the $\mathcal{R}$-matrices act in the space of irreducible representations.

The weighted trace~\cite{HOMFLY},\cite{Jones2} is defined so that the trace over all vectors in the irreducible representation $Q$ gives the character of the representation $S^*_{Q}(A,q)$ (Schur polynomial~\cite{Fulton}): $\mathrm{Tr}_{Q} I_{d_Q}=S^*_{Q}(A,q)$.
Here, the variable $A$ labels the choice of the gauge group $SU(N)$; it appears after the change of variables $(q,N)\ \rightarrow\ (q, A=q^N)$ and after the analytic continuation to arbitrary values of $A$. The properties of the $\mathcal{R}$-matrices and of the weighted trace lead to the character expansion formula for the HOMFLY polynomial~\cite{MMM2}:

\begin{equation}
H^{\mathcal{K}}_{T_1\otimes T_2\otimes\ldots}(A,q)=\sum\limits_{Q\vdash T_1\otimes T_2\otimes\ldots}h^{\mathcal{K},Q}_{T_1\otimes T_2\otimes\ldots}(q)S^*_{Q}(A,q).
\label{eq:charexp}
\end{equation}
The coefficients $h^{\mathcal{K},Q}_{T_1\otimes T_2\otimes\ldots}$ are Laurent polynomials in only one variable\footnote{In principle, $h^{\mathcal{K},Q}_{T_1\otimes T_2\otimes\ldots}$ can also depend on $A$ with some choices of the framing (see Sec.~\ref{framing}). This dependence is quite simple and can be derived from the fact that $h^{\mathcal{K},Q}_{T_1\otimes T_2\otimes\ldots}$  depend only on $q$ when the vertical framing is chosen.} $q$, i.e., they do not depend on $N$. The dependence of the HOMFLY polynomial on $N$ is totally described by the Schur polynomials. From the existence of the character expansion formula it follows that the coefficients corresponding to the different irreducible representations can be studied independently. The detailed description and analysis of this method can be found, e.g., in~\cite{MMM2},\cite{AMMM1}.

Let us emphasize that we slightly reformulate the approach presented in~\cite{MMM2}. Namely, if one considers the \textit{fundamental representation}, then there is a basis~(\ref{eq:nb}) where the form of \textit{all} the $\mathcal{R}$-matrices acting in the braid is easily defined. We find this form explicitly in Sec.~(\ref{sec:genR}) and then calculate the fundamental HOMFLY polynomials directly in terms of $\mathcal{R}$-matrices. Hence, there is no need for $U$-matrices (see~\cite{MMM2}) in the case of the fundamental representation. \textit{Inter alia}, this observation notably simplifies computer simulations of the fundamental HOMFLY polynomials.

The character expansion~(\ref{eq:charexp}) also leads to the construction of the so-called extended HOMFLY polynomial. To find them the Schur polynomials depending on the variables $A$ and $q$ should be replaced with the ones dependent on the set of the time variables $t_k$, while the coefficients $h^{\mathcal{K},Q}_{T_1\otimes T_2\otimes\ldots}$ remain intact. These extended HOMFLY polynomials~\cite{MMM1} are closely related to integrable systems~\cite{genexp1},\cite{genexp2}. Plain HOMFLY polynomials, which are knot invariants, recovers after the substitution $t_k=\frac{A^k-A{-k}}{q^k-q^{-k}}$ in the extended HOMFLY.

Calculations of the fundamental and of the colored HOMFLY polynomials using the Reshetikhin-Turaev formalism in a straightforward way were provided for several important examples in~\cite{IMMM1},\cite{IMMM3}. In the present paper, we  discuss slightly different topic. There is an approach, which allows one to study the colored knot invariants using the results in the fundamental representation. This is called cabling procedure~\cite{Adams}. In the present paper, we discuss how to use the cabling procedure within the Reshetikhin-Turaev formalism and interpret it from the group theory point of view.

\bigskip

The paper is organized as follows. In Sec.s~\ref{sec:CP}-\ref{framing}, we explain what is the cabling procedure, what elements are needed to use it and how to find them. Sec.s~\ref{sec:2str}-\ref{sec:links} provide some results evaluated using the cabling procedure. Then the group theory properties of the cabling procedure are discussed in Sec.~\ref{sec:repr}. Finally in Sec.s~\ref{sec:eig}-\ref{sec:ap} we give explanations and cabling descriptions for some conjectures suggested in previous papers~\cite{DMMSS},\cite{IMMM3}.

\

\bigskip

\

\noindent Through the paper we will use the following notations:

\

\noindent Irreducible representations of $sl_n$ algebra are enumerated by Young diagrams, i.e., by sets of natural numbers $Q=[q_1,\ldots, q_m]$ such that $q_1\geq q_2\geq \ldots \geq q_m$~\cite{Fulton}. Usually, we will denote a Young diagram by a string $Q=[q_1\ldots q_m]$ omitting the square brackets when it is clear what is implied. A Young diagram can be represented graphically by as the $k$ lines made of $q_1\leq q_2\leq \ldots \leq q_m$ boxes. The total number of boxes in the Young diagram $|Q|=\sum\limits_i k_i$ is called the size or the level of the corresponding representation.

\noindent The notation $Q\vdash T_1\otimes\ldots \otimes T_m$ implies that the representation $Q$ appears in the decomposition of the tensor product $T_1\otimes\ldots \otimes T_m$.

\noindent The notation $T_1\ldots T_m|Q$ is used for the component of the product of the representations $T_1,\ldots, T_m$ corresponding to the representation $Q$.

\noindent The $\mathcal{R}$-matrices are always considered as acting in the space of intertwining operators.

\noindent Describing a braid, by $i$-th strand we mean the strand, which occupies the $i$-th position in a certain piece of braid. When we say, that the $i$-th strand intersects the $i+1$-st one, we  imply the  piece of braid containing this crossing  unless otherwise is stated.

\noindent The colored $\mathcal{R}$-matrices corresponding to the crossing between representations $T_i$ and $T_{i+1}$ in the braid with representations $T_1,\ldots, T_k$ is denoted by $\mathcal{R}_{T_1\otimes\ldots(T_i\otimes T_{i+1})\ldots \otimes T_k}$. By definition~\cite{Tur}, this operator acts as the unit operator on all the representations but $T_i$ and $T_{i+1}$ but the concrete form of its matrix depends on all the representations $T_1,\ldots T_k$. We omit the sign $\otimes$ unless it leads to a confusion.

\noindent The fundamental $\mathcal{R}$-matrix corresponding to the crossing between the strands $i$ and $i+1$ in the fundamental braid is denoted by $R_i$.

\noindent The projector from representation $T$ onto representation $Q$ is denoted by $P^T_Q$. The notation $P_Q$ corresponds to the projector from representation $1^{|Q|}$ onto representation $Q$. The representation $Q$ can be reducible, e.g., $Q=Q_1\otimes Q_2$.

\noindent The Racah matrix~\cite{MMM2},\cite{Vil} describing the transition between the bases $T_1\otimes (T_2 \otimes T_3)$ and  $(T_1\otimes T_2) \otimes T_3$ is denoted by ${\cal U}_{T_1\otimes T_2\otimes T_3}$. This matrix is orthogonal, i.e., $UU^{\dagger}=1$ and becomes symmetric in a certain basis (at least in all cases of our interest), i.e., $U=U^{\dagger}$. Hence, $U^1=1$ or equivalently $U=U^{-1}$; by this reason we do not make a difference between straight and inverse Racah matrices.

\noindent The Schur polynomials $S_Q$ are usually defined as functions of an infinite set of time variables $\{t_k\}$~\cite{Fulton}. Those needed for the HOMFLY polynomials are functions of $A$ and $q$ only and are defined as
\begin{equation}
S^*_Q(A,q)=S_Q\ \ \mbox{ for  }\ t_k=\cfrac{A^k-A^{-k}}{q^k-q^{-k}}.
\end{equation}
Such choice of $\{t_k\}$ is called the topological locus. Transition from the usual HOMFLY polynomials to the extended ones corresponds to leaving the topological locus. For the $S^*_Q(A,q)$, there also exists a constructive hook formula, which allows one to calculate them easily:
\begin{equation}
S^*_Q(A,q)=\prod_{(i,j)\in Q}
\frac{Aq^{i-j}-A^{-1}q^{j-i}}{q^{h_{i,j}}-q^{-h_{i,j}}},
\begin{picture}(105,15)(-35,-15)
\put(0,0){\line(1,0){70}}
\put(0,-10){\line(1,0){70}}
\put(0,-20){\line(1,0){60}}
\put(0,-30){\line(1,0){40}}
\put(0,-40){\line(1,0){20}}
\put(0,-50){\line(1,0){20}}
\put(0,0){\line(0,-1){50}}
\put(10,0){\line(0,-1){50}}
\put(20,0){\line(0,-1){50}}
\put(30,0){\line(0,-1){30}}
\put(40,0){\line(0,-1){30}}
\put(50,0){\line(0,-1){20}}
\put(60,0){\line(0,-1){20}}
\put(70,0){\line(0,-1){10}}
\put(15,-15){\makebox(0,0)[cc]{\textbf{x}}}
\put(15,5){\makebox(0,0)[cc]{$i$}}
\put(-5,-15){\makebox(0,0)[cc]{$j$}}
\qbezier(19,-11)(45,20)(55,-15)
\put(40,10){\makebox(0,0)[cc]{$k$}}
\qbezier(11,-19)(-17,-40)(15,-45)
\put(60,-40){\makebox(0,0)[lc]{$h_{i,j}=k+l+1$.}}
\end{picture}
\end{equation}

\noindent $[n]_q$ denotes the quantum number $n$, i.e., $[n]_q\equiv\frac{q^n-q^{-n}}{q-q^{-1}}$.

\section{Cabling procedure\label{sec:CP}}

There are several ways to evaluate the colored HOMFLY polynomials. Since it is not clear yet how to use the direct approach for representations that are neither symmetric, nor antisymmetric another approach is often used (see, e.g.,~\cite{Morton1}). This approach uses the so-called cabling procedure~\cite{Adams}. The main idea behind this approach is the following. As was already mentioned, the HOMFLY polynomials can be represented as the character expansion~(\ref{eq:charexp}). The size of the representations $Q$ in the r.h.s of~(\ref{eq:charexp}) is equal to the sum product of the sizes of representation $T_1$, $T_2$ etc. at the l.h.s. It is therefore natural to reduce calculating of an invariant for a given knot/link in the representations $T_1$, $T_2$ etc. to the calculating of the same invariant in the fundamental representation but of the knon/link with $|T_1|+|T_2|+\ldots$ strands in the braid representation.

Hence the cabling procedure appears~\cite{Adams}. The initial knot is replaced with a satellite knot, i.e., with a braid placed along the initial knot. If this braid is just $|T_1|$ parallel strands, then the resulting HOMFLY in the representation $T_2$ will be equal to the HOMFLY polynomial of the initial knot in representation $T_2^{|T_1|}$:
\begin{equation}
\boxed{H^{\mathcal{K}}_{T_2^{|T_1|}}=H^{\mathcal{K}^{|T_1|}}_{T_2}}
\end{equation}
This is a basic relation, on which the cabling procedure is based. To get the HOMFLY (or Jones) polynomial in some other representation $Q$ of the size $|T_2|^{|T_1|}$ a certain linear combination of several different satellite knot polynomials should be taken\footnote{For given $T_1$ and $T_2$, one can obtain not an arbitrary representation $Q$, but only the ones that the reducible representation $T_2^{|T_1|}$ contains.}. Constructing this linear combination corresponds to the projection from the representation $T_2^{|T_1|}$ onto the representation $Q$.  In the Reshetikhin-Turaev formalism each crossing in the knot projection corresponds to the $\mathcal{R}$-matrix, hence the linear combination of the satellite knot polynomials (differing by a set of several crossings) corresponds to a combination of $\mathcal{R}$-matrices. Thus, the projection can be described as action of a linear operator called projector, which can be represented as a polynomial in $\mathcal{R}$-matrices (see Sec.~\ref{sec:proj} for details).

As a summary, to express the colored HOMFLY polynomials in terms of HOMFLY polynomials in the fundamental representation, one has to perform the following steps, which form the cabling procedure:
\begin{itemize}
\item If there is a representation $T_i$ at the strand then the strand is replaced with $|T_i|$ parallel strands. This step for the trefoil knot and for $|T_i|=2$ can be described with the following picture

\begin{picture}(360,100)(0,-37)
\put(0,0){\line(1,0){14}}
\put(0,24){\line(1,0){14}}
\put(14,0){\line(1,1){24}}
\put(14,24){\line(1,-1){10}}
\put(28,10){\line(1,-1){10}}
\put(38,0){\line(1,0){14}}
\put(38,24){\line(1,0){14}}
\put(52,0){\line(1,1){24}}
\put(52,24){\line(1,-1){10}}
\put(66,10){\line(1,-1){10}}
\put(76,0){\line(1,0){14}}
\put(76,24){\line(1,0){14}}
\put(90,0){\line(1,1){24}}
\put(90,24){\line(1,-1){10}}
\put(104,10){\line(1,-1){10}}
\put(114,0){\line(1,0){14}}
\put(114,24){\line(1,0){14}}
\put(145,12){\makebox(0,0)[cc]{$\longrightarrow$}}
\put(170,-24){\line(1,0){38}}
\put(170,0){\line(1,0){14}}
\put(170,24){\line(1,0){14}}
\put(170,48){\line(1,0){38}}
\put(208,-24){\line(1,1){48}}
\put(184,0){\line(1,1){48}}
\put(184,24){\line(1,-1){10}}
\put(198,10){\line(1,-1){20}}
\put(222,-14){\line(1,-1){10}}
\put(208,48){\line(1,-1){10}}
\put(222,34){\line(1,-1){20}}
\put(246,10){\line(1,-1){10}}
\put(232,-24){\line(1,0){62}}
\put(256,0){\line(1,0){14}}
\put(256,24){\line(1,0){14}}
\put(232,48){\line(1,0){62}}
\put(270,0){\line(1,1){48}}
\put(294,-24){\line(1,1){48}}
\put(270,24){\line(1,-1){10}}
\put(284,10){\line(1,-1){20}}
\put(308,-14){\line(1,-1){10}}
\put(294,48){\line(1,-1){10}}
\put(308,34){\line(1,-1){20}}
\put(332,10){\line(1,-1){10}}
\put(318,-24){\line(1,0){62}}
\put(342,0){\line(1,0){14}}
\put(342,24){\line(1,0){14}}
\put(318,48){\line(1,0){62}}
\put(356,0){\line(1,1){48}}
\put(380,-24){\line(1,1){48}}
\put(356,24){\line(1,-1){10}}
\put(370,10){\line(1,-1){20}}
\put(394,-14){\line(1,-1){10}}
\put(380,48){\line(1,-1){10}}
\put(394,34){\line(1,-1){20}}
\put(418,10){\line(1,-1){10}}
\put(404,-24){\line(1,0){38}}
\put(428,0){\line(1,0){14}}
\put(428,24){\line(1,0){14}}
\put(404,48){\line(1,0){38}}.
\end{picture}
\item The projector from representation $1^{|T_i|}$ onto representation $T_i$ is constructed.
\item The resulting knot/link polynomial is evaluated in the fundamental representation with the addition of the projector operator.
\end{itemize}

\newpage

\section{$\mathcal{R}$-matrices\label{sec:Rm}}

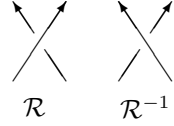
\begin{wrapfigure}{r}{80pt}
\begin{picture}(60,40)(0,-10)
\put(0,10){\vector(2,3){20}}
\put(20,10){\line(-2,3){8}}
\put(8,28){\vector(-2,3){8}}
\put(8,0){\makebox(0,0)[cc]{$\mathcal{R}$}}
\put(60,10){\vector(-2,3){20}}
\put(40,10){\line(2,3){8}}
\put(52,28){\vector(2,3){8}}
\put(50,0){\makebox(0,0)[cc]{$\mathcal{R}^{-1}$}}
\end{picture}
\label{invcr}
\caption{Direct and inverse crossings.}
\end{wrapfigure}

We use $\mathcal{R}$-matrices~\cite{Tur} as ``building blocks'' of the considered construction (see Introduction to the present paper for the details). Such an $\mathcal{R}$-matrix is located at each crossing in the planar projection of the knot. Generally speaking, the $\mathcal{R}$-matrices are some operators with the four indices corresponding to the crossings of strands in the planar projection of an oriented knot or link. To get the matrix form of these operators, it is convenient to consider a $2$-dimensional projection of a special form called a braid representation of the knot. Then appears the braid counterpart of usual $\mathcal{R}$-matrix that act not only on two strands which cross, but on the all strands in this section through the braid (it acts as the unity operator on all strands in the given section but those that cross). A matrix form of this operator is described in below. The $\mathcal{R}$-matrix acting in a section through a braid is still constructed using the operator with  four-indices. Therefore, different matrices correspond to crossings located in different places in a section through a braid. It can also be shown~\cite{Tur} that the inverse $\mathcal{R}$-matrix corresponds to the inverse crossing (see Fig.~\ref{invcr}).

\subsection{Diagonal $\mathcal{R}$-matrices \label{dRm}}

\begin{wrapfigure}{r}{80pt}
\begin{picture}(80,60)(0,-47)
\put(0,0){\line(1,0){14}}
\put(0,24){\line(1,0){14}}
\put(14,0){\line(1,1){24}}
\put(14,24){\line(1,-1){10}}
\put(28,10){\line(1,-1){10}}
\put(38,0){\line(1,0){14}}
\put(38,24){\line(1,0){14}}
\put(52,0){\line(1,1){10}}
\put(66,14){\line(1,1){10}}
\put(52,24){\line(1,-1){24}}
\put(76,0){\line(1,0){14}}
\put(76,24){\line(1,0){14}}
\put(25,-10){\makebox(0,0)[cc]{$\mathcal{R}^{-1}$}}
\put(65,-10){\makebox(0,0)[cc]{$\mathcal{R}$}}
\end{picture}
\label{invbr}
\caption{Direct and inverse crossings in a braid.}
\end{wrapfigure}
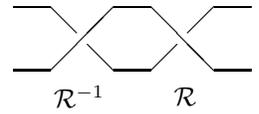
In the framework of the construction in which $\mathcal{R}$-matrices act on braids, the form of these $\mathcal{R}$-matrices depends not only on the representations located at strands crossing but also on the distribution of these strands relative to the remaining strands in the braid. Each $\mathcal{R}$-matrix can be diagonalized (but not all of them simultaneously) We will usually consider the basis in which the first $\mathcal{R}$-matrix (the one corresponding to the crossing of strands on the first and second positions in the section through the braid) is diagonal, unless the otherwise is stated.

The simplest case is the $\mathcal{R}$-matrix in the $2$-strand case with the representations $T_1$ and $T_2$. Then the $\mathcal{R}$-matrix is diagonal and the eigenvalues for each representation $Q_i\vdash T_1\otimes T_2$ are:
\begin{equation}
\begin{array}{c}
|\lambda_i|=q^{\varkappa_{Q_i}-\varkappa_{T_1}-\varkappa_{T_2}},
\\ \\
\varkappa_Q=\frac{1}{2}\sum\limits_{\{i,j\}\in Q} (j-i),\ \ i \text{ and } j \text{ enumerate all boxes in the Young diagram Q.}
\end{array}
\label{eq:Rmeig}
\end{equation}
The eigenvalues of the $\mathcal{R}$-matrix are defined up to a common factor since it does not change the Yang-Baxter equation~\cite{CSknots:MS}. This factor can be chosen differently depending on the studied quantities. The eigenvalues given in~(\ref{eq:Rmeig}) correspond to the vertical framing of the knot, we will discuss this in more details in Sec.~\ref{framing}.

When there are more than $2$ strands, the situation is a little bit more tricky. Let the representations be $T_1$, $T_2$,\ldots, $T_m$ and let us consider the diagonal $\mathcal{R}$-matrix corresponding to the crossing between the first two strands ($T_1$ and $T_2$). Then we should describe the irreducible representation expansion of $T_1\otimes T_2\otimes\ldots\otimes T_m=\sum\limits_j \bar{Q}_j$ as $(\sum\limits_i Q_i)\otimes\ldots\otimes T_m$. Then eigenvalues for $\bar{Q}_j$ are the same as for the representations $Q_i$ from which they are obtained~\cite{Tur},\cite{MMM2}.

The inverse $\mathcal{R}$-matrix whose eigenvalues are
\begin{equation}
|\lambda_i|^{-1}=q^{-\varkappa_{Q_i}+\varkappa_{T_1}+\varkappa_{T_2}}
\end{equation}
corresponds to the inverse crossing.

\subsection{General $\mathcal{R}$-matrices\label{sec:genR}}

The form of the general (non-diagonal) $\mathcal{R}$-matrix is a much
more difficult question. The general answer depending on the number of
strands and on the representations placed on them is not known. But still we know some
answers, such as a form of the $\mathcal{R}$-matrix in case of the fundamental
representations (we denote this $\mathcal{R}$-matrix by straight $R$).

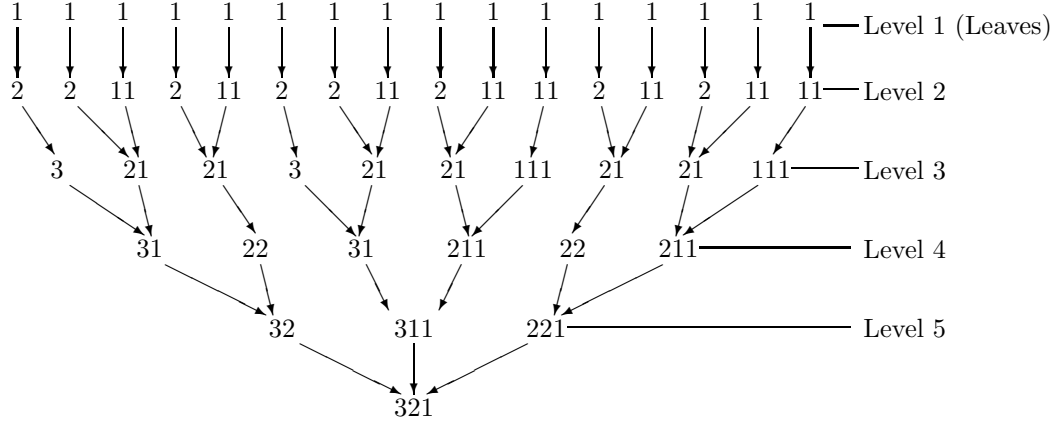
\begin{figure}
\begin{picture}(460,180)(-230,0)
\put(0,0){\makebox(0,0)[cc]{$321$}}
\put(43,24){\vector(-2,-1){38}}
\put(-43,24){\vector(2,-1){38}}
\put(0,24){\vector(0,-1){19}}
\put(-50,30){\makebox(0,0)[cc]{$32$}}
\put(0,30){\makebox(0,0)[cc]{$311$}}
\put(50,30){\makebox(0,0)[cc]{$221$}}
\put(58,30.5){\line(1,0){107}}
\put(170,30){\makebox(0,0)[ll]{Level $5$}}
\put(-94,54){\vector(2,-1){38}}
\put(-58,54){\vector(1,-4){4.75}}
\put(-18,54){\vector(1,-2){8.5}}
\put(18,54){\vector(-1,-2){8.5}}
\put(58,54){\vector(-1,-4){4.75}}
\put(94,54){\vector(-2,-1){38}}
\put(-100,60){\makebox(0,0)[cc]{$31$}}
\put(-60,60){\makebox(0,0)[cc]{$22$}}
\put(-20,60){\makebox(0,0)[cc]{$31$}}
\put(20,60){\makebox(0,0)[cc]{$211$}}
\put(60,60){\makebox(0,0)[cc]{$22$}}
\put(100,60){\makebox(0,0)[cc]{$211$}}
\put(108,60.5){\line(1,0){57}}
\put(170,60){\makebox(0,0)[ll]{Level $4$}}
\put(-130,84){\vector(3,-2){28}}
\put(-104,84){\vector(1,-4){4.75}}
\put(-72,84){\vector(2,-3){12}}
\put(-41,84){\vector(1,-1){19}}
\put(-16,84){\vector(-1,-4){4.75}}
\put(16,84){\vector(1,-4){4.75}}
\put(41,84){\vector(-1,-1){19}}
\put(72,84){\vector(-2,-3){12}}
\put(104,84){\vector(-1,-4){4.75}}
\put(130,84){\vector(-3,-2){28}}
\put(-135,90){\makebox(0,0)[cc]{$3$}}
\put(-105,90){\makebox(0,0)[cc]{$21$}}
\put(-75,90){\makebox(0,0)[cc]{$21$}}
\put(-45,90){\makebox(0,0)[cc]{$3$}}
\put(-15,90){\makebox(0,0)[cc]{$21$}}
\put(15,90){\makebox(0,0)[cc]{$21$}}
\put(45,90){\makebox(0,0)[cc]{$111$}}
\put(75,90){\makebox(0,0)[cc]{$21$}}
\put(105,90){\makebox(0,0)[cc]{$21$}}
\put(135,90){\makebox(0,0)[cc]{$111$}}
\put(143,90.5){\line(1,0){25}}
\put(170,90){\makebox(0,0)[ll]{Level $3$}}
\put(-148,114){\vector(2,-3){12}}
\put(-127,114){\vector(1,-1){19}}
\put(-109,114){\vector(1,-4){4.75}}
\put(-88,114){\vector(1,-2){9.5}}
\put(-71,114){\vector(-1,-4){4.75}}
\put(-49,114){\vector(1,-4){4.75}}
\put(-28,114){\vector(2,-3){12}}
\put(-9,114){\vector(-1,-4){4.75}}
\put(9,114){\vector(1,-4){4.75}}
\put(28,114){\vector(-2,-3){12}}
\put(49,114){\vector(-1,-4){4.75}}
\put(71,114){\vector(1,-4){4.75}}
\put(88,114){\vector(-1,-2){9.5}}
\put(109,114){\vector(-1,-4){4.75}}
\put(127,114){\vector(-1,-1){19}}
\put(148,114){\vector(-2,-3){12}}
\put(-150,120){\makebox(0,0)[cc]{$2$}}
\put(-130,120){\makebox(0,0)[cc]{$2$}}
\put(-110,120){\makebox(0,0)[cc]{$11$}}
\put(-90,120){\makebox(0,0)[cc]{$2$}}
\put(-70,120){\makebox(0,0)[cc]{$11$}}
\put(-50,120){\makebox(0,0)[cc]{$2$}}
\put(-30,120){\makebox(0,0)[cc]{$2$}}
\put(-10,120){\makebox(0,0)[cc]{$11$}}
\put(10,120){\makebox(0,0)[cc]{$2$}}
\put(30,120){\makebox(0,0)[cc]{$11$}}
\put(50,120){\makebox(0,0)[cc]{$11$}}
\put(70,120){\makebox(0,0)[cc]{$2$}}
\put(90,120){\makebox(0,0)[cc]{$11$}}
\put(110,120){\makebox(0,0)[cc]{$2$}}
\put(130,120){\makebox(0,0)[cc]{$11$}}
\put(150,120){\makebox(0,0)[cc]{$11$}}
\put(155,120.5){\line(1,0){13}}
\put(170,120){\makebox(0,0)[ll]{Level $2$}}
\put(-150,144){\vector(0,-1){19}}
\put(-130,144){\vector(0,-1){19}}
\put(-110,144){\vector(0,-1){19}}
\put(-90,144){\vector(0,-1){19}}
\put(-70,144){\vector(0,-1){19}}
\put(-50,144){\vector(0,-1){19}}
\put(-30,144){\vector(0,-1){19}}
\put(-10,144){\vector(0,-1){19}}
\put(10,144){\vector(0,-1){19}}
\put(30,144){\vector(0,-1){19}}
\put(50,144){\vector(0,-1){19}}
\put(70,144){\vector(0,-1){19}}
\put(90,144){\vector(0,-1){19}}
\put(110,144){\vector(0,-1){19}}
\put(130,144){\vector(0,-1){19}}
\put(150,144){\vector(0,-1){19}}
\put(-150,150){\makebox(0,0)[cc]{$1$}}
\put(-130,150){\makebox(0,0)[cc]{$1$}}
\put(-110,150){\makebox(0,0)[cc]{$1$}}
\put(-90,150){\makebox(0,0)[cc]{$1$}}
\put(-70,150){\makebox(0,0)[cc]{$1$}}
\put(-50,150){\makebox(0,0)[cc]{$1$}}
\put(-30,150){\makebox(0,0)[cc]{$1$}}
\put(-10,150){\makebox(0,0)[cc]{$1$}}
\put(10,150){\makebox(0,0)[cc]{$1$}}
\put(30,150){\makebox(0,0)[cc]{$1$}}
\put(50,150){\makebox(0,0)[cc]{$1$}}
\put(70,150){\makebox(0,0)[cc]{$1$}}
\put(90,150){\makebox(0,0)[cc]{$1$}}
\put(110,150){\makebox(0,0)[cc]{$1$}}
\put(130,150){\makebox(0,0)[cc]{$1$}}
\put(150,150){\makebox(0,0)[cc]{$1$}}
\put(155,144.5){\line(1,0){13}}
\put(170,144){\makebox(0,0)[ll]{Level $1$ (Leaves)}}
\end{picture}
\caption{\label{fig:321} Tree for representation $[321]$ for the $6$-strand braid in fundamental representation}
\end{figure}

The $R$-matrices have the simplest form in the basis of irreducible representations arising in the consecutive expansion of the tensor product with the nesting (denoted by parentheses):
\begin{equation}
\left((\ldots(1\otimes 1)\otimes\ldots\otimes 1) \otimes 1\right)
\label{eq:nb}
\end{equation}
We will call this basis the standard basis. The vectors of this basis are conveniently described in terms of pathes on a definite graph. For each irreducible representation, one can draw a tree depicting all ways to obtain that representation in the expansion of some tensor product. In in Fig.\ref{fig:321}, an example of such a tree for the representation $[321]$ arising in the expansion of the tensor product of $6$ fundamental representations. Each arrow in this tree corresponds to multiplication by one fundamental representation. Let us also introduce the $2\times 2$ block $b_j$:

\begin{equation}
b_j=
\left(\begin{array}{cc}
-\frac{1}{q^j[j]_q} & \frac{\sqrt{[j+1]_q[j-1]_q}}{[j]_q}
\\ \\
\frac{\sqrt{[j+1]_q[j-1]_q}}{[j]_q} & \frac{q^j}{[j]_q}
\end{array}\right).
\end{equation}

We claim that the matrix $R_{k-1}$ corresponding to the crossing of the $k-1$-th and $k$-th strands in some section through the braid consists only of $2\times 2$ blocks $b_{k}$ and $1\times 1$ blocks $q$ or $-q^{-1}$. It remains to describe the location of these blocks. For this, we use the example in Fig.\ref{fig:321}.

Each row and each column in the $R$-matrix corresponds to one of leaves of the tree, i.e., to one of points at which a path begins from the fundamental representation. Then $R_{k-1}$ is described by the level $k$ in the tree. Each path at the level $k$ is either a part of a doublet (i.e., is one of the pair of paths that coincide everywhere but the level $k$) or a singlet. If one uses the fundamental representation as the constructing block of the tree then there can be no triplets, quadruplets etc. If a pair of paths forms a doublet at the level $k$, then a block $b_j$, where $j$ is the length of the hook connecting the two boxes added to the Young diagram of the irreducible representation at the levels $k$ and $k-1$ minus one (see Fig.\ref{fig:Rdoub}), is placed in the intersection of the corresponding rows and columns. Strictly speaking, $b_j$ enters the matrix $R_{k-1}$ not as a block, but each of them (although not all simultaneously) can be transformed  into a block by exchanging rows and columns.

\begin{figure}
\begin{picture}(360,100)(-255,-80)
\put(-147,0){\line(0,-1){36}}
\put(-111,0){\line(0,-1){18}}
\put(-129,0){\line(0,-1){36}}
\put(-147,0){\line(1,0){36}}
\put(-147,-18){\line(1,0){36}}
\put(-147,-36){\line(1,0){18}}
\put(-102,-9){\circle{17}}
\put(-120,-27){\circle{17}}
\put(-120,-9){\line(1,0){18}}
\put(-120,-9){\line(0,-1){18}}
\put(-120,-70){\makebox(0,0)[cc]{$21\rightarrow 32$}}
\put(-120,-85){\makebox(0,0)[cc]{$b_2$ block in $R$}}
\put(46,0){\line(0,-1){36}}
\put(64,0){\line(0,-1){36}}
\put(46,0){\line(1,0){36}}
\put(46,-18){\line(1,0){36}}
\put(46,-36){\line(1,0){18}}
\put(82,0){\line(0,-1){18}}
\put(55,-45){\circle{17}}
\put(91,-9){\circle{17}}
\put(55,-9){\line(1,0){36}}
\put(55,-9){\line(0,-1){36}}
\put(73,-70){\makebox(0,0)[cc]{$21\rightarrow 311$}}
\put(73,-85){\makebox(0,0)[cc]{$b_4$ block in $R$}}
\end{picture}
\caption{\label{fig:Rdoub} To find out which $b_i$ blocks should be used, the diagram describing the transition at level $k$ should be drawn. If the length of a hook, connecting the two added boxes at the levels $k$ and $k+1$ (described by the circles in the picture) is equal to $j$ then the block $b_{j-1}$ should be used.}
\end{figure}

A path is a singlet if going from the level $k-1$ to the level $k+1$ corresponds to adding two boxes either in one row or in one column of the Young diagram (Fig. 6 corresponds to $k=4$). The diagonal element of the matrix $R_k$ corresponding to this path is $q$ in the first case and $-q^{-1}$ in the second.


\begin{figure}
\begin{picture}(360,100)(-255,-80)
\put(0,0){\line(0,-1){36}}
\put(18,0){\line(0,-1){18}}
\put(-18,0){\line(0,-1){36}}
\put(-18,0){\line(1,0){36}}
\put(-18,-18){\line(1,0){36}}
\put(-18,-36){\line(1,0){18}}
\put(27,-9){\circle{17}}
\put(-9,-45){\circle{17}}
\put(9,-70){\makebox(0,0)[cc]{$21\rightarrow 311$}}
\put(9,-85){\makebox(0,0)[cc]{$b_4$ block in $R$}}
\put(126,0){\line(0,-1){54}}
\put(144,0){\line(0,-1){54}}
\put(126,0){\line(1,0){18}}
\put(126,-18){\line(1,0){18}}
\put(126,-36){\line(1,0){18}}
\put(126,-54){\line(1,0){18}}
\put(153,-9){\circle{17}}
\put(171,-9){\circle{17}}
\put(153,-70){\makebox(0,0)[cc]{$111\rightarrow 311$}}
\put(153,-85){\makebox(0,0)[cc]{$q$ in $R$}}
\put(-164,0){\line(0,-1){18}}
\put(-144,0){\line(0,-1){18}}
\put(-126,0){\line(0,-1){18}}
\put(-108,0){\line(0,-1){18}}
\put(-162,0){\line(1,0){54}}
\put(-162,-18){\line(1,0){54}}
\put(-153,-27){\circle{17}}
\put(-153,-45){\circle{17}}
\put(-135,-70){\makebox(0,0)[cc]{$3\rightarrow 311$}}
\put(-135,-85){\makebox(0,0)[cc]{$-q^{-1}$ in $R$}}
\end{picture}
\caption{\label{fig:YDf} Description of $R$ in terms of added boxes to the Young diagrams for the level $4$ and for the representation $311$ as the final representation. Circles denote the boxes added to the initial diagrams.}
\end{figure}


Any any non-diagonal fundamental $R$-matrix can be constructed according to these rules. As an example, the block of the $R_4$-matrix corresponding to the representation $[321]$  (corresponding to the level $4$ in Fig.\ref{fig:321}) looks like:

\setlength{\arraycolsep}{1pt}

\begin{equation*}
\begin{array}{l}
R_{4|321}=
\\
\\
\left(
\begin{array}{cccccccccccccccc}
q &&&&&&&&&&&&&&&
\\
& -\frac{1}{[2]_q q^2} &0& \frac{\sqrt{[3]_q}}{[2]_q} &0&&&&&&&&&&&
\\
&0& -\frac{1}{[2]_q q^2} &0& \frac{\sqrt{[3]_q}}{[2]_q} &&&&&&&&&&&
\\
& \frac{\sqrt{[3]_q}}{[2]_q} &0& \frac{q^2}{[2]_q} &0&&&&&&&&&&&
\\
&0& \frac{\sqrt{[3]_q}}{[2]_q} &0& \frac{q^2}{[2]_q} &&&&&&&&&&&
\\
&&&&& -q^{-1} &&&&&&&&&&
\\
&&&&&& -\frac{1}{[4]_q q^4} &0& \frac{\sqrt{[3]_q[5]_q}}{[4]_q} &0&&&&&&
\\
&&&&&&0& -\frac{1}{[4]_q q^4} &0& \frac{\sqrt{[3]_q[5]_q}}{[4]_q} &&&&&&
\\
&&&&&& \frac{\sqrt{[3]_q[5]_q}}{[4]_q} &0& \frac{q^4}{[4]_q} &0&&&&&&
\\
&&&&&&0& \frac{\sqrt{[3]_q[5]_q}}{[4]_q} &0& \frac{q^4}{[4]_q} &&&&&&
\\
&&&&&&&&&& q &&&&&
\\
&&&&&&&&&&& -\frac{1}{[2]_q q^2} &0& \frac{\sqrt{[3]_q}}{[2]_q} &0&
\\
&&&&&&&&&&&0& -\frac{1}{[2]_q q^2} &0& \frac{\sqrt{[3]_q}}{[2]_q} &
\\
&&&&&&&&&&& \frac{\sqrt{[3]_q}}{[2]_q} &0& \frac{q^2}{[2]_q} &0&
\\
&&&&&&&&&&&0& \frac{\sqrt{[3]_q}}{[2]_q} &0& \frac{q^2}{[2]_q} &
\\
&&&&&&&&&&&&&&& -q^{-1}
\end{array}
\right).
\end{array}
\end{equation*}

\setlength{\arraycolsep}{6pt}

The inverse of the $R$-matrix described above is constructed in a similar way. It suffices to replace $q$ with $q^{-1}$ in all instances.

Let us explain the described form of the $R$-matrices.
First, we determine, which paths belong to the same block. Each point at the level $i\le k-1$ corresponds to a representation $T_i$ in the expansion of the tensor product $T_1\otimes \ldots \otimes T_{i-1}$ of the representations on which the matrix $R_k$ acts as the unit operator. Hence, any two paths mixed by the matrix $R_k$ coincides up to the level $k-1$. Then, the action of $R_k$ on $T_1\otimes 8 \ldots\otimes T_{k+1} \otimes T_m$ is obtained from its action on $T_1\otimes \ldots\otimes T_{k+1}$ where it does not mix the irreducible representations corresponding to different Young diagrams. Hence, the pair of paths mixed by $R_k$ has to arrive at the same Young diagram $Q$ at the level $k+1$. Again, the representations $T_{k+1},\ldots,T_m$ multiplied at the higher levels are unaffected by  $R_k$ and hence the corresponding parts of the mixing paths also have to coincide. Therefore all the paths in the same mixing block of $R_k$ coincide (i.e., go through the same Young diagrams) everywhere but at the level $k$.

The size of the blocks depends on the representations in the considered product. In the fundamental case ($T_i=\Box$), each arrow in the tree corresponds to the addition of one box to the Young diagram. The paths that coincide everywhere but at level $k$ correspond to the same Young diagram at level $k-1$ and to the same Young diagram at level $k+1$. Moreover, the diagram at the level $k+1$ differs from the diagram at the level $k-1$ by two boxes put at certain positions. Hence, there are either two such paths  (where two boxes are added at the levels $k-1$ and $k$ in one or in the other order) or only one (if the boxes are added at the same line or at the same column). The first case corresponds to $2\times 2$ block , and the second one to $1\times1$  block. and there is no other possibilities in the fundamental case.

The blocks $1\times1$ are merely the eigenvalues of the fundamental $R$-matrix. Since the addition of two boxes to the same line implies the symmetrization of the corresponding pair of fundamental representations and the addition of them to the same column corresponds to antisymmetrization, the first case corresponds to the eigenvalue $q$, and the second corresponds to  $-q^{-1}$. It is left to specify the form of the $2\times 2$ blocks, which is done in Sec.~\ref{sec:eig}.

\subsection{$R$-matrices properties and polynomial rings\label{sec:Rprop}}

There are several properties of $R$-matrices, which are important for our further calculations.

It is known that the $R$-matrices satisfy the same relations as the generators of the braid group~\cite{Pras}. This leads to the first two properties of the $R$-matrices:

\begin{equation}
R_iR_j-R_jR_i=0,\ \ \ |i-j|\neq 1
\label{Rcomm}
\end{equation}
\begin{equation}
R_iR_{i+1}R_i=R_{i+1}R_iR_{i+1},
\label{RYB}
\end{equation}
where $R_i$ corresponds to the crossing between strands $i$ and $i+1$. These relations are valid for the $\mathcal{R}$-matrices in \textbf{any} representation. Relation~(\ref{Rcomm}) is obvious if a braid is considered: if two crossings follow in order and have no common strand, then their order can be changed. This means that the $\mathcal{R}$-matrices that correspond to these crossings should commute. Relation~(\ref{RYB}) is the crucial property of $\mathcal{R}$-matrices, which in fact defines that crossings indeed correspond to the $\mathcal{R}$-matrices. It is the famous Yang-Baxter equation.

\begin{wrapfigure}{r}{180pt}
\begin{picture}(180,60)(0,-10)
\put(0,10){\vector(2,3){20}}
\put(20,10){\line(-2,3){8}}
\put(8,28){\vector(-2,3){8}}
\put(8,0){\makebox(0,0)[cc]{$R$}}
\put(30,25){\makebox(0,0)[cc]{$-$}}
\put(60,10){\vector(-2,3){20}}
\put(40,10){\line(2,3){8}}
\put(52,28){\vector(2,3){8}}
\put(50,0){\makebox(0,0)[cc]{$R^{-1}$}}
\put(100,25){\makebox(0,0)[cc]{=\ \ $\Big(q-q^{-1}\Big)$}}
\qbezier(140,10)(150,25)(140,40)
\put(142,37){\vector(-2,3){3}}
\qbezier(160,10)(150,25)(160,40)
\put(158,37){\vector(2,3){3}}
\put(150,0){\makebox(0,0)[cc]{$\mathds{1}$}}
\end{picture}
\caption{\label{skeinpic} Pictorial description of the fundamental skein relations in the vertical framing (see Sec.~\ref{framing})}
\end{wrapfigure}
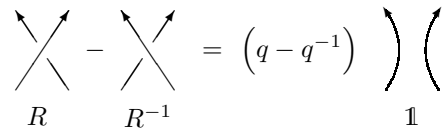

The third property of $\mathcal{R}$-matrices, which is important for our consideration, is quite different when different representations are considered. This differs the third property from the first two. This third property is in fact a characteristic equation on the $\mathcal{R}$-matrix, which in case of the fundamental representation is of the form
\begin{equation}
\Big(R_i-q\Big)\left(R_i+\frac{1}{q}\right)=0.
\label{Revals}
\end{equation}
The third property is equivalent to the skein relations of mathematical knot theory; this property is usually
described by the picture in Fig.\ref{skeinpic}. Since in the fundamental representation there are only two eigenvalues the characteristic equation is quite simple and can be used to find the knot polynomials as combinations of polynomials of simpler knots as it is done in the mathematical knot theory.
One can also treat characteristic equations for colored $\mathcal{R}$-matrices as the colored skein relations~\cite{MM}. However such colored skein relations are not as useful as the fundamental ones since the characteristic equation is of higher degree for higher representations and hence, no immediate way to untie knots with help of the colored skein relations is available. For example, for representation $[2]$ the colored skein relations are

\begin{equation}
\Big(\mathcal{R}_i-q^4\Big)\Big(\mathcal{R}_i+1\Big)\Big(\mathcal{R}_i-q^{-2}\Big)=0.
\end{equation}
This means that the skein relations now include three terms and not all of them are necessary simpler than the initial knot. It is not clear if one can use these colored skein relations to evaluate the colored HOMFLY polynomials.

The fundamental $R$-matrices satisfying relations~(\ref{Rcomm}),~(\ref{RYB}) and~(\ref{Revals}) generate the Hecke algebra. This means that the polynomial ring generated by the finite number of $R$-matrices is finite dimensional. Such polynomial rings will be quite important for our further calculations. Each $m$-strand braid can be represented as an element of such polynomial ring on $m-1$ $R$-matrices, which are $R_1,\ R_2,\ldots \ R_{m-1}$. However finite dimension of such polynomial ring does \textit{not} mean that there is only finite number of prime knots. The fact that the knot polynomial of some knot is equal to the sum of the knot polynomials of some other knots does not mean that the initial knot is a combination of the latter ones. In turn, a knot polynomial of a composite knot is equal to the product of knot polynomials of its parts.

It is known from the properties of Hecke algebra~\cite{Jones2} that the dimension of the described polynomial ring is equal to $m!$. The $m!$ basis elements can be constructed for example from the elementary blocks
\begin{equation}
\sigma_{l,0}=\mathds{1},\ \ \sigma_{l,k}=\prod_{i=k}^lR_{k+l-i},\ \ \  k=1,\ldots,m-1 ,\ l=k,\ldots,m-1,
\end{equation}
as
\begin{equation}
\Xi_{k_1k_2\ldots k_{m-1}}=\prod_{l=1}^{m-1}\sigma_{l,k_l},\ \ \ k_l=0,\ldots, l.
\end{equation}
In the case $q=1$, each $\sigma_{l,k}$ corresponds to the cyclic permutation of the elements from  $k$ to $l$. Each of the $\Xi_{k_1k_2\ldots k_{m-1}}$ realizes one of the $m!$ permutations. The elementary blocks and the simplest basis elements of the polynomial ring for the $m$-strand braid can be chosen as
\begin{equation}
\label{eq:prb}
\begin{array}{c}
\begin{array}{llllll}
\sigma_{1,0}=\mathds{1},&\sigma_{1,1}=R_1,\\
\sigma_{2,0}=\mathds{1},&\sigma_{2,1}=R_2R_1&\sigma_{2,2}=R_2,\\
\sigma_{3,0}=\mathds{1},&\sigma_{3,1}=R_3R_2R_1&\sigma_{3,2}=R_3R_2&\sigma_{3,3}=R_3,\\
\ldots&\ldots&\ldots&\ldots\\
\sigma_{m-1,0}=\mathds{1},&\sigma_{m-1,1}=R_{m-1}\ldots R_2R_1,&\sigma_{m-1,2}=R_{m-1}\ldots R_2,&\ldots&\sigma_{m-1,m-1}=R_{m-1},
\end{array}
\\
\\
\\
\begin{array}{lll}
\Xi_{00\ldots0}=    &\sigma_{1,0}\sigma_{2,0}\ldots\sigma_{m-1,0}=  &\mathds{1},
\\ \\
\Xi_{10\ldots0}=    &\sigma_{1,1}\sigma_{2,0}\ldots\sigma_{m-1,0}=  &R_1,
\\ \\
\Xi_{01\ldots0}=    &\sigma_{1,0}\sigma_{2,1}\ldots\sigma_{m-1,0}=  &R_2R_1,
\\ \\
\Xi_{11\ldots0}=    &\sigma_{1,1}\sigma_{2,1}\ldots\sigma_{m-1,0}=  &R_1R_2R_1,
\\ \\
\Xi_{02\ldots0}=    &\sigma_{1,0}\sigma_{2,2}\ldots\sigma_{m-1,0}=  &R_2,
\\ \\
\Xi_{12\ldots0}=    &\sigma_{1,1}\sigma_{2,2}\ldots\sigma_{m-1,0}=  &R_1R_2.
\end{array}
\end{array}
\end{equation}

\section{Projectors\label{sec:proj}}

If the strand in the knot is replaced with the cable consisting of,
e.g., two strands, this leads to the knot polynomial of the same knot but evaluated
in representation $[1]\otimes[1]$. Thus, if one asks for the
answers in irreducible representations (e.g., $[2]$ and $[11]$ in
this case) some operators that somehow project the answer in
reducible representation onto answers in irreducible representations
should be constructed. These operators are called projectors.

Generally speaking, each $\mathcal{R}$-matrix should be surrounded by four projectors (one for each colored strand that enters or leaves the corresponding crossing). From the standpoint of calculations, the crucial point in the method presented in this paper is that one can in fact use only one projector for the whole knot (or one per component of the link) or any number of projectors one wishes (but not less than one per component). This follows from the properties of the $\mathcal{R}$-matrix. Namely, $\mathcal{R}$-matrix cannot mix different irreducible representations with each other~\cite{Tur}. Thus, if one puts the projector onto some irreducible representation at one side of the $\mathcal{R}$-matrix then only the same representation would appear at the other side and thus at the whole connection component.

There are several approaches to construct the projectors.

\subsection{Path description of the projectors\label{sec:prpath}}

The easiest projectors to describe are the ones that project the
first $n$ strands onto some representations of the level $n$. In the
standard basis~(\ref{eq:nb}) used in the present paper, these ones are diagonal and can
be described in the similar way as the $R$-matrices. For
example, let us take the same tree as in Fig.\ref{fig:321}. With help of this tree, one can construct the
projectors needed for the calculation of $3$-strand knot in the representations
of size $2$ or a of $2$-strand knot in the representations of size $3$  (more precisely, the blocks corresponding to the representation $[321]$ of these projectors. E.g., to select the copies of representation $[321]$ arising from the expansion of $[2]\otimes [1]^4$, one has to keep only those paths in Fig.\ref{fig:321} that pass through the representation $[2]$ on the level $2$. By definition, each path Fig.\ref{fig:321} corresponds to a representation that is an eigenvector of $P_{2\otimes 1^{4}|321}$. Moreover, the eigenvalue $1$ corresponds to the retained paths (passing through $[2]$), and the eigenvalue $0$ corresponds to the rejected paths (passing through $[11]$). Therefore, in basis~(\ref{eq:nb}) all the non-diagonal elements of the projector are zero and each diagonal element is equal to the eigenvalue of the corresponding path. Hence, the matrix of the projector is:

\setlength{\arraycolsep}{1pt}

\begin{equation}
P_{2\otimes 1^{4}|321}=
\left(
\begin{array}{cccccccccccccccc}
1 &&&&&&&&&&&&&&&
\\
& 1 &&&&&&&&&&&&&&
\\
&& 0 &&&&&&&&&&&&&
\\
&&& 1 &&&&&&&&&&&&
\\
&&&& 0 &&&&&&&&&&&
\\
&&&&& 1 &&&&&&&&&&
\\
&&&&&& 1 &&&&&&&&&
\\
&&&&&&& 0 &&&&&&&&
\\
&&&&&&&& 1 &&&&&&&
\\
&&&&&&&&& 0 &&&&&&
\\
&&&&&&&&&& 0 &&&&&
\\
&&&&&&&&&&& 1 &&&&
\\
&&&&&&&&&&&& 0 &&&
\\
&&&&&&&&&&&&& 1 &&
\\
&&&&&&&&&&&&&& 0 &
\\
&&&&&&&&&&&&&&& 0
\end{array}
\right).
\end{equation}

\setlength{\arraycolsep}{6pt}

The described construction allows building only the projectors onto the first strand in the initial colored braid. Nevertheless, this suffices calculating the colored polynomial for an arbitrary knot: as already discussed, one projector that can be inserted in the first strand is used for this. Moreover, Reidemeister moves can be used to deform any braid such that each strand relating to one of the colored components turns out to be the first in some section through the braid. The projector for that strand can be placed in the corresponding section. To simplify the calculations, we can place several projectors in the first strand with two projectors for every colored crossings of the initial braid. Indeed, if each product of $\mathcal{R}$-matrices corresponding to one crossing of a cabled knot is multiplied from two sides by a diagonal projector, then some lines and columns becomes zeros. Then, one can obtain a knot polynomial multiplying the non-zero blocks, which are the matrices of a smaller size than the initial $R$-matrices.

\subsection{Projectors as polynomials of $R$-matrices}

The descriptions of the projectors in terms of paths is sufficient to calculated colored HOMFLY polynomials using the cabling procedure.
Nevertheless, it is interesting to understand what are expressions for the projectors via $R$-matrices. The very formulation of the cabling procedure implies that such a description should exist. The cabling procedure involves several satellite links differing by a set of crossings in the planar projection. In the construction considered, each crossing corresponds to the $R$-matrix. Thus, any linear combination of these link polynomials can be described as an element of the polynomial ring on the $R$-matrices. The $R$-matrix description of the projectors allows us to study connections between the colored HOMFLY polynomials and the fundamental ones and also to check the consistency of the cabling procedure (defined as in Sec.~\ref{sec:CP}) with the path description of the projectors.  In addition, projectors onto the same representation  placed in different strands in the braid and even in braids with different number of strands have the similar expressions via $R$-matrices. Hence, it is enough to find the $R$-matrix description, e.g., for the projector placed in the first cable, i.e., for the one that already has the path description.

\subsubsection{Projectors from the unknots}

The most straightforward method to construct the $R$-matrix description of the projectors (and the one that was introduced in~\cite{IMMM1}) is to use the idea that the form of the projector should depend not on the knot but only on the considered representation. Thus, to find the form of these projectors one can look at the simplest of the knots, at the unknot. The unknot in representation $Q$ can be represented in two ways. On one side, the corresponding colored HOMFLY polynomial is equal to $S^*_Q(A,q)$. On the other side, it can be represented as a sum of several knots and links in the fundamental representations with $|Q|$ strands in the braid representation with some coefficients. These two representations of the HOMFLY polynomial for the unknot provide constrains on the form of the projector.

\paragraph{Representations of size $\mathbf{|Q|=2}$.} This case is the simplest example where the cabling can be used. If $|Q|=2$, then  $2$-strand knots and links should be used to describe the unknot in representation $Q$.  There are no more than two linear independent HOMFLY polynomials among those of $2$-strand links since there are only two characters in the character expansion~(\ref{eq:charexp}) for the $2$-strand knots. The corresponding braids can be chosen as two strands without any crossings ($H_0=S_2^*(A,q)+S_{1,1}^*(A,q)$) and the representation of the unknot in the form of two strands with one crossing ($H_1=S_2^* q-S_{1,1}^* q^{-1}$). The HOMFLY polynomials of all other $2$-strand knots can be represented as à linear combinations of these two with the coefficients depending on $q$. The corresponding colored unknots are represented by the $1$-strand braid either in the representation $[2]$ or in the representation $[11]$ with the HOMFLY polynomials being $S^*_{[2]}(A,q)$ and $S^*_{[11]}(A,q)$ correspondingly. Hence, to construct the projectors, one has to solve the system of equations

\begin{equation}
\begin{array}{ll}
S^*_2(A,q)=p_{2}^{0}H_0+p_{2}^{1}H_1,&
S^*_{11}(A,q)=p_{11}^{0}H_0+p_{11}^{1}H_1.
\end{array}
\label{eq:p2sys}
\end{equation}
This system can be solved in two ways. The first one uses the exact form of the Schur polynomials in $A$ and $q$. If the coefficients $p$ are independent on $A$ there is only one solution for them. The second approach uses the idea that the cabling is applicable also to the extended HOMFLY polynomials~\cite{MMM1}. This means that the system~(\ref{eq:p2sys}) should be solved on the level of the coefficients in front of the Schur polynomials. Both approaches give the same solution:
\begin{equation}
\begin{array}{llll}
p_{2}^0= \frac{1}{q(q+q^{-1})},&p_{2}^1= \frac{1}{(q+q^{-1})},&
p_{11}^0= \frac{q}{(q+q^{-1})},&p_{11}^1= -\frac{1}{(q+q^{-1})}.
\end{array}
\end{equation}

With help of definition of $H_0$ and $H_1$, one can constructs the projectors in terms of $R$-matrices:
\begin{equation}\label{eq:projun}
\begin{array}{l}
P_{2}= p_{2}^{0}+p_{2}^{1}R_1=\cfrac{1}{q(q+q^{-1})}+\cfrac{1}{(q+q^{-1})}R_1,\\ \\
P_{11}= p_{11}^{0}+p_{11}^{1}R_1=\cfrac{q}{(q+q^{-1})}-\cfrac{1}{(q+q^{-1})}R_1.
\end{array}
\end{equation}
It can be easily checked that the projectors constructed this way are orthogonal:
\begin{equation}
\begin{array}{l}
P_{2}P_{11}=\cfrac{1}{(q+q^{-1})^2}\Big(1+R_1(q^{-1}-q)-R_1^2\Big)=
\\
=\cfrac{1}{(q+q^{-1})^2}\Big(1+R_1(q-q^{-1}))-\left(R_1(q-q^{-1})+1\right)\Big)=0.
\end{array}
\end{equation}
The check of the relations $P_{2,11}^2=P_{2,11}$ can be done in a similar way.

\paragraph{Higher representations.} Unfortunately this method does not provide answers for higher
representations. The reason is quite simple. For example, at
the level $3$ any $3$-strand knot is described by only $3$
$q$-dependent coefficients in the character expansion~(\ref{eq:charexp}). At the same
time the polynomial ring, as was described in Sec.~\ref{sec:Rprop}, in
this case is six dimensional, hence one has to define $6$ coefficients. This means that in fact $3$-strand
knots do not provide enough many equations to find the projectors. The same
applies to the higher representations. Thus, it is not clear how to
use the unknot approach to find the projectors onto representations
higher than $[2]$ and $[11]$.

\subsubsection{$R$-matrix description from the paths description}

Another method to find the desired formula for the projectors is to use the already known answers for the projectors from Sec.~\ref{sec:prpath}. One can try to look for a combination of the $R$-matrices (i.e., for an element of the polynomial ring) that gives the right matrix of the projector in the corresponding basis:

\begin{equation}
\sum_I\alpha_I\Xi_I=P_Q,\label{projrmgen}
\end{equation}
where $\Xi_I$ are the basis elements of the polynomial ring, $\alpha_I$ are the coefficients that have to be found and $I$ is the multi-index defined as in~(\ref{eq:prb}), which takes $|Q|!$ different values. At the r.h.s. of~(\ref{projrmgen}) stands the projector in the matrix form. This can be either the projector onto one specifically chosen representation $Q$ in the decomposition of $[1]^{|Q|}$ or on a sum of several such representations. Since all elements of the polynomial ring have the block structure with each block corresponding to an irreducible representation $Q$ from the decomposition $[1]^{|Q|}=\sum\limits_iN_{Q_i}Q_i$, the number of elements of the $\Xi_i$ that are not identically zero is equal to the sum of the squared multiplicities $\sum\limits_i \left(N_{Q_i}\right)^2$. This is the number of equations in~(\ref{projrmgen}) and one can see that it is exactly equal to $|Q|!$, which is the number of different $\Xi_i$ according to sec.\ref{sec:Rprop} and thus the number of variables $\alpha_I$. Indeed, it is well known in the representation theory~\cite{Fulton},\cite{Vil} that the multiplicities of the irreducible representations can be obtained from the expansion of the character of the fundamental representation to the power of $|Q|$, which equals $S_{1}^{|Q|}=t_1^{|Q|}$, over the characters $S_{Q_i}$ of the irreducible representations $Q_i\vdash 1^{|Q|}$. The characters satisfy the equations
\begin{equation}
\begin{array}{l}
t_1^{|Q|}=\sum\limits_{Q_i\vdash|Q|}N_{1^{|Q|}}^{Q_i}S_{Q_i}(t_k),\ \
\left(\frac{\partial}{\partial t_1}\right)^{|Q|}
=\sum\limits_{Q_i\vdash 1^{|Q|}}N_{1^{|Q|}}^{Q_i}S_{Q_i}(\frac{\partial}{k\partial t_k}).
\label{chidecomp}
\end{array}
\end{equation}
As well, one can define the scalar product for the characters as~\cite{Mac}:
\begin{equation}
S_{T}\left(\frac{\partial}{k\partial t_k}\right)S_{R}(t_k)=\delta_{Q,T}.
\label{chiscalprod}
\end{equation}
Applying~(\ref{chiscalprod}) to~(\ref{chidecomp}) one obtains the identity
\begin{equation}
|Q|!=\sum \left(N_{1^|Q|}^{Q_i}\right)^2\label{multsqid}.
\end{equation}
This proves that the number of equations in~(\ref{projrmgen}) is indeed equal to the number of defined coefficients $\alpha_I$.

\paragraph{Representations of size $\mathbf{|Q|=2}$.} The polynomial ring has the single generator $R_1$, which satisfying
\begin{equation}
(R_1-q)(R_1+q^{-1})=0.
\end{equation}
According to~(\ref{eq:prb}), the polynomial ring in this case is $2!=2$-dimensional and the basis can be chosen as
\begin{equation}
\Xi_0=\mathds{1}=
\left(\begin{array}{cc}
1 & \\ & 1
\end{array}\right),\
\Xi_1=R_1=
\left(\begin{array}{cc}
q & \\ & -q^{-1}
\end{array}\right).
\end{equation}
Then the equations for the projectors are
\setlength{\arraycolsep}{3pt}
\begin{equation}
\begin{array}{l}
P_{2}=\alpha_2^0  +\alpha_2^1 R_1=\left(\begin{array}{ccc}1&\\&0\end{array}\right),
\\
\\
P_{11}=\alpha_{11}^0  +\alpha_{11}^1 R_1=\left(\begin{array}{cc}0&\\&1\end{array}\right).
\end{array}
\end{equation}
\setlength{\arraycolsep}{6pt}
Since $[1]^2=[2]+[11]$, all elements of the ring are split into two  $1\times1$ blocks, one for $[2]$ the other one for $[11]$, so that there is exactly $1^2+1^2=2$ equations on $2!=2$ variables. The solutions of these equations coincide with~(\ref{eq:projun}):
\begin{equation}
\begin{array}{lcr}
P_2=\cfrac{1+qR_1}{q[2]_q},& \ & P_{11}=\cfrac{q-R_1}{[2]_q}.
\end{array}
\end{equation}

\

\paragraph{Representations of size $\mathbf{|Q|=3}$.} According to~(\ref{eq:prb}), the ring of $R$-matrices has two generators, $R_1$ and $R_2$. The polynomial ring is $3!=6$ dimensional with the basis that can be chosen as
\begin{equation}
\Xi_{00}=\mathds{1},\ \ \Xi_{10}=R_1,\ \ \Xi_{01}=R_2R_1,\ \ \Xi_{11}=R_1R_2R_1,\ \ \Xi_{02}=R_2,\ \ \Xi_{12}=R_1R_2.
\end{equation}
The equations for the projectors are
\setlength{\arraycolsep}{3pt}
\begin{equation}
\begin{array}{l}
P_3=\alpha_3^{00} +\alpha_3^{10}R_1+ \alpha_3^{02}R_2 +\alpha_3^{12}R_1R_2 +\alpha_3^{01} R_2R_1 +\alpha_3^{11}R_1R_2R_1 =\left(\begin{array}{cccc}1&&&\\&0&&\\&&0&\\&&&0\end{array}\right),
\\
\\
P_{\underline{21}}=\alpha_{\underline{21}}^{00} +\alpha_{\underline{21}}^{10}R_1+ \alpha_{\underline{21}}^{02}R_2+ \alpha_{\underline{21}}^{12}R_1R_2+ \alpha_{\underline{21}}^{01}R_2R_1+ \alpha_{\underline{21}}^{11}R_1R_2R_1 =\left(\begin{array}{cccc}0&&&\\&1&&\\&&0&\\&&&0\end{array}\right),
\\
\\
P_{\overline{21}}=\alpha_{\overline{21}}^{00} +\alpha_{\overline{21}}^{10}R_1+ \alpha_{\overline{21}}^{02}R_2 +\alpha_{\overline{21}}^{12}R_1R_2 +\alpha_{\overline{21}}^{01}R_2R_1+ \alpha_{\overline{21}}^{11}R_1R_2R_1 =\left(\begin{array}{cccc}0&&&\\&0&&\\&&1&\\&&&0\end{array}\right),
\\
\\
P_{111}=\alpha_{111}^{00} +\alpha_{111}^{10}R_1 +\alpha_{111}^{02}R_2 +\alpha_{111}^{12}R_1R_2 +\alpha_{111}^{01}R_2R_1 +\alpha_{111}^{11}R_1R_2R_1 =\left(\begin{array}{cccc}0&&&\\&0&&\\&&0&\\&&&1\end{array}\right).
\end{array}
\end{equation}
\setlength{\arraycolsep}{6pt}
Since $[1]^3=[3]+2\ [21]+[111]$, all the elements of the polynomial ring are split into three blocks, a $1\times 1$ block for representation $[3]$, a $2\times 2$ block for representation $[21]$, and a $1\times 1$ block for representation $[111]$. Thus, there are exactly $1^2+2^2+1^2=6$ equations on $3!=6$ variables. The solution is
\begin{equation}
\label{proj3def}
\begin{array}{l}
P_3=\cfrac{1}{q^3[2]_q[3]_q}\left(1+qR_1+qR_2+q^2R_1R_2+q^2R_2R_1+q^3R_1R_2R_1\right),
\\ \\
P_{\underline{21}}=\cfrac{1}{[3]_q}
\left(1+qR_1-\cfrac{1}{q^2[2]_q}R_2-\cfrac{1}{q[2]_q}(R_1R_2+R_2R_1)-\cfrac{1}{[2]_q}R_1R_2R_1\right),
\\ \\
P_{\overline{21}}=\cfrac{1}{[3]_q}
\left(1-q^{-1}R_1+\cfrac{q^2}{[2]_q}R_2-\cfrac{q}{[2]_q}(R_1R_2+R_2R_1)+\cfrac{1}{[2]_q}R_1R_2R_1\right),
\\ \\
P_{111}=\cfrac{q^3}{[2]_q[3]_q}
\left(1-q^{-1}R_1-q^{-1}R_2+q^{-2}R_1R_2+q^{-2}R_2R_1-q^{-3}R_1R_2R_1\right).
\end{array}
\end{equation}

The rank two projector onto the sum of two isomorphic representations $[21]$ can be obtained as a sum
\begin{equation}
P_{21}=P_{\underline{21}}+P_{\overline{21}}=\cfrac{1}{[3]_q}
\left(2+(q-q^{-1})(R_1+R_2)-R_1R_2-R_2R_1\right).
\end{equation}
Another way to obtain this projector is to solve the equation
\setlength{\arraycolsep}{3pt}
\begin{equation}
P_{21}=\alpha_{21}^{00} +\alpha_{21}^{10}R_1 +\alpha_{21}^{02}R_2 +\alpha_{21}^{12}R_1R_2 +\alpha_{21}^{01}R_2R_1 +\alpha_{21}^{11}R_1R_2R_1 =\left(\begin{array}{cccc}0&&&\\&1&&\\&&1&\\&&&0\end{array}\right),
\end{equation}
\setlength{\arraycolsep}{6pt}
whose solution gives the same result.

\paragraph{Representations of size $\mathbf{|Q|=4}$.} The polynomial ring of the $R$-matrices has three generators $R_1$, $R_2$ and $R_3$. There are $24$ basis elements, for example,

\begin{equation}
\begin{array}{llllll}
\Xi_{000}=\mathds{1},    &\Xi_{001}=R_3R_2R_1,            &\Xi_{002}=R_3R_2,           &\Xi_{003}=R_3,
\\ \\
\Xi_{100}=R_1,           &\Xi_{101}=R_1R_3R_2R_1,         &\Xi_{102}=R_1R_3R_2,        &\Xi_{103}=R_1R_3,
\\ \\
\Xi_{010}=R_2R_1,        &\Xi_{011}=R_2R_1R_3R_2R_1,      &\Xi_{012}=R_2R_1R_3R_2,     &\Xi_{013}=R_2R_1R_3,
\\ \\
\Xi_{110}=R_1R_2R_1,     &\Xi_{111}=R_1R_2R_1R_3R_2R_1,   &\Xi_{112}=R_1R_2R_1R_3R_2,  &\Xi_{113}=R_1R_2R_1R_3,
\\ \\
\Xi_{020}=R_2,           &\Xi_{021}=R_2R_3R_2R_1,         &\Xi_{022}=R_2R_3R_2,        &\Xi_{023}=R_2R_3,
\\ \\
\Xi_{120}=R_1R_2,        &\Xi_{121}=R_1R_2R_3R_2R_1,      &\Xi_{122}=R_1R_2R_3R_2,     &\Xi_{123}=R_1R_2R_3.
\end{array}
\end{equation}

Since $[1]^4=[4]+3\ [31]+2\ [22]+3\ [211]+[1111]$, each element of the ring are split into five blocks: a $1\times1$ block for $[4]$, a $3\times 3$ block for $[31]$, a $2\times 2$ block for $[22]$, a $3\times 3$ block for $[211]$, and a $1\times 1$ block for $[1111]$. Hence, there are exactly $1^2+3^2+2^2+3^2+1^2=24$ equations on $4!=24$ variables. The expressions for the projectors onto definite copies of representations are quite lengthy in this case thus we give them in Appendix~\ref{app:proj}, listing here only projectors onto representations $[4]$, $[1111]$ and onto the space of all representations $[31]$, $[22]$ and $[211]$:

\begin{equation}
\begin{array}{ll}
P_{4}=\cfrac{1}{q^6[4]_q!} &
\Bigl(1 +q(R_1+R_2+R_3) +q^2(R_1R_2+R_2R_1+R_2R_3+R_3R_2+R_1R_3)+
\\ & \\ &
+q^3(R_1R_2R_1+R_2R_3R_2+R_1R_2R_3+R_2R_1R_3+R_3R_1R_2+R_3R_2R_1)+
\\ & \\ &
+q^4(R_1R_2R_1R_3+R_1R_2R_3R_2+R_1R_3R_2R_1+R_2R_1R_3R_2+R_2R_3R_2R_1)+
\\ & \\ &
+q^5(R_1R_2R_1R_3R_2+R_1R_2R_3R_2R_1+R_2R_1R_3R_2R_1)+q^6R_1R_2R_1R_3R_2R_1\Bigr),
\\ & \\
P_{31}=\cfrac{1}{q^2[2]_q[4]_q} &
\Bigl(3+(2q-q^{-1})(R_1+R_2+R_3)+(q^2-1)(R_1R_2+R_2R_1+R_2R_3+R_3R_2)+
\\ & \\ &
+(q^2-2)R_1R_3+q^3(R_1R_2R_1+R_2R_3R_2)-
\\ & \\ &
-q(R_1R_2R_3+R_2R_1R_3+R_3R_1R_2+R_3R_2R_1)-R_2R_1R_3R_2-
\\ & \\ &
-q(R_1R_2R_1R_3R_2+R_2R_1R_3R_2R_1)+q^3R_1R_2R_3R_2R_1-q^2R_1R_2R_1R_3R_2R_1\Bigr),
\\ & \\
P_{22}=\cfrac{1}{[3]_q[2]_q^2} &
\Bigl(2+(q-q^{-1})(R_1+R_2+R_3)-(R_1R_2+R_2R_1+R_2R_3+R_3R_2)+
\\ & \\ &
+(q^2+q^{-2})R_1R_3-(R_1R_2R_1R_3+R_1R_2R_3R_2+R_1R_3R_2R_1+R_2R_3R_2R_1)+
\\ & \\ &
+(q^2+q^{-2})R_2R_1R_3R_2-
\\ & \\ &
-(q-q^{-1})(R_1R_2R_1R_3R_2+R_1R_2R_3R_2R_1+R_2R_1R_3R_2R_1)+2R_1R_2R_1R_3R_2R_1\Bigr),
\\ & \\ &
P_{211}(q)=P_{31}(-q^{-1}),
\\ & \\ &
P_{1111}(q)=P_{4}(-q^{-1}).
\end{array}
\end{equation}

\subsubsection{Projectors from characteristic equations.}

Here we consider one more approach, which allows one to construct projectors as a polynomials of the $R$-matrices. Unlike it was in the previous section, the obtained polynomials are neither of the minimal degree, nor with the minimal number of terms. Instead, this approach gives the answer with a more transparent structure what makes this method important for theoretical analysis.

The approach comes from the fact that if a characteristic equation for a linear operator is known,
\begin{equation}
\prod\limits_{i=1}^n(A - \lambda_i)=0,
\label{chareq}
\end{equation}
then it is easy to construct a projector onto the subspace corresponding to each eigenvalue,\footnote{The ordering of the factors is inessential both in~(\ref{chareq}) and~(\ref{projchareq}) since all these factors commute.}
\begin{equation}
P_{\lambda_j}=\prod\limits_{i\ne j}\cfrac{A-\lambda_i}{\lambda_j-\lambda_i},
\label{projchareq}
\end{equation}
or on a sum of such subspaces,
\begin{equation}
P_{\lambda_{j_1},\ldots,\lambda_{j_k}}=\sum\limits_{l=1}^kP_{\lambda_{j_k}}.
\label{projchareqmult}
\end{equation}
The property $P^2_{\lambda_j}=P_{\lambda_j}$ then follows directly from~(\ref{chareq}). A drawback of using this method when dealing with $R$-matrices is that many of their  eigenvalues coincide, e.g., in the fundamental case there are only two of them, namely $q$ and $-q^{-1}$. Therefore, the main task in this case is to find a combination of $R$-matrices that has enough different eigenvalues to distinguish all irreducible representations in the decomposition of $[1]^{|Q|}$.  To find such combinations and the corresponding characteristic equations, it suffices to consider $|Q|$-strand braids. It is natural to consider at first the particular cases $|Q|=2$ and $|Q|=3$ and then try to generalize the method.

Since the $R$-matrices are split into the blocks corresponding to different irreducible representations, the characteristic equations can be written in the form
\begin{equation}
\prod_{Q\vdash1^{|Q|}}F_Q(R)=0,
\end{equation}
where $F_Q(R)=0$ is the characteristic equation for block corresponding to the representation $Q$.

\paragraph{Level $\mathbf{|Q|=2}$.} In this case, there is one $R$-matrix satisfying the characteristic equation
\begin{equation}
F_2(R_1)F_{11}(R_1)\equiv(R_1 - q)(R_1 + q^{-1})=0.
\label{chareq2}
\end{equation}
This equation gives enough information to construct both projectors:
\begin{equation}
\begin{array}{lcr}
P_{2}=\cfrac{R_1 + q^{-1}}{q+q^{-1}}, & \ & P_{11}=\cfrac{R_1 - q}{-q^{-1}-q}.
\end{array}
\label{proj2chareq}
\end{equation}
This result coincides with the previously obtained formulas~(\ref{eq:projun}).

\paragraph{Level $\mathbf{|Q|=3}$.} In this case, there are the two matrices, $R_1$ and $R_2$, satisfying the same characteristic equation~(\ref{Revals}). The projectors, obtained from the characteristic equation for the first $R$-matrix distinguish the \emph{symmetric} in the first pair of strands representations $[3]$ and $\underline{[21]}$ from the \emph{antisymmetric} in the first pair of strands representations  $\overline{[21]}$ and $[111]$. The projectors obtained from the equation for the second $R$-matrix distinguish the similar two groups of representations with respect to the second pair of strands. This is not enough to construct the projectors on each of the representations $[3]$, $[21]$ and $[111]$. Thus one has to find some combinations of $R$-matrices that can provide all the needed projectors.
For example, for $|Q|=3$ a proper combination is $(R_1-R_2)^2$. From the explicit expressions for  $R_1$ and $R_2$ one can see that this combination satisfies the following characteristic equation,
\begin{equation}
(R_1-R_2)^2\Bigl((R_1-R_2)^2-(q^2+1+q^{-2})\Bigr)=0,
\label{chareq3}
\end{equation}
where the representation $[3]$ as well as the representation $[111]$ corresponds to the eigenvalue $0$ while the both representations $[21]$ correspond to the eigenvalue $q^2+1+q^{-2}$.
This gives the projector $P_{21}$:
\begin{equation}
P_{21}=\cfrac{(R_1-R_2)^2}{q^2+1+q^{-2}}.
\label{eq:P21}
\end{equation}
Together with~(\ref{chareq2}), it also gives all the 4 projectors:
\begin{equation}
\begin{array}{l}
P_3=\cfrac{(R_1+q^{-1})\Bigl((q^2+1+q^{-2})-(R_1-R_2)^2\Bigr)}{(q+q^{-1})(q^2+1+q^{-2})},
\\ \\
P_{\underline{21}}=\cfrac{(R_1+q^{-1})(R_1-R_2)^2}{(q+q^{-1})(q^2+1+q^{-2})},
\\ \\
P_{\overline{21}}=\cfrac{(q-R_1)(R_1-R_2)^2}{(q+q^{-1})(q^2+1+q^{-2})},
\\ \\
P_{111}=\cfrac{(q-R_1)\Bigl((q^2+1+q^{-2})-(R_1-R_2)^2\Bigr)}{(q+q^{-1})(q^2+1+q^{-2})}.
\label{proj3ce}
\end{array}
\end{equation}

\paragraph{Torus products of the $\mathbf{R}$-matrices.} The question remains if there is a universal method to construct the projectors onto arbitrary representations from the characteristic equations. It could be possible if one manages to find a combination of the $R$-matrices with eigenvalues known for any representation and with enough many different eigenvalues to distinguish all irreducible representations at the corresponding level. The obvious candidate for such a combination is the product
\begin{equation}
\mathfrak{R}_{|Q|}\equiv\prod\limits_{i=1}^{|Q|-1}R_{|Q|-i}.
\end{equation}
This is the product of the $R$-matrices that appears when the torus knots are studied. It is known from the Rosso-Jones formula that its eigenvalues are
\begin{equation}
\lambda_{Q,j}=q^{\frac{\varkappa_Q}{|Q|}}\Lambda_{Q,j},\ \ j=1\ldots N_Q,
\label{RossoJones}
\end{equation}
where the coefficients $\Lambda_{R,j}$ are $q$ independent numbers implicitly defined by the Adams rule~\cite{RJ}-\cite{DMMSS}. These coefficients are in fact roots of unity of the power $|Q|$, $\exp{\frac{2\pi ik}{|Q|}}$, and the multiplicity of each eigenvalue is non-trivially defined by  $Q$. Hence,
\begin{equation}
|\lambda_{Q,j}|=q^{\frac{\varkappa_Q}{|Q|}}\ \ j=1\ldots N_Q.
\label{RJmod}
\end{equation}
For such products of the $R$-matrices, sets of the eigenvalues corresponding to  different representations do not overlap up to the level $|Q|=5$. Thus, the projectors onto all the irreducible representations with $|Q|$ boxes can be obtained from the single equation for $\mathfrak{R}_{|Q|}$. This method stops working at the level $|Q|=6$ where $\varkappa_{411}=\varkappa_{33}=3$ and additional equations are required. Besides that fact, the explicit formulae for them are not known for arbitrary representation even when all the eigenvalues are different. One has either to extract them from the Adams rule~\cite{RJ}-\cite{DMMSS} or to straightforwardly calculate them for each $|Q|$ as it is done in~\cite{MMM2},\cite{AMMM1}. Moreover even in the simplest cases these eigenvalues are quite complicated and include fractional powers of $q$ and roots of unity. Due to these problems it is not clear how to find the general answer for the projectors onto arbitrary representations using the torus product $\mathfrak{R}_{|Q|}$. Thus, we only provide several examples of using these products. Since the $|Q|=2$ is trivial  ($\mathfrak{R}_2=R_1$), we start from the case $\mathbf{|Q|=3}$: one has
\begin{equation}
F_3(R_2R_1)F_{21}(R_2R_1)F_{111}(R_2R_1)\equiv(R_2R_1-q^2)\Bigl((R_2R_1)^2+R_2R_1+1\Bigr)(q^2R_2R_1-1)=0.
\label{chareq3toric}
\end{equation}
The corresponding projectors onto symmetric and onto antisymmetric representations are constructed immediately:
\begin{equation}
\begin{array}{lcr}
P_3=\cfrac{\Bigl((R_2R_1)^2+R_2R_1+1\Bigr)(R_2R_1-q^{-2})}{(q^4+q^2+1)(q^2-q^{-2})},
 & \  &
P_{111}=\cfrac{(R_2R_1-q^2)\Bigl((R_2R_1)^2+R_2R_1+1\Bigr)}{(q^{-2}-q^2)(q^{-4}+q^{-2}+1)}.
\end{array}
\label{projs3cet}
\end{equation}
But for the remaining representation, the simple construction like~(\ref{projchareq}) gives the operator $P\equiv(R_2R_1-q^2)(q^2R_2R_1-1)$ that does not satisfy the condition $P^2=const\cdot P$. Thus, it is not a projector even up to a normalization but a linear operator that annihilates representations $[3]$ and $[111]$ and somehow rotates two copies of representation $[21]$. To obtain the projector according to~(\ref{projchareqmult}) one has to further decompose~(\ref{chareq3toric}) in the following way
\begin{equation}
(R_2R_1-q^2)(R_2R_1- e^{\frac{2\pi i}{3}})(R_2R_1- e^{\frac{4\pi i}{3}})(q^2R_2R_1-1)=0.
\end{equation}
Then, $P_{21}$ can be obtained as a sum
\begin{equation}
\begin{array}{c}
P_{21}=\cfrac{(R_2R_1-q^2)(q^2R_2R_1-1)}{(e^{\frac{2\pi i}{3}}-e^{\frac{4\pi i}{3}})}\left(\cfrac{R_2R_1- e^{\frac{4\pi i}{3}}}{(e^{\frac{2\pi i}{3}}-q^2)(q^2e^{\frac{2\pi i}{3}}-1)}-\cfrac{R_2R_1- e^{\frac{2\pi i}{3}}}{(e^{\frac{4\pi i}{3}}-q^2)(q^2e^{\frac{4\pi i}{3}}-1)}\right)=
\\ \\
=-\cfrac{(R_2R_1-q^2)(q^2R_2R_1-1)(R_2R_1+1)}{q^4+q^2+1}.
\end{array}
\label{proj21cet}
\end{equation}
The summands are rank one projectors onto each of the isomorphic representations $[21]$, but these copies are \emph{not} $\underline{[21]}$, $\overline{[21]}$ of the standard basis ~(\ref{eq:nb}) used in this paper. It can be shown using the properties  of the $R$-matrices described in Sec.~\ref{sec:Rprop} that all three answers for the projectors~(\ref{proj3def}), ~(\ref{projs3cet},\ref{proj21cet}) and~(\ref{proj3ce}) are equivalent.

In the case of $\mathbf{|Q|=4}$, the relevant product now includes three $R$-matrices:
\begin{equation}
\mathfrak{R}_3\equiv R_3R_2R_1
\end{equation}
and satisfy the characteristic equation
\begin{equation}
F_4(\mathfrak{R}_3) F_{31}(\mathfrak{R}_3) F_{22}(\mathfrak{R}_3) F_{211}(\mathfrak{R}_3) F_{1111}(\mathfrak{R}_3)=0,
\end{equation}
where
\begin{equation}
\begin{array}{c}
\begin{array}{ccccc}
F_4(\mathfrak{R}_3)=q^3-\mathfrak{R}_3, & \ & F_{31}(\mathfrak{R}_3)=(q+\mathfrak{R}_3)(q^2+\mathfrak{R}_3^2), & \ & F_{22}(\mathfrak{R}_3)=\mathfrak{R}_3^2-1),
\end{array}
\\ \\
\begin{array}{ccc}
F_{211}(q|\mathfrak{R}_3)=F_{31}(-q^{-1}|\mathfrak{R}_3), & \ & F_{1111}(q|\mathfrak{R}_3)=F_4(-q^{-1}|\mathfrak{R}_3).
\end{array}
\end{array}
\end{equation}
Hence, the entire characteristic equation is
\begin{equation}
(q^3-\mathfrak{R}_3)\cdot(q+\mathfrak{R}_3)(q^2+\mathfrak{R}_3^2)\cdot(\mathfrak{R}_3^2-1)\cdot(1-q\mathfrak{R}_3)(1+q^2\mathfrak{R}_3^2)\cdot (1-q^3\mathfrak{R}_3)=0,
\end{equation}
where the ``$\cdot$'' separates factors corresponding to different irreducible representations. The projectors are constructed according to~(\ref{projchareq}) and~(\ref{projchareqmult}):
\begin{equation}
\begin{array}{rl}
P_4 = &
\prod\limits_{\substack{Q_i\vdash 1^4 \\ Q_i \ne 4}}
\cfrac{F_{Q_i}(\mathfrak{R}_3)}{F_{Q_i}(q^3)}=\cfrac{(q+\mathfrak{R}_3)(q^2+\mathfrak{R}_3^2)(\mathfrak{R}_3^2-1)(q\mathfrak{R}_3-1)(q^2\mathfrak{R}_3^2+1)(q^3\mathfrak{R}_3+1)}{q^3(q^2+1)(q^{12}-1)(q^{16}-1)},
\\ & \\
P_{31} = &
\prod\limits_{\substack{Q_i \vdash 1^4 \\ Q_i \ne 31}} F_{Q_i}(\mathfrak{R}_3) \left(
\frac{\mathfrak{R}_3+q^2}{F_{41}^{\prime}(-q)\prod\limits_{\substack{Q_i\vdash 1^4\\ Q_i \ne 31}}F_{Q_i}(-q)} +\frac{(\mathfrak{R}_3+q)(\mathfrak{R}_3+iq)}{F_{41}^{\prime}(iq)\prod\limits_{\substack{Q_i\vdash 1^4\\ Q_i \ne 31}}F_{Q_i}(iq)} +\frac{(\mathfrak{R}_3+q)(\mathfrak{R}_3-iq)}{F_{41}^{\prime}(-iq)\prod\limits_{\substack{Q_i\vdash 1^4\\ Q_i \ne 31}}F_{Q_i}(-iq)}\right)=
\\ & \\ = &
\cfrac{(\mathfrak{R}_3-q^3)(\mathfrak{R}_3^2-1)(q^8\mathfrak{R}_3^2-q^4\mathfrak{R}_3^2+\mathfrak{R}_3^2-q^5\mathfrak{R}_3+q\mathfrak{R}_3+q^2)(q\mathfrak{R}_3 -1) (q^2\mathfrak{R}_3^2+1)(q^3\mathfrak{R}_3^2+1)}{q^3(q^2+1)(q^4-1)(q^{16}-1)},
\\ & \\
P_{22} = &
\prod\limits_{\substack{Q_i\vdash 1^4\\ Q_i \ne 22}} F_{Q_i}(\mathfrak{R}_3)\left(\frac{\mathfrak{R}_3+1}{F_{22}^{\prime}(1)\prod\limits_{\substack{Q_i\vdash 1^4\\ Q_i \ne 22}}F_{Q_i}(1)} +\frac{\mathfrak{R}_3-1}{F_{22}^{\prime}(-1)\prod\limits_{\substack{Q_i\vdash 1^4\\ Q_i \ne 22}}F_{Q_i}(-1)}\right)=
\\ & \\ = &
\cfrac{(q^3-\mathfrak{R}_3)(q+\mathfrak{R}_3)(q^2+\mathfrak{R}_3^2)(1-q\mathfrak{R}_3)(1+q^2\mathfrak{R}_3^2)(1-q^3\mathfrak{R}_3)}{(q^4-1)^2(q^4+q^2+1)}.
\end{array}
\end{equation}

The same procedure can be repeated also for $|Q|=5$ and $|Q|=7$. However the method does not work for $|Q|=6$ and for the most of higher representations cases when there are different representations $Q$ with the same $\varkappa_{Q}$. E.g., in the case $|Q|=6$ such representations are $[411]$ and $[33]$. This means that for such representations eigenvalues of the torus product also coincide, namely $|\lambda_{411}|=|\lambda_{33}|=q$. Thus, the projectors constructed using the described method do not distinguish these two representations. Indeed,
\begin{equation}
\begin{array}{l}
F_{411}(\mathfrak{R}_3)=(\mathfrak{R}_3^2-q^2)^2(\mathfrak{R}_3^2-q\mathfrak{R}_3+q^2) (\mathfrak{R}_3^4+q^2\mathfrak{R}_3^2+q^4),
\\
F_{33}(\mathfrak{R}_3)=(\mathfrak{R}_3^2-q^2)(q^2-q\mathfrak{R}_3+\mathfrak{R}_3^2).
\end{array}
\label{toricdeg}
\end{equation}
Since all the eigenvalues of the representation $[33]$ are also the eigenvalues of the representation $[411]$ even the projector onto some particular copy of the representation $[33]$ cannot be constructed from these formulae. This means that some other equations are needed. It would be natural to use the equations from the same series but from lower levels, i.e., for $|Q|\le 6$, thus applying the already known answers for the projectors onto the lower representations. Yet, it cannot be done straightforwardly with the torus products since the eigenvectors of $\mathfrak{R}_{|Q|}$ are not in general the eigenvectors of $\mathfrak{R}_{|Q|+1}$.

\paragraph{Link products of the $\mathbf{R}$-matrices.} Another class of $R$-matrix combinations that have eigenvalues known in full generality, is the one we call link products,
\begin{equation}
\mathfrak{R}_{n,m}\mathfrak{R}_{n,m}^{\dagger}=
\left(\prod\limits_{i=1}^m \prod\limits_{j=1}^n R_{n+i-j}\right)
\left(\prod\limits_{i=1}^n \prod\limits_{j=1}^m R_{m+i-j}\right).
\label{lcd}
\end{equation}
These products emerge for the $2$-strand links. If one places a representation of size $n$ at the first strand and of size $m$ at the second one, then in the cabling procedure each pair $\mathcal{R}_{Q_1Q_2}\mathcal{R}_{Q_2Q_1}$ should be replaced with product~(\ref{lcd}). As will be discussed in details in Sec.~\ref{sec:2str}, such a product should satisfy the following characteristic equation
\begin{equation}
\prod_{\substack{T_1\vdash 1^m \\ T_2\vdash 1^n \\ Q \vdash 1^{m+n}}}
\left(\mathfrak{R}_{n,m}\mathfrak{R}_{n,m}^{\dagger}-q^{2\varkappa_Q-2\varkappa_{T_1}-2\varkappa_{T_2}}\right)=0.
\end{equation}

This link product is more lengthy than the torus one but for some reasons is more convenient. First of all, its eigenvalues are explicitly known for any number of strands and they are much simpler than the ones of the torus product since they are just integer powers of $q$. Also the standard basis~(\ref{eq:nb}) used in this paper  gives an eigenvectors of such product. These facts allows one to construct a recursive procedure to obtain the projectors.

The simplest case is when $T_1=[1]$. Then $\mathfrak{R}_{1,|T_2|}$ is  just a torus product:
\begin{equation}
\mathfrak{R}_{1,|T_2|}\mathfrak{R}_{1,|T_2|}^{\dagger}\equiv \mathfrak{R}_{|T_2|}\mathfrak{R}_{|T_2|}^{\dagger}.
\end{equation}
As before we will start with some examples and discuss the generalization possibilities afterwards. In the case of $\mathbf{|Q|=2}$, the characteristic equation is a quite simple deformation of its analog given above:
\begin{equation}
\begin{array}{c}
F_2=\mathfrak{R}_2\mathfrak{R}_2^{\dagger}-q^{2\varkappa_2}=\mathfrak{R}_2\mathfrak{R}_2^{\dagger}-q^2,
\\
F_{11}=\mathfrak{R}_2\mathfrak{R}_2^{\dagger}-q^{2\varkappa_{11}}=\mathfrak{R}_2\mathfrak{R}_2^{\dagger}-q^{-2}
\\
\Downarrow
\\
F_2F_{11}=\left(\mathfrak{R}_2\mathfrak{R}_2^{\dagger}-q^2\right)\left(\mathfrak{R}_2\mathfrak{R}_2^{\dagger}-q^{-2}\right)=0.
\end{array}
\end{equation}
It gives the following projectors
\begin{equation}
\begin{array}{ll}
P_2=\cfrac{\left(\mathfrak{R}_2\mathfrak{R}_2^{\dagger}-q^{-2}\right)}{q^2-q^{-2}},&
P_{11}=\cfrac{\left(\mathfrak{R}_2\mathfrak{R}_2^{\dagger}-q^2\right)}{q^{-2}-q^2}.
\end{array}
\end{equation}
This answer can be transformed into the answer~(\ref{proj2chareq}) obtained using the previous methods using that $\mathfrak{R}_2\mathfrak{R}_2^{\dagger}=R_1^2=\left(q-q^{-1}\right)R_1+1$.

In the first non-trivial case of $\mathbf{|Q|=3}$, the characteristic equation looks like
\begin{equation}
\begin{array}{c}
\begin{array}{l}
F_3=\mathfrak{R}_3\mathfrak{R}_3^{\dagger}-q^{2\varkappa_3-2\varkappa_2}=\mathfrak{R}_3\mathfrak{R}_3^{\dagger}-q^4,
\\
F_{21}=\left(\mathfrak{R}_3\mathfrak{R}_3^{\dagger}-q^{2\varkappa_{21}-2\varkappa_3}\right)\left(\mathfrak{R}_3\mathfrak{R}_3^{\dagger}-q^{2\varkappa_{21}-2\varkappa_{111}}\right)=\left(\mathfrak{R}_3\mathfrak{ R}_3^{\dagger}-q^{-2}\right)\left(\mathfrak{R}_3\mathfrak{R}_3^{\dagger}-q^2\right),
\\
F_{111}=\mathfrak{R}_3\mathfrak{R}_3^{\dagger}-q^{2\varkappa_{111}-2\varkappa_{11}}=\mathfrak{R}_3\mathfrak{R}_3^{\dagger}-q^{-4}
\end{array}
\\
\Downarrow
\\
F_3F_{21}F_{111}=\left(\mathfrak{R}_3\mathfrak{R}_3^{\dagger}-q^4\right)\left(\mathfrak{R}_3\mathfrak{R}_3^{\dagger}-q^{-2}\right)\left(\mathfrak{R}_3\mathfrak{R}_3^{\dagger}-q^2\right)\left(\mathfrak{R}_3\mathfrak{R}_3^{\dagger}-q^{-4}\right)=0.
\end{array}
\end{equation}
The rank one projectors onto symmetric and antisymmetric representations are
\begin{equation}
\begin{array}{c}
P_3=P^{(3)}_{q^4}=\cfrac{\left(\mathfrak{R}_3\mathfrak{R}_3^{\dagger}-q^{-2}\right) \left(\mathfrak{R}_3\mathfrak{R}_3^{\dagger}-q^2\right) \left(\mathfrak{R}_3\mathfrak{R}_3^{\dagger}-q^{-4}\right)} {\left(q^4-q^{-2}\right)\left(q^4-q^2\right)\left(q^4-q^{-4}\right)},
\\
\\
P_{111}=P^{(3)}_{q^{-4}}=\cfrac{\left(\mathfrak{R}_3\mathfrak{R}_3^{\dagger}-q^4\right) \left(\mathfrak{R}_3\mathfrak{R}_3^{\dagger}-q^{-2}\right) \left(q^{-4}-q^2\right)} {\left(q^{-4}-q^4\right)\left(q^{-4}-q^{-2}\right)\left(q^{-4}-q^2\right)},
\end{array}
\end{equation}
where $P^{(m)}_{\lambda}$ denotes the projector onto the eigenvalue $\lambda$ obtained from the $m$-strand equation. The remaining projector is of the rank two; it is obtained as the sum of the two rank one projectors:
\begin{equation}
\begin{array}{l}
P_{21}=P^{(3)}_{q^2}+P^{(3)}_{q^{-2}}=
\\
\\
=\cfrac{\left(\mathfrak{R}_3\mathfrak{R}_3^{\dagger}-q^4\right) \left(\mathfrak{R}_3\mathfrak{R}_3^{\dagger}-q^{-2}\right)\left(\mathfrak{R}_3\mathfrak{R}_3^{\dagger}-q^{-4}\right)} {\left(q^2-q^4\right)\left(q^2-q^{-2}\right)\left(q^2-q^{-4}\right)}
+\cfrac{\left(\mathfrak{R}_3\mathfrak{R}_3^{\dagger}-q^4\right) \left(\mathfrak{R}_3\mathfrak{R}_3^{\dagger}-q^2\right) \left(\mathfrak{R}_3\mathfrak{R}_3^{\dagger}-q^{-4}\right)} {\left(q^{-2}-q^4\right)\left(q^{-2}-q^2\right)\left(q^{-2}-q^{-4}\right)}=
\\
\\
=-\cfrac{\left(\mathfrak{R}_3\mathfrak{R}_3^{\dagger}-q^4\right) \left(\mathfrak{R}_3\mathfrak{R}_3^{\dagger}-q^{-4}\right)
\left(\mathfrak{R}_3\mathfrak{R}_3^{\dagger}-q^2-q^{-2}\right)}{(q-q^{-1})(q^3-q^{-3})}.
\end{array}
\end{equation}

For $\mathbf{|Q|=4}$, the characteristic equation is
\begin{equation}
\begin{array}{c}
\begin{array}{l}
F_4=\mathfrak{R}_4\mathfrak{ R}_4^{\dagger}-q^{2\varkappa_4-2\varkappa_3}=\mathfrak{R}_4\mathfrak{R}_4^{\dagger}-q^6,
\\
F_{31}=\left(\mathfrak{R}_4\mathfrak{R}_4^{\dagger}-q^{2\varkappa_{31}-2\varkappa_3}\right) \left(\mathfrak{R}_4\mathfrak{R}_4^{\dagger}-q^{2\varkappa_{31}-2\varkappa_{21}}\right)^2= \left(\mathfrak{R}_4\mathfrak{R}_4^{\dagger}-q^{-2}\right)\left(\mathfrak{R}_4\mathfrak{R}_4^{\dagger}-q^2\right)^2,
\\
F_{22}=\left(\mathfrak{R}_4\mathfrak{R}_4^{\dagger}-q^{2\varkappa_{22}-2\varkappa_{21}}\right)^2= \left(\mathfrak{R}_4\mathfrak{R}_4^{\dagger}-1\right)^2,
\\
F_{211}=\left(\mathfrak{R}_4\mathfrak{R}_4^{\dagger}-q^{2\varkappa_{211}-2\varkappa_{21}}\right)^2 \left(\mathfrak{R}_4\mathfrak{R}_4^{\dagger}-q^{2\varkappa_{211}-2\varkappa_{111}}\right)= \left(\mathfrak{R}_4\mathfrak{R}_4^{\dagger}-q^2\right)\left(\mathfrak{R}_4\mathfrak{R}_4^{\dagger}-q^2\right)^{-2},
\\
F_{1111}=\mathfrak{R}_4\mathfrak{R}_4^{\dagger}-q^{2\varkappa_4-2\varkappa_3}=\mathfrak{R}_4\mathfrak{ R}_4^{\dagger}-q^{-6}
\end{array}
\\
\Downarrow
\\
F_4F_{31}F_{22}F_{211}F_{1111}=\left(\mathfrak{R}_4\mathfrak{R}_4^{\dagger}-q^6\right) \left(\mathfrak{R}_4\mathfrak{R}_4^{\dagger}-q^{2}\right) \left(\mathfrak{R}_4\mathfrak{R}_4^{\dagger}-q^{-2}\right) \left(\mathfrak{R}_4\mathfrak{R}_4^{\dagger}-1\right)\left(\mathfrak{R}_4\mathfrak{R}_4^{\dagger}-q^{-6}\right)=0.
\end{array}
\label{cheqcabl4}
\end{equation}
In this case, the single equation is not enough to find all the projectors since there are different irreducible representations with the same set of eigenvalues. Namely, representations $[31]$ and $[211]$ both have two eigenvalues equal to $q^{\pm 2}$. This can be resolved using the observation that each factor in~(\ref{cheqcabl4}) corresponds to a certain path, i.e., to a certain sequence of irreducible representations at the previous levels. For example, in $F_{31}$ the double factor with $q^2$ corresponds to the pair of paths $[1]\rightarrow \alpha \rightarrow [21]\rightarrow [31]$, where $\alpha=[2]$ or $[11]$, and the factor with $q^{-2}$ corresponds to the path $[1]\rightarrow [2]\rightarrow [3]\rightarrow [31]$, in $F_{211}$ factor with $q^2$ corresponds to the path $[1]\rightarrow [11]\rightarrow [21]\rightarrow [31]$ and double factor with $q^{-2}$ corresponds to the pair of paths $[1]\rightarrow \alpha \rightarrow [21]\rightarrow [211]$, where $\alpha=[2]$ or $[11]$. This means that one can find the sought projectors using the already constructed projectors onto representations of the preceding levels:
\begin{equation}
\begin{array}{l}
P_{31}=P^{(4)}_{q^{-2}}P_{3}+P^{(4)}_{q^2}P_{21}=
\\
=\cfrac{\left(\mathfrak{R}_4\mathfrak{ R}_4^{\dagger}-q^6\right) \left(\mathfrak{R}_4\mathfrak{R}_4^{\dagger}-1\right) \left(\mathfrak{R}_4\mathfrak{R}_4^{\dagger}-q^{-6}\right)}{(q^4-q^{-4})(q^2-q^{-2})^2(q-q^{-1})} \left(q^3\left(\mathfrak{R}_4\mathfrak{R}_4^{\dagger}-q^2\right)P_3+q^{-3} \left(\mathfrak{R}_4\mathfrak{R}_4^{\dagger}-q^{-2}\right)P_{21}\right),
\\ \\
P_{211}=P^{(4)}_{q^{-2}}P_{21}+P^{(4)}_{q^2}P_{111}=
\\
=\cfrac{\left(\mathfrak{R}_4\mathfrak{ R}_4^{\dagger}-q^6\right) \left(\mathfrak{R}_4\mathfrak{R}_4^{\dagger}-1\right) \left(\mathfrak{R}_4\mathfrak{R}_4^{\dagger}-q^{-6}\right)} {(q^4-q^{-4})(q^2-q^{-2})^2(q-q^{-1})} \left(q^3\left(\mathfrak{R}_4\mathfrak{R}_4^{\dagger}-q^2\right)P_{21} +q^{-3}\left(\mathfrak{R}_4\mathfrak{R}_4^{\dagger}-q^{-2}\right)P_{111}\right).
\end{array}
\end{equation}
All the other projectors can be constructed as before
\begin{equation}
\begin{array}{l}
P_{4}=P^{(4)}_{q^6}= \cfrac{\left(\mathfrak{R}_4\mathfrak{R}_4^{\dagger}-q^6\right) \left(\mathfrak{R}_4\mathfrak{R}_4^{\dagger}-q^{2}\right) \left(\mathfrak{R}_4\mathfrak{R}_4^{\dagger}-q^{-2}\right) \left(\mathfrak{R}_4\mathfrak{R}_4^{\dagger}-1\right) \left(\mathfrak{R}_4\mathfrak{R}_4^{\dagger}-q^{-6}\right)} {\left(q^6-q^{2}\right)\left(q^6-q^{-2}\right)\left(q^6-1\right)\left(q^6-q^{-6}\right)},
\\ \\
P_{22}=P^{(4)}_1=\cfrac{\left(\mathfrak{R}_4\mathfrak{R}_4^{\dagger}-q^6\right)\left(\mathfrak{R}_4\mathfrak{R}_4^{\dagger}-q^{2}\right) \left(\mathfrak{R}_4\mathfrak{R}_4^{\dagger}-q^{-2}\right)\left(\mathfrak{R}_4\mathfrak{R}_4^{\dagger}-q^{-6}\right)} {\left(1-q^6\right)\left(1-q^{2}\right)\left(1-q^{-2}\right)\left(1-q^{-6}\right)},
\\ \\
P_{1111}=P^{(4)}_{q^{-6}}=\cfrac{\left(\mathfrak{R}_4\mathfrak{R}_4^{\dagger}-q^6\right) \left(\mathfrak{R}_4\mathfrak{R}_4^{\dagger}-q^{2}\right) \left(\mathfrak{R}_4\mathfrak{R}_4^{\dagger}-q^{-2}\right)\left(\mathfrak{R}_4\mathfrak{R}_4^{\dagger}-1\right)} {\left(q^{-6}-q^6\right)\left(q^{-6}-q^{2}\right)\left(q^{-6}-q^{-2}\right)}.
\end{array}
\end{equation}

\paragraph{Arbitrary representations.} In principle, the set of equations for the link products $\mathfrak{R}_{|Q|}\mathfrak{R}_{|Q|}^{\dagger}$ is enough to construct a projector onto any representation. If for example a representation $Q$ was made from a representation $T$ by adding one box with the coordinates $(i,j)$ to the Young diagram, then $\varkappa_Q=\varkappa_T+j-i$. Thus, for any $T$, all the $Q$ obtained in this way have different $\varkappa_Q$. If the representations $Q$ are chosen as eigenvectors of $\mathfrak{R}_{|Q|}\mathfrak{R}_{|Q|}^{\dagger}$, then the projectors can be constructed from the equation
\begin{equation}
\prod\limits_{i,j}\left(P_T\mathfrak{R}_{|Q|}\mathfrak{R}_{|Q|}^{\dagger}P_T-q^{2j-2i}\right)=0,
\label{evreqeq}
\end{equation}
and have the form
\begin{equation}
P_{Q=T\cup(k,l)}=\sum\limits_{i,j}\prod\limits_{(i,j)\ne (k,l)}\cfrac{P_T\mathfrak{R}_{|Q|}\mathfrak{R}_{|Q|}^{\dagger}P_T-q^{2j-2i}}{q^{2l-2k}-q^{2j-2i}} =\sum\limits_{i,j}\prod\limits_{(i,j)\ne(k,l)}\cfrac{\mathfrak{R}_{|Q|}\mathfrak{R}_{|Q|}^{\dagger}-q^{2j-2i}}{q^{2l-2k}-q^{2j-2i}}P_T,
\label{projreqsep}
\end{equation}
where all the products are calculated over all additions of the box (with the coordinates $(i,j)$) to the Young diagram $T$ that again yield a Young diagram. This formula gives the projector onto the entire space of the representations $Q$ simultaneously arising in the decompositions of $1^{|Q|}$ and $1\otimes T$. To obtain the projector onto one of representations $Q$ in the decomposition of $1^{|Q|}$, one has to substitute the projector onto one of representations $T$ in the decomposition of $1^{|T|}$ for $P_T$.  In contrast, to obtain the projector onto the entire space of representations $Q$ in the decomposition of $1^{|Q|}$, one has to take the sum over all possible $T$:
\begin{eqnarray}
P_{Q}=\sum\limits_{T:\ |T|=|Q|-1}P_{Q=T\cup(k,l)}.\label{projreqent}
\end{eqnarray}
The presented formulae give a recursive procedure to construct a projector onto any representation.

\subsubsection{(Anti-)symmetric representations.} The method described in the previous sections allows one in principle to find any projector but it is not clear how to write the general formula. Nevertheless such formula can be written for the simplest class of representations, namely for the symmetric ones. These projectors have the form
\begin{equation}
P_{[r]}=P_{[r-1]}\left(1+\sum\limits_{j=1}^k q^j\prod\limits^j_{i=1}R_{k-i+1}\right)=\prod\limits_{r-1}^{k=1}\left(1+\sum_{j=1}^k q^j\prod^j_{i=1}R_{k-i+1}\right)
\label{projsymm}
\end{equation}
or
\begin{equation}
P_{[r]}=\sum_{k=1}^{r!} q^{l_k}\Xi_k,
\label{qsymm}
\end{equation}
where the sum is over all $r!$ basis elements $\Xi_k$ (see (\ref{eq:prb})) of the polynomial ring and $l_k$ is the number of $R$-matrices in the $\Xi_k$.

\section{Framing in the cabling procedure\label{framing}}

The $\mathcal{R}$-matrices are defined up to a common factor because it would not affect the Yang-Baxter equation. This common factor can be chosen differently thus changing the answer for the HOMFLY polynomials.

From the standpoint of the Chern-Simons theory, the Wilson loop should correspond to the framed knot~\cite{KauffTB},\cite{Pras} and the common factor depends on the framing of the knot. Hence, the choices of the common factors are called framings. It can be shown that the different framings should differ by the factor with $q^{\varkappa_{T}+\frac{N|T|}{2}}$~\cite{Inds5}.

The framing can be chosen in different ways. One common choice comes from the representation theory where the eigenvalues of $\mathcal{R}$-matrix are equal to $q^{\varkappa_Q-\frac{N|Q|}{2}-\varkappa_{T_1}+\frac{N|T_1|}{2}-\varkappa_{T_2}+\frac{N|T_2|}{2}}$, where $Q$ describes corresponding irreducible representation and $T_1$ and $T_2$ describe the two crossing representations (i.e., this is the eigenvalue corresponding to the irreducible representation $Q$ in the decomposition of $T_1\otimes T_2$ thus actually $|Q|=|T_1|+|T_2|$ so there is no $N$ in this framing). This gives the eigenvalues $q$ and $-q^{-1}$ for the fundamental $\mathcal{R}$-matrix and $q^4$, $-1$ and $q^{-2}$ for those in the representation $[2]$. This framing is said to be vertical, and the cabling procedure in it gives the same answers as in a direct calculations in terms of colored $\mathcal{R}$-matrices.

Another framing should be chosen in order to provide the topological invariance (this framing was used, e.g., in~\cite{MMM2}-\cite{IMMM3} to evaluate knot polynomials). In this case, the eigenvalues of $\mathcal{R}$-matrices are $q^{\varkappa_Q-4\varkappa_T}A^{-|T|}=q^{\varkappa_Q-4\varkappa_T-N|T|}$ ($T_1=T_2=T$). This framing gives the eigenvalues $\frac{q}{A}$ and $-\frac{1}{qA}$ for the fundamental representations and $\frac{q^2}{A^2}$, $\frac{1}{A^2q^{2}}$ and $\frac{1}{A^2q^4}$ for representation $[2]$.

The topological framing is considered only in the case $T_1=T_2$, which corresponds to the sense of this framing. The topological framing is required to get the same polynomials for all braids representing a given knot, and, generally speaking, having different number of crossings. Precisely this is achieved by special choice of the common factor (presented above). If two strands carry different representations, then they necessarily belong to different components of the link. The algebraic number of crossings between different components of the link ia a topological invariant, i.e., it cannot be changed by any smooth transformations~\cite{Pras}. Hence, the topological framing must not include any factors for the crossings between different components of the link. The case where the different components carry the same representation is not an exception: in the topological framing, only self-crossings of connection components contribute.

\section{Cabling for two-strand knots\label{sec:2str}}

In the case of two-strand knots, the colored HOMFLY are known in any representation because only diagonal $\mathcal{R}$-matrices are needed to calculate them. Thus, the $2$-strand case is useful as a consistency check for the formulae for projectors and for eigenvalues of the $\mathcal{R}$-matrices in different representations. The general formula is

\begin{equation}
\label{eq:2strgen}
H_{T_1T_2}^{T[2,n]}=Tr_{T_1\otimes T_2}\left(\left(\mathcal{R}_{T_1T_2}\mathcal{R}_{T_2T_1}\right)^{\frac{n}{2}}\right)=
\frac{1}{q^{n\varkappa_{T_1}+n\varkappa_{T_2}}}\sum\limits_{Q_i\vdash T_1\otimes T_2}\left(\pm q^{\varkappa_{Q_i}}\right)^nN^{Q_i}_{T_1T_2}S^*_{Q_i},
\end{equation}
where $q^{\varkappa_{Q}}$ are eigenvalues of the $2$-strand $\mathcal{R}$-matrix described in Sec.~\ref{dRm}, and the sum is over all irreducible representations $Q_i$ in the expansion of the tensor product $T_1\otimes T_2$, their multiplicities are $N^{Q_i}_{T_1T_2}$. The coefficient in front of the sum gives the answer corresponding to the vertical framing (see Sec.~\ref{framing}).

The cabling procedure gives the formula
\begin{equation}
H_{T_1T_2}^{T[2,n]}=Tr_{1^{\otimes(|T_1|+|T_2|)}}\left(P_{T_1T_2}\left(
\prod\limits_{i=1}^{T_2}\prod\limits_{j=1}^{T_1}R_{|T_1|+i-j}
\prod\limits_{i=1}^{T_1}\prod\limits_{j=1}^{T_2}R_{|T_2|+i-j}
\right)^{\frac{n}{2}}\right).
\end{equation}

It is necessary to separate two cases: the two-strand links and the two-strand knots. Study of the links allows to find the form of eigenvalues for different representations on different components of the links, what can not be done while studying the knots. On the other side, since the $2$-strand links always have even number of crossings the signs of the eigenvalues cannot be found using the cabling procedure for the links, for this $2$-strand knots are needed.

\subsection{Two-strand links}

In this case, there are two $\mathcal{R}$-matrices, namely $\mathcal{R}_{T_1T_2}$ and $\mathcal{R}_{T_2T_1}$,  which are different from the operator point of view since they act on different spaces:

\begin{equation}
\begin{array}{ll}
\mathcal{R}_{T_1T_2}: T_1\otimes T_2\rightarrow T_2\otimes T_1,&
\mathcal{R}_{T_2T_1}: T_2\otimes T_1\rightarrow T_1\otimes T_2.
\end{array}
\end{equation}
But their matrices coincide up to transposition: $\mathcal{R}_{T_1T_2}=\mathcal{R}_{T_2T_1}^{\dagger}$. The answer~(\ref{eq:2strgen}) in the case of links with $2n$ crossings can be rewritten as
\begin{equation}
H^{T[2,2n]}_{T_1T_2}=\left(q^{\varkappa_{T_1}+\varkappa_{T_2}}\right)^{-2n}\sum\limits_{Q_i\vdash T_1\otimes T_2}
N^{Q_i}_{T_1T_2}q^{2n\varkappa_{Q_i}}S^*_{Q_i}=
\sum\limits_{Q_i\vdash T_1\otimes T_2} H^{[2,2n]}_{T_1T_2| Q_i} S^*_{Q_i}.
\end{equation}
It is impossible to find the signs of eigenvalues using the cabling procedure for the $2$-strand links since all the powers of the eigenvalues in this formula are even.

This answer can be reproduced using the cabling procedure even at the level of the coefficients $H^{[2,2n]}_{T_1Q_2| Q_i}$ in the character expansion for the HOMFLY polynomial:\footnote{It suffices to verify this equality for $2\le 2n\le \min ({\mathrm{ rank}\ P_{T_1},\ \mathrm{ rank}\ P_{T_2}})$}
\begin{equation}
H^{T[2,2n]}_{T_1T_2|Q_i}=N^{Q_i}_{T_1T_2} q^{2n\varkappa_{Q_i}-2n\varkappa_{T_1}-2n\varkappa_{T_2}}.
\label{2link}
\end{equation}

For the answers obtained using the cabling procedures, both the level-rank duality ($\tilde{T}$ here denotes the representation with the transposed $T$ diagram)
\begin{equation}
H^{T[2,2n]}_{T_1T_2|Q_i}(q)=H^{T[2,2n]}_{\tilde{T}_1\tilde{T}_2|\tilde{Q}_i}\left(-q^{-1}\right),
\label{mirr}
\end{equation}
and the symmetry under permutation of the representations
\begin{equation}
H^{T[2,2n]}_{T_2T_1|Q_i}=H^{T[2,2n]}_{T_1T_2|Q_i}
\label{trans}
\end{equation}
hold. These properties are described, e.g., in~\cite{MMM2},\cite{IMMM1}. They hold since $N^{Q_i}_{T_1T_2}=N^{Q_i}_{T_2T_1}=N^{\tilde{Q}_i}_{\tilde{T}_1\tilde{T}_2}$ and $\varkappa_Q=-\varkappa_{\tilde Q}$. It can also be checked that the answers would be the same no matter which copies of irreducible representations $T_1$ and $T_2$ from the decomposition of $1^{\otimes|T_1|+|T_2|}$ have been chosen.

The absolute values of the eigenvalues of the colored $\mathcal{R}$-matrices are listed in Appendix~\ref{app:eig}. The multiplicities of different irreducible representations can be found in a very simple way from representation theory using the following relation for the characters:
\begin{equation}
S_{T_1} S_{T_2}=\sum\limits_{Q_i\vdash T_1\otimes T_2} N^{Q_i}_{T_1T_2} S_{Q_i}.
\end{equation}
For example,
\begin{equation}
S_{31}S_{31}=S_{62}+S_{611}+S_{53}+2S_{521}+S_{5111}+S_{44}+2S_{431}+S_{422}+S_{4211}+S_{332}+S_{3311}.
\end{equation}
Most of the representations that we studied had trivial multiplicities $0$ or $1$. Non-trivial multiplicities for the $2$-strand knots and links start to appear for size $3$ of representations. The list of all representations with non-trivial multiplicities for the representations $T_1$ and $T_2$ of sizes $3$ and $4$ is given in Appendix~\ref{app:eig}.

The decomposition of $1^{\otimes |T_1|+|T_2|}$ includes some irreducible representations that do not appear in the decomposition of the $T_1\otimes T_2$. Thus, for all such representations the cabling procedure should somehow give zero coefficients. Since we place the projectors described in Sec. ~\ref{sec:prpath} on each side of each $R$-matrix it means that all the representations that do not appear in the decomposition of $T_1\otimes 1^{\otimes |T_2|}$ automatically vanish. It appears that the remaining parts of $R$-matrices are still degenerate so that all representations that do not appear in $1^{\otimes |T_1|}\otimes T_2$ also disappear as they should.

\subsection{Two-strand knots}

As was already mentioned, the study of the cabling procedure for $2$-strand knots allows one to find not only the absolute values of the eigenvalues of the colored $\mathcal{R}$-matrix but also their signs. The drawback is that only the cases where $T_1=T_2$ (since the strand in the knot can carry only one representation) can be studied. Using cabling methods it can be checked that in all multiplicity free cases (we checked this up to $|T_1|=|T_2|\leq4$) the eigenvalues of the colored $2$-strand $\mathcal{R}$-matrix satisfy the following rule:

\noindent {\emph {Eigenvalue with the maximal power of $q$ comes with the sign plus, the eigenvalue with the next to the maximum power
 comes with the sign minus, the next one with the sign plus, etc.}}

\noindent This rule was used, e.g., in~\cite{IMMM3}. The problems with this rule arise when there appear multiplicities in decomposition of the product of two representation. The simplest examples of such multiplicities are:
\begin{equation}
[21]\times[21]=[42]+[411]+[33]+\mathbf{2\ [321]}+[3111]+[222]+[2211].
\end{equation}
with the eigenvalues evaluated by the cabling procedure
\begin{equation}
q^5,\ -q^3,\ -q^3,\ 1,\ -1,\ q^{-3},\ q^{-3},\ -q^{-5},
\end{equation}
and
\begin{equation}
\begin{array}{r}
[31]\times[31]\ =\ [62]+[611]+[53]+\mathbf{2\ [521]}+[5111]+[44]+
\\
+\mathbf{2\ [431]}+[422]+[4211]+[332]+[3311]
\end{array}
\end{equation}
with the eigenvalues evaluated by the cabling procedure
\begin{equation}
q^{10},\ -q^8,\ -q^6,\ q^3,\ -q^3,\ 1,\ q^4,\ 1,\ -1,\ q^{-2},\ -q^{-4},\ -q^{-4},\ q^{-6}.
\end{equation}
Thus, the general rule for the choice of signs in the $2$-strand $\mathcal{R}$-matrix is not clear.

The results for the $\mathcal{R}$-matrix eigenvalues evaluated by the cabling procedure are presented in Appendix~\ref{app:eig} for all representations up to the level $4$.

As a check, one can substitute the calculated eigenvalues into formula~(\ref{eq:2strgen}) and make sure that the following two properties of the colored HOMFLY are satisfied.
First, these HOMFLY should indeed be \emph{polynomials}. Second, polynomial of the 2-strand knot with only one crossing should be equal to polynomial of the unknot up to the framing coefficient. Indeed, such a knot is transformed into the unknot by the first Reidemeister move.

Thus, the eigenvalues should satisfy the following equations\footnote{The first equation is valid only in the topological framing (see Sec.~\ref{framing}).}:
\begin{equation}
\begin{array}{lll}
S^*_R=\sum \lambda_QS^*_Q, &\ \ & \sum (\lambda_Q)^{2n+1}S^*_Q\ \vdots\ S^*_R.
\end{array}
\end{equation}
These requirements seem to fix all the signs, but it is not clear how to find the signs from these equations.

\section{Cabling for three- and four-strand knots}

\subsection{Three-strand knots}

Results for the $3$-strand knots in representations $[2]$ and $[11]$ were obtained in~\cite{IMMM1}. These results can be evaluated also using the cabling methods. To evaluate these answers one should replace each colored $\mathcal{R}$-matrix in the $3$-stand braid with a following product of the fundamental ones in the $6$-strand braid:
\begin{equation}
\begin{array}{lcr}
\mathcal{R}_{(1^2\otimes 1^2)\otimes 1^2}=R_2R_1R_3R_2,
& \ &
\mathcal{R}_{1^2\otimes (1^2\otimes 1^2)}=R_4R_3R_5R_4.
\end{array}
\label{eq:cb3}\end{equation}
Then the projectors onto the corresponding representations should be inserted. As was already explained in Sec.~\ref{sec:proj} any number of projectors but not less than one per component can be inserted and any form of those described in Sec.~\ref{sec:proj} can be used. The most convenient way to simplify the calculations is to use the path description of the projectors and to insert two of them for each $\mathcal{R}$-matrix, surrounding each of combinations~(\ref{eq:cb3}) with them.

In the case of level $3$ representations, i.e., $[3]$, $[21]$, and $[111]$, the $3$-strand braid should be replaced with the fundamental $9$-strand braid with the following replacement for the $\mathcal{R}$-matrices:
\begin{equation}
\begin{array}{lcr}
\mathcal{R}_{(1^3\otimes 1^3)\otimes 1^3}=R_3R_2R_1R_4R_3R_2R_5R_4R_3,
& \ &
\mathcal{R}_{1^3\otimes (1^3\otimes 1^3)}=R_6R_5R_4R_7R_6R_5R_8R_7R_6.
\end{array}
\end{equation}
All the needed projectors are described in Sec.~\ref{sec:proj}. The HOMFLY polynomials for  $3$-strand knots in representations $[3]$ and $[111]$ were provided in~\cite{IMMM3} thus in the present paper we provide only the results for representation $[21]$, see Appendix~\ref{app:3str}.

This procedure can be repeated for higher representations but a higher computational power is needed.

\subsection{Four-strand knots}

For the $4$-strand knots the scheme is the same as for the $3$-strand knots. For representations $[2]$ and $[11]$ the colored $\mathcal{R}$-matrices should be replaced with the following fundamental ones:
\begin{equation}
\begin{array}{lcccr}
\mathcal{R}_{(1^2\otimes 1^2)\otimes 1^2 \otimes 1^2}=R_2R_1R_3R_2,
& \ &
\mathcal{R}_{1^2\otimes (1^2\otimes 1^2) \otimes 1^2}=R_4R_3R_5R_4,
& \ &
\mathcal{R}_{1^2 \otimes 1^2 \otimes (1^2\otimes 1^2)}=R_6R_5R_7R_6.
\end{array}
\end{equation}
In addition, the corresponding projectors in Sec.~\ref{sec:proj} should be inserted.
The HOMFLY polynomials of the $4$-strand knots for the level-two representations are provided in Appendix~\ref{app:4str}. This procedure can be repeated for higher representations.

\section{Multicolored three-strand links \label{sec:links}}

The colored links are much more involved than the colored knots since they have several components and different representations can be placed on each of these components. The links with all the components of the same color are evaluated in the same way as the colored knots thus we will concentrate on the multicolored links in this section. Another property, which even fundamental knots and links possess, is an orientation. There is only one possible orientation for the knots (in fact two but they pass into each other under a symmetry transformation), but in the case of links different components can have different relative orientation. The orientation is automatically accounted for when the braid representation is used since differently oriented links have different braid representations. Also it is important to note that the HOMFLY polynomials for the links are in fact not polynomials, there is always some denominator dependent on the studied representations.

In the present section, we provide the simplest and mostly known examples to illustrate the approach that is used.
\subsection{Colored $\mathcal{R}$-matrices approach}

The main problem with the calculations following the Reshetikhin-Turaev approach in this case is the following one. If all strands in the braid are of the same color, then all the $\mathcal{R}$-matrices are labeled by the ordinal numbers of the strands crossing, namely, $\mathcal{R}_k$ corresponds to the crossing of $k$-th and $k+1$-th strands of the braid. Unlike that, the $\mathcal{R}$-matrices for the multicolored braid are labeled \textit{both} by the strands which cross, and by their colors (see the concrete examples in the below). Also one should take into account that $\mathcal{R}$-matrices includes the permutation operators and thus change the placement of representation on the strands, i.e:

\begin{equation}
\begin{array}{l}
\mathcal{R}_{T_1\ldots (T_iT_{i+1})\ldots T_{m-1}}:\ T_1\otimes\ldots (T_i\otimes T_{i+1})\ldots T_{m-1}\ \rightarrow\ T_1\otimes\ldots (T_{i+1}\otimes T_i)\ldots T_{m-1},
\\
\\
\mathcal{U}_{T_1\ldots (T_{i-1}T_iT_{i+1})\ldots T_{m-1}}:\ T_1\otimes\ldots \Bigl((T_{i-1}\otimes T_i)\otimes T_{i+1}\Bigr)\ldots T_{m-1}\ \rightarrow\ T_1\otimes\ldots \Bigl(T_{i-1}\otimes (T_i\otimes T_{i+1})\Bigr)\ldots T_{m-1}.
\end{array}\label{Rmulc}
\end{equation}
With help of the rule~(\ref{Rmulc}), one can find all possible sequences of $\mathcal{R}$-matrices that appear in a given multicolored braid.

The eigenvalues of the $\mathcal{R}$-matrices are known: they are the same as for the diagonal $\mathcal{R}$-matrix described in Sec.~\ref{dRm} and are equal to $\mathcal{R}_{T_1T_2|Q}=\pm q^{\varkappa_Q-\varkappa_{T_1}-\varkappa_{T_2}}$ in the vertical framing. The signs are essential only for crossings of stands in the same representation strands since the number of the crossings between every pair of strands in different representations is always even. In the considered examples, these signs simply alternate in each $\mathcal{R}$-matrix, i.e., the eigenvalue with the highest power of $q$ has the sign plus, the next one the size minus etc. The non-diagonal matrices can be constructed from the diagonal ones with the help of Racah matrices~\cite{MMM2}, which can be evaluated using the methods of representation theory~\cite{Vil}.

We consider three particular examples of the multicolored links, they are $[1]\otimes[1]\otimes[2]$, $[1]\otimes[2]\otimes[2]$, and $[1]\otimes[2]\otimes[3]$. In all these cases, representation theory of the $SU_q(2)$ group is enough to find the Racah matrices.

\subsection{The case $[1]\otimes[1]\otimes[2]$\label{sec:112}}

\paragraph{Calculations using the colored $\mathbf{\mathcal{R}}$-matrices.}
The tensor product of the representations is decomposed as
\begin{equation}
[1]\otimes [1]\otimes [2] = [4]+2\ [31]+[22]+[211].
\end{equation}
For the singlets there are no mixing matrices and the corresponding components of the $\mathcal{R}$-matrices are entirely described by their eigenvalues:
\begin{equation}
\begin{array}{c}
\mathcal{R}_{(1\otimes2)\otimes1|4}=\mathcal{R}_{(2\otimes1)\otimes1|4}=\mathcal{R}_{1\otimes(1\otimes2)|4}= \mathcal{R}_{1\otimes(2\otimes1)|4}=q^{\varkappa_3-\varkappa_2}=q^2,
\\ \\
\mathcal{R}_{(1\otimes2)\otimes1|22}=\mathcal{R}_{(2\otimes1)\otimes1|22}=\mathcal{R}_{1(12)|22}= \mathcal{R}_{1\otimes(2\otimes1)|22}=q^{\varkappa_{21}-\varkappa_2}=-q^{-1},
\\ \\
\mathcal{R}_{(1\otimes2)\otimes1|211}=\mathcal{R}_{(2\otimes1)\otimes1|211}=\mathcal{R}_{1\otimes(1\otimes2)|211}= \mathcal{R}_{1\otimes(2\otimes1)|211}=q^{\varkappa_{21}-\varkappa_2}=-q^{-1},
\\ \\
\mathcal{R}_{(1\otimes1)\otimes2|4}=\mathcal{R}_{2\otimes(1\otimes1)|4}=q^{\varkappa_2}=q,
\\ \\
\mathcal{R}_{(1\otimes1)\otimes2|22}=\mathcal{R}_{2\otimes(1\otimes1)|22}=q^{\varkappa_2}=q,
\\ \\
\mathcal{R}_{(1\otimes1)\otimes2|211}=\mathcal{R}_{2\otimes(1\otimes1)|211}=-q^{\varkappa_{11}}=-q^{-1}.
\end{array}
\end{equation}
For the doublets both the diagonal and the non-diagonal $\mathcal{R}$-matrices are needed, the latter being calculated with help of the mixing matrices. The doublet components of the diagonal $\mathcal{R}$-matrices are
\setlength{\arraycolsep}{3pt}
\begin{equation}
\begin{array}{l}
\mathcal{R}_{(1\otimes1)\otimes2|31}=\left(
\begin{array}{cc}
q^{\varkappa_2} &
\\
&-q^{\varkappa_{11}}
\end{array}\right)
=\left(
\begin{array}{cc}
q &
\\
&-q^{-1}
\end{array}
\right),
\\ \\
\mathcal{R}_{(1\otimes2)\otimes1|31}= \mathcal{R}^{\dagger}_{(2\otimes1)\otimes1|31}=\left(
\begin{array}{cc}
q^{\varkappa_3-\varkappa_2} &
\\
&-q^{\varkappa_{21}-\varkappa_2}
\end{array}\right)
=\left(
\begin{array}{cc}
q^2&
\\
&-q^{-1}
\end{array}
\right).
\end{array}
\end{equation}
\setlength{\arraycolsep}{6pt}
The mixing matrices evaluated using the representation theory for the group $SU_q(2)$ are\footnote{See, e.g.,~\cite{MMM2} for the detailed derivation of the Racah matrices from representation theory.}:
\begin{equation}
\begin{array}{l}
{\mathcal{U}}_{1\otimes2\otimes1|31}=\left(
\begin{array}{cc}
\frac{1}{[3]_q} & \frac{\sqrt{[2]_q[4]_q}}{[3]_q}
\\
\frac{\sqrt{[2]_q[4]_q}}{[3]_q} & -\frac{1}{[3]_q}
\end{array}
\right),\ \
{\mathcal{U}}_{1\otimes1\otimes2|31}={\mathcal{U}}_{2\otimes1\otimes1|31}=\left(
\begin{array}{cc}
\frac{1}{\sqrt{[3]_q}} & \sqrt{\frac{[4]_q}{[2]_q[3]_q}}
\\
\sqrt{\frac{[4]_q}{[2]_q[3]_q}} & -\frac{1}{\sqrt{[3]_q}}
\end{array}
\right).
\end{array}\end{equation}
This gives the following non-diagonal $\mathcal{R}$-matrices:
\begin{equation}
\begin{array}{l}
\mathcal{R}_{2\otimes(1\otimes1)|31}={\mathcal{U}}_{2\otimes1\otimes1|31}\left(
\setlength{\arraycolsep}{3pt}
\begin{array}{cc}
q&
\\
&-q^{-1}
\end{array}
\setlength{\arraycolsep}{6pt}
\right)\mathcal{U}^{\dagger}_{2\otimes1\otimes1|31}=\left(
\begin{array}{cc}
\frac{1}{q^3[3]_q} & \frac{\sqrt{[2]_q[4]_q}}{[3]_q}
\\
\frac{\sqrt{[2]_q[4]_q}}{[3]_q} & -\frac{q^3}{[3]_q}
\end{array}
\right),
\\
\mathcal{R}_{1\otimes(1\otimes2)|31}=\mathcal{R}_{1\otimes(2\otimes1)|31}^{\dagger}= \mathcal{U}_{1\otimes1\otimes2|31}\left(
\setlength{\arraycolsep}{3pt}
\begin{array}{cc}
q^2&
\\
&-q^{-1}
\end{array}
\setlength{\arraycolsep}{6pt}
\right){\mathcal{U}}^{\dagger}_{1\otimes2\otimes1|31}=\left(
\begin{array}{cc}
-\frac{1}{q\sqrt{[3]_q}} & \frac{q\sqrt{[4]_q}}{\sqrt{[2]_q[3]_q}}
\\
\frac{\sqrt{[4]_q}}{\sqrt{[2]_q[3]_q}} & \frac{q^3}{\sqrt{[3]_q}}
\end{array}
\right).
\end{array}
\end{equation}

To give an illustrative example of how to deal with the multicolored links, we compute the HOMFLY polynomials for several simplest $2$-component $3$-strand links in the representation $[1]\otimes[1]\otimes[2]$. If one studies such links then the $2$-strand component should carry the fundamental representation and the $1$-strand component should carry the representation $[2]$. The simplest link of this type is the $3$-strand representation of the pair of spit unknots, which has the braid word $\sigma_1$. Then the HOMFLY in the vertical framing is equal to
\begin{equation}
H^{\bigcirc^2}_{1\otimes2}=\mathrm{Tr}_{1\otimes1\otimes2}\,\mathcal{R}_{(1\otimes1)\otimes2}=qS^*_4+(q-q^{-1})S^*_{31}+qS^*_{22}-q^{-1}S^*_{211}=S^*_1S^*_2.
\end{equation}

Next in simplicity example is the $3$-strand version of the torus link $T[2,4]$ with the braid word $\sigma_{1}\sigma_{2}\sigma_{1}\sigma_{2}\sigma_{1}$ or $\sigma_{2}\sigma_{1}\sigma_{1}\sigma_{2}\sigma_{1}$
The corresponding HOMFLY is
\begin{equation}
H^{T[2,4]}_{2\otimes1\otimes1}=\mathrm{Tr}_{2\otimes1\otimes1}\, \mathcal{R}_{(2\otimes1)\otimes1}\mathcal{R}_{1\otimes(2\otimes1)}\mathcal{R}_{(1\otimes1)\otimes2} \mathcal{R}_{1\otimes(1\otimes2)}\mathcal{R}_{(1\otimes2)\otimes1} =q^9S^*_4+(q-q^3)S^*_{31}+q^3S^*_{22}-q^{-5}S^*_{311}.
\end{equation}
This answer coincides with the one for the $2$-strand version of the same link (if the topological framing is used, which differs from the vertical one by factor of $A^{-1}$):
\begin{equation}
A^{-1}H^{T[2,4]-3strand}_{2\otimes1\otimes1}=H^{T[2,4]-2strand}_{1\otimes2} =q^{4\varkappa_3-8\varkappa_2}S^*_3+q^{4\varkappa_{21}-8\varkappa_2}S^*_{21}=q^8S^*_3+q^{-4}S^*_{21}.
\end{equation}

The simplest non-torus $3$-strand $2$-component link is $5_1^2$ or the Whitehead link\footnote{The Rolfsen notation of a link $c^n_k$ generalizes the Rolfsen notation $c_k$ of a knot. In this notation, $c$ is the crossing number, i.e., is the minimal number of the intersections in the planar diagram of the knot or link and $n$ is the number of components.} with the braid word $\sigma^{-1}_1\sigma_{2}\sigma^{-1}_{1}\sigma_{2}\sigma^{-1}_{1}$. It differs from the torus link $T[2,4]$ by inverting all $\mathcal{R}$-matrices acting on the first pair of strands. The HOMFLY polynomial is given by the following formula:
\begin{equation}
\begin{array}{l}
H^{5_1^2}_{2\otimes 1\otimes 1}=\mathrm{Tr}_{2\otimes1\otimes1}\, \mathcal{R}^{-1}_{(2\otimes1)\otimes1}\mathcal{R}_{1\otimes(2\otimes1)}\mathcal{R}^{-1}_{(1\otimes1)\otimes2} \mathcal{R}_{1\otimes(1\otimes2)}\mathcal{R}_{(1\otimes2)\otimes1}=
\\ \\
=q^{-1}S^*_4+(-q^{7}+q^{5}+q^{3}-2\,q+2\,q^{-1}-q^{-3}-q^{-5}+q^{-7})S^*_{31}+q^{-1}S^*_{22}-qS^*_{221}=
\\ \\
=\cfrac{S^*_2}{q^2-1}\Bigl((-q^{3}+q^{-1}-q^{-3})A^{-2}+
\\ \\
+(q^{7}-q^{5}-q^{3}+3\,q-q^{-1}-q^{-3}+q^{-5})+(-q^{5}+q^{3}+q-q^{-1})A^2\Bigr).
\end{array}
\end{equation}

\paragraph{Calculations using the cabling procedure.} To perform the calculations of the same polynomials in the cabling approach the colored $\mathcal{R}$-matrices should be substituted by the corresponding products of the fundamental ones, in this case in the $4$-strand braid (since $1+1+2=4$):
\begin{equation}
\begin{array}{rcl}
\mathcal{R}_{(1\otimes1)\otimes2} & \rightarrow & R_1,
\\
\mathcal{R}_{(1\otimes2)\otimes1}=\mathcal{R}_{(2\otimes1)\otimes1}^{\dagger} & \rightarrow & R_2R_1,
\\
\mathcal{R}_{1\otimes(1\otimes2)}=\mathcal{R}_{1\otimes(2\otimes1)}^{\dagger} & \rightarrow & R_3R_2,
\\
\mathcal{R}_{2\otimes(1\otimes1)} & \rightarrow & R_3.
\end{array}
\label{R112}
\end{equation}
with the corresponding substitution for the inverse crossings
\begin{equation}
\begin{array}{rcl}
\mathcal{R}^{-1}_{(1\otimes1)\otimes2} & \rightarrow& R^{-1}_1,
\\
\mathcal{R}^{-1}_{(1\otimes2)\otimes1}=\mathcal{R}_{(2\otimes1)\otimes1}^{\dagger} & \rightarrow & R^{-1}_2R^{-1}_1,
\\
\mathcal{R}^{-1}_{1\otimes(1\otimes2)}=\mathcal{R}_{1\otimes(2\otimes1)}^{\dagger} & \rightarrow & R^{-1}_3R^{-1}_2,
\\
\mathcal{R}^{-1}_{2\otimes(1\otimes1)} & \rightarrow & R^{-1}_3.
\end{array}
\end{equation}
In addition, the projector should be inserted. Depending on the place where the projector is put and on the used placement of the representations, one can use one of the three projectors\footnote{As was described in Sec.~\ref{sec:proj} different placements of the projectors and different numbers of projectors can be used. In calculations here, we use the projector placed at the beginning of the braid.}
\begin{equation}
\begin{array}{lll}
P_{2\otimes1\otimes1}=\cfrac{1+qR_1}{1+q^2},&
P_{1\otimes2\otimes1}=\cfrac{1+qR_2}{1+q^2},&
P_{1\otimes1\otimes2}=\cfrac{1+qR_3}{1+q^2}.
\end{array}
\label{P112}
\end{equation}
The answers for particular links can be obtained by applying relations~(\ref{R112}) and~(\ref{P112}) to the braid word and then evaluating the HOMFLY for the resulting fundamental braids. The answers obtained in this way coincide with the ones obtained using the colored $\mathcal{R}$-matrices.

In the case of split union of two unknots with the braid word $\sigma_1$ (the $3$-strand representation is implied), the HOMFLY obtained using the cabling procedure is equal to
\begin{equation}
H^{\bigcirc^2}_{1\otimes2}=\mathrm{Tr}_{1^4}\ P_{1\otimes1\otimes2}R_1=qS^*_4+(q-q^{-1})S^*_{31}+qS^*_{22}-q^{-1}S^*_{211}=A\ S^*_1S^*_2.
\end{equation}
The torus link $T[2,4]$ in the 3-strand representation has the HOMFLY polynomial
\begin{equation}
\begin{array}{l}
H^{T[2,4]-3strand}_{2\otimes1\otimes 1}=\mathrm{Tr}_{1^4}\, P_{2\otimes1\otimes 1}R_1R_2\cdot R_2R_3\cdot R_1\cdot R_3R_2\cdot R_2R_1 =
\\ \\
=q^9S^*_4+(q-q^3)S^*_{31}+q^3S^*_{22}-q^{-5}S^*_{311}=AH^{T[2,4]-2strand}_{1\otimes2}.
\end{array}
\end{equation}
Finally, for the Whitehead link, the answer obtained by the cabling procedure is
\begin{equation}
\begin{array}{l}
H^{5^2_1}_{2\otimes1\otimes 1}=\mathrm{Tr}_{2\otimes1\otimes 1}\, P_{2\otimes1\otimes 1}R^{-1}_1R^{-1}_2\cdot R_2R_3\cdot R^{-1}_1\cdot R_3R_2\cdot R_2R_1=
\\ \\
=q^{-1}S^*_4+(-q^{7}+q^{5}+q^{3}-2\,q+2\,q^{-1}-q^{-3}-q^{-5}+q^{-7})S^*_{31}+q^{-1}S^*_{22}-qS^*_{221}.
\end{array}
\end{equation}

\subsection{The case $[2]\otimes[2]\otimes[1]$}

\paragraph{Calculations using the colored $\mathbf{\mathcal{R}}$-matrices.}
The tensor product of representations in this case decomposes as
\begin{equation}
[2]\otimes [2]\otimes [1]=[5]+2\ [41]+2\ [32]+[311]+[221].
\end{equation}
The singlet components of the colored $\mathcal{R}$-matrices are
\begin{equation}
\begin{array}{c}
\mathcal{R}_{(1\otimes2)\otimes2|5}=\mathcal{R}_{(2\otimes1)\otimes2|5}=\mathcal{R}_{2\otimes(1\otimes2)|5}= \mathcal{R}_{2\otimes(2\otimes1)|5}=q^{\varkappa_3-\varkappa_2}=q^2,
\\ \\
\mathcal{R}_{(1\otimes2)\otimes2|311}=\mathcal{R}_{(2\otimes1)\otimes2|311}= \mathcal{R}_{2\otimes(1\otimes2)|311}=\mathcal{R}_{2\otimes(2\otimes1)|311}= q^{\varkappa_{21}-\varkappa_2}=-q^{-1},
\\ \\
\mathcal{R}_{(1\otimes2)\otimes2|221}=\mathcal{R}_{(2\otimes1)\otimes2|221}=\mathcal{R}_{2\otimes(1\otimes2)|221}= \mathcal{R}_{2\otimes(2\otimes1)|221}=q^{\varkappa_{21}-\varkappa_2}=-q^{-1},
\\ \\
\mathcal{R}_{(2\otimes2)\otimes1|5}=\mathcal{R}_{1\otimes(2\otimes2)|5}= q^{\varkappa_4-2\varkappa_2}=q^4,
\\ \\
\mathcal{R}_{(2\otimes2)\otimes1|311}=\mathcal{R}_{1\otimes(2\otimes2)|311}= -q^{\varkappa_{31}-2\varkappa_2}=-1,
\\ \\
\mathcal{R}_{(2\otimes2)\otimes1|221}=\mathcal{R}_{1\otimes(2\otimes2)|221}= q^{\varkappa_{22}-2\varkappa_2}=q^{-2}.
\end{array}
\end{equation}
The doublet components of the diagonal colored $\mathcal{R}$-matrices in this case are:
\setlength{\arraycolsep}{2pt}
\begin{equation}
\begin{array}{l}
\mathcal{R}_{(1\otimes2)\otimes2|41}=\mathcal{R}_{(1\otimes2)\otimes2|32}= \mathcal{R}^{\dagger}_{(2\otimes1)\otimes2|41}=\mathcal{R}^{\dagger}_{(2\otimes1)\otimes2|32}=\left(
\begin{array}{cc}
q^{\varkappa_3-\varkappa_2} &
\\
&-q^{\varkappa_{21}-\varkappa_2}
\end{array}
\right)=\left(
\begin{array}{cc}
q^2 &
\\
&-q^{-1}
\end{array}
\right),
\\ \\
\mathcal{R}_{(2\otimes2)\otimes1|41}=\left(
\begin{array}{cc}
q^{\varkappa_4-2\varkappa_2} &
\\
&-q^{\varkappa_{31}-2\varkappa_2}
\end{array}
\right)=\left(
\begin{array}{cc}
q^4&
\\
&-1
\end{array}
\right),
\\ \\
\mathcal{R}_{(2\otimes2)\otimes1|32}=\left(
\begin{array}{cc}
-q^{\varkappa_{31}-2\varkappa_2} &
\\
&q^{\varkappa_{22}-2\varkappa_2}
\end{array}
\right)=\left(
\begin{array}{cc}
-1 &
\\
&q^{-2}
\end{array}
\right).
\end{array}
\end{equation}
\setlength{\arraycolsep}{6pt}
The corresponding Racah matrices are
\begin{equation}
\begin{array}{l}
\mathcal{U}_{2\otimes2\otimes1|41}=\mathcal{U}_{1\otimes2\otimes2|41}=\left(
\begin{array}{cc}
\sqrt{\frac{[2]_q}{[3]_q[4]_q}} & \sqrt{\frac{[2]_q[5]_q}{[3]_q[4]_q}}
\\
\sqrt{\frac{[2]_q[5]_q}{[3]_q[4]_q}} & -\sqrt{\frac{[2]_q}{[3]_q[4]_q}}
\end{array}
\right),
\\ \\
\mathcal{U}_{2\otimes2\otimes1|32}=\mathcal{U}_{1\otimes2\otimes2|32}=\left(
\begin{array}{cc}
\frac{1}{\sqrt{[3]_q}} & \sqrt{\frac{[4]_q}{[2]_q[3]_q}}
\\
\sqrt{\frac{[4]_q}{[2]_q[3]_q}} & -\frac{1}{\sqrt{[3]_q}}
\end{array}
\right),
\\ \\
\mathcal{U}_{2\otimes1\otimes2|41}=\left(
\begin{array}{cc}
\frac{[2]_q}{[3]_q} & \frac{\sqrt{[5]_q}}{[3]_q}
\\
\frac{\sqrt{[5]_q}}{[3]_q} & -\frac{[2]_q}{[3]_q}
\end{array}
\right),
\ \
\mathcal{U}_{2\otimes1\otimes2|32}=\left(
\begin{array}{cc}
\frac{1}{[3]_q} & \frac{\sqrt{[2]_q[4]_q}}{[3]_q}
\\
\frac{\sqrt{[2]_q[4]_q}}{[3]_q} & -\frac{1}{[3]_q}
\end{array}
\right).
\end{array}
\end{equation}
This gives the following doublet components of the non-diagonal colored $\mathcal{R}$-matrices:
\begin{equation}
\begin{array}{l}
\mathcal{R}_{2\otimes(1\otimes2)|41}=\mathcal{R}_{2\otimes(1\otimes2)|41}^{\dagger}= \mathcal{U}_{2\otimes1\otimes2|41}\left(
\setlength{\arraycolsep}{3pt}
\begin{array}{cc}
q^2 &
\\
&-q^{-1}
\end{array}
\setlength{\arraycolsep}{6pt}
\right)\mathcal{U}^{\dagger}_{2\otimes2\otimes1|41}=\left(
\begin{array}{cc}
-\frac{\sqrt{[2]_q}}{q^3\sqrt{[3]_q[4]_q}} & \frac{q\sqrt{[2]_q[5]_q}}{\sqrt{[3]_q[4]_q}}
\\
\frac{\sqrt{[2]_q[5]_q}}{\sqrt{[3]_q[4]_q}} & \frac{q^4\sqrt{[2]_q}}{\sqrt{[3]_q[4]_q}}
\end{array}
\right),
\\ \\
\mathcal{R}_{2\otimes(1\otimes2)|32}=\mathcal{R}^{\dagger}_{2\otimes(1\otimes2)|32}= \mathcal{U}_{2\otimes1\otimes2|32}\left(
\setlength{\arraycolsep}{3pt}
\begin{array}{cc}
q^2 &
\\
&-q^{-1}
\end{array}
\setlength{\arraycolsep}{6pt}
\right)\mathcal{U}^{\dagger}_{2\otimes2\otimes1|32}=\left(
\begin{array}{cc}
-\frac{1}{q^2\sqrt{[3]_q}} & \frac{\sqrt{[4]_q}}{\sqrt{[2]_q[3]_q}}
\\
\frac{q\sqrt{[4]_q}}{\sqrt{[2]_q[3]_q}} & \frac{q^3}{\sqrt{[3]_q}}
\end{array}
\right),
\\ \\
\mathcal{R}_{1\otimes(2\otimes2)|41}=\mathcal{U}_{2\otimes2\otimes1|41}\left(
\begin{array}{cc}
q^4 &
\\
&-1
\end{array}
\right)\mathcal{U}^{\dagger}_{2\otimes2\otimes1|41}=\left(
\begin{array}{cc}
-\frac{[2]_q}{q[3]_q} & \frac{q^2\sqrt{[5]_q}}{[3]_q}
\\
\frac{q^2\sqrt{[5]_q}}{[3]_q} & \frac{q^5[2]_q}{[3]_q}
\end{array}
\right),
\\ \\
\mathcal{R}_{1\otimes(2\otimes2)|32}={\mathcal{U}}_{2\otimes2\otimes1|32}\left(
\begin{array}{cc}
-1 &
\\
&q^{-2}
\end{array}
\right)\mathcal{U}^{\dagger}_{2\otimes2\otimes1|32}=\left(
\begin{array}{cc}
\frac{1}{q^4[3]_q} & -\frac{\sqrt{[2]_q[4]_q}}{q[3]_q}
\\
-\frac{\sqrt{[2]_q[4]_q}}{q[3]_q} & -\frac{q^2}{[3]_q}
\end{array}
\right).\end{array}
\end{equation}
As examples, we use the same links as in Sec.~\ref{sec:112} with representation $[2]$ on the $2$-strand component and the fundamental one on the $1$-strand component.

For the split union of two unknots the colored $\mathcal{R}$-matrices approach gives:
\begin{equation}
H^{\bigcirc^2}_{1\otimes2}=q^{4}S^*_5+(q^4-1)S^*_{41}+(q^{-2}-1)S^*_{32}-S^*_{311}+q^{-2}S^*_{221}=A^2q^2S^*_1S^*_2.
\end{equation}
The factor $A^2q^2$ relates the vertical and the topological framings (see Sec.~\ref{framing}). For the 3-strand version of the torus link $T[2,4]$ the answer is
\begin{equation}
\begin{array}{l}
\mathcal{H}^{T[2,4]-3strand}_{1\otimes2\otimes2}=\mathrm{Tr}_{1\otimes2\otimes2}\, \mathcal{R}_{(1\otimes2)\otimes2}\mathcal{R}_{2\otimes(1\otimes2)}\mathcal{R}_{(2\otimes2)\otimes1} \mathcal{R}_{2\otimes(2\otimes1)}\mathcal{R}^{\dagger}_{(2\otimes1)\otimes2}
\\ \\
= q^{12}S^*_5+(q^2-q^6)S^*_{41}+(q^2-1)S^*_{32}-q^{-4}S^*_{311}+q^{-6}S^*_{221}=
\\ \\
=A^2q^2H^{T[2,4]-2strand}_{2\otimes1}= A^2q^2\left(q^{4\varkappa_3-8\varkappa_2}S^*_3+q^{4\varkappa_{21}-8\varkappa_2}S^*_{21}\right)= A^2q^{10}S^*_3+A^2q^{-2}S^*_{21}.
\end{array}
\end{equation}
The Whitehead link gives the following HOMFLY polynomial:
\begin{equation}
\begin{array}{l}
H^{5_1^2}_{1\otimes 2\otimes 2}=\mathrm{Tr}_{1\otimes2\otimes2}\, \mathcal{R}_{(1\otimes2)\otimes2}^{-1}\mathcal{R}_{2\otimes(1\otimes2)} \mathcal{R}_{(2\otimes2)\otimes1}^{-1}\mathcal{R}_{2\otimes(2\otimes1)}\mathcal{R}^{\dagger}_{(2\otimes1)\otimes2}=
\\ \\
=q^{-4}S^*_5+(-q^{6}+q^{4}+q^{2}-2+2\,q^{-4}-q^{-6}-q^{-8}+q^{-10})S^*_{41}+\\ \\+(-q^{6}+q^{4}+q^{2}-2+2\,q^{-4}-q^{-6}-q^{-8}+q^{-10})S^*_{32}-S^*_{311}+q^2S^*_{221}=A^{-1}q^{-1}H^{5_1^2}_{2\otimes 1\otimes 1},
\end{array}
\end{equation}
i.e., the polynomials $H^{5_1^2}_{2\otimes 1\otimes 1}$ and $H^{5_1^2}_{1\otimes 2\otimes 2}$ are equal up to a factor. This factor appears due to the difference between the vertical framing, which was used in these calculations, and the topological framing, which should give the same answers for these two polynomials. The relation $H^{5_1^2}_{2\otimes 1\otimes 1}=H^{5_1^2}_{1\otimes 2\otimes 2}$ in the topological framing should exist because using the Reidemeister moves the components of the Whitehead link can be exchanged.

One more consistency check is that the answer possess the factorization property~\cite{DMMSS},\cite{chinsp1},\cite{chinsp2} (see Sec.~\ref{sec:sp} for the details). When untied, both components of the Whitehead link are isomorphic to the unknots. This is consistent with the fact that
\begin{equation}
\cfrac{H^{5_1^2}_{1\otimes 2\otimes 2}}{S^*_1S^*_2}\Biggr|_{q\rightarrow 1}=A^{-2}.
\end{equation}

\paragraph{Calculations using the cabling procedure.}
The cabling procedure is analogous to the one in Sec.~\ref{sec:112}. The colored $\mathcal{R}$-matrices are substituted in the following way:
\begin{equation}
\begin{array}{rcl}
\mathcal{R}_{(2\otimes2)\otimes1} & \longrightarrow & R_2R_1R_3R_2,
\\ \\
\mathcal{R}_{(1\otimes2)\otimes2}=\mathcal{R}_{(2\otimes1)\otimes2}^{\dagger} & \longrightarrow & R_1R_2,
\\ \\
\mathcal{R}_{2\otimes(1\otimes2)}=\mathcal{R}_{2\otimes(2\otimes1)}^{\dagger} & \longrightarrow & R_3R_4,
\\ \\
\mathcal{R}_{1\otimes(2\otimes2)} & \longrightarrow & R_3R_2R_4R_3.
\end{array}
\label{R122}
\end{equation}
For the inverse crossings the same products of the inverted $R$-matrices are substituted. The relevant projectors are
\begin{equation}
\begin{array}{lll}
\mathcal{P}_{1\otimes 2\otimes 2}=\cfrac{1+qR_2}{1+q^2}\,\cdot\cfrac{1+qR_4}{1+q^2},
&
\mathcal{P}_{2\otimes 1\otimes 2}=\cfrac{1+qR_1}{1+q^2}\,\cdot\cfrac{1+qR_4}{1+q^2},
&
\mathcal{P}_{2\otimes 2\otimes 1}=\cfrac{1+qR_1}{1+q^2}\,\cdot\cfrac{1+qR_3}{1+q^2}.
\end{array}
\label{P122}
\end{equation}
The HOMFLY polynomial for the split union of two unknots is
\begin{equation}
H^{\bigcirc^2}_{1\otimes2}=\mathrm{Tr}_{1^5}\ P_{2\otimes2\otimes1}R_1=qS^*_4+(q-q^{-1})S^*_{31}+qS^*_{22}-q^{-1}S^*_{211}=AS^*_1S^*_2.
\end{equation}
For the $3$-strand version of the torus link $T^{2,4}$ one has
\begin{equation}
\begin{array}{l}
\mathcal{H}^{T[2,4]-3strand}_{1\otimes2\otimes 2}=\mathrm{Tr}_{1^5}\, P_{1\otimes2\otimes 2}
R_1R_2 \cdot R_3R_4 \cdot R_2R_1R_3R_2 \cdot R_4R_3 \cdot R_2R_1=
\\ \\
=q^{12}S^*_5+(q^2-q^6)S^*_{41}+(q^2-1)S^*_{32}-q^{-4}S^*_{311}+q^{-6}S^*_{221}.
\end{array}
\end{equation}
For the non-torus Whitehead link the HOMFLY polynomial is
\begin{equation}
\begin{array}{l}
\mathcal{H}^{5^2_1}_{1\otimes2\otimes 2}=\mathrm{Tr}_{1\otimes2\otimes2}\, P_{1\otimes2\otimes 2}
R^{-1}_1R^{-1}_2 \cdot R_3R_4 \cdot R^{-1}_2R^{-1}_1R^{-1}_3R^{-1}_2 \cdot R_4R_3 \cdot R_2R_1=
\\ \\
=q^{-4}S^*_5+(-{q}^{6}+{q}^{4}+{q}^{2}-2+2\,{q}^{-4}-{q}^{-6}-{q}^{-8}+{q}^{-10})S^*_{41}+\\\\
+({q}^{8}-{q}^{6}-{q}^{4}+2\,{q}^{2}-2+{q}^{-2}+{q}^{-4}-{q}^{-6})S^*_{32}-S^*_{311}+q^2S^*_{221}.
\end{array}
\end{equation}
These answers coincide with the ones obtained by the colored $\mathcal{R}$-matrix calculations.

\subsection{The case $[1]\otimes [2]\otimes [3]$}

\paragraph{Calculations using the colored $\mathbf{\mathcal{R}}$-matrices.}
The tensor product of these representations decomposes as
\begin{equation}
[1]\otimes [2]\otimes [3] = [6]+2\ [51]+2\ [42]+[411]+[33]+[321].
\end{equation}
The singlet components of the $\mathcal{R}$-matrices are equal to
\begin{equation}
\begin{array}{l}
\mathcal{R}_{(1\otimes2)\otimes3|6}=\mathcal{R}_{(2\otimes1)\otimes3|6}=\mathcal{R}_{3\otimes(1\otimes2)|6} =\mathcal{R}_{3\otimes(2\otimes1)|6}=q^{\varkappa_3-\varkappa_2}=q^2,
\\ \\
\mathcal{R}_{(1\otimes2)\otimes3|411}=\mathcal{R}_{(2\otimes1)\otimes3|411}=\mathcal{R}_{3\otimes(1\otimes2)|411} =\mathcal{R}_{3\otimes(2\otimes1)|411}=-q^{\varkappa_{21}-\varkappa_2}=-q^{-1},
\\ \\
\mathcal{R}_{(1\otimes2)\otimes3|33}=\mathcal{R}_{(2\otimes1)\otimes3|33}=\mathcal{R}_{3\otimes(1\otimes2)|33} =\mathcal{R}_{3\otimes(2\otimes1)|33}=q^{\varkappa_3-\varkappa_2}=q^2,
\\ \\
\mathcal{R}_{(1\otimes2)\otimes3|321}=\mathcal{R}_{(2\otimes1)\otimes3|321}=\mathcal{R}_{3\otimes(1\otimes2)|321} =\mathcal{R}_{3\otimes(2\otimes1)|321}=-q^{\varkappa_{21}-\varkappa_2}=-q^{-1},
\\ \\
\mathcal{R}_{(1\otimes3)\otimes2|6}=\mathcal{R}_{(3\otimes1)\otimes2|6}=\mathcal{R}_{2\otimes(1\otimes3)|6} =\mathcal{R}_{2\otimes(3\otimes1)|6}=q^{\varkappa_4-\varkappa_3}=q^3,
\\ \\
\mathcal{R}_{(1\otimes3)\otimes2|411}=\mathcal{R}_{(3\otimes1)\otimes2|411}=\mathcal{R}_{2\otimes(1\otimes3)|411} =\mathcal{R}_{2\otimes(3\otimes1)|411}=-q^{\varkappa_{31}-\varkappa_3}=-q^{-1},
\\ \\
\mathcal{R}_{(1\otimes3)\otimes2|33}=\mathcal{R}_{(3\otimes1)\otimes2|33}=\mathcal{R}_{2\otimes(1\otimes3)|33} =\mathcal{R}_{2\otimes(3\otimes1)|33}=-q^{\varkappa_{31}-\varkappa_3}=-q^{-1},
\\ \\
\mathcal{R}_{(1\otimes3)\otimes2|321}=\mathcal{R}_{(3\otimes1)\otimes2|321}=\mathcal{R}_{2\otimes(1\otimes3)|321} =\mathcal{R}_{2\otimes(3\otimes1)|321}=q^{\varkappa_{31}-\varkappa_3}=q^{-3},
\\ \\
\mathcal{R}_{(2\otimes3)\otimes1|6}=\mathcal{R}_{(3\otimes2)\otimes1|6}=\mathcal{R}_{1\otimes(2\otimes3)|6} =\mathcal{R}_{1\otimes(3\otimes2)|6}=q^{\varkappa_5-\varkappa_2-\varkappa_3}=q^6,
\\ \\
\mathcal{R}_{(2\otimes3)\otimes1|411}=\mathcal{R}_{(3\otimes2)\otimes1|411}=\mathcal{R}_{1\otimes(2\otimes3)|411} =\mathcal{R}_{1\otimes(3\otimes2)|411}=-q^{\varkappa_{41}-\varkappa_2-\varkappa_3}=-q,
\\ \\
\mathcal{R}_{(2\otimes3)\otimes1|33}=\mathcal{R}_{(3\otimes2)\otimes1|33}=\mathcal{R}_{1\otimes(2\otimes3)|33} =\mathcal{R}_{1\otimes(3\otimes2)|33}=q^{\varkappa_{32}-\varkappa_2-\varkappa_3}=q^{-2},
\\ \\
\mathcal{R}_{(2\otimes3)\otimes1|321}=\mathcal{R}_{(3\otimes2)\otimes1|321}=\mathcal{R}_{1\otimes(2\otimes3)|321} =\mathcal{R}_{1\otimes(3\otimes2)|321}=q^{\varkappa_{32}-\varkappa_2-\varkappa_3}=q^{-2}.
\end{array}
\end{equation}
The doublet components of the diagonal $\mathcal{R}$-matrices are
\setlength{\arraycolsep}{2pt}
\begin{equation}
\begin{array}{l}
\mathcal{R}_{(1\otimes2)\otimes3|51}=\mathcal{R}_{(1\otimes2)\otimes3|42}= \mathcal{R}^{\dagger}_{(2\otimes1)\otimes3|51}=\mathcal{R}^{\dagger}_{(2\otimes1)\otimes3|42}=\left(
\begin{array}{cc}
q^{\varkappa_3-\varkappa_2} & \\
&-q^{\varkappa_{21}-\varkappa_2}
\end{array}
\right)=\left(
\begin{array}{cc}
q^2 & \\
&-q^{-1}
\end{array}
\right),
\\ \\
\mathcal{R}_{(1\otimes3)\otimes2|51}=\mathcal{R}_{(1\otimes3)\otimes2|42}= \mathcal{R}^{\dagger}_{(3\otimes1)\otimes2|51}=\mathcal{R}^{\dagger}_{(3\otimes1)\otimes2|42}=\left(
\begin{array}{cc}
q^{\varkappa_4-\varkappa_3} & \\
&-q^{\varkappa_{31}-\varkappa_3}
\end{array}
\right)=\left(
\begin{array}{cc}
q^3 & \\
&-q^{-1}
\end{array}
\right),
\\ \\
\mathcal{R}_{(2\otimes3)\otimes1|51}=\mathcal{R}^{\dagger}_{(3\otimes2)\otimes1|51}=\left(
\begin{array}{cc}
q^{\varkappa_5-\varkappa_2-\varkappa_3} & \\
&-q^{\varkappa_{41}-\varkappa_2-\varkappa_3}
\end{array}
\right)=\left(
\begin{array}{cc}
q^6 & \\
&-q
\end{array}
\right),\\\\
\mathcal{R}_{(2\otimes3)\otimes1|42}=\mathcal{R}^{\dagger}_{(3\otimes2)\otimes1|42}=\left(
\begin{array}{cc}
q^{\varkappa_{41}-\varkappa_2-\varkappa_3}\\
&-q^{\varkappa_{32}-\varkappa_2-\varkappa_3}
\end{array}
\right)=\left(
\begin{array}{cc}
q & \\
&-q^2
\end{array}
\right).
\end{array}
\end{equation}
\setlength{\arraycolsep}{6pt}
The corresponding Racah matrices are
\setlength{\arraycolsep}{1pt}
\begin{equation}
\begin{array}{l}
\mathcal{U}_{1\otimes2\otimes3|51}=\mathcal{U}_{3\otimes2\otimes1|51}=\left(
\begin{array}{cc}
\frac{1}{\sqrt{[5]_q}} & \sqrt{\frac{[2]_q[6]_q}{[3]_q[5]_q}} \\
\sqrt{\frac{[2]_q[6]_q}{[3]_q[5]_q}} & -\frac{1}{\sqrt{[5]_q}}
\end{array}
\right)
,\ \
\mathcal{U}_{1\otimes2\otimes3|42}=\mathcal{U}_{3\otimes2\otimes1|42}=\left(
\begin{array}{cc}
\frac{[2]_q}{[3]_q} & \frac{\sqrt{[5]_q}}{[3]_q} \\
\frac{\sqrt{[5]_q}}{[3]_q} & -\frac{[2]_q}{[3]_q}
\end{array}
\right),
\\ \\
\mathcal{U}_{1\otimes3\otimes2|51}=\mathcal{U}_{2\otimes3\otimes1|51}=\left(
\begin{array}{cc}
\sqrt{\frac{[2]_q}{[4]_q[5]_q}} & \sqrt{\frac{[3]_q[6]_q}{[4]_q[5]_q}}\\
\sqrt{\frac{[3]_q[6]_q}{[4]_q[5]_q}} & -\sqrt{\frac{[2]_q}{[4]_q[5]_q}}
\end{array}
\right)
,\ \
\mathcal{U}_{1\otimes3\otimes2|42}=\mathcal{U}_{2\otimes3\otimes1|42}=\left(
\begin{array}{cc}
\sqrt{\frac{[2]_q}{[3]_q[4]_q}} & \sqrt{\frac{[2]_q[5]_q}{[3]_q[4]_q}} \\
\sqrt{\frac{[2]_q[5]_q}{[3]_q[4]_q}} & -\sqrt{\frac{[2]_q}{[3]_q[4]_q}}
\end{array}
\right),
\\ \\
\mathcal{U}_{2\otimes1\otimes3|51}=\mathcal{U}_{3\otimes1\otimes2|51}=\left(
\begin{array}{cc}
\sqrt{\frac{[2]_q}{[4]_q}} & \sqrt{\frac{[6]_q}{[3]_q[4]_q}} \\
\sqrt{\frac{[6]_q}{[3]_q[4]_q}} & -\sqrt{\frac{[2]_q}{[4]_q}}
\end{array}
\right)
,\ \
\mathcal{U}_{2\otimes1\otimes3|42}=\mathcal{U}_{3\otimes1\otimes2|42}=\left(
\begin{array}{cc}
\sqrt{\frac{[2]_q}{[3]_q[4]_q}} & \sqrt{\frac{[2]_q[5]_q}{[3]_q[4]_q}} \\
\sqrt{\frac{[2]_q[5]_q}{[3]_q[4]_q}} & -\sqrt{\frac{[2]_q}{[3]_q[4]_q}}
\end{array}
\right).\end{array}
\end{equation}
\setlength{\arraycolsep}{6pt}
This gives the following formulae for the doublet components of the non-diagonal $\mathcal{R}$-matrices:
\setlength{\arraycolsep}{3pt}
\begin{equation}
\begin{array}{l}
\mathcal{R}_{3\otimes(1\otimes2)|51}=\mathcal{R}_{3\otimes(2\otimes1)|51}^{\dagger}= \mathcal{U}_{3\otimes1\otimes2|51}\left(
\begin{array}{cc}
q^2 & \\
&-q^{-1}
\end{array}
\right)\mathcal{U}^{\dagger}_{3\otimes2\otimes1|51}=\left(
\begin{array}{cc}
-\frac{\sqrt{[2]_q}}{q^4\sqrt{[4]_q[5]_q}} & \frac{q\sqrt{[3]_q[6]_q}}{\sqrt{[4]_q[5]_q}} \\
\frac{\sqrt{[3]_q[6]_q}}{\sqrt{[4]_q[5]_q}} & \frac{q^5\sqrt{[2]_q}}{\sqrt{[4]_q[5]_q}}
\end{array}
\right),
\\ \\
\mathcal{R}_{3\otimes(1\otimes2)|42}=\mathcal{R}_{3\otimes(2\otimes1)|42}^{\dagger}=\mathcal{U}_{3\otimes1\otimes2|42}\left(
\begin{array}{cc}
q^2 & \\
&-q^{-1}
\end{array}
\right)\mathcal{U}^{\dagger}_{3\otimes2\otimes1|42}=\left(
\begin{array}{cc}
-\frac{\sqrt{[2]_q}}{q^3\sqrt{[3]_q[4]_q}} & \frac{\sqrt{[2]_q[5]_q}}{\sqrt{[3]_q[4]_q}} \\
\frac{q\sqrt{[2]_q[5]_q}}{\sqrt{[3]_q[4]_q}} & \frac{q^4\sqrt{[2]_q}}{\sqrt{[3]_q[4]_q}}
\end{array}
\right),
\\ \\
\mathcal{R}_{2\otimes(1\otimes3)|51}=\mathcal{R}_{2\otimes(3\otimes1)|51}^{\dagger}= \mathcal{U}_{2\otimes1\otimes3|51}\left(
\begin{array}{cc}
q^3 &\\
&-q^{-1}
\end{array}
\right)\mathcal{U}^{\dagger}_{2\otimes3\otimes1|51}=\left(
\begin{array}{cc}
-\frac{1}{q^3\sqrt{[5]_q}} & \frac{q^2\sqrt{[2]_q[6]_q}}{\sqrt{[3]_q[5]_q}} \\
\frac{\sqrt{[2]_q[6]_q}}{\sqrt{[3]_q[5]_q}} & \frac{q^5}{\sqrt{[5]_q}}
\end{array}
\right),
\\ \\
\mathcal{R}_{2\otimes(1\otimes3)|42}=\mathcal{R}_{2\otimes(3\otimes1)|42}^{\dagger}=\mathcal{U}_{2\otimes1\otimes3|42}\left(
\begin{array}{cc}
q^3 & \\
&-q^{-1}
\end{array}
\right)\mathcal{U}^{\dagger}_{2\otimes3\otimes1|42}=\left(
\begin{array}{cc}
-\frac{1}{q^2[3]_q} & \frac{q\sqrt{[5]_q}}{[3]_q} \\
\frac{q\sqrt{[5]_q}}{[3]_q} & \frac{q^4}{[3]_q}
\end{array}
\right),
\\ \\
\mathcal{R}_{1\otimes(2\otimes3)|51}=\mathcal{R}_{1\otimes(3\otimes2)|51}^{\dagger}= \mathcal{U}_{1\otimes2\otimes3|51}\left(
\begin{array}{cc}
q^6 & \\
&-q
\end{array}
\right)\mathcal{U}^{\dagger}_{1\otimes3\otimes2|51}=\left(
\begin{array}{cc}
-\frac{\sqrt{[2]_q}}{\sqrt{[4]_q}} & \frac{q^4\sqrt{[6]_q}}{\sqrt{[3]_q[4]_q}} \\
\frac{q^3\sqrt{[6]_q}}{\sqrt{[3]_q[4]_4}} & \frac{q^7\sqrt{[2]_q}}{\sqrt{[4]_q}}
\end{array}
\right),
\\ \\
\mathcal{R}_{1\otimes(2\otimes3)|42}=\mathcal{R}_{1\otimes(3\otimes2)|42}^{\dagger}= {\mathcal{U}}_{1\otimes2\otimes3|42}\left(
\begin{array}{cc}
q & \\
&-q^2
\end{array}
\right)\mathcal{U}^{\dagger}_{1\otimes3\otimes2|42}=\left(
\begin{array}{cc}
-\frac{\sqrt{[2]_q}}{\sqrt{q^4[3]_q[4]_q}} & \frac{\sqrt{[2]_q[5]_q}}{\sqrt{[3]_q[4]_q}}
\\\frac{\sqrt{[2]_q[5]_q}}{q\sqrt{[3]_q[4]_q}} & \frac{q^3\sqrt{[2]_q}}{\sqrt{[3]_q[4]_q}}
\end{array}
\right).
\end{array}
\end{equation}
\setlength{\arraycolsep}{6pt}

Now, let us provide several examples of computation of the HOMFLY polynomials for the simplest 3-colored 3-strand links. The simplest nontrivial $3$-component link is the split union of the Hopf link and of the unknot represented by the $3$-strand braid with the braid word $\sigma_1^2$. There are $3!=6$ possible placements of three representations on the $3$-strand braid, which easily can be reduced to the $3$ different links, since each link of this type has two braid representations corresponding to the permutations of the $\mathcal{R}$-matrices within the trace. The three remaining links correspond to the different choice of representation of the unknot:
\begin{equation}
\begin{array}{l}
\mathrm{Tr}\,\mathcal{R}_{(1\otimes2)\otimes3}\mathcal{R}_{(2\otimes1)\otimes3}=H^{T[2,2]}_{1\otimes2}\,S^*_3,
\\
\mathrm{Tr}\,\mathcal{R}_{(1\otimes3)\otimes2}\mathcal{R}_{(3\otimes1)\otimes2}=H^{T[2,2]}_{1\otimes3}\,S^*_2,
\\
\mathrm{Tr}\,\mathcal{R}_{(2\otimes3)\otimes1}\mathcal{R}_{(3\otimes2)\otimes1}=H^{T[2,2]}_{2\otimes3}\,S^*_1,
\end{array}
\end{equation}
it can be checked that the answers for the HOMFLY of the Hopf link at the r.h.s. obtained from the $3$-strand calculations are the same as the one obtained from the $2$-strand calculation:
\begin{equation}
\begin{array}{l}
H^{T[2,2]}_{1\otimes2}=q^{\varkappa_3-\varkappa_2}S^*_3+q^{\varkappa_{21}-\varkappa_2}S^*_{21}=q^4S^*_3+q^{-2}S^*_{21},
\\
H^{T[2,2]}_{1\otimes3}=q^{\varkappa_4-\varkappa_3}S^*_4+q^{\varkappa_{31}-\varkappa_3}S^*_{21}=q^6S^*_4+q^{-2}S^*_{21},
\\
H^{T[2,2]}_{2\otimes3}=q^{\varkappa_5-\varkappa_3-\varkappa_2}S^*_5+q^{\varkappa_{41}-\varkappa_3-\varkappa_2}S^*_{41} +q^{\varkappa_{32}-\varkappa_3-\varkappa_2}S^*_{32}=q^{12}S^*_5+q^2S^*_{41}+q^{-4}S^*_{32}.\nn
\end{array}
\end{equation}
The next to the simplest example is the ``Double Hopf'' composite link represented by the braid
$\sigma_{1}\sigma_{1}\sigma_{2}\sigma_{2}$. There are again six different placements of the representations on the braid with three different HOMFLY polynomials:
\begin{equation}
\begin{array}{l}
\mathrm{Tr}_{1\otimes2\otimes3}\,\mathcal{R}_{1\otimes(2\otimes3)}\mathcal{R}_{(1\otimes3)\otimes2} \mathcal{R}_{(3\otimes1)\otimes2}\mathcal{R}_{1\otimes(3\otimes2)}=
\\
=\mathrm{Tr}_{2\otimes1\otimes3}\,\mathcal{R}_{2\otimes(1\otimes3)}\mathcal{R}_{(2\otimes3)\otimes1} \mathcal{R}_{(3\otimes2)\otimes1}\mathcal{R}_{2\otimes(3\otimes1)}= \cfrac{H^{T[2,2]}_{2\otimes3}H^{T[2,2]}_{1\otimes3}}{S^*_3},
\\
\mathrm{Tr}_{3\otimes1\otimes2}\, \mathcal{R}_{3\otimes(1\otimes2)}\mathcal{R}_{(3\otimes2)\otimes1} \mathcal{R}_{(2\otimes3)\otimes1}\mathcal{R}_{3\otimes(2\otimes1)}=
\\
=\mathrm{Tr}_{1\otimes3\otimes2}\,\mathcal{R}_{1\otimes(3\otimes2)}\mathcal{R}_{(1\otimes2)\otimes3} \mathcal{R}_{(2\otimes1)\otimes3}\mathcal{R}_{1\otimes(2\otimes3)}= \cfrac{H^{T[2,2]}_{1\otimes2}H^{T[2,2]}_{2\otimes3}}{S^*_2},
\\
\mathrm{Tr}_{3\otimes2\otimes1}\,\mathcal{R}_{3\otimes(2\otimes1)}\mathcal{R}_{(3\otimes1)\otimes2} \mathcal{R}_{(1\otimes3)\otimes2}\mathcal{R}_{3\otimes(1\otimes2)}=
\\
=\mathrm{Tr}_{2\otimes3\otimes1}\,\mathcal{R}_{2\otimes(3\otimes1)}\mathcal{R}_{(2\otimes1)\otimes3} \mathcal{R}_{(1\otimes2)\otimes3}\mathcal{R}_{2\otimes(1\otimes3)}= \cfrac{H^{T[2,2]}_{1\otimes2}H^{T[2,2]}_{1\otimes3}}{S^*_1}.
\end{array}
\end{equation}

It is known~\cite{Pras} that the HOMFLY of the composite knot or link is proportional to the product of HOMFLY of its components and this is indeed satisfied in case considered.

The simplest prime $3$-strand $3$-component link is the torus link $T[3,3]=6_3^3$ with the braid word $\sigma_{1}\sigma_{2}\sigma_{1}\sigma_{2}\sigma_{1}\sigma_{2}$. In this case, one can apply the evolution method~\cite{DMMSS},~\cite{MMM2},\cite{MMM3}, which allows one to find an answer not for the single link, but for the whole series at the same time. In this case, the series is $T[3,3n]$. In the corresponding calculation, we need the eigenvalues of the $\mathcal{R}$-matrices product
\begin{equation}
\mathfrak{R}\equiv \mathcal{R}_{(1\otimes2)\otimes3} \mathcal{R}_{2\otimes(1\otimes3)} \mathcal{R}_{(2\otimes3)\otimes1} \mathcal{R}_{3\otimes(2\otimes1)} \mathcal{R}_{(3\otimes1)\otimes2} \mathcal{R}_{1\otimes(3\otimes2)}.
\end{equation}
Then coefficients in the character expansion of the HOMFLY are given by the $n$-th powers of the these eigenvalues as
\begin{equation}
H^{T[3,3n]}_{1\otimes2\otimes3}=\mathrm{Tr}_{1\otimes2\otimes3}\,\mathfrak{R}^n= q^{22n}S^*_6+2q^{10n}S^*_{51}+2q^{2n}S^*_{42}+q^{-2n}S^*_{411}+q^{-2n}S^*_{33}+q^{-8n}S^*_{321}.
\end{equation}
This answer is obviously invariant under the permutations of the representations on the strands since it corresponds to the cyclic permutation of the $\mathcal{R}$-matrices under the trace.

The simplest non-torus $3$-component $3$-strand link is the Borromean rings link $6_2^3$ with the braid word \\$\sigma_{1}^{-1}\sigma_{2}\sigma_{1}^{-1}\sigma_{2}\sigma_{1}^{-1}\sigma_{2}$.
The corresponding HOMFLY polynomial is:
\begin{equation}
\begin{array}{l}
H^{6_2^3}_{1\otimes2\otimes3}=\mathrm{Tr}_{1\otimes2\otimes3}\, \mathcal{R}^{-1}_{(1\otimes2)\otimes3}\mathcal{R}_{2\otimes(1\otimes3)} \mathcal{R}^{-1}_{(2\otimes3)\otimes1}\mathcal{R}_{3\otimes(2\otimes1)} \mathcal{R}^{-1}_{(3\otimes1)\otimes2}\mathcal{R}_{1\otimes(3\otimes2)}=
\\ \\
=S^*_6+\left(-q^{12}+q^{10}+q^{8}-q^{4}-q^{2}+4-q^{-2}-q^{-4}+q^{-8}+q^{-10}-q^{-12}\right)S^*_{51}+
\\
+\left(-q^{10}+q^{8}+2\,q^{6}-2\,q^{4}-q^{2}+4-q^{-2}-2\,q^{-4}+2\,q^{-6}+q^{-8}-q^{-10}\right)S^*_{42} +S^*_{411}+S^*_{33}+S^*_{321}=
\\ \\
=\cfrac{S_3^*}{(q^2-1)^2(q^4-1)}\Bigl((q^{15}-2\,q^{13}+q^{9}+q^{7}-q^{5}-2\,q^{3}+q)A^3+
\\
+(-q^{17}+2\,q^{15}-2\,q^{11}+2\,q^{7}+3\,q^{5}-2\,q^{3}-q+2\,q^{-1}+q^{-3}-q^{-5})A+
\\
+(q^{13}-q^{11}-2\,q^{9}+q^{7}+2\,q^{5}-3\,q^{3}-2\,q+2\,q^{-3}-2\,q^{-7}+q^{-9})A^{-1}+
\\
+(-q^{7}+2\,q^{5}+q^{3}-q-q^{-1}+2\,q^{-5}-q^{-7})A^{-3}.
\end{array}
\end{equation}
The answer is again symmetric under the permutation of the representations.

The Borromean rings link consists of 3 intertwined unknots. Hence, the factorization property (\cite{DMMSS},~\cite{DMMSS},\cite{chinsp1},\cite{chinsp2}, Sec.~\ref{sec:sp}) has to be
\begin{equation}
\cfrac{H^{6_2^3}_{1\otimes2\otimes3}}{S^*_1S^*_2S^*_3}\Biggr|_{q\rightarrow 1}=1.
\end{equation}
This relation is indeed satisfied.

\paragraph{Calculations using the cabling procedure.}
There are $6$ different substitutions corresponding to different colored crossings:
\begin{equation}
\begin{array}{rcl}
\mathcal{R}_{(1\otimes2)\otimes3}=\mathcal{R}_{(2\otimes1)\otimes3}^{\dagger} & \longrightarrow & R_1R_2,
\\
\mathcal{R}_{(1\otimes3)\otimes2}=\mathcal{R}_{(3\otimes1)\otimes2}^{\dagger} & \longrightarrow &R_1R_2R_3,
\\
\mathcal{R}_{(2\otimes3)\otimes1}=\mathcal{R}_{(3\otimes2)\otimes1}^{\dagger} & \longrightarrow & R_2R_1R_3R_2R_4R_3,
\\
\mathcal{R}_{1\otimes(2\otimes3)}=\mathcal{R}_{1\otimes(3\otimes2)}^{\dagger} & \longrightarrow & R_3R_2R_4R_3R_5R_4,
\\
\mathcal{R}_{2\otimes(1\otimes3)}=\mathcal{R}_{2\otimes(3\otimes1)}^{\dagger} & \longrightarrow & R_3R_4R_5,
\\
\mathcal{R}_{3\otimes(1\otimes2)}=\mathcal{R}_{3\otimes(2\otimes1)}^{\dagger} & \longrightarrow & R_4R_5.
\end{array}
\label{R123}
\end{equation}
The inverse crossings are substituted by the same products of the inverted $R$-matrices. There are $6$ projectors corresponding to the $3!=6$ permutations of the representations placed on the strands:
\begin{equation}
\begin{array}{l}
P_{1\otimes2\otimes3}= \cfrac{1+qR_2}{1+q^2}\,\cdot\cfrac{1+qR_4+qR_5+qR_4R_5+qR_5R_4+q^2R_4R_5R_4}{(1+q^2)(1+q^2+q^4)},
\\
P_{2\otimes 1\otimes 3}= \cfrac{1+qR_1}{1+q^2}\,\cdot\cfrac{1+qR_4+qR_5+qR_4R_5+qR_5R_4+q^2R_4R_5R_4}{(1+q^2)(1+q^2+q^4)},
\\
P_{1\otimes 3\otimes 2}= \cfrac{1+qR_5}{1+q^2}\,\cdot\cfrac{1+qR_2+qR_3+qR_2R_3+qR_3R_2+q^2R_2R_3R_2}{(1+q^2)(1+q^2+q^4)},
\\
P_{3\otimes 1\otimes 2}= \cfrac{1+qR_5}{1+q^2}\,\cdot\cfrac{1+qR_1+qR_2+qR_1R_2+qR_2R_1+q^2R_1R_2R_1}{(1+q^2)(1+q^2+q^4)},
\\
P_{2\otimes 3\otimes 1}= \cfrac{1+qR_1}{1+q^2}\,\cdot\cfrac{1+qR_3+qR_4+qR_3R_4+qR_4R_3+q^2R_3R_4R_3}{(1+q^2)(1+q^2+q^4)},
\\
P_{3\otimes 2\otimes 1}= \cfrac{1+qR_4}{1+q^2}\,\cdot\cfrac{1+qR_1+qR_2+qR_1R_2+qR_2R_1+q^2R_1R_2R_1}{(1+q^2)(1+q^2+q^4)}.
\end{array}
\label{P123}
\end{equation}
Combining these relations all the HOMFLY for the multicolored $3$-strand $3$-component links in the representation $[1]\otimes[2]\otimes[3]$ can be computed. This way, we obtain the HOMFLY polynomial for the split union of the Hopf link
\begin{equation}
\begin{array}{l}
\mathrm{Tr}_{1^6}\,P_{1\otimes 2\otimes 3}R_2R_1 \cdot R_1R_2 = H^{T[2,2]}_{1\otimes2}\,S^*_3,
\\
\mathrm{Tr}_{1^6}\,P_{1\otimes 3\otimes 2}R_3R_2R_1 \cdot R_1R_2R_3=H^{T[2,2]}_{1\otimes3}\,S^*_2,
\\
\mathrm{Tr}_{1^6}\,P_{2\otimes 3\otimes 1}R_2R_1R_3R_2R_4R_3\cdot R_3R_4R_2R_3R_1R_2=H^{T[2,2]}_{2\otimes3}\,S^*_1,
\end{array}
\end{equation}
for the ``Double Hopf'' link
\begin{equation}
\begin{array}{l}
\mathrm{Tr}_{1^6}\, P_{1\otimes 2\otimes 3} R_3R_2R_4R_3R_5R_4 \cdot R_3R_2R_1 \cdot R_1R_2R_3 \cdot R_4R_5R_3R_4R_2R_3=
\\
=\mathrm{Tr}_{1^6}\,P_{2\otimes 1\otimes 3} R_5R_4R_3 \cdot R_2R_1R_3R_2R_4R_3 \cdot R_3R_4R_2R_3R_1R_2 \cdot R_3R_4R_5= \cfrac{H^{T[2,2]}_{2\otimes3}H^{T[2,2]}_{1\otimes3}}{S^*_3},
\\ \\
\mathrm{Tr}_{1^6}\,P_{3\otimes 1\otimes 2} R_4R_5 \cdot R_3R_4R_2R_3R_1R_2 \cdot R_2R_1R_3R_2R_4R_3 \cdot R_4R_5=
\\
=\mathrm{Tr}_{1^6}\,P_{1\otimes 3\otimes 2} R_3R_2R_4R_3R_5R_4 \cdot R_2R_1 \cdot R_1R_2 \cdot R_4R_5R_3R_4R_2R_3= \cfrac{H^{T[2,2]}_{1\otimes2}H^{T[2,2]}_{2\otimes3}}{S^*_2},
\\ \\
\mathrm{Tr}_{1^6}\,P_{3\otimes 2\otimes 1} R_5R_4 \cdot R_1R_2R_3 \cdot R_3R_2R_1 \cdot R_4R_5=
\\
=\mathrm{Tr}_{1^6}\,P_{2\otimes 3\otimes 1} R_3R_4R_5 \cdot R_1R_2 \cdot R_2R_1 \cdot R_5R_4R_3= \cfrac{H^{T[2,2]}_{1\otimes2}H^{T[2,2]}_{1\otimes3}}{S^*_1},
\end{array}
\end{equation}
for the series of torus links
\begin{equation}
\begin{array}{c}
H^{T[3,3n]}_{1\otimes2\otimes3}= \mathrm{Tr}_{1^6}\,P_{1\otimes 2\otimes 3}\ \biggl(R_2R_1\cdot R_5R_4R_3\cdot R_2R_1R_3R_2R_4R_3\cdot R_4R_5\cdot R_3R_2R_1\cdot R_4R_5R_3R_4R_2R_3\biggr)^n=
\\ \\
=q^{22n}S^*_6+2q^{10n}S^*_{51}+2q^{2n}S^*_{42}+q^{-2n}S^*_{411}+q^{-2n}S^*_{33}+q^{-8n}S^*_{321},
\end{array}
\end{equation}
and for the Borromean rings link
\begin{equation}
\begin{array}{l}
H^{6_2^3}_{1\otimes2\otimes3}=\mathrm{Tr}_{1^6}\,P_{1\otimes 2\otimes 3}\ R^{-1}_2R^{-1}_1\cdot R_5R_4R_3\cdot R^{-1}_2R^{-1}_1R^{-1}_3R^{-1}_2R^{-1}_4R^{-1}_3\cdot
\\
\cdot R_4R_5\cdot R^{-1}_3R^{-1}_2R^{-1}_1\cdot R_4R_5R_3R_4R_2R_3=
\\
=S^*_6+(-q^{12}+q^{10}+q^{8}-q^{4}-q^{2}+4-q^{-2}-q^{-4}+q^{-8}+q^{-10}-q^{-12})S^*_{51}+
\\
+(-q^{10}+q^{8}+2\,q^{6}-2\,q^{4}-q^{2}+4-q^{-2}-2\,q^{-4}+2\,q^{-6}+q^{-8}-q^{-10})S^*_{42} +S^*_{411}+S^*_{33}+S^*_{321}.
\end{array}
\end{equation}
In all these cases, the results of these calculations are the same as those obtained using the colored $\mathcal{R}$-matrices.

\section{Cabling procedure from the representation theory\label{sec:repr}}

In the previous sections, we described the cabling procedure with all necessary elements and provided
some examples of calculations using this procedure but the procedure itself was so far treated as a postulate.
In the present section, we discuss why this cabling procedure should work from the representation theory point of view.
We point out that the cabling procedure arises not from topology, but from representation theory.
Hence, it can be used not only for constructing topological invariants, i.e., for the HOMFLY polynomials,
but also for the objects closely related with these but not topologically invariant themselves such as the extended
HOMFLY polynomials~\cite{MMM1} and the $\mathcal{R}$-matrices.

In representation theory, higher representations can be introduced with help of the co-product operation.
The co-product defines an action of an algebra on the tensor product of the representations and by that dictates
the decomposition of the tensor product of representations into the sum of irreducible ones.
This is exactly what the cabling approach does. Let us consider the $2$-strand $\mathcal{R}$-matrix corresponding
to the crossing between the strands with the representations $T_1$ and $T_2$. We would like to describe it using
the fundamental $R$-matrices for the $|T_1|+|T_2|$-strand braid. The corresponding cabling procedure can be schematically
described with the following two steps:

\begin{itemize}
\item{The colored $\mathcal{R}$-matrix is substituted with the fundamental ones,
\begin{equation}
\mathcal{R}_{T_1\otimes T_2}\ \longrightarrow\ \prod\limits^{|T_2|}_{i=1}\prod\limits^{|T_1|}_{j=1}R_{|T_1|+i-j}.
\label{cablcshem}
\end{equation}}
\item{The projectors $P_R$ are inserted.}
\end{itemize}
The first step is based on the statement
\begin{equation}
\mathcal{R}_{1^{|T_1|}\otimes 1^{|T_2|}}=\prod\limits^{|T_2|}_{i=1}\prod\limits^{|T_1|}_{j=1}R_{|T_1|+i-j},
\label{cabl1b}
\end{equation}
which in turn relies on the definition of the co-product for the
$\mathcal{R}$-matrices
\cite{Rfus:KR}-\cite{Rfus:JMP}
\begin{equation}
\begin{array}{c}
\mathcal{R}_{(T_1\otimes T_2)\otimes T_3}=\left(I_{T_1}\otimes\mathcal{R}_{T_2\otimes T_3}\right)\cdot\left(\mathcal{R}_{T_1\otimes T_3}\otimes I_{T_2}\right),
\\ \\
\mathcal{R}_{T_1\otimes (T_2\otimes T_3)}=\left(\mathcal{R}_{T_1\otimes T_2}\otimes I_{T_3}\right)\cdot\left(I_{T_2}\otimes\mathcal{R}_{T_1\otimes T_3}\right),
\end{array}
\label{Rmcoprod}
\end{equation}
where $T_1$, $T_2$ and $T_2$ are arbitrary representations, not necessary the irreducible ones.

Relation~(\ref{cabl1b}) follows from~(\ref{Rmcoprod}) by induction. First, one applies the first of relations~(\ref{Rmcoprod})  $|T_1|$ times,
\begin{equation}
\mathcal{R}_{1^{m+1}\otimes 1}=\mathcal{R}_{(1^m\otimes 1)\otimes 1}=\left(I_{1^m}\otimes\mathcal{R}_{1\otimes 1}\right)\cdot\left(\mathcal{R}_{1^m\otimes 1}\otimes I_{1}\right) = R_{m+1}\left(\mathcal{R}_{1^m\otimes 1}\otimes I_{1}\right)=\prod\limits_{i=1}^{m+1}R_{m-i+1},
\end{equation}
then one applies the second of relations~(\ref{Rmcoprod}) $|T_2|$ times in a similar way.

The second step is based on the definition of the projectors. This definition relies on the expansion of the tensor power of the fundamental representation (or of a higher one) into the irreducible ones $1^m=\sum_{T\vdash 1^m}T$, see Sec.~\ref{sec:proj} for details. Once the projectors are defined, the second step of~(\ref{cablcshem}) is based on the simple identity of linear algebra
\begin{equation}
\mathrm{Tr}_{1^{|T_1|}\otimes 1^{|T_2|}}P_{T_1} P_{T_2}\cdot(\ldots)=\mathrm{Tr}_{T_1\otimes T_2}\cdot(\ldots).
\label{cabl2b}
\end{equation}

Summarizing what is said above, the cabling procedure relies on the claim: \textit{colored $\mathcal{R}$-matrix is equal to the product of the fundamental $R$-matrices and of the projectors}\footnote{The exact equality takes place in the vertical framing (see Sec.~\ref{framing}). If another framing is used, framing factors must be also taken into account.}. In particular, this means that in the basis where the corresponding projectors are diagonal the cabling product of the fundamental $R$-matrices has to split into the colored blocks corresponding to the different irreducible representations. This is a nontrivial statement from the point of view of Sec.~\ref{sec:Rm} where the fundamental and colored $\mathcal{R}$-matrices are considered as two independent quantities defined via their eigenvalues.

The main technical obstacle here is changing the standard basis~(\ref{eq:nb}) where the form of the fundamental $\mathcal{R}$-matrices is known, for the basis
\begin{equation}
(1\otimes1\otimes\ldots\otimes1)\otimes(1\otimes1\otimes\ldots1)\otimes\ldots\otimes(1\otimes1\otimes\ldots\otimes1),
\end{equation}
where the colored $\mathcal{R}$-matrices should be defined. The Racah coefficients~\cite{Vil} describing this transition in a straightforward way are not known in the general case. However the necessary transition matrix can be found in a number of cases. To simplify the calculations, one can use that the sought transition matrix diagonalizes the projectors
\begin{equation}
P_Q:\
1\otimes1\otimes\ldots\otimes(1\otimes1\otimes\ldots\otimes1)\otimes\ldots\otimes 1\ \rightarrow\ 1\otimes1\otimes\ldots\otimes Q\otimes\ldots\otimes 1,
\end{equation}
which are the known polynomials of the fundamental $R$-matrices (see Sec.~\ref{sec:proj} for the details). The fundamental $R$-matrices are also known explicitly (see Sec.~\ref{sec:Rprop}).

\subsection{Colored $\mathcal{R}$-matrices from the fundamental ones\label{sec:ftoc}}

As was described in Sec.~\ref{sec:Rprop} the fundamental $R$-matrices are built from the blocks of sizes $1\times 1$ and $2\times 2$. The $1\times 1$ blocks are just the eigenvalues $\pm q^{\pm 1}$ while the  $2\times 2$ blocks have the form
\begin{equation}
b_k=\left(
\begin{array}{cc}
-\cfrac{1}{q^k[k]_q} & \cfrac{\sqrt{[k-1]_q[k+1]_q}}{[k]_q}
\\ \\
\cfrac{\sqrt{[k-1]_q[k+1]_q}}{[k]_q} & \cfrac{q^k}{[k]_q}
\end{array}\right)
\end{equation}
and are diagonalized by conjugation with the block
\begin{equation}
\left(
\begin{array}{cc}
\sqrt{\cfrac{[k-1]_q}{[2]_q[k]_q}} & -\sqrt{\cfrac{[k+1]_q}{[2]_q[k]_q}}
\\ \\
\sqrt{\cfrac{[k+1]_q}{[2]_q[k]_q}} & \sqrt{\cfrac{[k-1]_q}{[2]_q[k]_q}}
\end{array}\right).
\end{equation}
Thus, it is easy to diagonalize each fundamental $R$-matrix separately. This leads to a way of constructing the $\mathcal{R}$-matrices for level-two representations from the fundamental ones.

\subsubsection{Level $|Q|=2$}
The colored $\mathcal{R}$-matrices for the level-two representations $2$ (i.e., the representations $[2]$ and $[11]$) are diagonal in the bases of the form
\begin{equation}
(1\otimes 1)\otimes(1\otimes 1)\otimes(1\otimes1)\otimes\ldots.
\label{basiscol2}
\end{equation}
In these bases, all the projectors
\begin{equation}
\begin{array}{l}
P^{(j)}_2:\ (1\otimes 1)\otimes(1\otimes 1)\ldots\otimes(1\otimes 1)\otimes\ldots\rightarrow(1\otimes 1)\otimes(1\otimes 1)\ldots\otimes 2\otimes\ldots,
\\ \\
P^{(j)}_{11}:\ (1\otimes 1)\otimes(1\otimes1)\ldots\times(1\otimes 1)\otimes\ldots\rightarrow(1\otimes1)\otimes(1\otimes 1)\ldots\otimes11\otimes\ldots,
\end{array}
\end{equation}
where $j$ denotes the number of the projected pair of the fundamental representations, are diagonal. Since (see Sec.~\ref{sec:proj})
\begin{equation}
\begin{array}{lcr}
P^{(j)}_2=\cfrac{1+qR_{2j-1}}{1+q^2},&\ &P^{(j)}_{11}=\cfrac{q^2- qR_{2j-1}}{1+q^2},
\end{array}
\end{equation}
diagonalizing all the projectors $P^{(j)}_{2,11}$ (which is equivalent to diagonalizing all the $P^{(j)}_{2}$) means diagonalizing all the ``odd'' $R_{2j-1}$-matrices. This actually can be done since the only $R$-matrices that do not commute are the adjacent ones $R_i$ and $R_{i+1}$. For the form of the $R$-matrices described above it follows that when there is a $2\times 2$ block in one of the matrices, in all other
there would be either a diagonal block with equal diagonal elements or the same $2\times 2$ block.

\subsubsection{Level $|Q|=3$\label{sec:level3col}}

The bases where the $\mathcal{R}$-matrices for the level-two representations of are diagonal, look like
\begin{equation}
(1\otimes1\otimes1)\otimes (1\otimes 1\otimes 1)\otimes\ldots\times(1\otimes1\otimes 1).
\label{basiscol3}
\end{equation}
In these bases, all the projectors
\begin{equation}
\begin{array}{l}
P^{(j)}_3:\ (1\otimes 1\otimes 1)\otimes(1\otimes 1\otimes 1)\ldots\otimes(1\otimes 1\otimes 1)\otimes\ldots\rightarrow(1\otimes 1\otimes 1)\otimes(1\otimes 1\otimes 1)\ldots\otimes 3\otimes\ldots,
\\
P^{(j)}_{21}:\ (1\otimes 1\otimes 1)\otimes(1\otimes 1\otimes 1)\ldots\otimes(1\otimes 1\otimes 1)\otimes\ldots\rightarrow(1\otimes 1\otimes 1)\otimes(1\otimes 1\otimes 1)\ldots\otimes 21\otimes\ldots,
\\
P^{(j)}_{111}:\ (1\otimes 1\otimes 1)\otimes(1\otimes1\otimes 1)\ldots\otimes(1\otimes 1\otimes 1)\otimes\ldots\rightarrow(1\otimes1\otimes 1)\otimes(1\otimes 1\otimes 1)\ldots\otimes 111\otimes\ldots
\end{array}
\end{equation}
are diagonal.
In what follows, we show that it suffices to diagonalize only the projectors $P^{(j)}_{21}$. As follows from~(\ref{eq:P21}), the
corresponding $R$-matrix expression is
\begin{equation}
P^{(j)}_{21}=\cfrac{(R_{3j-2}-R_{3j-1})^2}{q^{-2}+1+q^2}.
\end{equation}
Basis~(\ref{basiscol3}) is not well defined since there is an ambiguity in the choice of bases  in the two-dimensional spaces of the representations $[21]$ coming from the expansion of  each bracket\footnote{
There is also a more general ambiguity in choosing basis since the order of multiplying the brackets is not defined. In all cases, we choose the basis $\biggl(\Bigl(\bigl((\ldots)\otimes(\ldots)\bigr)\otimes(\ldots)\Bigr)\ldots\biggr)\otimes\ldots$, which is generalization of the standard basis~(\ref{eq:nb}).}
$[1]\otimes[1]\otimes [1]=[3]+2\ [21]+[111]$. This ambiguity can be fixed by the definition of the order of multiplying the fundamental representations in each bracket, e.g., as:
\begin{equation}
\biggl(\Bigl((1\otimes 1)\otimes1\Bigl)\otimes\Bigl((1\otimes 1)\otimes1\Bigr)\otimes\ldots\biggr)\otimes\Bigl((1\otimes 1)\otimes1\Bigr).
\label{basiscol3up}
\end{equation}
This means that additionally all the projectors
\begin{equation}
^{*}P^{(j)}_2=\cfrac{1+qR_{3j-2}}{1+q^2}
\end{equation}
turn diagonal. The projectors $^{*}P^{(j)}_2$ can be diagonalized in the same way as the projectors $P^{(j)}_2$ were diagonalized in the previous section. Then the matrices of $P^{(j)}_{21}$ in the  basis obtained can be calculated. The non-diagonal elements in this matrix will appear only where there are unit or zero blocks in the already diagonalized projectors $P^{(j)}_{21}$ and $^{*}P^{(l)}_21$ since these projectors commute for any $j$ and $l$. In the considered $6$-strand case, all the $P^{(j)}_{21}$ consist of $2\times 2$ blocks except for the single $3\times 3$ block appearing several times in the component  $P^{(j)}_{21|321}$ corresponding to the representation $[321]$.  In this case, we calculated the $3\times 3$ blocks and present the explicit expressions in Appendix~\ref{app:projbl}. The universal form of the blocks in the projectors onto level-three, in contrast to the blocks in the fundamental $R$-matrices and the projectors onto level-two representations, is unknown.

\subsection{The permutation operators in $\mathcal{R}$ and $U$-matrices}

The mixing matrices $\mathcal{U}$ can be defined in two ways. The first way is to define them as the matrices that just switch from one basis (e.g., where the $\mathcal{R}$-matrix corresponding to the crossing between the first pair of strands is diagonal) to the other (e.g., where the $\mathcal{R}$-matrix corresponding to the crossing between the second pair of strands is diagonal). All the $\mathcal{U}$-matrices described in Sec.~\ref{sec:links} are of this type. The approach, which uses the mixing matrices of this type, implies that the diagonal form of $\mathcal{R}$-matrix is known irrespective to where the corresponding crossing stands. There is another way to define the  $\mathcal{U}$-matrices. The difference is that the operator $\mathcal{U}$ now includes a permutation operator and consequently changes the placement of the representations in the braid. We denote these operators by straight $U$. $U$-operators discussed in this section allow one to define any $\mathcal{R}$-matrix using only the diagonal $\mathcal{R}$-matrices corresponding to the crossing of the first two strands. The $U$-operators can be defined by the following pictures (hereafter, $m$ denotes the number of the strands):

\paragraph{for $\mathbf{m=2}$,}
\begin{center}
\begin{picture}(150,20)
\put(0,0){$T_1\otimes T_2$} \put(120,0){$T_2\otimes T_1$,}
\put(45,5){\vector(1,0){60}} \put(60,10){$\mathcal{R}_{T_1T_2}$}
\end{picture}
\end{center}

\

\paragraph{$\mathbf{m=3}$,}
\begin{center}
\begin{picture}(200,100)
\put(0,80){$(T_1\otimes T_2)\otimes T_3$} \put(140,80){$(T_2\otimes T_1)\otimes
T_3$} \put(0,0){$T_3\otimes(T_2\otimes T_1)$} \put(140,0){$T_3\otimes(T_1\otimes
T_2)$} \put(70,85){\vector(1,0){60}} \put(70,5){\vector(1,0){60}}
\put(30,70){\vector(0,-1){50}} \put(190,70){\vector(0,-1){50}}
\put(80,90){$\mathcal{R}_{(T_1T_2)T_3}$}
\put(80,10){$\mathcal{R}_{T_3(T_2T_1)}$}
\put(35,45){$U_{\,T_1T_2T_3}$}
\put(195,45){$U_{\,T_2T_1T_3}$,}
\end{picture}
\end{center}

\begin{equation}
\mathcal{R}_{T_3(T_2T_1)}=U_{\,T_1T_2T_3}\ \mathcal{R}_{(T_1T_2)T_3}\ U^{\,\dagger}_{\,T_2T_1T_3},
\label{Ucolored}
\end{equation}

\

In what follows, we consider some particular examples and describe the form of the products of the fundamental $R$-matrices corresponding to the colored $\mathcal{R}$-matrices. In all the examples given in this section, as well as everywhere else in the present paper, the vertical framing is used (see Sec.~\ref{framing}).

\subsection{Level $|T|=2$, $m=2$ strands}

In this case, there is the single $\mathcal{R}$-matrix, which should be substituted according to~(\ref{cablcshem}) with the following product
\begin{equation}
\mathcal{R}_1\ \rightarrow\ \mathfrak{R}_1 \equiv R_2 R_1 R_3 R_2.
\end{equation}
In the basis~(\ref{basiscol2}), the matrix $\mathfrak{R}$ splits into the blocks corresponding to the colored $\mathcal{R}$-matrices of the type $\mathcal{R}_{QQ^{\prime}}$ with $|Q|=|Q^{\prime}|=2$ (Hereafter the solid lines denote the symmetric representation and the dashed lines the antisymmetric one):

\begin{flushright}
\begin{picture}(50,45)
\put(0,20){\line(1,1){20}} \put(20,20){\line(-1,1){20}}
\put(0,0){$ \mathcal{R}_{2\otimes2}$}
\end{picture}
\begin{picture}(50,45)
\put(0,20){\line(1,1){20}} \multiput(20,20)(-2,2){11}{.}
\put(0,0){$ \mathcal{R}_{11\otimes2}$}
\end{picture}
\begin{picture}(50,45)
\multiput(0,20)(2,2){11}{.} \put(20,20){\line(-1,1){20}}
\put(0,0){$ \mathcal{R}_{2\otimes11}$}
\end{picture}
\begin{picture}(25,45)
\multiput(0,20)(2,2){11}{.} \multiput(20,20)(-2,2){11}{.}
\put(0,0){$ \mathcal{R}_{11\otimes11}$}
\end{picture}
\begin{picture}(130,40)
\put(100,35){\line(1,0){10}} \multiput(100,20)(2,0){6}{.}
\put(117,35){$[2]$} \put(115,18){$[11]$}
\end{picture}
\end{flushright}

\paragraph{Representation $\mathbf{Q=[4]}$} arises from the decomposition of $2\otimes2$ and does \emph{not} appear in the decompositions of $2\otimes11$, $11\otimes2$, $11\otimes11$. Hence, the $\mathfrak{R}$-matrix contains one out of the four possible blocks:
\begin{equation}
{\mathfrak{R}}_{1|4}=q^4=
\mathcal{R}_{2\otimes2|4}.
\end{equation}

\paragraph{Representation $\mathbf{Q=[31]}$} arises from the decompositions of $2\otimes2$, $2\otimes11$, $11\otimes2$ and does \emph{not} appear in the decomposition of $11\otimes11$. Hence, the $\mathfrak{R}$-matrix contains three out of the four possible blocks:
\begin{equation}
{\mathfrak{R}}_{1|31}=\left(\begin{array}{ccc}-1\\&&\pm q^2\\&\pm q^2\end{array}\right)=\left(\begin{array}{ccc}
\mathcal{R}_{2\otimes2|31}\\&&
\mathcal{R}_{2\otimes11|31}\\&
\mathcal{R}_{11\otimes2|31}\end{array}\right).
\end{equation}
\paragraph{Representation $\mathbf{Q=[22]}$} arises from the decompositions of $2\otimes2$, $11\otimes11$ and does \emph{not} appear in the decompositions of $2\otimes11$, $11\otimes2$. Hence, the $\mathfrak{R}$-matrix contains two out of the four possible blocks:
\begin{equation}
{\mathfrak{R}}_{1|22}=\left(\begin{array}{ccc}q^{-2}\\&q^2\end{array}\right)
=\left(\begin{array}{cc}
\mathcal{R}_{2\otimes2|22}\\&
\mathcal{R}_{11\otimes11|22}\end{array}\right).
\end{equation}
\paragraph{Representation $\mathbf{Q=[211]}$} arises from the decompositions of $2\otimes11$, $11\otimes22$, $11\otimes2$ and does \emph{not} appear in the decomposition of $2\otimes2$. Hence, the $\mathfrak{R}$-matrix contains three out of the four possible blocks:
\begin{equation}
{\mathfrak{R}}_{1|211}=\left(\begin{array}{ccc}&\pm q^{-2}\\\pm q^{-2}\\&&-1\end{array}\right)=\left(\begin{array}{ccc}
&\mathcal{R}_{2\otimes11|211}\\
\mathcal{R}_{11\otimes2|211}\\
&&\mathcal{R}_{11\otimes11|211}\end{array}\right).
\end{equation}
We write  $\pm$ before the  off-diagonal elements to emphasize that they are defined up to a sign. 

\subsection{Level $|T|=2$, $m=3$ strands\label{sec:lev2str3}}

For the $3$-strand knots two different $\mathcal{R}$-matrices are needed:
\begin{equation}
\begin{array}{l}
\mathcal{R}_1\ \rightarrow\ \mathfrak{R}_1\equiv R_2 R_1 R_3 R_2,
\\ \\
\mathcal{R}_2\ \rightarrow\ \mathfrak{R}_2\equiv R_4 R_3 R_5 R_4.
\end{array}
\end{equation}
The colored mixing matrix $U$ also should be replaced with the product of the fundamental ones, $\mathfrak{U}$. There are $8$ various
types of crossings corresponding the $\mathfrak{R}$-matrices:
\begin{center}
\begin{picture}(50,70)
\put(0,30){\line(1,1){20}} \put(20,30){\line(-1,1){20}}
\put(0,20){\line(1,0){20}} \put(0,0){$ \mathcal{R}_{(2\otimes2)\otimes2}$}
\end{picture}
\begin{picture}(50,70)
\put(0,30){\line(1,1){20}} \put(20,30){\line(-1,1){20}}
\multiput(0,20)(3,0){8}{.} \put(0,0){$ \mathcal{R}_{(2\otimes2)\otimes11}$}
\end{picture}
\begin{picture}(50,70)
\multiput(0,30)(2,2){11}{.} \put(20,30){\line(-1,1){20}}
\put(0,20){\line(1,0){20}} \put(0,0){$ \mathcal{R}_{(2\otimes11)\otimes2}$}
\end{picture}
\begin{picture}(50,70)
\put(0,30){\line(1,1){20}} \multiput(20,30)(-2,2){11}{.}
\put(0,20){\line(1,0){20}} \put(0,0){$ \mathcal{R}_{(11\otimes2)\otimes2}$}
\end{picture}
\begin{picture}(50,70)
\put(0,30){\line(1,1){20}} \multiput(20,30)(-2,2){11}{.}
\multiput(0,20)(3,0){8}{.} \put(0,0){$ \mathcal{R}_{(2\otimes11)\otimes11}$}
\end{picture}
\begin{picture}(50,70)
\multiput(0,30)(2,2){11}{.} \put(20,30){\line(-1,1){20}}
\multiput(0,20)(3,0){8}{.} \put(0,0){$ \mathcal{R}_{(11\otimes2)\otimes11}$}
\end{picture}
\begin{picture}(50,70)
\multiput(0,30)(2,2){11}{.} \multiput(20,30)(-2,2){11}{.}
\put(0,20){\line(1,0){20}} \put(0,0){$ \mathcal{R}_{(11\otimes11)\otimes2}$}
\end{picture}
\begin{picture}(25,70)
\multiput(0,30)(2,2){11}{.} \multiput(20,30)(-2,2){11}{.}
\multiput(0,20)(3,0){8}{.} \put(0,0){$ \mathcal{R}_{(11\otimes11)\otimes11}$}
\end{picture}
\end{center}
\begin{center}
\begin{picture}(50,70)
\put(0,20){\line(1,1){20}} \put(20,20){\line(-1,1){20}}
\put(0,50){\line(1,0){20}} \put(0,0){$\mathcal{R}_{2\otimes(2\otimes2)}$}
\end{picture}
\begin{picture}(50,70)
\multiput(0,20)(2,2){11}{.} \put(20,20){\line(-1,1){20}}
\put(0,50){\line(1,0){20}} \put(0,0){$\mathcal{R}_{2\otimes(2\otimes11)}$}
\end{picture}
\begin{picture}(50,70)
\put(0,20){\line(1,1){20}} \multiput(20,20)(-2,2){11}{.}
\put(0,50){\line(1,0){20}} \put(0,0){$\mathcal{R}_{2\otimes(11\otimes2)}$}
\end{picture}
\begin{picture}(50,70)
\put(0,20){\line(1,1){20}} \put(20,20){\line(-1,1){20}}
\multiput(0,50)(3,0){7}{.} \put(0,0){$\mathcal{R}_{11\otimes(2\otimes2)}$}
\end{picture}
\begin{picture}(50,70)
\multiput(0,20)(2,2){11}{.} \multiput(20,20)(-2,2){11}{.}
\put(0,50){\line(1,0){20}} \put(0,0){$\mathcal{R}_{2\otimes(11\otimes11)}$}
\end{picture}
\begin{picture}(50,70)
\multiput(0,20)(2,2){11}{.} \put(20,20){\line(-1,1){20}}
\multiput(0,50)(3,0){7}{.} \put(0,0){$\mathcal{R}_{11\otimes(2\otimes11)}$}
\end{picture}
\begin{picture}(50,70)
\put(0,20){\line(1,1){20}} \multiput(20,20)(-2,2){11}{.}
\multiput(0,50)(3,0){7}{.} \put(0,0){$\mathcal{R}_{11\otimes(11\otimes2)}$}
\end{picture}
\begin{picture}(25,70)
\multiput(0,20)(2,2){11}{.} \multiput(20,20)(-2,2){11}{.}
\multiput(0,50)(3,0){7}{.} \put(0,0){$\mathcal{R}_{11\otimes(11\otimes11)}$}
\end{picture}
\end{center}

This gives 8 transitions described by the operator $\mathfrak{U}=UVWY\cdot UVW$ (the notations as in~\cite{MMM2},\cite{AMMM1}):
\begin{center}
\begin{picture}(120,80)
\put(0,45){\line(1,-1){30}} \put(30,15){\line(1,1){30}}
\put(15,30){\line(1,1){15}} \put(75,45){\line(1,-1){30}}
\put(105,15){\line(1,1){30}} \put(120,30){\line(-1,1){15}}
\put(50,20){\vector(1,0){35}} \put(50,0){$U_{2\otimes2\otimes2}$}
\end{picture}
\end{center}

\begin{center}
\begin{picture}(150,60)
\put(0,45){\line(1,-1){30}} \multiput(30,15)(2,2){16}{.}
\put(15,30){\line(1,1){15}} \multiput(75,45)(2,-2){16}{.}
\put(105,15){\line(1,1){30}} \put(120,30){\line(-1,1){15}}
\put(50,20){\vector(1,0){35}} \put(50,0){$U_{2\otimes2\otimes11}$}
\end{picture}
\begin{picture}(150,60)
\put(0,45){\line(1,-1){30}} \put(30,15){\line(1,1){30}}
\multiput(15,30)(2,2){8}{.} \put(75,45){\line(1,-1){30}}
\put(105,15){\line(1,1){30}} \multiput(120,30)(-2,2){8}{.}
\put(50,20){\vector(1,0){35}} \put(50,0){$U_{2\otimes11\otimes2}$}
\end{picture}
\begin{picture}(120,60)
\multiput(0,45)(2,-2){16}{.} \put(30,15){\line(1,1){30}}
\put(15,30){\line(1,1){15}} \put(75,45){\line(1,-1){30}}
\multiput(105,15)(2,2){15}{.} \put(120,30){\line(-1,1){15}}
\put(50,20){\vector(1,0){35}} \put(50,0){$U_{11\otimes2\otimes2}$}
\end{picture}
\end{center}

\begin{center}
\begin{picture}(150,60)
\put(0,45){\line(1,-1){30}} \multiput(30,15)(2,2){16}{.}
\multiput(15,30)(2,2){8}{.} \multiput(75,45)(2,-2){16}{.}
\put(105,15){\line(1,1){30}} \multiput(120,30)(-2,2){8}{.}
\put(50,20){\vector(1,0){35}} \put(50,0){$U_{2\otimes11\otimes11}$}
\end{picture}
\begin{picture}(150,60)
\multiput(0,45)(2,-2){16}{.} \multiput(30,15)(2,2){16}{.}
\put(15,30){\line(1,1){15}} \multiput(75,45)(2,-2){16}{.}
\multiput(105,15)(2,2){16}{.} \put(120,30){\line(-1,1){15}}
\put(50,20){\vector(1,0){35}} \put(50,0){$U_{11\otimes2\otimes11}$}
\end{picture}
\begin{picture}(120,60)
\multiput(0,45)(2,-2){16}{.} \put(30,15){\line(1,1){30}}
\multiput(15,30)(2,2){8}{.} \put(75,45){\line(1,-1){30}}
\multiput(105,15)(2,2){16}{.} \multiput(120,30)(-2,2){8}{.}
\put(50,20){\vector(1,0){35}} \put(50,0){$U_{11\otimes11\otimes2}$}
\end{picture}
\end{center}

\begin{center}
\begin{picture}(150,60)
\multiput(0,45)(2,-2){16}{.} \multiput(30,15)(2,2){16}{.}
\multiput(15,30)(2,2){8}{.} \multiput(75,45)(2,-2){16}{.}
\multiput(105,15)(2,2){16}{.} \multiput(120,30)(-2,2){8}{.}
\put(50,20){\vector(1,0){35}} \put(50,0){$U_{11\otimes11\otimes11}$}
\end{picture}
\end{center}

In the basis $\Bigl((1\otimes 1)\otimes (1\otimes 1)\Bigr)\otimes(1\otimes 1)$, the matrices $\mathfrak{R}$ and $\mathfrak{U}$ indeed have the block structure:
\setlength{\arraycolsep}{-1pt}
\begin{equation*}
\mathfrak{R}_1=
\left(\begin{array}{cccccccc}
\mathcal{R}_{(2\otimes2)\otimes2}
\\
&\mathcal{R}_{(2\otimes2)\otimes11}&0&0
\\
&0&0&\mathcal{R}_{(2\otimes11)\otimes2}
\\
&0&\mathcal{R}_{(11\otimes2)\otimes2}&0&
\\
&&&&0&\mathcal{R}_{(2\otimes11)\otimes11}&0
\\
&&&&\mathcal{R}_{(11\otimes2)\otimes11}&0&0
\\
&&&&0&0&\mathcal{R}_{(11\otimes11)\otimes2}
\\
&&&&&&&\mathcal{R}_{(11\otimes11)\otimes11}
\end{array}\right),
\end{equation*}
\begin{equation*}
\mathfrak{R}_2=
\left(\begin{array}{cccccccc}
\mathcal{R}_{2\otimes(2\otimes2)}
\\
&0&\mathcal{R}_{2\otimes(2\otimes11)}&0
\\
&\mathcal{R}_{2\otimes(11\otimes2)}&0&0
\\
&0&0&\mathcal{R}_{11\otimes(2\otimes2)}
\\
&&&&\mathcal{R}_{2\otimes(11\otimes11)}&0&0
\\
&&&&0&0&\mathcal{R}_{11\otimes(2\otimes11)}
\\
&&&&0&\mathcal{R}_{11\otimes(11\otimes2)}&0
\\
&&&&&&&\mathcal{R}_{11\otimes(11\otimes11)}
\end{array}\right),
\end{equation*}
\setlength{\arraycolsep}{1pt}
\begin{equation*}
\mathfrak{U}=
\left(\begin{array}{cccccccc}
U_{2\otimes2\otimes2}
\\
&0&U_{11\otimes2\otimes2}&0
\\
&0&0&U_{2\otimes11\otimes2}
\\
&U_{2\otimes2\otimes11}&0&0
\\
&&&&0&0&U_{11\otimes11\otimes2}
\\
&&&&U_{11\otimes2\otimes11}&0&0
\\
&&&&0&U_{2\otimes11\otimes11}&0
\\
&&&&&&&U_{11\otimes11\otimes11}
\end{array}\right).
\end{equation*}
\setlength{\arraycolsep}{6pt}
This means that the equality $\mathfrak{R}_2=\mathfrak{U}\,\mathfrak{R}_1\mathfrak{U}^{\dagger}$ can be split into separate parts:
\begin{equation}
\begin{array}{ll}
\mathcal{R}_{2\otimes(2\otimes2)}=U_{2\otimes2\otimes2}\mathcal{R}_{(2\otimes2)\otimes2}U^{\,\dagger}_{2\otimes2\otimes2},
&
\mathcal{R}_{11\otimes(11\otimes11)}=U_{11\otimes11\otimes11}\mathcal{R}_{(11\otimes11)\otimes11}U^{\,\dagger}_{11\otimes11\otimes11},
\\ \\
\mathcal{R}_{11\otimes(2\otimes2)}=U_{2\otimes2\otimes11}\mathcal{R}_{(2\otimes2)\otimes11}U^{\,\dagger}_{11\otimes2\otimes2},
&
\mathcal{R}_{2\otimes(11\otimes2)}=U_{2\otimes11\otimes2}\mathcal{R}_{(2\otimes11)\otimes2}U^{\,\dagger}_{2\otimes11\otimes2},
\\ \\
\mathcal{R}_{2\otimes(2\otimes11)}=U_{11\otimes2\otimes2}\mathcal{R}_{(11\otimes2)\otimes2}U^{\,\dagger}_{2\otimes11\otimes11},
&
\mathcal{R}_{2\otimes(11\otimes11)}=U_{11\otimes11\otimes2}\mathcal{R}_{(11\otimes11)\otimes2}U^{\,\dagger}_{2\otimes11\otimes11},
\\ \\
\mathcal{R}_{11\otimes(2\otimes11)}=U_{11\otimes2\otimes11}\mathcal{R}_{(11\otimes2)\otimes11}U^{\,\dagger}_{11\otimes2\otimes11},
&
\mathcal{R}_{22\otimes(11\otimes11)}=U_{11\otimes11\otimes2}\mathcal{R}_{(11\otimes11)\otimes2}U^{\,\dagger}_{2\otimes11\otimes11}.
\end{array}
\end{equation}
The explicit form of the blocks is given in Appendix~\ref{app:colbl2}.

\subsection{Level $|T|=3$, $m=2$ strands\label{sec:lev3str2}}

Due to technical complexity of the calculations, we provide only the calculation for the $2$-strand knots for the representations of the size $3$. This case requires $2\cdot 3=6$-strand calculation of the knots/links in the fundamental representation (and so do the $3$-strand knots in the representations of size $2$). The basis that should be considered in this case is of the type $(1\otimes 1\otimes 1)\otimes(1\otimes 1\otimes 1)$.

According to~(\ref{cablcshem}), the single $\mathcal{R}$-matrix should be substituted with the product of six fundamental $R$-matrices:
\begin{equation}
\mathcal{R}_1\ \rightarrow\ \mathfrak{R}=R_3R_2R_1R_4R_3R_2R_5R_4R_3.
\end{equation}
In the basis~(\ref{basiscol3up}), this matrix decomposes into the blocks corresponding to the irreducible representations coming from the decomposition $(1\otimes 1)\otimes 1=3+\underline{21}+\overline{21}+111$. We have
\setlength{\arraycolsep}{3pt}
\begin{equation}
\begin{array}{ll}
\multicolumn{2}{c}{
\mathfrak{R}=\left(\begin{array}{cccc}
\mathfrak{r}_{3\otimes111}
\\
&\mathfrak{r}_{3\otimes21}
\\
&&\mathfrak{r}_{21\otimes111}
\\
&&&\mathfrak{r}_{21\otimes21}
\end{array}\right),}
\end{array}
\end{equation}
where
\begin{equation}
\begin{array}{ll}
\mathfrak{r}_{3\otimes111}=\small\left(\begin{array}{cccc}
\mathcal{R}_{3\otimes3}
\\
&0&\mathcal{R}_{3\otimes111}
\\
&\mathcal{R}_{111\otimes3}&0
\\
&&&\mathcal{R}_{111\otimes111}
\end{array}\right),
&
\mathfrak{r}_{3\otimes21}=\small\left(\begin{array}{cccc}
0&\mathcal{R}_{3\otimes\underline{21}}
\\
\mathcal{R}_{\underline{21}\otimes3}&0
\\
&&0&\mathcal{R}_{3\otimes\overline{21}}
\\
&&\mathcal{ R}_{\overline{21}\otimes3}&0
\end{array}\right),
\\ \\
\mathfrak{r}_{21\otimes111}=\small\left(\begin{array}{cccc}
0&\mathcal{R}_{111\otimes\underline{21}}
\\
\mathcal{R}_{\underline{21}\otimes111}&0
\\
&&0&\mathcal{R}_{111\otimes\overline{21}}
\\
&&\mathcal{R}_{\overline{21}\otimes111}&0
\end{array}\right),
&
\mathfrak{r}_{21\otimes21}=\small\left(\begin{array}{cccc}
\mathcal{R}_{\underline{21}\otimes\underline{21}}
\\
&0&\mathcal{R}_{\underline{21}\otimes\overline{21}}
\\
&\mathcal{R}_{\overline{21}\otimes\underline{21}}&0
\\
&&&\mathcal{ R}_{\overline{21}\otimes\overline{21}}\end{array}\right).
\end{array}
\end{equation}
\setlength{\arraycolsep}{6pt}
The explicit form of the blocks is given in Appendix~\ref{app:colbl3}.

\section{Eigenvalue conjecture and cabling.\label{sec:eig}}

It was proposed in paper~\cite{IMMM3} that there exists a certain relation between eigenvalues of $\mathcal{R}$-matrix and the corresponding Racah-coefficients.
It was asserted that all elements of the $3$-strand $U$-matrices depend only on the (somehow normalized) eigenvalues of the $\mathcal{R}$-matrix; it was supposed that all three strands are in the same representation. The cabling procedure provides an explanation, why the $U$-matrix elements are expressed though $R$-matrix eigenvalues.

\subsection{Constraints on the $\mathcal{R}$-matrix elements following from the cabling procedure\label{sec:conU}}
The existence of the cabling procedure implies that there are severe constraints on the form of the $\mathcal{R}$-matrices. As was discussed in Sec.~\ref{sec:repr} the cabling procedure in fact describes the fusion (co-product) rule for the $\mathcal{R}$-matrices. This implies that the specific products of $\mathcal{R}$-matrices should in certain bases split into blocks corresponding to the $\mathcal{R}$-matrices in higher representations (see examples in Sec.~\ref{sec:repr}). This means that in fact co-product rule~(\ref{Rmcoprod}) can be considered as a set of constraints on the form of $\mathcal{R}$-matrices:

\begin{equation}
\mathcal{U}_{xxx}\mathcal{R}_{(xx)x}\mathcal{U}_{xxx}^{\dagger}\mathcal{R}_{(xx)x}\mathcal{U}_{xxx}=\mathrm{ diag}\Bigl(\mathcal{R}_{wx}\Bigr)_{w\vdash x\otimes x},
\label{RURUR}
\end{equation}
where the diagonalized $\mathcal{R}$-matrix is in the r.h.s.; each possible $w$ corresponds to one of the eigenvalues of the matrix. The appearance of the three transition matrices in this relation can be explained in the following way. The $\mathcal{R}$-matrix

\begin{equation}
\mathcal{R}_{wx}:\ w\otimes x\ \rightarrow\ x\otimes w
\end{equation}
is diagonal if it acts from the space $w\otimes x$ to the space $x\otimes w$. Thus, if one wants the matrix on the r.h.s. of~(\ref{RURUR}) to be diagonal three transition matrices are needed.
Relation~(\ref{RURUR}) as always splits into separate colored blocks, each one corresponding to the set of irreducible representations $Q\vdash x\otimes x\otimes x$ with the same Young diagram. We imply that all the identities in the present section are written for the separate colored blocks $\mathcal{U}_{xxx|Q}$ though the index $Q$ is omitted. In particular, $w$ in~(\ref{RURUR}) runs not over all irreducible representations from the decomposition of $x\otimes x$ but only over those that contain $Q$ in the decomposition of $w\otimes x$.

If all elements of the diagonal matrix $\mathcal{R}_{xx}$ are known, the system~(\ref{RURUR}) gives severe constraints on the form of the Racah matrix $U$ and on the eigenvalues of the matrix $\mathcal{R}_{wx}$ for the representations $w$ higher than $x$ (i.e., $|w|>|x|$). At a first glance, system~(\ref{RURUR}) is overdetermined since the condition of vanishing of the off-diagonal elements in the l.h.s. gives $N^2-N$ constraints while there are only $\frac{1}{2}(N^2-N)$ free parameters in the orthogonal matrix $\mathcal{U}$. This contradiction is resolved since the Racah matrix $\mathcal{U}$ is not only orthogonal but also satisfies $\mathcal{U}\sigma=\sigma \mathcal{U}^{\dagger}$ for some matrix sigma such that $\sigma_{ij}=0$ for $i\ne j$ and $\sigma_{ii}=\pm 1$. We call such matrices $\mathcal{U}$ pseudosymmetric. Assuming $\mathcal{U}$ to be  pseudosymmetric, one can see that the number of independent equations in~(\ref{RURUR}) reduces; in particular cases, we show that the number of independent equations exactly equals to the number of variables. In fact, it is enough to consider only the symmetric matrices since each pseudosymmetric solution $\mathcal{U}$ of~(\ref{RURUR}) corresponds to a symmetric one\footnote{If the matrix $\mathcal{U}$ is pseudosymmetric, i.e., $\mathcal{U}\sigma=\sigma \mathcal{U}^{\dagger}$, then the matrix $U\sigma$ is symmetric: $(U\sigma)^{\dagger}=U\sigma$. Hence, a symmetric solution can be constructed from a pseudosymmetric one by substituting $\mathcal{U}$ with $\mathcal{U}\sigma$ in~(\ref{RURUR}).}.

\subsection{Elements of $\mathcal{U}$-matrix via eigenvalues of $\mathcal{R}$-matrix}

Let us sightly simplify~(\ref{RURUR}) for further analysis. From relation~(\ref{RURUR}) and from the orthogonal matrix property $\det\mathcal{U}=\pm 1$ it follows that $\left(\det \mathcal{R}_{(xx)x}\right)^2=\det \mathcal{R}_{wx}$. Hence, a direct substitution shows that~(\ref{RURUR}) is still satisfied if the $\mathcal{R}$-matrices are rescaled at the both sides as
\begin{equation}
\mathcal{R}\ \rightarrow\ \cfrac{\mathcal{R}}{(\pm\det\mathcal{R})^{1/n}}\ ,
\label{eq:norm}
\end{equation}
where $n$ is the size of the matrices and the sign in front of the determinant is chosen so that the expression in the brackets is positive for $q$ being the positive real (this choice slightly simplifies the formulae below). Then,~(\ref{RURUR}) can be rewritten in components as
\begin{equation}
\sum_{j,k=1}^n\mathcal{U}_{ij}\mathcal{U}_{jk}\mathcal{U}_{lk} \xi_j\xi_k = \delta_{il}\eta_i,
\label{RURURc}
\end{equation}
where  $\xi$ and $\eta$ are the normalized eigenvalues introduced in~\cite{IMMM3}:\footnote{In~\cite{IMMM3}, the normalized eigenvalues are denoted as $\tilde{\xi}$ while $\xi$ is reserved for the ordinary eigenvalues. We on the other side use $\xi$ for the normalized  eigenvalues since the ordinary ones do not appear in the calculations in the present section.}
\begin{equation}
\cfrac{\mathcal{R}_{(xx)x}}{(\pm\det \mathcal{R})^{1/n}}\equiv(\mathrm{diag}\ \xi_1,\ldots,\xi_n),\ \ \ \cfrac{\mathrm{diag}\Bigl(\mathcal{R}_{wx}\Bigr)_{w\vdash x\otimes x}}{(\pm\det \mathcal{R})^{1/n}}\equiv(\mathrm{ diag}\ \eta_1,\ldots,\eta_n).
\end{equation}
Equations~(\ref{RURURc}) together with the orthogonality condition
\begin{equation}
\sum_{k=1}^n\mathcal{U}_{ik}\mathcal{U}_{lk}=\delta_{il}
\label{orth}
\end{equation}
are the constraints for the $\mathcal{U}$-matrix. At a first glance, the explicit solution is not available since the obtained equations are highly non-linear. Yet, they can be reduced to the linear form in some particular examples.

Since $\mathcal{U}$ is assumed to be orthogonal and symmetric, it can be expressed via a symmetric linear projector $\mathcal{P}$:
\begin{equation}
\mathcal{U}=1-2\mathcal{P},\ \ \mathcal{P}^2=\mathcal{P},\ \ \mathcal{P}^{\dagger}=\mathcal{P}\ \ \Rightarrow\ \mathcal{U}^2=1,\ \ \mathcal{U}^{\dagger}=\mathcal{U}.
\label{UP}
\end{equation}
The simplest possible form of the projector is
\begin{equation}
\mathrm{rank}\ \mathcal{P}=1\ \Rightarrow\ \mathcal{P}=u_iu_j\ \Rightarrow\ \mathcal{ U}_{ij}=\delta_{ij}-2u_iu_j.
\label{UP1}
\end{equation}
The substitution of~(\ref{UP1}) in~(\ref{RURURc}) then gives
\begin{equation}
\delta_{il}\xi_i^2-2u_iu_l(\xi_l^2+\xi_i^2+\xi_i\xi_l) +4u_iu_l\Bigl((\xi_i+\xi_l)\sum_{j=1}^n u_j^2\xi_j+\sum_{j=1}^n u_j^2\xi_j^2\Bigr) -8u_iu_l(\sum_{j=1}^n u_j^2\xi_j)^2=\delta_{ij}\eta_{i}.
\label{RURUR1}
\end{equation}
For the off-diagonal components this gives
\begin{equation}
-(\xi_l^2+\xi_i^2+\xi_i\xi_l) +2\left((\xi_i+\xi_l)\sum_{j=1}^n u_j^2\xi_j+\sum_{j=1}^n u_j^2\xi_j^2\right) -4\left(\sum_{j=1}^n u_j^2\xi_j\right)^2=0,\ \ \ i\ne l.
\label{RURURh}
\end{equation}
Condition~(\ref{UP}) then reduces to
\begin{equation}
\sum_{i=1}^n u_i^2=1.
\label{P1}
\end{equation}
The projector in the form~(\ref{UP1}) is the only possible choice for the mixing matrix blocks of size $n=2$. When $n=3$, there is also the case of the $\mathrm{rank}\ \mathcal{P}=2$ projectors, which can be reduced to the previous case using the relations
\begin{equation}
\mathcal{U}=1-2\mathcal{P}=2(\mathcal{P}-1)-1,\ \ (1-\mathcal{P})^2=1-\mathcal{P},\ \ \mathrm{ rank}(1-\mathcal{P})=n-\mathrm{rank} \mathcal{P}.
\end{equation}
The explicit solutions for the $\mathcal{U}$-matrix for $n=2,3$ is calculated in Sec.s~\ref{sec:2x2mb},~\ref{sec:3x3mb}. For $n\ge 4$ for the $\mathrm{rank} \mathcal{P}=1$, system~(\ref{RURURh},\ref{P1}) turns to be incompatible and $\mathrm{rank} \mathcal{P}=2$ cannot be reduced to~(\ref{UP1}). Hence, there is no solutions of the form~(\ref{UP1}) for $n\ge 4$.

Once the $\mathcal{U}$-matrices are found, the normalized eigenvalues $\eta$ for the higher representations can be immediately determined from~(\ref{RURUR1}). Another way is to determine $\eta$ straight from~(\ref{RURUR}) supposing that $\mathcal{U}^{\dagger}=\mathcal{U}$. Indeed, introducing the notation $\mathrm{diag}\Bigl(\mathcal{R}_{wx}\Bigr)_{w\vdash x\otimes x}\equiv \tilde{\mathcal{R}}$, we have:
\begin{equation}
\mathcal{U}\mathcal{R}\mathcal{U}\mathcal{R}\mathcal{U}=\tilde{\mathcal{R}} \ \ \Rightarrow\ \ \left\{\begin{array}{l}\mathcal{U}\mathcal{R}\mathcal{U}=\tilde{\mathcal{R}}\mathcal{U}\mathcal{R}^{-1}\\\mathcal{U}\mathcal{ R}\mathcal{U}=\mathcal{R}^{-1}\mathcal{U}\tilde{\mathcal{R}}\end{array}\right.\ \ \Rightarrow\ \tilde{\mathcal{R}}\mathcal{ U}\mathcal{R}^{-1}-\mathcal{R}^{-1}\mathcal{U}\tilde{\mathcal{R}}=0.
\end{equation}
If $\mathcal{R}$ and $\tilde{\mathcal{R}}$ are rescaled according to~(\ref{eq:norm}), then the last equation rewritten in components gives
\begin{equation}
(\xi^{-1}_i\eta_j-\xi^{-1}_j\eta_i)\mathcal{U}_{ij}=0.
\end{equation}
wherefrom $\xi_i\eta_i=C$ provided that $\mathcal{U}_{ij}\ne 0$ for all $i,j$ (as it is for all known $\mathcal{U}$-matrices). From the relation on the determinants for the matrices from~(\ref{RURUR}) it follows that $C^3=1$ and one can set $C=1$ since the common phase factor is inessential for the definition of the normalized eigenvalues. Thus, the normalized eigenvalues should satisfy
\begin{equation}
\eta_i=\xi_i^{-1}.
\end{equation}
The same relation can be obtained from the general formula $\mathcal{R}_{xy|Q}=q^{\varkappa_Q-\varkappa_x-\varkappa_y}$ (according to~(\ref{eq:Rmeig})).

\subsubsection{Mixing matrices of size $2\times 2$ \label{sec:2x2mb}}

Constraints~(\ref{RURURh},~\ref{P1}) provide two equations for $u_1^2$ and $u_2^2$, one quadratic and one linear:
\begin{equation}
\begin{array}{c}
4(u_1^4\xi_1^2+2\xi_1\xi_2u_1^2u_2^2+u_2^4\xi_2^2)-2(2\xi_1^2+\xi_1\xi_2)u_1^2 -2(2\xi_2^2+\xi_1\xi_2)u_1^2+\xi+1^2+\xi_1\xi_2+\xi+2^2=0,
\\
u_1^2+u_2^2=1.
\end{array}
\end{equation}
Eliminating $u_2^2$ and using that $\xi_2\xi_1=-1$, one obtains
\begin{equation}
\left(2u_1^2(\xi_1-\xi_2)-\xi_1+\xi_2+1\right)\left(2u_1^2(\xi_1-\xi_2)-\xi_1+\xi_2-1\right)=0.
\end{equation}
Hence, system~(\ref{RURURh},~\ref{P1}) possesses a single solution (up to permutation of indices), and this solution is rational w.r.t $u_1^2$ and $u_2^2$.
Substitution of this solution to~(\ref{UP}) yields the $\mathcal{U}$-matrix
\begin{equation}
\left(
\begin{array}{cc}
\mathcal{U}_{11}&\mathcal{U}_{12}\\
\mathcal{U}_{12}&-\mathcal{U}_{11}
\end{array}
\right)=\left(
\begin{array}{cc}
\frac{1}{\xi_2-\xi_1}&\frac{\sqrt{\xi_1^2+1+\xi_2^2}}{\xi_1-\xi_2}\\
\frac{\sqrt{\xi_1^2+1+\xi_2^2}}{\xi_1-\xi_2}&\frac{1}{\xi_1-\xi_2}
\end{array}
\right).
\end{equation}
This expression coincides with the one in~\cite{IMMM3}.

\subsubsection{Mixing matrices of size $3\times 3$ \label{sec:3x3mb}}

For $n>2$~(\ref{RURURh}) provides not one but several relations (one per each off-diagonal element of the symmetric matrix) and the difference of any two of them gives a linear equation for $u_i^2$:
\begin{equation}
-(\xi_l^2-\xi_m^2+\xi_i(\xi_l-\xi_m))+2(\xi_l-\xi_m)\sum_{j=1}^n u_j^2\xi_j=0,\ \ \ i\ne l\ne m.
\label{RURURd}
\end{equation}
For $n=3$~(\ref{RURURd}) provides 3 different equations, each of which factors as
\begin{equation}
\begin{array}{lcl}
i=1,\ l=2,\ m=3 & & (\xi_2-\xi_3)(2\xi_1u_1^2+2\xi_2u_2^2+2\xi_3u_3^2-\xi_1-\xi_2-\xi_3)=0,
\\
i=2,\ l=1,\ m=3 & & (\xi_1-\xi_3)(2\xi_1u_1^2+2\xi_2u_2^2+2\xi_3u_3^2-\xi_1-\xi_2-\xi_3)=0,
\\
i=3,\ l=1,\ m=2 & & (\xi_2-\xi_3)(2\xi_1u_1^2+2\xi_2u_2^2+2\xi_3u_3^2-\xi_1-\xi_2-\xi_3)=0.
\end{array}
\label{d3}
\end{equation}
All of them have the same factor (the second one) that depend on $u$. This factor is linear in $u_1^2$, $u_2^2$ and $u_3^2$ as well as~(\ref{P1}), which for $n=3$ looks like:
\begin{equation}
u_1^2+u_2^2+u_3^2=1.
\label{orth3}
\end{equation}
Expressing $u_2^2$ and $u_3^2$ from~(\ref{d3},\ref{orth3}) and substituting the results in~(\ref{RURURh}) for $i=1$, $l=2$, for example, one unexpectedly obtains a linear equation for $u_1^2$:
\begin{equation}
2u_1^2(\xi_1^2-\xi_1\xi_2-\xi_1\xi_3+\xi_2\xi_3)-\xi_1^2-\xi_2\xi_3=0.
\end{equation}
The solutions of this equation together with the corresponding solutions for $u_2^2$ and $u_3^2$ after substituting in~(\ref{UP}) reproduce the formula in~\cite{IMMM3} for the $3\times 3$ mixing block:
\begin{equation}
\begin{array}{cccccc}
\multicolumn{2}{c}{\mathcal{U}_{11}=-\cfrac{\xi_1(\xi_2+\xi_3)}{(\xi_1-\xi_2)(\xi_1-\xi_3)}\ ,}
&
\multicolumn{2}{c}{\mathcal{ U}_{22}=-\cfrac{\xi_2(\xi_1+\xi_3)}{(\xi_2-\xi_1)(\xi_2-\xi_3)}\ ,}
&
\multicolumn{2}{c}{\mathcal{U}_{33}=-\cfrac{\xi_3(\xi_1+\xi_2)}{(\xi_3-\xi_1)(\xi_3-\xi_2)}\ ,}
\\ &&&&& \\
\multicolumn{3}{c}{\mathcal{U}^2_{12}=-\cfrac{(\xi_1^2-1)(\xi_2^2-1)}{\xi_1\xi_2(\xi_1-\xi_2)^2(\xi_1-\xi_3)(\xi_2-\xi_3)}\ ,}
&
\multicolumn{3}{c}{\mathcal{ U}^2_{23}=-\cfrac{(\xi_2^2-1)(\xi_3^2-1)}{\xi_2\xi_3(\xi_2-\xi_3)^2(\xi_3-\xi_2)(\xi_3-\xi_1)}\ ,}
\\ &&&&& \\
\multicolumn{6}{c}{\mathcal{U}^2_{13}=-\cfrac{(\xi_1^2-1)(\xi_3^2-1)}{\xi_1\xi_3(\xi_1-\xi_3)^2(\xi_1-\xi_2)(\xi_3-\xi_2)}\ .}
\end{array}
\label{U3xidiag}
\end{equation}

\subsection{Eigenvalue conjecture for different representations\label{sec:eigdiff}}

With help of the methods discussed in Sec.~\ref{sec:conU}, the eigenvalue conjecture can be extended to the case with different representations on the different connection components of the link. This generalization can potentially give a possibility of using the Reshetikhin-Turaev straightforwardly to calculate invariants of colored links. Our generalization of the eigenvalue conjecture, which we illustrate below, is:

\emph{The mixing matrices of the multicolored braid are completely defined by the normalized eigenvalues of all the $\mathcal{R}$-matrices that can appear in this braid.}

\subsubsection{Two different representations}

\underline{The conjecture states}: \emph{each} of the mixing matrices $\mathcal{U}_{QSQ}$ and $\mathcal{U}_{QQS}=\mathcal{U}_{SQQ}^{\dagger}$  is expressed through \emph{two} sets of the normalized eigenvalues: that of $\mathcal{R}_{QQ}$ \emph{and} that of $\mathcal{ R}_{SQ}$.

\

In this case, one can write two relations instead of~(\ref{RURUR}):
\begin{equation}
\begin{array}{l}
\mathcal{U}_{xyx}\mathcal{R}_{(yx)x}\mathcal{U}_{xxy}\mathcal{R}_{(xx)y} \mathcal{U}_{xxy}=\mathcal{R}_{wx},
\\
\mathcal{U}_{xxy}\mathcal{R}_{(xy)x}\mathcal{U}_{xyx}\mathcal{R}_{(xy)x} \mathcal{U}_{yxx}=\mathcal{R}_{wy}.
\end{array}
\label{Rcolxxy}
\end{equation}
though the first of them is enough to define \emph{both} the mixing matrices and the eigenvalues of the corresponding matrix at the r.h.s.

\paragraph{Case of $\mathbf{2\times 2}$ mixing blocks.} Writing $\mathcal{R}$ and $\mathcal{U}$ in matrix form and substituting them in~(\ref{Rcolxxy}), one obtains two matrix equations:
\begin{equation}
\setlength{\arraycolsep}{3pt}
\begin{array}{l}
\left(\begin{array}{ccc}-c_x&s_x\\s_x&c_x\end{array}\right)\left(\begin{array}{ccc}\xi_{yx}&\\&-\xi^{-1}_{yx}\end{array}\right)
\left(\begin{array}{ccc}-c_y&s_y\\s_y&c_y\end{array}\right)\left(\begin{array}{ccc}\xi_{yy}&\\&-\xi^{-1}_{yy}\end{array}\right)
\left(\begin{array}{ccc}-c_y&s_y\\s_y&c_y\end{array}\right)=\left(\begin{array}{ccc}\xi_{wy}&\\&-\xi^{-1}_{wy}\end{array}\right),
\\ \\
\left(\begin{array}{ccc}-c_y&s_y\\s_y&c_y\end{array}\right)\left(\begin{array}{ccc}\xi_{xy}&\\&-\xi^{-1}_{xy}\end{array}\right)
\left(\begin{array}{ccc}-c_x&s_x\\s_x&c_x\end{array}\right)\left(\begin{array}{ccc}\xi_{yx}&\\&-\xi^{-1}_{yx}\end{array}\right)
\left(\begin{array}{ccc}-c_y&s_y\\s_y&c_y\end{array}\right)=\left(\begin{array}{ccc}\xi_{wx}&\\&-\xi^{-1}_{wx}\end{array}\right).
\end{array}
\label{2yyxeqs}
\setlength{\arraycolsep}{6pt}
\end{equation}
The solution of this system is quite simple.
\begin{itemize}
 \item{The compatibility condition on the off-diagonal components of the first equation can be reduced to the equation linear w.r.t. $c_y^2$ and $s_y^2=1-c_y^2$. From this equation $c_y^2$ can be found.}
 \item{After $c_y^2$ is evaluated, $c_x,\ s_x$ can be found as a solution of the obtained degenerate system.}
 \item{Substituting the obtained $c_x^2$ and $c_y^2$ back into~(\ref{Rcolxxy}) one can obtain $\xi_{xx}$ and $\xi_{xy}$.}
\end{itemize}
The solutions constructed in this way are
\begin{equation}
\begin{array}{cc}
c^2_x=\cfrac{(\xi_{xx}^2-\xi_{xy}^4)}{(1-\xi_{xy}^4)(1+\xi_{xx}^2)}\ ,
&
s^2_x=\cfrac{(1-\xi_{xx}^2\xi_{xy}^4)}{(1-\xi_{xy}^4)(1+\xi_{xx}^2)}\ ,
\\ \\
c^2_y=\cfrac{\xi_{xy}^2(1-\xi_{xx}^2)^2}{\xi_{xx}^2(1-\xi_{yx}^4)^2}\ ,
&
s^2_y=\cfrac{(\xi_{xx}^2-\xi_{yx}^4)(1-\xi_{xy}^4\xi_{xx}^2)}{\xi_{xx}^2(1-\xi_{yx}^4)^2}\ ,
\\ \\
\xi_{wx}=-\cfrac{1}{\xi_{yx}}\ ,
&
\xi_{wy}=-\cfrac{1}{\xi_{xx}}\ .
\end{array}
\label{2xxyanswer}
\end{equation}

\subsubsection{Three different representations}

\underline{The conjecture states:} \emph{each} of the mixing matrices $\mathcal{U}_{zxy}=\mathcal{U}_{yxz}^{\dagger}$, $\mathcal{U}_{xzy}=\mathcal{ U}_{yxz}^{\dagger}$ and $\mathcal{U}_{zyx}=\mathcal{U}_{xyz}^{\dagger}$ is expressed through \emph{three} sets of the normalized eigenvalues: that of $\mathcal{R}_{yz}$, that of $\mathcal{R}_{xz}$ \emph{and} that of $\mathcal{ R}_{yx}$.

\

In this case, there are three equations instead of one~(\ref{RURUR}):
\begin{equation}
\begin{array}{l}
\mathcal{U}_{xyz}\mathcal{R}_{(yz)x}\mathcal{U}_{xzy}\mathcal{R}_{(xz)y} \mathcal{U}_{zxy}=\mathcal{ R}_{wz},
\\
\mathcal{U}_{xzy}\mathcal{R}_{(zy)x}\mathcal{U}_{xyz}\mathcal{R}_{(xy)z} \mathcal{U}_{yxz}=\mathcal{R}_{wy},
\\
\mathcal{U}_{zyx}\mathcal{R}_{(yx)z}\mathcal{U}_{zxy}\mathcal{R}_{(zx)y} \mathcal{U}_{xzy}=\mathcal{R}_{wx}.
\end{array}
\label{Rcolxyz}
\end{equation}
In fact, any \emph{two} of them are enough to define \emph{all} the mixing matrices and the eigenvalues of the corresponding matrices at the r.h.s.

\paragraph{Case of $\mathbf{2\times 2}$ mixing blocks.}
The first of equations~(\ref{Rcolxyz}) this case becomes
\begin{equation}
\setlength{\arraycolsep}{3pt}
\left(\begin{array}{ccc}-c_y&s_y\\s_y&c_y\end{array}\right)\left(\begin{array}{ccc}\xi_{zy}&\\&-\xi^{-1}_{zy}\end{array}\right)
\left(\begin{array}{ccc}-c_z&s_z\\s_z&c_z\end{array}\right)\left(\begin{array}{ccc}\xi_{xz}&\\&-\xi^{-1}_{xz}\end{array}\right)
\left(\begin{array}{ccc}-c_x&s_x\\s_x&c_x\end{array}\right)=\left(\begin{array}{ccc}\xi_{wz}&\\&-\xi^{-1}_{wz}\end{array}\right).
\label{2xyzeqs}
\setlength{\arraycolsep}{6pt}
\end{equation}
There are four equations for the off-diagonal components of two of the equations in system~(\ref{Rcolxyz}). The solutions of this system can be constructed by studying the compatibility conditions  of these four equations (these conditions can be reduced to the linear equations for squared parameters of the mixing matrices). The answer reads
\begin{equation}
\begin{array}{lr}
c^2_z=\cfrac{(\xi_{xz}^2-\xi_{xy}^2\xi_{zy}^2)(\xi_{zy}^2-\xi_{xz}^2\xi_{xy}^2)}{\xi_{xy}^2(1-\xi_{zy}^4)(1-\xi_{xz}^4)}\ ,
&
s^2_z=\cfrac{(\xi_{xy}^2-\xi_{xz}^2\xi_{zy}^2)(1-\xi_{zy}^2\xi_{xz}^2\xi_{xy}^2)}{\xi_{xy}^2(1-\xi_{zy}^4)(1-\xi_{xz}^4)}\ ,
\\
\multicolumn{2}{c}{\xi_{wz}=-\cfrac{1}{\xi_{yx}}\ .}
\end{array}
\label{2yxzanswer}
\end{equation}
The signs in~(\ref{2yxzanswer}) can be restored by substituting the obtained expressions in~(\ref{2xyzeqs}).

\subsection{Blocks in the non-diagonal $\mathcal{R}$-matrices\label{sec:blfm}}
Following the approach of Sec.s~\ref{sec:conU}-\ref{sec:eigdiff}, one can also determine the form of blocks in the $\mathcal{R}$-matrices, which was discussed in Sec.~\ref{sec:genR}.
From consistency of the cabling approach, one can write the constrains directly on the elements of the non-diagonal $R$-matrices:

\begin{equation}
\mathrm{diag}\Bigl(\mathcal{R}^2_{ux|w}\Bigr)_{\substack{w\vdash u\otimes x\\w\otimes x\dashv Q}}=\mathcal{ R}_{x^{m+1}\otimes x}\mathcal{R}_{x\otimes x^{m+1}}.
\label{Rbl}
\end{equation}

Then, from relation~(\ref{RURUR}) and co-product rule~(\ref{Rmcoprod}) it follows that
\begin{equation}
\begin{array}{r}
\mathrm{diag}\Bigl(\mathcal{R}^2_{ux|w}\Bigr)_{\substack{w\vdash u\otimes x\\w\otimes x\dashv Q}}=\mathcal{R}_{x^m\otimes (x\otimes x)}\mathcal{R}_{(x^m\otimes x)\otimes x}\mathcal{R}_{(x\otimes x^m)\otimes x}\mathcal{R}_{x^m \otimes (x\otimes x)}=
\\
=\mathcal{R}_{x^m\otimes (x\otimes x)}\mathrm{ diag}\Bigl(\mathcal{R}^2_{wx|Q}\Bigr)_{\substack{w\vdash u\otimes x\\w\otimes x\dashv Q}}\mathcal{R}_{x^m\otimes (x\otimes x)},
\end{array}
\end{equation}
where $u$ is an irreducible representation such that $u\vdash x^{m-1}$ and $Q\vdash u\otimes x^2$. Different $u$ never appear in the same mixing block due to properties of the $\mathcal{R}$-matrices. The size $n$ of the non-diagonal block is equal to the number of representations $w$ that satisfy both $w\vdash u\otimes x$ and $w\otimes x\dashv Q$. In the following, we define the explicit form of the blocks in the non-diagonal $\mathcal{R}$ matrices in two cases where the mixing blocks are of the size $n=2$. These cases are the fundamental representation case $x=\Box$ and case where the representations $x$, $u$, $Q$ and hence all $w$ are described by the hook diagrams.

Introducing the notations $r_{11}$, $r_{12}$ and $r_{22}$ for the elements of the unknown block and $\lambda_1$ and $\lambda_2$ for the corresponding eigenvalues of $\mathcal{R}_{ux|w}$, equation~(\ref{Rbl}) for the $2\times 2$ block can be rewritten in the following form:
\begin{equation}
\left(\begin{array}{cc}r_{11}&r_{12}\\r_{12}&r_{22}\end{array}\right)
\left(\begin{array}{cc}\lambda_1^2&0\\0&\lambda_2^2\end{array}\right)
\left(\begin{array}{cc}r_{11}&r_{12}\\r_{12}&r_{22}\end{array}\right)=
\left(\begin{array}{cc}\tilde\lambda_1^2&0\\0&\tilde\lambda_2^2\end{array}\right).
\label{Rbl2}
\end{equation}
The equations for the off-diagonal elements can be written as
\begin{equation}
r_{12}(\lambda_1^2r_{11}+\lambda_2^2r_{22})=0.
\label{Rbloff}
\end{equation}
We also use the well known property of the $2\times 2$ matrix:
\begin{equation}
\begin{array}{l}
r_{11}+r_{22}=\mu_1+\mu_2,
\\ \\
r_{11}r_{22}-r_{12}^2=\mu_1\mu_2.\nn
\end{array}
\label{Rblev}
\end{equation}
where $\mu_1$ and $\mu_2$ are the eigenvalues of the non-diagonal block, which are supposed to be known. The solutions of these equations are
\begin{equation}
\begin{array}{lcr}
r_{11}=-\cfrac{(\mu_1+\mu_2)\lambda_2^2}{\lambda_1^2-\lambda_2^2}\ ,
&
r_{22}=-\cfrac{(\mu_1+\mu_2)\lambda_1^2}{\lambda_1^2-\lambda_2^2}\ ,
&
r_{12}^2=\cfrac{(\mu_1\lambda_1^2+\mu_2\lambda_2^2)(\mu_1\lambda_2^2+\mu_2\lambda_1^2)}{(\lambda_1^2-\lambda_2^2)^2}\ .
\end{array}
\label{Rblans}
\end{equation}

\paragraph{The fundamental representation.} In the case of $x=\square$, $Q$ is $u$ with the pair of added boxes $(i_1,j_1)$ and  $(i_2,j_2)$ hence there are two $w$ obtained from $u$ by addition of one or the other of the two boxes. According to the general formula given by~(\ref{eq:Rmeig}),
\begin{equation}
\mu_1=q,\ \ \mu_2=-q^{-1},\ \ \lambda_{1,2}=q^{\varkappa_{w_{1,2}}-\varkappa_u}= q^{j_{1,2}-i_{1,2}}.
\label{fblev}
\end{equation}
The substitution of~(\ref{fblev}) into~(\ref{Rblans}) gives the formulae from Sec.~\ref{sec:genR} for the  blocks in the fundamental $R$-matrices:
\begin{equation}
\begin{array}{c}
\mathcal{R}_{1^m\otimes (1\otimes 1)|Q}=\left(\begin{array}{ccc}-\cfrac{q-q^{-1}}{q^n(q^{n}-q^{-n})}&&\cfrac{\sqrt{\left(q^{n+1}-q^{-n-1}\right)\left(q^{n-1}-q^{1-n}\right)}}{q^{n}-q^{-n}}\\\\
\cfrac{\sqrt{\left(q^{n+1}-q^{-n-1}\right)\left(q^{n-1}-q^{1-n}\right)}}{q^{n}-q^{-n}}&&\cfrac{q^n\left(q-q^{-1}\right)}{q^{n}-q^{-n}}\end{array}\right),
\\ \\
Q=w_1\cup(i_1,j_1)=w_2\cup(i_2,j_2),\ \ \ n=j_2-i_2+j_1-i_1.
\end{array}
\end{equation}

\paragraph{Hook representations.} If representations $x$ and $u$ are the hook ones,
\begin{equation}
x=[r_1,1^{s_1}],\ \ \ u=[r_2, 1^{s_2}],
\end{equation}
then there are only three possible hook representations $Q$ since
\begin{equation}
\begin{array}{c}
[r_2,1^{s_2}]\times[r_1,1^{s_1}]\times[r_1,1^{s_1}]=
\\
=\left([r_2+r_1, 1^{s_2+s_1}]+[r_2+r_1-1, 1^{s_2+s_1+1}]+\mbox{not one-hook}\right)\times[r_1,1^{s_1}]=
\\
=[2r_1+r_2, 1^{2s_1+s_2}]+2[2r_1+r_2-1, 1^{2s_1+s_2+1}]+[2r_1+r_2-2, 1^{2s_1+s_2+2}]+\mbox{not one-hook}.
\end{array}
\end{equation}
The three possible hook representations are:
\begin{equation}
Q=[2r_1+r_2, 1^{2s_1+s_2}],\ \ Q^{\prime}=[2r_1+r_2-1, 1^{2s_1+s_2+1}],\ \ Q^{\prime\prime}=[2r_1+r_2-2, 1^{2s_1+s_2+2}],
\end{equation}
where the first and the last representations are the singlets and the second one is a doublet with the intermediate representations being
\begin{equation}
w_1=[r_1+r_2,1^{s_1+s_2}],\ \ w_2=[r_1+r_2-1,1^{s_1+s_2+1}].
\end{equation}
Since a path that ends at the hook representation $Q$ goes only through the hook representations, the relevant block of the matrix $\mathcal{R}_{x^m\otimes x\otimes x}$ corresponds to one of the two representations $y$ from the decomposition of $x\otimes x$:
\begin{equation}
y_1=[2r_1,1^{2s_1}],\ \ y_2=[2r_1-1,1^{2s_1+1}].
\end{equation}
For these representations, the eigenvalues in~(\ref{Rblans}) are given by~(\ref{eq:Rmeig}):
\begin{equation}
\begin{array}{cc}
\mu_1=q^{\varkappa_{y_1}-\varkappa_x}=q^{r_1(2r_1-1)-s_1(2s_1+1)}\equiv q^{r_1+s_1}\mu,
&
\mu_2=-q^{\varkappa_{y_2}-\varkappa_x}=-q^{-r_1-s_1}\mu,
\\
\lambda_1=q^{\varkappa_{w_1}-\varkappa_u}\equiv q^{r_2+s_2}\lambda,
&
\lambda_2=q^{\varkappa_{w_2}-\varkappa_u}=q^{-r_2-s_2}\lambda.
\end{array}
\end{equation}
Substituting these eigenvalues in~(\ref{Rblans}), one gets the blocks of the form:
\begin{equation}
\begin{array}{l}
\mathcal{R}_{x^m\otimes (x\otimes x)|Q}=q^{2r_1(r_1-1)-2s_1(s_1
+1)}\times
\\
\times\left(\begin{array}{ccc}
-\cfrac{q^r-q^{-r}}{q^{rn}(q^{rn}-q^{-rn})}&&\cfrac{\sqrt{\left(q^{rn+r}-q^{-rn-r}\right)\left(q^{rn-r}-q^{r-rn}\right)}}{q^{rn}-q^{-rn}}
\\ \\
\cfrac{\sqrt{\left(q^{rn+r}-q^{-rn-r}\right)\left(q^{rn-r}-q^{r-rn}\right)}}{q^{rn}-q^{-rn}}&&\cfrac{q^{rn}(q^r-q^{-r})}{q^{rn}-q^{-rn}}
\end{array}\right),
\end{array}
\label{eq:1hans}
\end{equation}
where
\begin{equation}
\begin{array}{c}
x=[r_1,1^{s_1}],\ \ w_1=[r_1+r_2,1^{s_1+s_2}],\ \ w_2=[r_1+r_2-1,1^{s_1+s_2+1}],\\ \\
Q=[2r_1+r_2-1, 1^{2s_1+s_2+1}],\ \ n=s_2-r_2+1,\ \ r=r_1+s_1.
\end{array}
\end{equation}

\section{Special polynomials\label{sec:sp}}

Special polynomials describe the double scaling limit of the HOMFLY polynomial with the large $N$ and small $\hbar$ (i.e., $q\equiv e^{\hbar} \rightarrow 1$ and $A=q^N=const$). The whole expansion in this limit is called a genus expansion~\cite{genexp1},~\cite{genexp2}. But here we are more interested in the zero-order term of this expansion, which is called special polynomial,~\cite{DMMSS}
\begin{equation}
\sigma^{\mathcal{K}}_Q(A)=H^{\mathcal{K}}_Q(A=q^N=const,q=1).
\end{equation}
Behavior of the colored special polynomials in this limit is obtained in~\cite{DMMSS} and studied in~\cite{genexp1},\cite{genexp2},\cite{chinsp1}-\cite{AntM2}  describes the connections between fundamental and colored special polynomials:
\begin{equation}
\sigma^{\mathcal{K}}_Q\ =\ \sigma^{\mathcal{K}^{|Q|}}_1\ =\ \left(\sigma^{K}_q\right)^{|Q|}.
\label{special}
\end{equation}
This property is called \emph{the factorization property}. It follows from the basic properties of HOMFLY polynomials, i.e., from skein-relations~(\ref{skeinpic}), from the possibility to apply cabling procedure, from the factorization of the HOMFLY polynomial for the split union of knots, and from the theorem that the fundamental unreduced (not divided by the unknot) HOMFLY of the $n$-component link diverges in the $q\rightarrow 1$ limit as $(q-q^{-1})^{-n}$~\cite{linksing}.

It follows from the cabling procedure that the colored HOMFLY polynomial in representation $Q$ diverges in the limit $q \rightarrow 1$ as fundamental HOMFLY the $|Q|$-component link, i.e., as $(q-q^{-1})^{-|Q|}$. Indeed, there cannot be more than $|Q|$ components and there is always a term with $|Q|$ components in the cabling for the representation $Q$, thus $\sigma^{\mathcal{K}}_Q\ \sim\ (q-q^{-1})^{-|Q|}$. Then, using the skein relations the $|Q|$-component link can be transformed into the split union of $|Q|$  components; in the $q\rightarrow 1$ limit, this will not change the HOMFLY polynomial. Indeed, if the skein relations are applied to a crossing of two different components then the term corresponding to resolving this crossing contains one connection component less and thus diverges as $(q-q^{-1})^{-|Q|+1}$. In account for the coefficient $(q-q^{-1})$ before this term, the corresponding expression diverges as $(q-q^{-1})^{-|Q|+2 }$ and hence is less singular as $q \rightarrow 1$ then the rest two terms and can be set as zero in this limit. The resulting identity reads that HOMFLY polynomials of the two links with the linking number differing by one are equal in the $q \rightarrow 1$ limit. Hence, $\sigma^{\mathcal{K}^{|Q|}}_1\ =\ \left(\sigma^{K}_q\right)^{|Q|}$. Note, that if the vertical framing of the knot is used then there is no $A$ in the skein relations; hence relation~(\ref{special}) do not contain $A$ explicitly.

Our considerations were done for the HOMFLY polynomials evaluated in the vertical framing (Sec.~\ref{framing}). However, one can easily see that~(\ref{special}) is as well correct in the topological framing.

\section{Alexander polynomials\label{sec:ap}}

Alexander polynomial is a HOMFLY polynomial in the $A\rightarrow 1$ limit,
\begin{equation}
A_Q^{\mathcal{K}}(q)=H^{\mathcal{K}}_Q(A=q^N=1,q=const).
\end{equation}
For hook representations $Q$, there is the conjecture concerning the representation dependence of the Alexander polynomial~\cite{DMMSS},~\cite{IMMM1}. This conjecture asserts that
\begin{equation}
A^{\mathcal{K}}_{Q}(q)=A^{\mathcal{K}}_1(q^{|Q|}).
\label{Alexander}
\end{equation}
The proof of this fact in the case can be performed using the method similar to one described in Sec.~\ref{sec:Rm}. The answers for the $R$-matrices presented in that section can be generalized to the hook diagrams.

In the limit $A\rightarrow 1$, only the factor $\left(A-A^{-1}\right)$ in the Schur polynomials is essential. The power of this term in $S_Q^*$ is equal to the number of hooks in the diagram $Q$. This means that if one studies the representation with the hook diagram $[r,1^s]$, then all the diagrams in the character expansion for HOMFLY polynomial that are non-hook ones vanish from the answer for the reduced HOMFLY polynomial in the limit $A\rightarrow 1$. For all diagrams appearing in the answer, the $\mathcal{R}$-matrices consist of blocks of a size not exceeding $2\times 2$ (see Sec.~\ref{sec:blfm}), similarly to what happens in the fundamental case as. These blocks has form~(\ref{eq:1hans}), and it can be straightforward checked that they satisfy the following property with respect to the substitution $q\rightarrow q^{r+s}$:
\begin{equation}
\mathcal{R}_{\square}^i\ \stackrel{q\rightarrow q^{r+s}}{\longrightarrow}\ \mathcal{ R}_{[r,1^s]}^i.\label{AlexanderRm}
\end{equation}
This means that property~(\ref{Alexander}) is satisfied for the individual coefficients of the Schur polynomials in character expansion~(\ref{eq:charexp}). It can be checked that the characters (Schur polynomials) themselves in the limit $A\rightarrow 1$ also satisfy the needed relation. Hence, the Alexander polynomials should also satisfy relation~(\ref{Alexander}).

\section{Conclusion}

In the present paper, we studied the method of evaluating the colored HOMFLY polynomials via the fundamental ones. This method is called the cabling procedure and it consists of three steps. The first step is to construct the cabled knot from the initial one. This step can be described by a simple picture (see Sec.~\ref{sec:CP}). The second step is to find the projector, which describes the combination of the fundamental HOMFLY polynomials into which the colored HOMFLY polynomial decomposes. The third step is to evaluate the colored HOMFLY polynomials.

To evaluate the fundamental HOMFLY polynomials using the Reshetikhin-Turaev approach (the one that we studied) the fundamental $\mathcal{R}$-matrices are needed. Their matrix form can be described using quite simple paths construction provided in Sec.~\ref{sec:Rm}. This answer in principle allows one to construct the fundamental HOMFLY polynomial of \emph{any} knot. The main problem is the calculational difficulty of evaluation of the knot polynomials, which increases rapidly with increasing the minimal number of strands in the braid representing the knot.

The paths description also allows one to construct the matrix form of the projectors onto any representation (see Sec.~\ref{sec:proj}). Though this matrix form describes only the projector onto the first cable it is enough to evaluate in principle the HOMFLY polynomial in \emph{any} representation of \emph{any} knot. The main problem is again the computational difficulties. Since the cabling procedure implies that the HOMFLY polynomials in representation $Q$ are expressed through the fundamental ones for the knots with $|Q|$ times more strands in the braid representation then there is in the braid representation of the initial knot, the difficulties are much more severe for the colored HOMFLY polynomials. At the moment the explicit calculation can be done up to 12 strands in the braid. This allows to evaluate the required fundamental HOMFLY in various representations for a number of knots and links. The answers are listed in Appendices~\ref{app:3str}-\ref{app:FigWh}.

We also discussed the group theory description of the cabling procedure. This description allows one to derive the form of the fundamental $\mathcal{R}$-matrices described in Sec.~\ref{sec:Rm}. Other applications of the cabling procedure include explanations of the conjectures stated in the previous papers: of the eigenvalue conjecture~\cite{IMMM3} and of the dependence of the Alexander and special polynomials on the representation~\cite{DMMSS}.

Though there is a principal method to construct any colored HOMFLY polynomial a lot of studies remain to be done. An internal structure of the HOMFLY polynomials in different representations for different knots remains to be found. There are several examples when the general formula (or at least more general than just one particular HOMFLY polynomial) is already known; they are torus knots~\cite{RJ}-\cite{DMMSS}, twist knots~\cite{IMMM2},\cite{Inds6}, and double braid knots~\cite{MMM3}.
But there are much more examples that are completely mysterious. Obtaining the general formulas for the series of knots and representations would be very helpful not only for the exploring of HOMFLY polynomials themselves, but also for the studies of some adjacent topics like difference equations and $\tau$-functions~\cite{Gaur}-\cite{FGSS},~\cite{MMM1},\cite{genexp1},\cite{genexp2},  superpolynomials~\cite{sup1}-\cite{Art},~\cite{DMMSS}-\cite{GorNeg} and Khovanov homologies~\cite{KhR:Kh}-\cite{KhR:DM2}.

For the sake of obtaining the general formulae for the HOMFLY polynomials the eigenvalue conjecture~\cite{IMMM3} may turn very fruitful. This conjecture in fact implies that the HOMFLY polynomial is fully described in terms of the ${\cal R}$-matrix eigenvalues although the straightforward expression involves several ${\cal R}$-matrices that do not commute. The cabling description of the eigenvalue conjecture (Sec.~\ref{sec:eig}) gives a possible tool for further studies in this direction.

Another direction of studies is the complexification of the topology of the space (see, e.g.,~\cite{Pras} and references therein). Nearly all the known answers represent the theories on the $S^3$ manifold and almost nothing is known even about the next in simplicity case of $S^1\times S^2$.

\section*{Acknowledgements}

The authors would like to thank G.Aminov, I.Danilenko, D.Diakonov, H.Itoyama, P.Dunin-Barkowski, A.Mironov, A.Morozov, A.Popolitov, A.Sleptsov, D.Vasiliev, Ye.Zenkevich and all the participants of the ITEP weekly seminars for the very useful discussions and criticism. The authors are also indebted to S.Arthamonov for the help with calculations. Our work is partly supported by Ministry of Education and Science of the Russian Federation, by NSh-3349.2012.2, by RFBR grants 12-01-00482 and 12-02-31078\_young\_a, by joint grants 12-02-92108-Yaf-a (An.Mor.) and 13-02-91371-ST-a and by the Dynasty Foundation (An.Mor.).

\newpage

\appendix

\section{Rank-one projectors onto level-four representations\label{app:proj}}
See Sec.~\ref{sec:proj} for the definitions.
We label the multiple representations as
\begin{equation}
1\otimes1\otimes1\otimes1=4+\underline{\underline{31}}+\underline{\overline{31}}+\overline{\overline{31}}+\underline{22}+
\overline{22}
+\underline{\underline{211}}+\overline{\overline{211}}+\underline{\overline{211}}+1111,
\end{equation}
where the summands are ordered in the same way as the branches in the representation tree (see Sec.~\ref{sec:genR}).
\begin{equation}
\begin{array}{l}
P_{\underline{\underline{31}}}=
\cfrac{1}{q^2[2]_q[4]_q}\Bigl(1+q(R_1+R_2)-\cfrac{1}{q^3[3]_q}R_3+
q^{-2}(R_1R_2+R_2R_1)-\cfrac{1}{q^2[3]_q}(R_2R_3+
R_3R_2+R_1R_3)
+\\ \\
\phantom{P_{\underline{\underline{31}}}=
\cfrac{1}{q^2[2]_q[4]_q}}+q^3R_1R
_2R_1-\cfrac{1}{q[3]_q}(R_2R_3R_2
+R_1R_2R_3+R_2R_1R_3
+R_3R_1R_2+R_3R_2R_1)
-\\ \\
\phantom{P_{\underline{\underline{31}}}=
\cfrac{1}{q^2[2]_q[4]_q}}-\cfrac{1}{[3]_q}(R_1R_2R_1R_3+R_1R_2R_3R_2
+R_1R_3R_2R_1
+R_2R_1R_3R_2+R_2R_3R_2R_1)
-\\ \\\phantom{P_{\underline{\underline{31}}}=
\cfrac{1}{q^2[2]_q[4]_q}}-
\cfrac{q}{[3]_q}(\mathcal{R}_1R_2R_1R_3R_2
+R_1R_2R_3R_2R_1
+R_2R_1R_3R_2R_1)
-\cfrac{q^2}{[3]_q}R_1R_2R_1R_3R_2R_1\Bigr),
\\ \\
P_{\overline{\underline{31}}}=
\cfrac{1}{q^2[2]_q[4]_q}\Bigl(1+qR_1-\cfrac{1}{q^2[2]_q}R_2+\cfrac{q^3}{[3]_q}R_3-
\cfrac{1}{q[2]_q}(R_1R_2+R_2R_1)+q^2(R_2R_3
+R_3R_2)-\cfrac{q}{[2]_q}\mathcal{ R}_1R_3
-\\ \\-
\cfrac{1}{[2]_q}R_1R_2R_1
+\cfrac{q^4+q^2-2+q^{-2}}{[2]_q[3]_q}R_2R_3R_2
-\cfrac{q^2}{[2]_q[3]_q}(R_1R_2R_3+R_2R_1R_3
+R_3R_1R_2+R_3R_2R_1)
-\\ \\-
\cfrac{q^2}{[2]_q[3]_q}(R_1R_2R_1R_3
+R_1R_3R_2R_1)
+\cfrac{q(q^4+q^2-2+q^{-2})}{[2]_q[3]_q}(R_1R_2R_3R_2
+R_2R_1R_3R_2+R_2R_3R_2R_1)
+\\ \\-
\cfrac{q[4]_q}{[2]_q[3]_q}(R_1R_2R_1R_3R_2
+R_2R_1R_3R_2R_1)+ \cfrac{q^2(q^4+q^2-2+q^{-2})}{[2]_q[3]_q}R_1R_2R_3R_2R_1
-\cfrac{q^2[4]_q}{[2]_q[3]_q}R_1R_2R_1R_3R_2R_1\Bigr),
\\ \\
P_{\overline{\overline{31}}}=
\cfrac{1}{q^2[2]_q[4]_q}\Bigl(1-q^{-1}R_1+\cfrac{q^2}{[2]_q}R_2+qR_3
-
\cfrac{q}{[2]_q}(R_1R_2+R_2R_1)
-R_2R_3-q(R_3R_2+R_1R_3)
+\\ \\\phantom{P_{\underline{\underline{31}}}=
\cfrac{1}{q^2[2]_q[4]_q}}+
\cfrac{q^3}{[2]_q}R_1R_2R_1+\cfrac{1}{[2]_q}R_2R_3R_2
+\cfrac{q^4}{[2]_q}R_1R_2R_3+\cfrac{q^2}{[2]_q}(R_2R_1R_3
+R_3R_1R_2+R_3R_2R_1)
+\\ \\\phantom{P_{\underline{\underline{31}}}=
\cfrac{1}{q^2[2]_q[4]_q}}+
\cfrac{q}{[2]_q}(R_1R_2R_1R_3+R_1R_3R_2R_1)
-\cfrac{q^3}{[2]_q}(R_1R_2R_3R_2+R_2R_1R_3R_2)
+\cfrac{q}{[2]_q}R_1R_2R_1R_3R_2\Bigr),
\\ \\
P_{\underline{22}}=
\cfrac{1}{[2]_q^2[3]_q}\Bigl(1+q(R_1+R_3)-\cfrac{1}{q^2[2]_q}R_2
-\cfrac{1}{q[2]_q}(R_1R_2+R_2R_1+R_3R_2
+R_1R_3)+q^2R_2R_3
+\\ \\\phantom{P_{\underline{22}}=
\cfrac{1}{[2]_q^2[3]_q}}-
\cfrac{1}{[2]_q}(R_1R_2R_1+R_2R_3R_2
+R_1R_2R_3+R_2R_1R_3
+R_3R_1R_2+R_3R_2R_1)
-\\ \\\phantom{P_{\underline{22}}=
\cfrac{1}{[2]_q^2[3]_q}}
-\cfrac{q}{[2]_q}(R_1R_2R_1R_3
+R_1R_2R_3R_2+R_1R_3R_2R_1
+R_2R_1R_3R_2)+q^{-2}R_2R_3R_2R_1
+\\ \\\phantom{P_{\underline{22}}=
\cfrac{1}{[2]_q^2[3]_q}}+
q^{-1}(R_1R_2R_1R_3R_2
+R_2R_1R_3R_2R_1)
+\cfrac{1}{q^2[2]_q}R_1R_2R_3R_2R_1
+R_1R_2R_1R_3R_2R_1\Bigr),
\\ \\
P_{\overline{22}}(q)=P_{\underline{22}}(-q^{-1}),\\ \\
P_{\overline{\overline{211}}}(q)=P_{\overline{\overline{31}}}(-q^{-1}),\ \ \
P_{\overline{\underline{211}}}(q)=P_{\overline{\underline{31}}}(-q^{-1}),\ \ \
P_{\underline{\underline{211}}}(q)=P_{\underline{\underline{31}}}(-q^{-1}).
\end{array}
\nn\end{equation}
\newpage
\section{Eigenvalues of colored $\mathcal{R}$-matrices from $2$-strand cabling\label{app:eig}}

We used the $2$-strand knots and links to find the eigenvalues of the colored $\mathcal{R}$-matrices. In the table below, we provide the answers for the squares of the eigenvalues for the colored $\mathcal{R}$-matrices corresponding to the crossings between different representations with the summarized size up to $|T_1|+|T_2|=8$. These eigenvalues were obtained by studying the cabling for the $2$-strand links with representations $T_1$ on one component and $T_2$ on the other one, $Q$ are the irreducible representations in the expansion of $T_1\otimes T_2$. The listed answers are obtained using the vertical framing, i.e., $\lambda_Q=q^{\varkappa_Q-\varkappa_{T_1}-\varkappa_{T_2}}$, see Sec.~\ref{framing} for details. In these tables with the symbol $1^n$ we denote $n$ $1$-s in the Young diagram of the representation.

\begin{equation}
\begin{array}{|c|c|c|c||lll|l}
\cline{1-7}
\multicolumn{4}{|c||}{Q/\varkappa_Q}&[3]&[21]&\\
\hhline{|----||---|~} T_1&\varkappa_{T_1}&T_2&\varkappa_{T_2}&3&0&&\\
\hhline{|----||---|~}{}
[1]&0&[2]&1&q^4&q^{-2}&&\\{}
&&[11]&-1&&q^2&&\\
\hhline{|====||===|~}
\multicolumn{4}{|c||}{}&[4]&[31]&[22]&\\
\hhline{|----||---|~}
T_1&\varkappa_{T_1}&T_2&\varkappa_{T_2}&6&2&0&\\
\hhline{|----||---|~}
[1]&0&[3]&3&q^6&q^{-2}&&\\{}
&&[21]&0&&q^4&1&\\{}
&&[111]&-3&&&&\\
\hhline{|====||===|~}
[2]&1&[2]&1&q^8&1&q^{-4}&\\{}
&&[11]&-1&&q^4&&\\
\hhline{|----||---|~}{}
[11]&-1&[11]&-1&&&q^4&\\
\cline{1-7}
\end{array}
\begin{array}{|c|c|c|c||llll|}
\hline
\multicolumn{4}{|c||}{Q/\varkappa_Q}&[5]&[41]&[32]&[311]\\
\hhline{|----||----|}
T_1&\varkappa_{T_1}&T_2&\varkappa_{T_2}&10&5&2&0\\
\hhline{|----||----|}
\hhline{|----||----|}
[1]&0&[4]&6&q^8&q^{-2}&&\\{}
&&[31]&2&&q^6&1&q^{-4}\\{}
&&[22]&0&&&q^4&\\{}
&&[211]&-2&&&&q^4\\{}
&&[1^4]&-6&&&&\\
\hhline{|====||====|}
[2]&1&[3]&3&q^{12}&q^2&q^{-4}&\\{}
&&[21]&0&&q^8&q^2&q^{-2}\\{}
&&[111]&-3&&&&q^4\\
\hhline{|----||----|}{}
[11]&-1&[3]&3&&q^6&&q^{-4}\\{}
&&[21]&0&&&q^6&q^2\\{}
&&[111]&-3&&&&\\
\hline
\end{array}
\nn\end{equation}
\begin{equation}
\begin{array}{|c|c|c|c||llllll|}
\hline
\multicolumn{4}{|c||}{Q/\varkappa_Q}&[6]&[51]&[42]&[411]&[33]&[321]\\
\hhline{|----||------|}
T_1&\varkappa_{T_1}&T_2&\varkappa_{T_2}&15&9&5&3&3&0\\
\hhline{|----||------|}
[1]&0&[5]&10&q^{10}&q^{-2}&&&&\\{}
&&[41]&5&&q^8&1&q^{-4}&&\\{}
&&[32]&2&&&q^6&&q^2&q^{-4}\\{}
&&[311]&0&&&&q^6&&1\\{}
&&[221]&-2&&&&&&q^4\\{}
&&[2111]&-5&&&&&&\\{}
&&[1^5]&-10&&&&&&\\
\hhline{|====||======|}
[2]&1&[4]&6&q^{16}&q^4&q^{-4}&&&\\{}
&&[31]&2&&q^{12}&q^4&1&1&q^{-6}\\{}
&&[22]&0&&&q^8&&&q^{-2}\\{}
&&[211]&-2&&&&q^8&&q^2\\{}
&&[1^4]&-6&&&&&&\\
\hhline{|----||------|}
[11]&-1&[4]&6&&q^8&&q^{-4}&&\\{}
&&[31]&2&&&q^8&q^4&&q^{-2}\\{}
&&[22]&0&&&&&q^8&q^2\\{}
&&[211]&-2&&&&&&q^6\\{}
&&[1^4]&-6&&&&&&\\
\hhline{|====||======|}
[3]&3&[3]&3&q^{18}&q^6&q^{-2}&&q^{-6}&\\{}
&&[21]&0&&q^{12}&q^4&1&&q^{-6}\\{}
&&[111]&-3&&&&q^6&&\\
\hhline{|----||------|}
[21]&0&[21]&0&&&q^{10}&q^6&q^6&2\\{}
&&[111]&-3&&&&&&q^6\\
\hhline{|----||------|}
[111]&-3&[111]&-3&&&&&&\\
\hline
\end{array}
\nn\end{equation}

\begin{equation}
\begin{array}{|c|c|c|c||llllllll|}
\hline
\multicolumn{4}{|c||}{}&\multicolumn{8}{|c|}{Q,\ \varkappa_Q}\\
\cline{5-12}
\multicolumn{4}{|c||}{}&[7]&[61]&[52]&[511]&[43]&[421]&[4111]&[331]\\
\hhline{|----||--------|} T_1&\varkappa_{T_1}&T_2&\varkappa_{T_2}&21&14&9&7&6&3&0&1\\
\hhline{|----||--------|}{}
[1]&0&[6]&15&q^{12}&q^{-2}&&&&&&\\{}
&&[51]&9&&q^{10}&1&q^{-4}&&&&\\{}
&&[42]&5&&&q^8&&q^2&q^{-4}&&\\{}
&&[411]&3&&&&q^8&&1&q^{-6}&\\{}
&&[33]&3&&&&&q^6&&&q^{-4}\\{}
&&[321]&0&&&&&&q^6&&q^2\\{}
&&[3111]&-3&&&&&&&q^6&\\{}
&&[222]&-3&&&&&&&&\\{}
&&[2211]&-5&&&&&&&&\\{}
&&[2 1^4]&-9&&&&&&&&\\{}
&&[1^6]&-15&&&&&&&&\\
\hhline{|====||========|}{}
[2]&1&[5]&10&q^{20}&q^6&q^{-4}&&&&&\\{}
&&[41]&2&&q^{16}&q^6&q^2&1&q^{-6}&&\\{}
&&[32]&5&&&q^{12}&&q^6&1&&q^{-4}\\{}
&&[311]&0&&&&q^{12}&&q^4&q^{-2}&1\\{}
&&[221]&-5&&&&&&q^8&&\\{}
&&[2111]&-2&&&&&&&q^8&\\{}
&&[1^5]&-10&&&&&&&&\\
\hhline{|----||--------|}{}
[11]&-1&[5]&10&&q^{10}&&q^{-4}&&&&\\{}
&&[41]&2&&&q^{10}&q^6&&q^{-2}&q^{-8}&\\{}
&&[32]&5&&&&&q^{10}&q^4&&1\\{}
&&[311]&0&&&&&&q^8&q^2&\\{}
&&[221]&-2&&&&&&&&q^8\\{}
&&[2111]&-5&&&&&&&&\\{}
&&[1^5]&-10&&&&&&&&\\
\hhline{|====||========|}{}
[3]&3&[4]&6&q^{24}&q^{10}&1&&q^{-6}&&&\\{}
&&[31]&2&&q^{18}&q^8&q^4&q^2&q^{-4}&&q^{-8}\\{}
&&[22]&0&&&q^{12}&&&1&&\\{}
&&[211]&-2&&&&q^{12}&&q^4&q^{-2}&\\{}
&&[1^4]&-6&&&&&&&q^6&\\
\hhline{|----||--------|}{}
[21]&0&[4]&6&&q^{16}&q^6&q^2&&q^{-6}&&\\{}
&&[31]&2&&&q^{14}&q^{10}&q^8&2q^2&q^{-4}&q^{-2}\\{}
&&[22]&0&&&&&q^{12}&q^6&&q^2\\{}
&&[211]&-2&&&&&&q^{10}&q^4&q^6\\{}
&&[1^4]&-6&&&&&&&&\\
\hhline{|====||========|}{}
[111]&-3&[4]&6&&&&q^8&&&q^{-6}&\\{}
&&[31]&2&&&&&&q^8&q^2&\\{}
&&[22]&0&&&&&&&&q^8\\{}
&&[211]&-2&&&&&&&&\\{}
&&[1^4]&-6&&&&&&&&\\
\hline
\end{array}
\nn\end{equation}

\setlength{\arraycolsep}{4.5pt}
\begin{equation}
\begin{array}{|c|c|c|c||llllllllllll|}
\hline
\multicolumn{4}{|c||}{}&\multicolumn{12}{|c|}{Q,\ \varkappa_Q}\\
\cline{5-16}
\multicolumn{4}{|c||}{}&[8]&[71]&[62]&[611]&[53]&[521]&[5111]&[44]&[431]&[422]&[4211]&[332]\\
\hhline{|----||------------|} T_1&\varkappa_{T_1}&T_2&\varkappa_{T_2}&28&20&14&12&10&7&4&8&4&2&0&0\\
\hhline{|----||------------|}{}
[1]&0&[7]&21&q^{14}&q^{-2}&&&&&&&&&&\\{}
&&[61]&14&&q^{12}&1&q^{-4}&&&&&&&&\\{}
&&[52]&9&&&q^{10}&&q^2&q^{-4}&&&&&&\\{}
&&[511]&7&&&&q^{10}&&1&q^{-6}&&&&&\\{}
&&[43]&6&&&&&q^8&&&q^4&q^{-4}&&&\\{}
&&[421]&3&&&&&&q^8&&&q^2&q^{-2}&q^{-6}&\\{}
&&[4111]&0&&&&&&&q^8&&&&1&\\{}
&&[331]&1&&&&&&&&&q^6&&&q^{-2}\\{}
&&[322]&-1&&&&&&&&&&q^6&&q^2\\{}
&&[3211]&-3&&&&&&&&&&&q^6&\\{}
&&[3 1^4]&-7&&&&&&&&&&&&\\{}
&&[22111]&-6&&&&&&&&&&&&\\{}
&&[2 1^5]&-9&&&&&&&&&&&&\\{}
&&[1^7]&-14&&&&&&&&&&&&\\
\hhline{|====||============|}
[2]&1&[6]&15&q^{24}&q^8&q^{-4}&&&&&&&&&\\{}
&&[51]&9&&q^{20}&q^8&q^4&1&q^{-6}&&&&&&\\{}
&&[42]&5&&&q^{16}&&q^8&q^2&&q^4&q^{-4}&q^{-8}&&\\{}
&&[411]&3&&&&q^{16}&&q^6&1&&1&&q^{-8}&\\{}
&&[33]&3&&&&&q^{12}&&&&1&&&q^{-8}\\{}
&&[321]&0&&&&&&q^{12}&&&q^6&q^2&q^{-2}&q^{-2}\\{}
&&[3111]&-3&&&&&&&q^{12}&&&&q^4&\\{}
&&[222]&-3&&&&&&&&&&q^8&&\\{}
&&[2211]&-5&&&&&&&&&&&q^8&\\{}
&&[2 1^4]&-9&&&&&&&&&&&&\\{}
&&[1^6]&-15&&&&&&&&&&&&\\
\hhline{|----||------------|}{}
[11]&-1&[6]&14&&q^{12}&&q^{-4}&&&&&&&&\\{}
&&[51]&9&&&q^{12}&q^8&&q^{-2}&q^{-8}&&&&&\\{}
&&[42]&5&&&&&q^{12}&q^6&&&1&&q^{-8}&\\{}
&&[411]&3&&&&&&q^{10}&q^4&&&1&q^{-4}&\\{}
&&[33]&3&&&&&&&&q^{12}&q^4&&&\\{}
&&[321]&0&&&&&&&&&q^{10}&q^6&q^2&q^2\\{}
&&[3111]&-3&&&&&&&&&&&q^8&\\{}
&&[222]&-3&&&&&&&&&&&&q^8\\{}
&&[2211]&-5&&&&&&&&&&&&\\{}
&&[2 1^4]&-9&&&&&&&&&&&&\\{}
&&[1^6]&-15&&&&&&&&&&&&\\
\hline
\end{array}
\nn\end{equation}
\begin{equation}
\begin{array}{|c|c|c|c||llllllllllll|}
\hline
\multicolumn{4}{|c||}{}&\multicolumn{12}{|c|}{Q,\ \varkappa_Q}\\
\cline{5-16}
\multicolumn{4}{|c||}{}&[8]&[71]&[62]&[611]&[53]&[521]&[5111]&[44]&[431]&[422]&[4211]&[332]\\
\hhline{|----||------------|} T_1&\varkappa_{T_1}&T_2&\varkappa_{T_2}&28&20&14&12&10&7&4&8&4&2&0&0\\
\hhline{|----||------------|}{}
[3]&3&[5]&10&q^{30}&q^{14}&q^2&&q^{-6}&&&&&&&\\{}
&&[41]&2&&q^{24}&q^{12}&q^8&q^4&q^{-2}&&1&q^{-8}&&&\\{}
&&[32]&5&&&q^{18}&&q^{10}&q^4&&&q^{-2}&q^{-6}&&q^{-10}\\{}
&&[311]&0&&&&q^{18}&&q^8&q^2&&q^2&&q^{-6}&\\{}
&&[221]&-2&&&&&&q^{12}&&&&q^2&q^{-2}&\\{}
&&[2111]&-5&&&&&&&q^{12}&&&&q^4&\\{}
&&[1^5]&-10&&&&&&&&&&&&\\
\hhline{|----||------------|}
[21]&0&[5]&10&&q^{20}&q^8&q^4&&q^{-6}&&&&&&\\{}
&&[41]&2&&&q^{18}&q^{14}&q^{10}&2q^4&q^{-2}&&q^{-2}&q^{-6}&q^{-10}&\\{}
&&[32]&5&&&&&q^{16}&q^{10}&&q^{12}&2q^4&1&q^{-4}&q^{-4}\\{}
&&[311]&0&&&&&&q^{14}&q^8&&q^8&q^4&q^2&1\\{}
&&[221]&-2&&&&&&&&&q^{12}&q^8&q^4&q^4\\{}
&&[2111]&-5&&&&&&&&&&&q^{10}&\\{}
&&[1^5]&-10&&&&&&&&&&&&\\
\hhline{|----||------------|}
[111]&-3&[5]&10&&&&q^{10}&&&q^{-6}&&&&&\\{}
&&[41]&2&&&&&&q^{10}&q^4&&&&q^{-4}&\\{}
&&[32]&5&&&&&&&&&q^{10}&&q^2&\\{}
&&[311]&0&&&&&&&&&&q^{10}&q^6&\\{}
&&[221]&-2&&&&&&&&&&&&q^{10}\\{}
&&[2111]&-5&&&&&&&&&&&&\\{}
&&[1^5]&-10&&&&&&&&&&&&\\
\hhline{|====||============|}
[4]&6&[4]&6&q^{32}&q^{16}&q^4&&q^{-4}&&&q^{-8}&&&&\\{}
&&[31]&2&&q^{24}&q^{12}&q^8&q^4&q^{-2}&&&q^{-8}&&&\\{}
&&[22]&0&&&&q^{16}&&q^2&&&&q^{-8}&&\\{}
&&[211]&-2&&&&q^{16}&&q^6&1&&&&q^{-8}&\\{}
&&[1^4]&-6&&&&&&&q^8&&&&&\\
\hhline{|----||------------|}{}
[31]&2&[31]&2&&&q^{20}&q^{16}&q^{12}&2q^6&1&q^8&2&q^{-4}&q^{-8}&q^{-8}\\{}
&&[22]&0&&&&&q^{16}&q^{10}&&&q^4&1&q^{-4}&q^{-4}\\{}
&&[211]&-2&&&&&&q^{14}&q^8&&q^8&q^4&2&\\{}
&&[1^4]&-6&&&&&&&&&&&q^8&\\
\hhline{|----||------------|}{}
[22]&0&[22]&0&&&&&&&&q^{16}&q^8&q^4&&\\{}
&&[211]&-2&&&&&&&&&&q^{12}&q^4&q^4\\{}
&&[1^4]&-6&&&&&&&&&&&&\\
\hhline{|----||------------|}{}
[211]&-2&[211]&-2&&&&&&&&&&q^{12}&q^8&q^8\\{}
&&[1^4]&-6&&&&&&&&&&&&\\
\hhline{|----||------------|}{}
[1^4]&-6&[1^4]&-6&&&&&&&&&&&&\\
\hline
\end{array}
\nn\end{equation}
\setlength{\arraycolsep}{6pt}

\newpage

The study of $2$-strand knots allows one to find not only the absolute values of the eigenvalues of the colored $\mathcal{R}$-matrices but also their signs. The drawback is that one study only the case of equal representations on the crossing strands, $T_1=T_2=T$. In the table below, the eigenvalues for the colored $\mathcal{R}$-matrices corresponding to the crossings between $T\otimes T$ with for $|T|\le 4$ are listed. In these tables with the symbol $1^n$ we denote $n$ $1$-s in the Young diagram of the representation.

\begin{equation*}
\begin{array}{|c|c||ccccccccccc|}
\hline
\multicolumn{2}{|c||}{}&\multicolumn{11}{|c|}{Q,\ \varkappa_Q}\\
\cline{3-13}
\multicolumn{2}{|c||}{}&[4]&[31]&[22]&[211]&[1^4]&&&&&&\\
\hhline{|--||-----------|}
T&\varkappa_{T}&6&2&0&-2&-6&&&&&&\\
\hhline{|--||-----------|}{}
[2]&1&q^4&-1&q^{-2}&&&&&&&&\\
\hhline{|~~||~~~~~~~~~~~|}{}
[11]&-1&&&q^2&-1&q^{-4}&&&&&&\\
\hhline{|==||===========|}
\multicolumn{2}{|c||}{}&[6]&[51]&[42]&[411]&[33]&[321]&[3111]&[222]&[2211]&[2 1^4]&[1^6]\\
\cline{3-13}
\hhline{|--||-----------|}
T&\varkappa_{T}&15&9&5&3&3&0&-3&-3&-5&-9&-15\\
\hhline{|--||-----------|}{}
[3]&3&q^{9}&-q^3&q^{-1}&&-q^{-3}&&&&&&\\
\hhline{|~~||~~~~~~~~~~~|}
[21]&0&&&q^5&-q^3&-q^3&\pm 1&q^{-3}&q^{-3}&-q^{-5}&&\\
\hhline{|~~||~~~~~~~~~~~|}
[111]&-3&&&&&&&&q^3&-q&q^{-3}&-q^{-9}\\
\hhline{|==||===========|}
\multicolumn{2}{|c||}{}&[8]&[71]&[62]&[611]&[53]&[521]&[5111]&[44]&[431]&[422]&[4211]\\
\cline{3-13}
\hhline{|--||-----------|}
T&\varkappa_{T}&28&20&14&12&10&7&4&8&4&2&0\\
\hhline{|--||-----------|}
[4]&6&q^{16}&-q^8&q^2&&-q^{-2}&&&q^{-4}&&&\\
\hhline{|~~||~~~~~~~~~~~|}
[31]&2&&&q^{10}&-q^8&-q^6&\pm q^3&1&q^4&\pm 1&q^{-2}&-q^{-4}\\
\hhline{|~~||~~~~~~~~~~~|}
[22]&0&&&&&&&&q^8&-q^4&q^2&\\
\hhline{|~~||~~~~~~~~~~~|}
[211]&-2&&&&&&&&&&q^6&-q^4\\
\hhline{|~~||~~~~~~~~~~~|}
[1^4]&-6&&&&&&&&&&&\\
\hhline{|==||===========|}
\multicolumn{2}{|c||}{}&[4 1^4]&[332]&[3311]&[3221]&[32111]&[3 1^5]&[2222]&[22211]&[22 1^4]&[2 1^6]&[1^8]
\\
\cline{3-13}
\hhline{|--||-----------|}
T&\varkappa_{T}&-4&0&-2&-4&-7&-12&-8&-10&-14&-20&-28
\\
\cline{3-13}
\hhline{|--||-----------|}
[4]&6&&&&&&&&&&&\\
\hhline{|~~||~~~~~~~~~~~|}
[31]&2&&-q^{-4}&q^{-6}&&&&&&&&\\
\hhline{|~~||~~~~~~~~~~~|}
[22]&0&&&q^{-2}&-q^{-4}&&&q^{-8}&&&&\\
\hhline{|~~||~~~~~~~~~~~|}
[211]&-2&1&-q^4&q^2&\pm1&\pm q^{-3}&-q^{-8}&q^{-4}&-q^{-6}&q^{-10}&&\\
\hhline{|~~||~~~~~~~~~~~|}
[1^4]&-6&&&&&&&q^4&-q^2&q^{-2}&-q^{-8}&q^{-16}\\
\hline
\end{array}
\end{equation*}

Another important property of the expansion into irreducible representations is the multiplicity. It appears while studying higher representations. Starting from the size $6$, some irreducible representations in the decomposition of the tensor product of two representations appear several times. Though most representations have multiplicity equal to one (or zero if they do not appear at all), the ones with the multiplicities higher than one have the most nontrivial behavior. In the table below, all the representations with the multiplicity $2$ from the size $6$ to the size $8$ are listed; there are no other nontrivial multiplicities in this range.

\begin{equation}
\begin{array}{c}
[2,1] \times [2,1] \rightarrow [3,2,1],
\\
\\
\begin{array}{cc}
[2,1] \times [3,1] \rightarrow [4,2,1],
&
[2,1] \times [2,1,1] \rightarrow [3,2,2,1],
\end{array}
\\
\\
\begin{array}{lll}
[2,1] \times [4,1] \rightarrow [5,2,1],
&
[2,1] \times [3,2] \rightarrow [4,3,1],
&
[2,1] \times [2,2,1] \rightarrow [3,2,2,1],
\\{}
[2,1] \times [2,1,1,1] \rightarrow [3,2,1,1,1],
&
[3,1] \times [3,1] \rightarrow [5,2,1],
&
[3,1] \times [3,1] \rightarrow [4,3,1],
\\{}
[3,1] \times [2,1,1] \rightarrow [4,2,1,1],
&
[2,1,1] \times [2,1,1] \rightarrow [3,2,2,1],
&
[2,1,1] \times [2,1,1] \rightarrow [3,2,1,1,1].
\end{array}
\end{array}
\end{equation}

\newpage

\section{Three-strand knots in representation $[21]$\label{app:3str}}

In this Appendix, we present HOMFLY polynomials for the $3$-strand knots with up to
eight crossings in representation $[21]$. The answers for the same
knots in lower representations and in level-three symmetric and
antisymmetric representations can be found in~\cite{IMMM1,IMMM3}. All the answers are given in the vertical framing, see~Sec.~\ref{framing}.

We use here the notation used in~\cite{Inds8}. The matrix describes
the coefficients of a polynomial in $A^2$ and $q^2$ as

\setlength{\arraycolsep}{3pt}

\begin{equation}
\nonumber q^{10}A^{16} \left(\begin{array}{rr}
3 & 4 \\
& \\
1 & 2 \\
\end{array}\right)
= q^{10}A^{16}+2q^{12}A^{16}+3q^{10}A^{18}+4q^{12}A^{18}.
\end{equation}

\textbf{Knot $3_1$}

\begin{equation}
\nonumber H_{[2,1]}= \frac{1}{q^{10}A^{3}}
\left(\begin{array}{rrrrrrrrrrr}
0 & 0 & 0 & 0 & 0 & -1 & 0 & 0 & 0 & 0 & 0 \\
&&&&&&&&&& \\
0 & 0 & 1 & 1 & 1 & 0 & 1 & 1 & 1 & 0 & 0 \\
&&&&&&&&&& \\
-1 & 0 & -2 & 0 & -3 & 0 & -3 & 0 & -2 & 0 & -1 \\
&&&&&&&&&& \\
1 & 0 & 2 & -1 & 2 & 0 & 2 & -1 & 2 & 0 & 1
\end{array}\right)
\end{equation}
\textbf{Knot $4_1$}

\begin{equation}
\nonumber H_{[2,1]}= \frac{1}{q^{10}A^{6}}
\left(\begin{array}{rrrrrrrrrrr}
0 & 0 & 0 & 0 & 0 & 1 & 0 & 0 & 0 & 0 & 0 \\
&&&&&&&&&& \\
0 & 0 & -1 & 0 & -1 & 1 & -1 & 0 & -1 & 0 & 0 \\
&&&&&&&&&& \\
1 & -1 & 3 & -3 & 5 & -4 & 5 & -3 & 3 & -1 & 1 \\
&&&&&&&&&& \\
-2 & 2 & -5 & 6 & -8 & 7 & -8 & 6 & -5 & 2 & -2 \\
&&&&&&&&&& \\
1 & -1 & 3 & -3 & 5 & -4 & 5 & -3 & 3 & -1 & 1 \\
&&&&&&&&&& \\
0 & 0 & -1 & 0 & -1 & 1 & -1 & 0 & -1 & 0 & 0 \\
&&&&&&&&&& \\
0 & 0 & 0 & 0 & 0 & 1 & 0 & 0 & 0 & 0 & 0
\end{array}\right)
\end{equation}
\textbf{Knot $5_2$}

\begin{equation}
\nonumber
\begin{array}{l}
H_{[2,1]}=\frac{1}{q^{14}A^{6}}
\\
\\
\left(\begin{array}{rrrrrrrrrrrrrrr}
0 & 0 & 1 & -2 & 3 & -4 & 5 & -5 & 5 & -4 & 3 & -2 & 1 & 0 & 0 \\
&&&&&&&&&&&&&& \\
0 & 1 & -2 & 5 & -7 & 9 & -11 & 13 & -11 & 9 & -7 & 5 & -2 & 1 & 0 \\
&&&&&&&&&&&&&& \\
1 & -2 & 4 & -6 & 9 & -12 & 13 & -14 & 13 & -12 & 9 & -6 & 4 & -2 & 1 \\
&&&&&&&&&&&&&& \\
-1 & 1 & -3 & 3 & -5 & 7 & -8 & 7 & -8 & 7 & -5 & 3 & -3 & 1 & -1 \\
&&&&&&&&&&&&&& \\
0 & 0 & 0 & 0 & 1 & 0 & 0 & -2 & 0 & 0 & 1 & 0 & 0 & 0 & 0 \\
&&&&&&&&&&&&&& \\
0 & 0 & 0 & 0 & 1 & 0 & 1 & -1 & 1 & 0 & 1 & 0 & 0 & 0 & 0 \\
&&&&&&&&&&&&&& \\
0 & 0 & 0 & 0 & 0 & 0 & 0 & -1 & 0 & 0 & 0 & 0 & 0 & 0 & 0
\end{array}\right)
\end{array}
\end{equation}

\begin{landscape}

\noindent\textbf{Knot $6_2$}

\begin{equation}
\nonumber
H_{[2,1]}=\frac{1}{q^{20}A^{6}}
\left(\begin{array}{rrrrrrrrrrrrrrrrrrrrr}
0 & 0 & 0 & 0 & 0 & 1 & -2 & 3 & -4 & 5 & -5 & 5 & -4 & 3 & -2 & 1 & 0 & 0 & 0 & 0 & 0 \\
&&&&&&&&&&&&&&&&&&&& \\
0 & 0 & -1 & 1 & -1 & 0 & -1 & -2 & 5 & -6 & 4 & -6 & 5 & -2 & -1 & 0 & -1 & 1 & -1 & 0 & 0 \\
&&&&&&&&&&&&&&&&&&&& \\
1 & -2 & 5 & -6 & 12 & -14 & 17 & -16 & 21 & -18 & 18 & -18 & 21 & -16 & 17 & -14 & 12 & -6 & 5 & -2 & 1 \\
&&&&&&&&&&&&&&&&&&&& \\
-2 & 3 & -8 & 12 & -22 & 24 & -33 & 35 & -42 & 39 & -44 & 39 & -42 & 35 & -33 & 24 & -22 & 12 & -8 & 3 & -2 \\
&&&&&&&&&&&&&&&&&&&& \\
1 & -1 & 5 & -7 & 13 & -14 & 24 & -22 & 29 & -26 & 32 & -26 & 29 & -22 & 24 & -14 & 13 & -7 & 5 & -1 & 1 \\
&&&&&&&&&&&&&&&&&&&& \\
0 & 0 & -1 & 0 & -3 & 2 & -5 & 3 & -8 & 4 & -8 & 4 & -8 & 3 & -5 & 2 & -3 & 0 & -1 & 0 & 0 \\
&&&&&&&&&&&&&&&&&&&& \\
0 & 0 & 0 & 0 & 0 & 1 & 0 & 2 & -1 & 2 & 0 & 2 & -1 & 2 & 0 & 1 & 0
& 0 & 0 & 0 & 0
\end{array}\right)
\end{equation}
\textbf{Knot $6_3$}

\begin{equation}
\nonumber H_{[2,1]}=\frac{1}{q^{20}A^{6}}
\left(\begin{array}{rrrrrrrrrrrrrrrrrrrrr}
0 & 0 & 0 & 0 & 0 & -1 & 2 & -3 & 4 & -5 & 5 & -5 & 4 & -3 & 2 & -1 & 0 & 0 & 0 & 0 & 0 \\
&&&&&&&&&&&&&&&&&&&& \\
0 & 0 & 1 & -1 & 2 & -2 & 4 & -4 & 6 & -5 & 7 & -5 & 6 & -4 & 4 & -2 & 2 & -1 & 1 & 0 & 0 \\
&&&&&&&&&&&&&&&&&&&& \\
-1 & 2 & -6 & 9 & -17 & 23 & -36 & 41 & -55 & 56 & -62 & 56 & -55 & 41 & -36 & 23 & -17 & 9 & -6 & 2 & -1 \\
&&&&&&&&&&&&&&&&&&&& \\
2 & -4 & 10 & -16 & 31 & -40 & 60 & -71 & 90 & -92 & 105 & -92 & 90 & -71 & 60 & -40 & 31 & -16 & 10 & -4 & 2 \\
&&&&&&&&&&&&&&&&&&&& \\
-1 & 2 & -6 & 9 & -17 & 23 & -36 & 41 & -55 & 56 & -62 & 56 & -55 & 41 & -36 & 23 & -17 & 9 & -6 & 2 & -1 \\
&&&&&&&&&&&&&&&&&&&& \\
0 & 0 & 1 & -1 & 2 & -2 & 4 & -4 & 6 & -5 & 7 & -5 & 6 & -4 & 4 & -2 & 2 & -1 & 1 & 0 & 0 \\
&&&&&&&&&&&&&&&&&&&& \\
0 & 0 & 0 & 0 & 0 & -1 & 2 & -3 & 4 & -5 & 5 & -5 & 4 & -3 & 2 & -1
& 0 & 0 & 0 & 0 & 0
\end{array}\right)
\end{equation}

\newpage

\noindent\textbf{Knot $7_2$}

\begin{equation}
\nonumber
\begin{array}{l}
H_{[2,1]}=\frac{1}{q^{24}A^{6}}
\\
\\
\left(\begin{array}{rrrrrrrrrrrrrrrrrrrrrrrrr}
0 & 0 & 1 & -2 & 3 & -6 & 10 & -11 & 12 & -15 & 16 & -15 & 15 & -15 & 16 & -15 & 12 & -11 & 10 & -6 & 3 & -2 & 1 & 0 & 0 \\
&&&&&&&&&&&&&&&&&&&&&&&& \\
0 & 1 & -2 & 6 & -11 & 17 & -23 & 31 & -35 & 38 & -41 & 42 & -40 & 42 & -41 & 38 & -35 & 31 & -23 & 17 & -11 & 6 & -2 & 1 & 0 \\
&&&&&&&&&&&&&&&&&&&&&&&& \\
1 & -2 & 5 & -9 & 16 & -22 & 28 & -33 & 40 & -41 & 39 & -41 & 44 & -41 & 39 & -41 & 40 & -33 & 28 & -22 & 16 & -9 & 5 & -2 & 1 \\
&&&&&&&&&&&&&&&&&&&&&&&& \\
-1 & 1 & -4 & 5 & -9 & 10 & -14 & 13 & -14 & 11 & -10 & 7 & -6 & 7 & -10 & 11 & -14 & 13 & -14 & 10 & -9 & 5 & -4 & 1 & -1 \\
&&&&&&&&&&&&&&&&&&&&&&&& \\
0 & 0 & 0 & 0 & 0 & 1 & -2 & 3 & -8 & 9 & -13 & 13 & -18 & 13 & -13 & 9 & -8 & 3 & -2 & 1 & 0 & 0 & 0 & 0 & 0 \\
&&&&&&&&&&&&&&&&&&&&&&&& \\
0 & 0 & 0 & 0 & 1 & 0 & 3 & -2 & 5 & -3 & 8 & -4 & 8 & -4 & 8 & -3 & 5 & -2 & 3 & 0 & 1 & 0 & 0 & 0 & 0 \\
&&&&&&&&&&&&&&&&&&&&&&&& \\
0 & 0 & 0 & 0 & 0 & 0 & 0 & -1 & 0 & -2 & 1 & -2 & 0 & -2 & 1 & -2 &
0 & -1 & 0 & 0 & 0 & 0 & 0 & 0 & 0
\end{array}\right)
\end{array}
\end{equation}
\textbf{Knot $7_5$}

\begin{equation}
\nonumber
\begin{array}{l}
H_{[2,1]}=\frac{1}{q^{24}A^{6}}
\\
\\
\begin{footnotesize}
\left(\begin{array}{rrrrrrrrrrrrrrrrrrrrrrrrr}
0 & 0 & 0 & 0 & 0 & 0 & 0 & -1 & 2 & -3 & 4 & -5 & 5 & -5 & 4 & -3 & 2 & -1 & 0 & 0 & 0 & 0 & 0 & 0 & 0 \\
&&&&&&&&&&&&&&&&&&&&&&&& \\
0 & 0 & 0 & 0 & 1 & -2 & 3 & -6 & 9 & -11 & 12 & -14 & 16 & -14 & 12 & -11 & 9 & -6 & 3 & -2 & 1 & 0 & 0 & 0 & 0 \\
&&&&&&&&&&&&&&&&&&&&&&&& \\
0 & 0 & 0 & 1 & 0 & 2 & -4 & 9 & -12 & 17 & -24 & 27 & -26 & 27 & -24 & 17 & -12 & 9 & -4 & 2 & 0 & 1 & 0 & 0 & 0 \\
&&&&&&&&&&&&&&&&&&&&&&&& \\
-1 & 2 & -6 & 10 & -17 & 24 & -32 & 41 & -47 & 51 & -55 & 58 & -56 & 58 & -55 & 51 & -47 & 41 & -32 & 24 & -17 & 10 & -6 & 2 & -1 \\
&&&&&&&&&&&&&&&&&&&&&&&& \\
1 & -3 & 8 & -18 & 30 & -48 & 69 & -92 & 111 & -133 & 146 & -156 & 158 & -156 & 146 & -133 & 111 & -92 & 69 & -48 & 30 & -18 & 8 & -3 & 1 \\
&&&&&&&&&&&&&&&&&&&&&&&& \\
0 & 1 & -3 & 9 & -19 & 33 & -49 & 69 & -90 & 107 & -121 & 130 & -134 & 130 & -121 & 107 & -90 & 69 & -49 & 33 & -19 & 9 & -3 & 1 & 0 \\
&&&&&&&&&&&&&&&&&&&&&&&& \\
0 & 0 & 1 & -2 & 5 & -9 & 15 & -20 & 27 & -32 & 38 & -40 & 42 & -40
& 38 & -32 & 27 & -20 & 15 & -9 & 5 & -2 & 1 & 0 & 0
\end{array}\right)
\end{footnotesize}
\end{array}
\end{equation}

\newpage

\noindent\textbf{Knot $8_2$}

\begin{equation}
\nonumber
\begin{array}{l}
H_{[2,1]}=\frac{1}{q^{30}A^{6}}
\\
\\
{\tiny \left(\begin{array}{rrrrrrrrrrrrrrrrrrrrrrrrrrrrrrr}
0 & 0 & 0 & 0 & 0 & 1 & -2 & 3 & -6 & 10 & -11 & 12 & -15 & 16 & -15 & 15 & -15 & 16 & -15 & 12 & -11 & 10 & -6 & 3 & -2 & 1 & 0 & 0 & 0 & 0 & 0 \\
&&&&&&&&&&&&&&&&&&&&&&&&&&&&&& \\
0 & 0 & -1 & 1 & -1 & 1 & -1 & -5 & 9 & -13 & 20 & -28 & 29 & -31 & 33 & -35 & 33 & -31 & 29 & -28 & 20 & -13 & 9 & -5 & -1 & 1 & -1 & 1 & -1 & 0 & 0 \\
&&&&&&&&&&&&&&&&&&&&&&&&&&&&&& \\
1 & -2 & 5 & -7 & 14 & -17 & 22 & -22 & 26 & -20 & 19 & -12 & 12 & -7 & 8 & -4 & 8 & -7 & 12 & -12 & 19 & -20 & 26 & -22 & 22 & -17 & 14 & -7 & 5 & -2 & 1 \\
&&&&&&&&&&&&&&&&&&&&&&&&&&&&&& \\
-2 & 3 & -8 & 13 & -25 & 30 & -45 & 50 & -62 & 58 & -69 & 64 & -71 & 62 & -71 & 65 & -71 & 62 & -71 & 64 & -69 & 58 & -62 & 50 & -45 & 30 & -25 & 13 & -8 & 3 & -2 \\
&&&&&&&&&&&&&&&&&&&&&&&&&&&&&& \\
1 & -1 & 5 & -7 & 15 & -18 & 32 & -32 & 46 & -43 & 57 & -48 & 61 & -52 & 64 & -52 & 64 & -52 & 61 & -48 & 57 & -43 & 46 & -32 & 32 & -18 & 15 & -7 & 5 & -1 & 1 \\
&&&&&&&&&&&&&&&&&&&&&&&&&&&&&& \\
0 & 0 & -1 & 0 & -3 & 2 & -7 & 4 & -12 & 7 & -15 & 8 & -18 & 8 & -18 & 9 & -18 & 8 & -18 & 8 & -15 & 7 & -12 & 4 & -7 & 2 & -3 & 0 & -1 & 0 & 0 \\
&&&&&&&&&&&&&&&&&&&&&&&&&&&&&& \\
0 & 0 & 0 & 0 & 0 & 1 & 0 & 2 & -1 & 4 & -1 & 4 & -1 & 4 & -1 & 5 &
-1 & 4 & -1 & 4 & -1 & 4 & -1 & 2 & 0 & 1 & 0 & 0 & 0 & 0 & 0
\end{array}\right)
}
\end{array}
\end{equation}

\noindent\textbf{Knot $8_5$}

\begin{equation}
\nonumber
\begin{array}{l}
H_{[2,1]}=\frac{1}{q^{30}A^{6}}
\\
\\
{\tiny \left(\begin{array}{rrrrrrrrrrrrrrrrrrrrrrrrrrrrrrr}
0 & \!\! 0 & \!\! 0 & \!\! 0 & \!\! 0 & \!\! 1 & \!\! 0 & \!\! 4 & \!\! -2 & \!\! 8 & \!\! -4 & 13 & \!\! -6 & \!\! 16 & \!\! -8 & \!\! 20 & \!\! -8 & \!\! 16 & \!\! -6 & \!\! 13 & \!\! -4 & \!\! 8 & \!\! -2 & \!\! 4 & \!\! 0 & \!\! 1 & \!\! 0 & \!\! 0 & \!\! 0 & \!\! 0 & \!\! 0 \\
&&&&&&&&&&&&&&&&&&&&&&&&&&&&&& \\
0 & \!\! 0 & \!\! -1 & \!\! 0 & \!\! -5 & \!\! 3 & \!\! -14 & \!\! 10 & \!\! -29 & \!\! 18 & \!\! -46 & \!\! 27 & \!\! -62 & \!\! 31 & \!\! -70 & \!\! 36 & \!\! -70 & \!\! 31 & \!\! -62 & \!\! 27 & \!\! -46 & \!\! 18 & \!\! -29 & \!\! 10 & \!\! -14 & \!\! 3 & \!\! -5 & \!\! 0 & \!\! -1 & \!\! 0 & \!\! 0 \\
&&&&&&&&&&&&&&&&&&&&&&&&&&&&&& \\
1 & \!\! -1 & \!\! 7 & \!\! -10 & \!\! 26 & \!\! -31 & \!\! 61 & \!\! -61 & \!\! 102 & \!\! -91 & \!\! 143 & \!\! -114 & \!\! 173 & \!\! -133 & \!\! 194 & \!\! -136 & \!\! 194 & \!\! -133 & \!\! 173 & \!\! -114 & \!\! 143 & \!\! -91 & \!\! 102 & \!\! -61 & \!\! 61 & \!\! -31 & \!\! 26 & \!\! -10 & \!\! 7 & \!\! -1 & \!\! 1 \\
&&&&&&&&&&&&&&&&&&&&&&&&&&&&&& \\
-2 & \!\! 3 & \!\! -11 & \!\! 16 & \!\! -35 & \!\! 40 & \!\! -68 & \!\! 64 & \!\! -97 & \!\! 77 & \!\! -115 & \!\! 83 & \!\! -130 & \!\! 84 & \!\! -136 & \!\! 89 & \!\! -136 & \!\! 84 & \!\! -130 & \!\! 83 & \!\! -115 & \!\! 77 & \!\! -97 & \!\! 64 & \!\! -68 & \!\! 40 & \!\! -35 & \!\! 16 & \!\! -11 & \!\! 3 & \!\! -2 \\
&&&&&&&&&&&&&&&&&&&&&&&&&&&&&& \\
1 & \!\! -2 & \!\! 6 & \!\! -7 & \!\! 16 & \!\! -14 & \!\! 22 & \!\! -11 & \!\! 18 & \!\! 8 & \!\! 0 & \!\! 30 & \!\! -16 & \!\! 46 & \!\! -23 & \!\! 50 & \!\! -23 & \!\! 46 & \!\! -16 & \!\! 30 & \!\! 0 & \!\! 8 & \!\! 18 & \!\! -11 & \!\! 22 & \!\! -14 & \!\! 16 & \!\! -7 & \!\! 6 & \!\! -2 & \!\! 1 \\
&&&&&&&&&&&&&&&&&&&&&&&&&&&&&& \\
0 & \!\! 0 & \!\! -1 & \!\! 1 & \!\! -2 & \!\! 0 & \!\! 0 & \!\! -10 & \!\! 14 & \!\! -26 & \!\! 30 & \!\! -46 & \!\! 42 & \!\! -49 & \!\! 44 & \!\! -54 & \!\! 44 & \!\! -49 & \!\! 42 & \!\! -46 & \!\! 30 & \!\! -26 & \!\! 14 & \!\! -10 & \!\! 0 & \!\! 0 & \!\! -2 & \!\! 1 & \!\! -1 & \!\! 0 & \!\! 0 \\
&&&&&&&&&&&&&&&&&&&&&&&&&&&&&& \\
0 & \!\! 0 & \!\! 0 & \!\! 0 & \!\! 0 & \!\! 1 & \!\! -2 & \!\! 4 &
\!\! -6 & \!\! 9 & \!\! -8 & \!\! 7 & \!\! -5 & \!\! 5 & \!\! -1 &
\!\! 0 & \!\! -1 & \!\! 5 & \!\! -5 & \!\! 7 & \!\! -8 & \!\! 9 &
\!\! -6 & \!\! 4 & \!\! -2 & \!\! 1 & \!\! 0 & \!\! 0 & \!\! 0 &
\!\! 0 & \!\! 0
\end{array}\right)
}
\end{array}
\end{equation}

\noindent\textbf{Knot $8_7$}

\begin{equation}
\nonumber
\begin{array}{l}
H_{[2,1]}=\frac{1}{q^{30}A^{6}}
\\
\\
{\tiny \left(\begin{array}{rrrrrrrrrrrrrrrrrrrrrrrrrrrrrrr}
0 & \!\!\! 0 & \!\!\! 0 & \!\!\! 0 & \!\!\! 0 & \!\!\! -1 & \!\!\! 2 & \!\!\! -3 & \!\!\! 6 & \!\!\! -10 & \!\!\! 11 & \!\!\! -12 & \!\!\! 15 & \!\!\! -16 & \!\!\! 15 & \!\!\! -15 & 15 & \!\!\! -16 & \!\!\! 15 & \!\!\! -12 & \!\!\! 11 & \!\!\! -10 & \!\!\! 6 & \!\!\! -3 & \!\!\! 2 & \!\!\! -1 & \!\!\! 0 & \!\!\! 0 & \!\!\! 0 & \!\!\! 0 & \!\!\! 0 \\
&&&&&&&&&&&&&&&&&&&&&&&&&&&&&& \\
0 & \!\!\! 0 & \!\!\! 1 & \!\!\! -1 & 2 & \!\!\! -4 & \!\!\! 7 & \!\!\! -6 & \!\!\! 11 & \!\!\! -17 & \!\!\! 19 & \!\!\! -21 & \!\!\! 29 & \!\!\! -30 & \!\!\! 32 & \!\!\! -32 & \!\!\! 32 & \!\!\! -30 & \!\!\! 29 & \!\!\! -21 & \!\!\! 19 & \!\!\! -17 & \!\!\! 11 & \!\!\! -6 & \!\!\! 7 & \!\!\! -4 & \!\!\! 2 & \!\!\! -1 & \!\!\! 1 & \!\!\! 0 & \!\!\! 0 \\
&&&&&&&&&&&&&&&&&&&&&&&&&&&&&& \\
-1 & \!\!\! 2 & \!\!\! -6 & \!\!\! 11 & \!\!\! -23 & \!\!\! 34 & \!\!\! -55 & \!\!\! 75 & \!\!\! -104 & \!\!\! 124 & \!\!\! -155 & \!\!\! 171 & \!\!\! -190 & \!\!\! 198 & \!\!\! -210 & \!\!\! 204 & \!\!\! -210 & \!\!\! 198 & \!\!\! -190 & \!\!\! 171 & \!\!\! -155 & \!\!\! 124 & \!\!\! -104 & \!\!\! 75 & \!\!\! -55 & \!\!\! 34 & \!\!\! -23 & \!\!\! 11 & \!\!\! -6 & \!\!\! 2 & \!\!\! -1 \\
&&&&&&&&&&&&&&&&&&&&&&&&&&&&&& \\
2 & \!\!\! -4 & \!\!\! 11 & \!\!\! -20 & \!\!\! 40 & \!\!\! -57 & \!\!\! 91 & \!\!\! -118 & \!\!\! 159 & \!\!\! -181 & \!\!\! 220 & \!\!\! -233 & \!\!\! 260 & \!\!\! -255 & \!\!\! 273 & \!\!\! -264 & \!\!\! 273 & \!\!\! -255 & \!\!\! 260 & \!\!\! -233 & \!\!\! 220 & \!\!\! -181 & \!\!\! 159 & \!\!\! -118 & \!\!\! 91 & \!\!\! -57 & \!\!\! 40 & \!\!\! -20 & \!\!\! 11 & \!\!\! -4 & \!\!\! 2 \\
&&&&&&&&&&&&&&&&&&&&&&&&&&&&&& \\
-1 & \!\!\! 2 & \!\!\! -7 & \!\!\! 11 & \!\!\! -22 & \!\!\! 31 & \!\!\! -49 & \!\!\! 55 & \!\!\! -76 & \!\!\! 78 & \!\!\! -90 & \!\!\! 77 & \!\!\! -85 & \!\!\! 66 & \!\!\! -72 & \!\!\! 56 & \!\!\! -72 & \!\!\! 66 & \!\!\! -85 & \!\!\! 77 & \!\!\! -90 & \!\!\! 78 & \!\!\! -76 & \!\!\! 55 & \!\!\! -49 & \!\!\! 31 & \!\!\! -22 & \!\!\! 11 & \!\!\! -7 & \!\!\! 2 & \!\!\! -1 \\
&&&&&&&&&&&&&&&&&&&&&&&&&&&&&& \\
0 & \!\!\! 0 & \!\!\! 1 & \!\!\! -1 & \!\!\! 3 & \!\!\! -2 & \!\!\! 3 & \!\!\! 2 & \!\!\! -5 & \!\!\! 18 & \!\!\! -25 & \!\!\! 45 & \!\!\! -56 & \!\!\! 75 & \!\!\! -78 & \!\!\! 88 & \!\!\! -78 & \!\!\! 75 & \!\!\! -56 & \!\!\! 45 & \!\!\! -25 & 18 & \!\!\! -5 & \!\!\! 2 & \!\!\! 3 & \!\!\! -2 & \!\!\! 3 & \!\!\! -1 & \!\!\! 1 & \!\!\! 0 & \!\!\! 0 \\
&&&&&&&&&&&&&&&&&&&&&&&&&&&&&& \\
0 & \!\!\! 0 & \!\!\! 0 & \!\!\! 0 & \!\!\! 0 & \!\!\! -1 & \!\!\! 2
& \!\!\! -5 & \!\!\! 9 & \!\!\! -15 & \!\!\! 20 & \!\!\! -27 &
\!\!\! 32 & \!\!\! -38 & \!\!\! 40 & \!\!\! -42 & \!\!\! 40 & \!\!\!
-38 & \!\!\! 32 & \!\!\! -27 & \!\!\! 20 & -15 & \!\!\! 9 & \!\!\!
-5 & \!\!\! 2 & \!\!\! -1 & \!\!\! 0 & \!\!\! 0 & \!\!\! 0 & \!\!\!
0 & \!\!\! 0
\end{array}\right)
}
\end{array}
\end{equation}

\newpage

\noindent\textbf{Knot $8_9$}

\begin{equation}
\nonumber
\begin{array}{l}
H_{[2,1]}=\frac{1}{q^{30}A^{6}}
\\
\\
{\tiny \left(\begin{array}{rrrrrrrrrrrrrrrrrrrrrrrrrrrrrrr} 0 &
\!\!\! 0 & \!\!\! 0 & \!\!\! 0 & \!\!\! 0 & \!\!\! 1 & \!\!\! -2 &
\!\!\! 5 & \!\!\! -9 & \!\!\! 15 & \!\!\! -20 & \!\!\! 27 & \!\!\!
-32 & \!\!\! 38 & \!\!\! -40 & \!\!\!
42 & \!\!\! -40 & \!\!\! 38 & \!\!\! -32 & \!\!\! 27 & \!\!\! -20 & \!\!\! 15 & \!\!\! -9 & \!\!\! 5 & \!\!\! -2 & \!\!\! 1 & \!\!\! 0 & \!\!\! 0 & \!\!\! 0 & \!\!\! 0 & \!\!\! 0 \\
&&&&&&&&&&&&&&&&&&&&&&&&&&&&&& \\
0 & \!\!\! 0 & \!\!\! -1 & \!\!\! 1 & \!\!\! -3 & \!\!\! 3 & \!\!\! -5 & \!\!\! 1 & \!\!\! -1 & \!\!\! -5 & \!\!\! 8 & \!\!\! -18 & \!\!\! 16 & \!\!\! -21 & \!\!\! 20 & \!\!\! -26 & \!\!\! 20 & \!\!\! -21 & \!\!\! 16 & \!\!\! -18 & 8 & \!\!\! -5 & -1 & \!\!\! 1 & \!\!\! -5 & \!\!\! 3 & \!\!\! -3 & \!\!\! 1 & \!\!\! -1 & \!\!\! 0 & \!\!\! 0 \\
&&&&&&&&&&&&&&&&&&&&&&&&&&&&&& \\
1 & \!\!\! -2 & \!\!\! 7 & \!\!\! -12 & \!\!\! 25 & \!\!\! -36 &
\!\!\! 58 & \!\!\! -74 & \!\!\! 103 & \!\!\! -121 & \!\!\! 155 &
\!\!\! -175 & 208 & \!\!\! -224 & \!\!\! 247 & \!\!\! -242 & \!\!\!
247 & \!\!\! -224 & \!\!\! 208 & \!\!\! -175 & \!\!\!
155 & \!\!\! -121 & \!\!\! 103 & -74 & \!\!\! 58 & \!\!\! -36 & \!\!\! 25 & \!\!\! -12 & \!\!\! 7 & \!\!\! -2 & \!\!\! 1 \\
&&&&&&&&&&&&&&&&&&&&&&&&&&&&&& \\
-2 & \!\!\! 4 & \!\!\! -12 & \!\!\! 22 & \!\!\! -44 & \!\!\! 64 & \!\!\! -103 & \!\!\! 136 & -186 & \!\!\! 225 & \!\!\! -286 & \!\!\! 332 & \!\!\! -389 & \!\!\! 414 & \!\!\! -454 & \!\!\! 459 & \!\!\! -454 & \!\!\! 414 & \!\!\! -389 & \!\!\! 332 & \!\!\! -286 & \!\!\! 225 & \!\!\! -186 & \!\!\! 136 & \!\!\! -103 & \!\!\! 64 & \!\!\! -44 & \!\!\! 22 & \!\!\! -12 & \!\!\! 4 & \!\!\! -2 \\
&&&&&&&&&&&&&&&&&&&&&&&&&&&&&& \\
1 & \!\!\! -2 & \!\!\! 7 & \!\!\! -12 & \!\!\! 25 & \!\!\! -36 & \!\!\! 58 & \!\!\! -74 & \!\!\! 103 & \!\!\! -121 & \!\!\! 155 & \!\!\! -175 & \!\!\! 208 & \!\!\! -224 & \!\!\! 247 & \!\!\! -242 & \!\!\! 247 & \!\!\! -224 & \!\!\! 208 & \!\!\! -175 & \!\!\! 155 & \!\!\! -121 & \!\!\! 103 & \!\!\! -74 & \!\!\! 58 & \!\!\! -36 & 25 & \!\!\! -12 & \!\!\! 7 & \!\!\! -2 & \!\!\! 1 \\
&&&&&&&&&&&&&&&&&&&&&&&&&&&&&& \\
0 & \!\!\! 0 & \!\!\! -1 & \!\!\! 1 & \!\!\! -3 & \!\!\! 3 & \!\!\! -5 & \!\!\! 1 & \!\!\! -1 & \!\!\! -5 & \!\!\! 8 & \!\!\! -18 & \!\!\! 16 & \!\!\! -21 & \!\!\! 20 & \!\!\! -26 & \!\!\! 20 & \!\!\! -21 & \!\!\! 16 & \!\!\! -18 & \!\!\! 8 & \!\!\! -5 & \!\!\! -1 & \!\!\! 1 & \!\!\! -5 & \!\!\! 3 & \!\!\! -3 & \!\!\! 1 & \!\!\! -1 & \!\!\! 0 & \!\!\! 0 \\
&&&&&&&&&&&&&&&&&&&&&&&&&&&&&& \\
0 & \!\!\! 0 & \!\!\! 0 & \!\!\! 0 & \!\!\! 0 & \!\!\! 1 & \!\!\! -2
& \!\!\! 5 & \!\!\! -9 & \!\!\! 15 & \!\!\! -20 & \!\!\! 27 & \!\!\!
-32 & \!\!\! 38 & \!\!\! -40 & \!\!\! 42 & \!\!\! -40 & \!\!\! 38 &
\!\!\! -32 & \!\!\! 27 & \!\!\! -20 & \!\!\! 15 & \!\!\! -9 & \!\!\!
5 & \!\!\! -2 & \!\!\! 1 & \!\!\! 0 & \!\!\! 0 & \!\!\! 0 & \!\!\! 0
& \!\!\! 0
\end{array}\right)
}
\end{array}
\end{equation}

\noindent\textbf{Knot $8_{10}$}

\begin{equation}
\nonumber
\begin{array}{l}
H_{[2,1]}=\frac{1}{q^{30}A^{6}}
\\
\\
{\tiny \left(\begin{array}{rrrrrrrrrrrrrrrrrrrrrrrrrrrrrrr}
0 & \!\!\! 0 & \!\!\! 0 & \!\!\! 0 & \!\!\! 0 & \!\!\! -1 & \!\!\! 2 & \!\!\! -5 & \!\!\! 9 & \!\!\! -15 & \!\!\! 20 & \!\!\! -27 & \!\!\! 32 & \!\!\! -38 & \!\!\! 40 & \!\!\! -42 & \!\!\! 40 & \!\!\! -38 & \!\!\! 32 & \!\!\! -27 & \!\!\! 20 & \!\!\! -15 & \!\!\! 9 & \!\!\! -5 & \!\!\! 2 & \!\!\! -1 & \!\!\! 0 & \!\!\! 0 & \!\!\! 0 & \!\!\! 0 & \!\!\! 0 \\
&&&&&&&&&&&&&&&&&&&&&&&&&&&&&& \\
0 & \!\!\! 0 & \!\!\! 1 & \!\!\! -1 & \!\!\! 4 & \!\!\! -5 & \!\!\! 11 & \!\!\! -13 & \!\!\! 26 & \!\!\! -31 & \!\!\! 50 & \!\!\! -56 & \!\!\! 77 & \!\!\! -78 & \!\!\! 95 & \!\!\! -88 & \!\!\! 95 & \!\!\! -78 & \!\!\! 77 & \!\!\! -56 & \!\!\! 50 & \!\!\! -31 & \!\!\! 26 & \!\!\! -13 & \!\!\! 11 & \!\!\! -5 & \!\!\! 4 & \!\!\! -1 & \!\!\! 1 & \!\!\! 0 & \!\!\! 0 \\
&&&&&&&&&&&&&&&&&&&&&&&&&&&&&& \\
-1 & \!\!\! 2 & \!\!\! -8 & \!\!\! 14 & \!\!\! -33 & \!\!\! 49 & \!\!\! -89 & \!\!\! 119 & \!\!\! -186 & \!\!\! 222 & \!\!\! -303 & \!\!\! 333 & \!\!\! -410 & \!\!\! 414 & \!\!\! -471 & \!\!\! 444 & \!\!\! -471 & \!\!\! 414 & \!\!\! -410 & \!\!\! 333 & \!\!\! -303 & \!\!\! 222 & \!\!\! -186 & \!\!\! 119 & \!\!\! -89 & \!\!\! 49 & \!\!\! -33 & \!\!\! 14 & \!\!\! -8 & \!\!\! 2 & \!\!\! -1 \\
&&&&&&&&&&&&&&&&&&&&&&&&&&&&&& \\
2 & \!\!\! -4 & \!\!\! 14 & \!\!\! -23 & \!\!\! 51 & \!\!\! -72 & \!\!\! 129 & \!\!\! -159 & \!\!\! 242 & \!\!\! -273 & \!\!\! 371 & \!\!\! -377 & \!\!\! 476 & \!\!\! -454 & \!\!\! 530 & \!\!\! -474 & \!\!\! 530 & \!\!\! -454 & \!\!\! 476 & \!\!\! -377 & \!\!\! 371 & \!\!\! -273 & \!\!\! 242 & \!\!\! -159 & \!\!\! 129 & \!\!\! -72 & \!\!\! 51 & \!\!\! -23 & \!\!\! 14 & \!\!\! -4 & \!\!\! 2 \\
&&&&&&&&&&&&&&&&&&&&&&&&&&&&&& \\
-1 & \!\!\! 2 & \!\!\! -8 & \!\!\! 11 & \!\!\! -26 & \!\!\! 31 & \!\!\! -58 & \!\!\! 58 & \!\!\! -95 & \!\!\! 80 & \!\!\! -124 & \!\!\! 87 & \!\!\! -129 & \!\!\! 76 & \!\!\! -127 & \!\!\! 68 & \!\!\! -127 & \!\!\! 76 & \!\!\! -129 & \!\!\! 87 & \!\!\! -124 & \!\!\! 80 & \!\!\! -95 & \!\!\! 58 & \!\!\! -58 & \!\!\! 31 & \!\!\! -26 & \!\!\! 11 & \!\!\! -8 & \!\!\! 2 & \!\!\! -1 \\
&&&&&&&&&&&&&&&&&&&&&&&&&&&&&& \\
0 & \!\!\! 0 & \!\!\! 1 & \!\!\! -1 & \!\!\! 4 & \!\!\! -1 & \!\!\! 4 & \!\!\! 6 & \!\!\! -5 & \!\!\! 30 & \!\!\! -34 & \!\!\! 70 & \!\!\! -72 & \!\!\! 119 & \!\!\! -106 & \!\!\! 132 & \!\!\! -106 & \!\!\! 119 & \!\!\! -72 & \!\!\! 70 & \!\!\! -34 & \!\!\! 30 & \!\!\! -5 & \!\!\! 6 & \!\!\! 4 & \!\!\! -1 & \!\!\! 4 & \!\!\! -1 & \!\!\! 1 & \!\!\! 0 & \!\!\! 0 \\
&&&&&&&&&&&&&&&&&&&&&&&&&&&&&& \\
0 & \!\!\! 0 & \!\!\! 0 & \!\!\! 0 & \!\!\! 0 & \!\!\! -1 & \!\!\! 2
& \!\!\! -6 & \!\!\! 9 & \!\!\! -16 & \!\!\! 20 & \!\!\! -30 &
\!\!\! 32 & \!\!\! -39 & \!\!\! 39 & \!\!\! -47 & \!\!\! 39 & \!\!\!
-39 & \!\!\! 32 & \!\!\! -30 & \!\!\! 20 & \!\!\! -16 & \!\!\! 9 &
\!\!\! -6 & \!\!\! 2 & \!\!\! -1 & \!\!\! 0 & \!\!\! 0 & \!\!\! 0 &
\!\!\! 0 & \!\!\! 0
\end{array}\right)
}
\end{array}
\end{equation}

\noindent\textbf{Knot $8_{16}$}

\begin{equation}
\nonumber
\begin{array}{l}
H_{[2,1]}=\frac{1}{q^{30}A^{6}}
\\
\\
{\tiny \left(\begin{array}{rrrrrrrrrrrrrrrrrrrrrrrrrrrrrrr}
0 & \!\!\! 0 & \!\!\! 0 & \!\!\! 0 & 0 & \!\!\! -1 & \!\!\! 4 & \!\!\! -9 & \!\!\! 16 & \!\!\! -27 & \!\!\! 40 & \!\!\! -51 & \!\!\! 61 & \!\!\! -71 & \!\!\! 76 & \!\!\! -77 & \!\!\! 76 & \!\!\! -71 & \!\!\! 61 & \!\!\! -51 & \!\!\! 40 & \!\!\! -27 & \!\!\! 16 & \!\!\! -9 & \!\!\! 4 & \!\!\! -1 & \!\!\! 0 & \!\!\! 0 & \!\!\! 0 & \!\!\! 0 & \!\!\! 0 \\
&&&&&&&&&&&&&&&&&&&&&&&&&&&&&& \\
0 & \!\!\! 0 & \!\!\! 1 & \!\!\! -3 & \!\!\! 5 & \!\!\! -5 & \!\!\! 4 & \!\!\! 2 & \!\!\! -19 & \!\!\! 48 & \!\!\! -84 & \!\!\! 130 & \!\!\! -183 & \!\!\! 232 & \!\!\! -260 & \!\!\! 270 & \!\!\! -260 & \!\!\! 232 & \!\!\! -183 & \!\!\! 130 & \!\!\! -84 & \!\!\! 48 & \!\!\! -19 & \!\!\! 2 & \!\!\! 4 & \!\!\! -5 & \!\!\! 5 & \!\!\! -3 & \!\!\! 1 & \!\!\! 0 & \!\!\! 0 \\
&&&&&&&&&&&&&&&&&&&&&&&&&&&&&& \\
-1 & \!\!\! 4 & \!\!\! -11 & \!\!\! 22 & \!\!\! -40 & \!\!\! 67 & \!\!\! -100 & \!\!\! 133 & \!\!\! -170 & \!\!\! 206 & \!\!\! -229 & \!\!\! 236 & \!\!\! -240 & \!\!\! 238 & \!\!\! -232 & \!\!\! 222 & \!\!\! -232 & 238 & -240 & 236 & -229 & \!\!\! 206 & \!\!\! -170 & \!\!\! 133 & \!\!\! -100 & \!\!\! 67 & \!\!\! -40 & \!\!\! 22 & \!\!\! -11 & \!\!\! 4 & \!\!\! -1 \\
&&&&&&&&&&&&&&&&&&&&&&&&&&&&&& \\
2 & \!\!\! -8 & \!\!\! 21 & \!\!\! -46 & \!\!\! 91 & \!\!\! -160 & \!\!\! 254 & \!\!\! -375 & \!\!\! 518 & \!\!\! -674 & \!\!\! 827 & \!\!\! -971 & \!\!\! 1098 & \!\!\! -1187 & \!\!\! 1246 & \!\!\! -1264 & \!\!\! 1246 & \!\!\! -1187 & \!\!\! 1098 & \!\!\! -971 & \!\!\! 827 & \!\!\! -674 & \!\!\! 518 & \!\!\! -375 & \!\!\! 254 & \!\!\! -160 & \!\!\! 91 & \!\!\! -46 & \!\!\! 21 & \!\!\! -8 & \!\!\! 2 \\
&&&&&&&&&&&&&&&&&&&&&&&&&&&&&& \\
-1 & \!\!\! 4 & \!\!\! -12 & \!\!\! 30 & \!\!\! -62 & \!\!\! 113 & \!\!\! -188 & \!\!\! 296 & \!\!\! -426 & \!\!\! 573 & \!\!\! -733 & \!\!\! 890 & \!\!\! -1023 & \!\!\! 1128 & \!\!\! -1198 & \!\!\! 1218 & \!\!\! -1198 & \!\!\! 1128 & \!\!\! -1023 & \!\!\! 890 & \!\!\! -733 & \!\!\! 573 & \!\!\! -426 & \!\!\! 296 & \!\!\! -188 & \!\!\! 113 & \!\!\! -62 & \!\!\! 30 & \!\!\! -12 & \!\!\! 4 & \!\!\! -1 \\
&&&&&&&&&&&&&&&&&&&&&&&&&&&&&& \\
0 & \!\!\! 0 & \!\!\! 1 & \!\!\! -3 & \!\!\! 6 & \!\!\! -13 & \!\!\! 23 & \!\!\! -38 & \!\!\! 62 & \!\!\! -97 & \!\!\! 133 & \!\!\! -173 & \!\!\! 217 & \!\!\! -250 & \!\!\! 272 & \!\!\! -280 & \!\!\! 272 & \!\!\! -250 & \!\!\! 217 & \!\!\! -173 & 133 & \!\!\! -97 & \!\!\! 62 & \!\!\! -38 & \!\!\! 23 & \!\!\! -13 & \!\!\! 6 & \!\!\! -3 & \!\!\! 1 & \!\!\! 0 & \!\!\! 0 \\
&&&&&&&&&&&&&&&&&&&&&&&&&&&&&& \\
0 & \!\!\! 0 & \!\!\! 0 & \!\!\! 0 & \!\!\! 0 & \!\!\! -1 & \!\!\! 4
& \!\!\! -9 & \!\!\! 19 & \!\!\! -33 & \!\!\! 46 & \!\!\! -61 &
\!\!\! 78 & \!\!\! -90 & \!\!\! 96 & \!\!\! -98 & \!\!\! 96 & \!\!\!
-90 & \!\!\! 78 & \!\!\! -61 & \!\!\! 46 & \!\!\! -33 & \!\!\! 19 &
-9 & \!\!\! 4 & \!\!\! -1 & \!\!\! 0 & \!\!\! 0 & \!\!\! 0 & \!\!\!
0 & \!\!\! 0
\end{array}\right)
}
\end{array}
\end{equation}

\newpage

\noindent\textbf{Knot $8_{17}$}

\begin{equation}
\nonumber
\begin{array}{l}
H_{[2,1]}=\frac{1}{q^{30}A^{6}}
\\
\\
{\tiny \left(\begin{array}{rrrrrrrrrrrrrrrrrrrrrrrrrrrrrrr}
0 & \!\!\! 0 & \!\!\! 0 & \!\!\! 0 & \!\!\! 0 & \!\!\! 1 & \!\!\! -4 & \!\!\! 10 & \!\!\! -20 & \!\!\! 36 & \!\!\! -58 & \!\!\! 85 & \!\!\! -112 & \!\!\! 136 & \!\!\! -154 & \!\!\! 161 & \!\!\! -154 & \!\!\! 136 & \!\!\! -112 & \!\!\! 85 & \!\!\! -58 & \!\!\! 36 & \!\!\! -20 & \!\!\! 10 & \!\!\! -4 & \!\!\! 1 & \!\!\! 0 & \!\!\! 0 & \!\!\! 0 & \!\!\! 0 & \!\!\! 0 \\
&&&&&&&&&&&&&&&&&&&&&&&&&&&&&& \\
0 & \!\!\! 0 & \!\!\! -1 & \!\!\! 3 & \!\!\! -6 & \!\!\! 9 & \!\!\! -13 & \!\!\! 16 & \!\!\! -17 & \!\!\! 13 & \!\!\! -7 & \!\!\! 1 & \!\!\! 3 & \!\!\! -9 & \!\!\! 13 & \!\!\! -13 & \!\!\! 13 & \!\!\! -9 & \!\!\! 3 & \!\!\! 1 & \!\!\! -7 & \!\!\! 13 & \!\!\! -17 & \!\!\! 16 & \!\!\! -13 & \!\!\! 9 & \!\!\! -6 & \!\!\! 3 & \!\!\! -1 & \!\!\! 0 & \!\!\! 0 \\
&&&&&&&&&&&&&&&&&&&&&&&&&&&&&& \\
1 & \!\!\! -4 & \!\!\! 12 & \!\!\! -27 & \!\!\! 54 & \!\!\! -96 & \!\!\! 159 & \!\!\! -243 & \!\!\! 351 & \!\!\! -478 & \!\!\! 626 & \!\!\! -780 & \!\!\! 925 & \!\!\! -1049 & \!\!\! 1131 & \!\!\! -1158 & \!\!\! 1131 & \!\!\! -1049 & \!\!\! 925 & \!\!\! -780 & \!\!\! 626 & \!\!\! -478 & \!\!\! 351 & \!\!\! -243 & \!\!\! 159 & \!\!\! -96 & \!\!\! 54 & \!\!\! -27 & \!\!\! 12 & \!\!\! -4 & \!\!\! 1 \\
&&&&&&&&&&&&&&&&&&&&&&&&&&&&&& \\
-2 & \!\!\! 8 & \!\!\! -22 & \!\!\! 48 & \!\!\! -96 & \!\!\! 172 & \!\!\! -285 & \!\!\! 434 & \!\!\! -628 & \!\!\! 862 & \!\!\! -1122 & \!\!\! 1388 & \!\!\! -1640 & \!\!\! 1844 & \!\!\! -1980 & \!\!\! 2031 & \!\!\! -1980 & \!\!\! 1844 & \!\!\! -1640 & \!\!\! 1388 & \!\!\! -1122 & \!\!\! 862 & \!\!\! -628 & \!\!\! 434 & \!\!\! -285 & \!\!\! 172 & \!\!\! -96 & \!\!\! 48 & \!\!\! -22 & \!\!\! 8 & \!\!\! -2 \\
&&&&&&&&&&&&&&&&&&&&&&&&&&&&&& \\
1 & \!\!\! -4 & \!\!\! 12 & \!\!\! -27 & \!\!\! 54 & \!\!\! -96 & \!\!\! 159 & \!\!\! -243 & \!\!\! 351 & \!\!\! -478 & \!\!\! 626 & \!\!\! -780 & \!\!\! 925 & \!\!\! -1049 & \!\!\! 1131 & \!\!\! -1158 & \!\!\! 1131 & \!\!\! -1049 & \!\!\! 925 & \!\!\! -780 & \!\!\! 626 & \!\!\! -478 & \!\!\! 351 & \!\!\! -243 & \!\!\! 159 & \!\!\! -96 & \!\!\! 54 & \!\!\! -27 & \!\!\! 12 & \!\!\! -4 & \!\!\! 1 \\
&&&&&&&&&&&&&&&&&&&&&&&&&&&&&& \\
0 & \!\!\! 0 & \!\!\! -1 & \!\!\! 3 & \!\!\! -6 & \!\!\! 9 & \!\!\! -13 & \!\!\! 16 & \!\!\! -17 & \!\!\! 13 & \!\!\! -7 & \!\!\! 1 & \!\!\! 3 & \!\!\! -9 & \!\!\! 13 & \!\!\! -13 & \!\!\! 13 & \!\!\! -9 & \!\!\! 3 & \!\!\! 1 & \!\!\! -7 & \!\!\! 13 & \!\!\! -17 & \!\!\! 16 & \!\!\! -13 & \!\!\! 9 & \!\!\! -6 & \!\!\! 3 & \!\!\! -1 & \!\!\! 0 & \!\!\! 0 \\
&&&&&&&&&&&&&&&&&&&&&&&&&&&&&& \\
0 & \!\!\! 0 & \!\!\! 0 & \!\!\! 0 & \!\!\! 0 & \!\!\! 1 & \!\!\! -4
& \!\!\! 10 & \!\!\! -20 & \!\!\! 36 & \!\!\! -58 & \!\!\! 85 &
\!\!\! -112 & \!\!\! 136 & \!\!\! -154 & \!\!\! 161 & \!\!\! -154 &
\!\!\! 136 & \!\!\! -112 & \!\!\! 85 & \!\!\! -58 & \!\!\! 36 &
\!\!\! -20 & \!\!\! 10 & \!\!\! -4 & \!\!\! 1 & \!\!\! 0 & \!\!\! 0
& \!\!\! 0 & \!\!\! 0 & \!\!\! 0
\end{array}\right)
}
\end{array}
\end{equation}

\noindent\textbf{Knot $8_{18}$}

\begin{equation}
\nonumber
\begin{array}{l}
H_{[2,1]}=\frac{1}{q^{30}A^{6}}
\\
\\
{\tiny \left(\begin{array}{rrrrrrrrrrrrrrrrrrrrrrrrrrrrrrrr}
0 & \!\!\!\! 0 & \!\!\!\! 0 & \!\!\!\! 0 & \!\!\!\! 0 & \!\!\!\! 1 & \!\!\!\! -6 & \!\!\!\! 16 & \!\!\!\! -32 & \!\!\!\! 62 & \!\!\!\! -106 & \!\!\!\! 155 & \!\!\!\! -210 & \!\!\!\! 264 & \!\!\!\! -298 & \!\!\!\! 307 & \!\!\!\! -298 & \!\!\!\! 264 & \!\!\!\! -210 & \!\!\!\! 155 & \!\!\!\! -106 & \!\!\!\! 62 & \!\!\!\! -32 & \!\!\!\! 16 & \!\!\!\! -6 & \!\!\!\! 1 & \!\!\!\! 0 & \!\!\!\! 0 & \!\!\!\! 0 & \!\!\!\! 0 & \!\!\!\! 0 \\
&&&&&&&&&&&&&&&&&&&&&&&&&&&&&& \\
0 & \!\!\!\! 0 & \!\!\!\! -1 & \!\!\!\! 5 & \!\!\!\! -10 & \!\!\!\!
15 & \!\!\!\! -25 & \!\!\!\! 36 & \!\!\!\! -41 & \!\!\!\!
42 & \!\!\!\! -40 & \!\!\!\! 34 & \!\!\!\! -26 & \!\!\!\! 23 & \!\!\!\! -13 & \!\!\!\! 11 & \!\!\!\! -13 & \!\!\!\! 23 & \!\!\!\! -26 & \!\!\!\! 34 & \!\!\!\! -40 & \!\!\!\! 42 & \!\!\!\! -41 & \!\!\!\! 36 & \!\!\!\! -25 & \!\!\!\! 15 & \!\!\!\! -10 & \!\!\!\! 5 & \!\!\!\! -1 & \!\!\!\! 0 & \!\!\!\! 0 \\
&&&&&&&&&&&&&&&&&&&&&&&&&&&&&& \\
1 & \!\!\!\! -6 & \!\!\!\! 18 & \!\!\!\! -43 & 90 & \!\!\!\! -168 & \!\!\!\! 289 & \!\!\!\! -458 & \!\!\!\! 678 & \!\!\!\! -947 & \!\!\!\! 1255 & \!\!\!\! -1578 & \!\!\!\! 1876 & \!\!\!\! -2142 & \!\!\!\! 2301 & \!\!\!\! -2362 & \!\!\!\! 2301 & \!\!\!\! -2142 & \!\!\!\! 1876 & \!\!\!\! -1578 & \!\!\!\! 1255 & \!\!\!\! -947 & \!\!\!\! 678 & \!\!\!\! -458 & \!\!\!\! 289 & \!\!\!\! -168 & \!\!\!\! 90 & \!\!\!\! -43 & \!\!\!\! 18 & \!\!\!\! -6 & \!\!\!\! 1 \\
&&&&&&&&&&&&&&&&&&&&&&&&&&&&&& \\
-2 & \!\!\!\! 12 & \!\!\!\! -34 & \!\!\!\! 76 & \!\!\!\! -160 & \!\!\!\! 304 & \!\!\!\! -517 & \!\!\!\! 812 & \!\!\!\! -1210 & \!\!\!\! 1691 & \!\!\!\! -2218 & \!\!\!\! 2778 & \!\!\!\! -3290 & \!\!\!\! 3710 & \!\!\!\! -3980 & \!\!\!\! 4101 & \!\!\!\! -3980 & \!\!\!\! 3710 & \!\!\!\! -3290 & \!\!\!\! 2778 & \!\!\!\! -2218 & \!\!\!\! 1691 & \!\!\!\! -1210 & \!\!\!\! 812 & \!\!\!\! -517 & \!\!\!\! 304 & \!\!\!\! -160 & \!\!\!\! 76 & \!\!\!\! -34 & \!\!\!\! 12 & \!\!\!\! -2 \\
&&&&&&&&&&&&&&&&&&&&&&&&&&&&&& \\
1 & \!\!\!\! -6 & \!\!\!\! 18 & \!\!\!\! -43 & \!\!\!\! 90 & \!\!\!\! -168 & \!\!\!\! 289 & \!\!\!\! -458 & \!\!\!\! 678 & \!\!\!\! -947 & \!\!\!\! 1255 & \!\!\!\! -1578 & \!\!\!\! 1876 & \!\!\!\! -2142 & \!\!\!\! 2301 & \!\!\!\! -2362 & \!\!\!\! 2301 & \!\!\!\! -2142 & \!\!\!\! 1876 & \!\!\!\! -1578 & \!\!\!\! 1255 & \!\!\!\! -947 & \!\!\!\! 678 & \!\!\!\! -458 & \!\!\!\! 289 & \!\!\!\! -168 & \!\!\!\! 90 & \!\!\!\! -43 & \!\!\!\! 18 & \!\!\!\! -6 & \!\!\!\! 1 \\
&&&&&&&&&&&&&&&&&&&&&&&&&&&&&& \\
0 & \!\!\!\! 0 & \!\!\!\! -1 & \!\!\!\! 5 & \!\!\!\! -10 & \!\!\!\! 15 & \!\!\!\! -25 & \!\!\!\! 36 & \!\!\!\! -41 & \!\!\!\! 42 & \!\!\!\! -40 & \!\!\!\! 34 & \!\!\!\! -26 & \!\!\!\! 23 & \!\!\!\! -13 & \!\!\!\! 11 & -13 & \!\!\!\! 23 & \!\!\!\! -26 & \!\!\!\! 34 & \!\!\!\! -40 & \!\!\!\! 42 & \!\!\!\! -41 & \!\!\!\! 36 & \!\!\!\! -25 & \!\!\!\! 15 & \!\!\!\! -10 & \!\!\!\! 5 & \!\!\!\! -1 & \!\!\!\! 0 & \!\!\!\! 0 \\
&&&&&&&&&&&&&&&&&&&&&&&&&&&&&& \\
0 & \!\!\!\! 0 & \!\!\!\! 0 & \!\!\!\! 0 & \!\!\!\! 0 & \!\!\!\! 1 &
\!\!\!\! -6 & \!\!\!\! 16 & \!\!\!\! -32 & \!\!\!\! 62 & \!\!\!\!
-106 & \!\!\!\! 155 & \!\!\!\! -210 & \!\!\!\! 264 & \!\!\!\! -298 &
\!\!\!\! 307 & \!\!\!\! -298 & \!\!\!\! 264 & \!\!\!\! -210 &
\!\!\!\! 155 & \!\!\!\! -106 & \!\!\!\! 62 & \!\!\!\! -32 & \!\!\!\!
16 & \!\!\!\! -6 & \!\!\!\! 1 & \!\!\!\! 0 & \!\!\!\! 0 & \!\!\!\! 0
& \!\!\!\! 0 & \!\!\!\! 0
\end{array}\right)
}
\end{array}
\end{equation}

\noindent\textbf{Knot $8_{19}$}

\begin{equation}
\nonumber
\begin{array}{l}
H_{[2,1]}=\frac{1}{q^{30}A^{6}}
\\
\\
{\tiny \left(\begin{array}{rrrrrrrrrrrrrrrrrrrrrrrrrrrrrrr}
1 & 0 & 2 & 1 & 3 & 1 & 5 & 3 & 6 & 3 & 8 & 4 & 10 & 3 & 10 & 5 & 10 & 3 & 10 & 4 & 8 & 3 & 6 & 3 & 5 & 1 & 3 & 1 & 2 & 0 & 1 \\
&&&&&&&&&&&&&&&&&&&&&&&&&&&&&& \\
-1 & -1 & -3 & -3 & -6 & -6 & -10 & -10 & -16 & -14 & -21 & -16 & -25 & -19 & -27 & -19 & -27 & -19 & -25 & -16 & -21 & -14 & -16 & -10 & -10 & -6 & -6 & -3 & -3 & -1 & -1 \\
&&&&&&&&&&&&&&&&&&&&&&&&&&&&&& \\
0 & 1 & 1 & 3 & 4 & 8 & 9 & 13 & 15 & 20 & 22 & 26 & 27 & 31 & 29 & 32 & 29 & 31 & 27 & 26 & 22 & 20 & 15 & 13 & 9 & 8 & 4 & 3 & 1 & 1 & 0 \\
&&&&&&&&&&&&&&&&&&&&&&&&&&&&&& \\
0 & 0 & 0 & -1 & -1 & -3 & -4 & -7 & -8 & -13 & -13 & -18 & -17 & -22 & -18 & -25 & -18 & -22 & -17 & -18 & -13 & -13 & -8 & -7 & -4 & -3 & -1 & -1 & 0 & 0 & 0 \\
&&&&&&&&&&&&&&&&&&&&&&&&&&&&&& \\
0 & 0 & 0 & 0 & 0 & 0 & 1 & 1 & 3 & 3 & 5 & 5 & 7 & 8 & 8 & 8 & 8 & 8 & 7 & 5 & 5 & 3 & 3 & 1 & 1 & 0 & 0 & 0 & 0 & 0 & 0 \\
&&&&&&&&&&&&&&&&&&&&&&&&&&&&&& \\
0 & 0 & 0 & 0 & 0 & 0 & 0 & 0 & 0 & 0 & -1 & -1 & -2 & -1 & -2 & -1 & -2 & -1 & -2 & -1 & -1 & 0 & 0 & 0 & 0 & 0 & 0 & 0 & 0 & 0 & 0 \\
&&&&&&&&&&&&&&&&&&&&&&&&&&&&&& \\
0 & 0 & 0 & 0 & 0 & 0 & 0 & 0 & 0 & 0 & 0 & 0 & 0 & 0 & 0 & 1 & 0 &
0 & 0 & 0 & 0 & 0 & 0 & 0 & 0 & 0 & 0 & 0 & 0 & 0 & 0
\end{array}\right)
}
\end{array}
\end{equation}

\newpage

\noindent\textbf{Knot $8_{20}$}

\begin{equation}
\nonumber
\begin{array}{l}
H_{[2,1]}=\frac{1}{q^{22}A^{6}}
\\
\\
\left(\begin{array}{rrrrrrrrrrrrrrrrrrrrrrr}
0 & 0 & 0 & 0 & 0 & 0 & -1 & 0 & -2 & 1 & -2 & 0 & -2 & 1 & -2 & 0 & -1 & 0 & 0 & 0 & 0 & 0 & 0 \\
&&&&&&&&&&&&&&&&&&&&&& \\
0 & 0 & 0 & 1 & 1 & 3 & 1 & 3 & 3 & 5 & 5 & 4 & 5 & 5 & 3 & 3 & 1 & 3 & 1 & 1 & 0 & 0 & 0 \\
&&&&&&&&&&&&&&&&&&&&&& \\
0 & -1 & -1 & -2 & -2 & -3 & -7 & -5 & -11 & -5 & -14 & -6 & -14 & -5 & -11 & -5 & -7 & -3 & -2 & -2 & -1 & -1 & 0 \\
&&&&&&&&&&&&&&&&&&&&&& \\
1 & -1 & 4 & -3 & 10 & -5 & 17 & -8 & 24 & -6 & 26 & -6 & 26 & -6 & 24 & -8 & 17 & -5 & 10 & -3 & 4 & -1 & 1 \\
&&&&&&&&&&&&&&&&&&&&&& \\
-1 & 2 & -4 & 5 & -10 & 9 & -14 & 9 & -16 & 6 & -15 & 4 & -15 & 6 & -16 & 9 & -14 & 9 & -10 & 5 & -4 & 2 & -1 \\
&&&&&&&&&&&&&&&&&&&&&& \\
0 & 0 & 1 & -1 & 1 & -2 & 2 & 3 & -4 & 7 & -8 & 14 & -8 & 7 & -4 & 3 & 2 & -2 & 1 & -1 & 1 & 0 & 0 \\
&&&&&&&&&&&&&&&&&&&&&& \\
0 & 0 & 0 & 0 & 0 & -1 & 2 & -2 & 4 & -8 & 8 & -7 & 8 & -8 & 4 & -2
& 2 & -1 & 0 & 0 & 0 & 0 & 0
\end{array}\right)
\end{array}
\end{equation}

\noindent\textbf{Knot $8_{21}$}

\begin{equation}
\nonumber
\begin{array}{l}
H_{[2,1]}=\frac{1}{q^{22}A^{6}}
\\
\\
\left(\begin{array}{rrrrrrrrrrrrrrrrrrrrrrr}
0 & 0 & 0 & 0 & 0 & 0 & 1 & -2 & 3 & -4 & 5 & -5 & 5 & -4 & 3 & -2 & 1 & 0 & 0 & 0 & 0 & 0 & 0 \\
&&&&&&&&&&&&&&&&&&&&&& \\
0 & 0 & 0 & -1 & 1 & -3 & 4 & -6 & 7 & -12 & 11 & -11 & 11 & -12 & 7 & -6 & 4 & -3 & 1 & -1 & 0 & 0 & 0 \\
&&&&&&&&&&&&&&&&&&&&&& \\
0 & 1 & 0 & 2 & -1 & 2 & 3 & 0 & 8 & -7 & 14 & -8 & 14 & -7 & 8 & 0 & 3 & 2 & -1 & 2 & 0 & 1 & 0 \\
&&&&&&&&&&&&&&&&&&&&&& \\
-2 & 3 & -8 & 12 & -25 & 30 & -51 & 58 & -83 & 85 & -106 & 93 & -106 & 85 & -83 & 58 & -51 & 30 & -25 & 12 & -8 & 3 & -2 \\
&&&&&&&&&&&&&&&&&&&&&& \\
2 & -4 & 10 & -13 & 29 & -36 & 59 & -65 & 98 & -93 & 121 & -108 & 121 & -93 & 98 & -65 & 59 & -36 & 29 & -13 & 10 & -4 & 2 \\
&&&&&&&&&&&&&&&&&&&&&& \\
0 & 0 & -2 & 0 & -4 & 2 & -14 & 9 & -25 & 20 & -38 & 23 & -38 & 20 & -25 & 9 & -14 & 2 & -4 & 0 & -2 & 0 & 0 \\
&&&&&&&&&&&&&&&&&&&&&& \\
0 & 0 & 0 & 0 & 0 & 4 & -2 & 6 & -4 & 11 & -7 & 11 & -7 & 11 & -4 &
6 & -2 & 4 & 0 & 0 & 0 & 0 & 0
\end{array}\right)
\end{array}
\end{equation}

\end{landscape}

\newpage

\section{$4$-strand knots in representation $[2]$\label{app:4str}}

In this Appendix, we present answers for $4$-strand knots with up to
seven crossings in representation $[2]$. The answers for
representation $[11]$ can be obtained from these by the simple
substitution $q\rightarrow-q^{-1}$.  All the answers are given in the vertical framing, see Sec.~\ref{framing}. We use here the same notation as in Appendix~\ref{app:3str}.

\

\noindent\textbf{Knot $6_{1}$}

\begin{equation}
\nonumber H_{[2]}=\frac{1}{q^{8}A^{6}}
\left(\begin{array}{rrrrrrrrr}
0 & 0 & 0 & 0 & 0 & 0 & 0 & 1 & 0 \\
&&&&&&&& \\
0 & 0 & 0 & 0 & -1 & 0 & 1 & -1 & -1 \\
&&&&&&&& \\
0 & 0 & 0 & -1 & 1 & 2 & -1 & -1 & 1 \\
&&&&&&&& \\
1 & 0 & -2 & 2 & 3 & -2 & -1 & 1 & 0 \\
&&&&&&&& \\
0 & -3 & 0 & 4 & -2 & -2 & 1 & 0 & 0 \\
&&&&&&&& \\
-1 & 0 & 2 & 0 & -1 & 0 & 0 & 0 & 0 \\
&&&&&&&& \\
0 & 1 & 0 & 0 & 0 & 0 & 0 & 0 & 0
\end{array}\right)
\end{equation}

\noindent\textbf{Knot $7_{2}$}

\begin{equation}
\nonumber H_{[2]}=\frac{1}{q^{14}A^{6}}
\left(\begin{array}{rrrrrrrrrrr}
0 & 0 & 0 & 0 & 0 & 0 & 0 & 0 & 0 & 1 & 0 \\
&&&&&&&&&& \\
0 & 0 & 0 & 0 & 0 & 0 & -1 & 0 & 1 & -1 & -1 \\
&&&&&&&&&& \\
0 & 0 & 0 & 0 & 0 & -1 & 1 & 2 & -1 & -1 & 1 \\
&&&&&&&&&& \\
0 & 0 & 0 & -1 & -2 & 1 & 2 & -2 & -1 & 1 & 0 \\
&&&&&&&&&& \\
1 & 1 & -1 & -2 & 2 & 3 & -2 & -1 & 1 & 0 & 0 \\
&&&&&&&&&& \\
1 & 0 & -3 & 1 & 3 & -2 & -1 & 1 & 0 & 0 & 0 \\
&&&&&&&&&& \\
1 & -1 & -1 & 2 & 0 & -1 & 1 & 0 & 0 & 0 & 0
\end{array}\right)
\end{equation}

\noindent\textbf{Knot $7_{4}$}

\begin{equation}
\nonumber H_{[2]}=\frac{1}{q^{6}A^{6}}
\left(\begin{array}{rrrrrrrrrrr}
0 & 0 & 0 & 0 & 1 & -2 & -1 & 4 & -1 & -2 & 1 \\
&&&&&&&&&& \\
0 & 0 & 0 & 2 & -2 & -4 & 6 & 2 & -6 & 0 & 2 \\
&&&&&&&&&& \\
0 & 0 & 3 & -2 & -4 & 6 & 4 & -5 & -1 & 2 & 1 \\
&&&&&&&&&& \\
0 & 2 & -2 & -4 & 5 & 3 & -4 & -1 & 1 & 0 & 0 \\
&&&&&&&&&& \\
1 & -2 & -3 & 3 & 1 & -3 & -1 & 0 & 0 & 0 & 0 \\
&&&&&&&&&& \\
-1 & -1 & 2 & 1 & -1 & 0 & 0 & 0 & 0 & 0 & 0 \\
&&&&&&&&&& \\
0 & 1 & 0 & 0 & 0 & 0 & 0 & 0 & 0 & 0 & 0
\end{array}\right)
\end{equation}

\newpage

\noindent\textbf{Knot $7_{6}$}

\begin{equation}
\nonumber H_{[2]}=\frac{1}{q^{16}A^{6}}
\left(\begin{array}{rrrrrrrrrrrrr}
0 & 0 & 0 & 0 & 0 & 0 & 0 & 0 & 0 & 0 & 1 & 0 & 0 \\
&&&&&&&&&&&& \\
0 & 0 & 0 & 0 & 0 & 0 & 0 & -2 & -1 & 2 & -1 & -2 & 0 \\
&&&&&&&&&&&& \\
0 & 0 & 0 & 0 & 1 & 3 & -4 & -2 & 8 & 0 & -4 & 3 & 1 \\
&&&&&&&&&&&& \\
0 & 0 & -2 & 1 & 5 & -7 & -6 & 10 & -1 & -8 & 3 & 1 & -2 \\
&&&&&&&&&&&& \\
1 & -2 & 0 & 7 & -4 & -7 & 10 & 2 & -7 & 4 & 2 & -2 & 1 \\
&&&&&&&&&&&& \\
-1 & 0 & 3 & -2 & -4 & 4 & 1 & -4 & 1 & 1 & -1 & 0 & 0 \\
&&&&&&&&&&&& \\
0 & 1 & -1 & -1 & 2 & 0 & -1 & 1 & 0 & 0 & 0 & 0 & 0
\end{array}\right)
\end{equation}

\noindent\textbf{Knot $7_{7}$}

\begin{equation}
\nonumber H_{[2]}=\frac{1}{q^{10}A^{6}}
\left(\begin{array}{rrrrrrrrrrrrr}
0 & 0 & 0 & 0 & 0 & 1 & -2 & -1 & 4 & -1 & -2 & 1 & 0 \\
&&&&&&&&&&&& \\
0 & 0 & -1 & 1 & 3 & -6 & -2 & 10 & -2 & -6 & 3 & 1 & -1 \\
&&&&&&&&&&&& \\
1 & -2 & 1 & 7 & -8 & -6 & 16 & -1 & -10 & 5 & 2 & -2 & 1 \\
&&&&&&&&&&&& \\
-2 & 0 & 5 & -7 & -8 & 11 & 1 & -10 & 2 & 2 & -2 & 0 & 0 \\
&&&&&&&&&&&& \\
1 & 4 & -3 & -3 & 8 & 2 & -4 & 2 & 1 & 0 & 0 & 0 & 0 \\
&&&&&&&&&&&& \\
0 & -2 & -2 & 2 & 0 & -2 & 0 & 0 & 0 & 0 & 0 & 0 & 0 \\
&&&&&&&&&&&& \\
0 & 0 & 1 & 0 & 0 & 0 & 0 & 0 & 0 & 0 & 0 & 0 & 0
\end{array}\right)
\end{equation}


\section{The knot $4_1$ and the link $5^2_1$\label{app:FigWh}}

In this Appendix, the results for the colored HOMFLY polynomials for the figure-eight knot $4_1$ and for the Whitehead link $5_1^2$ are provided. The HOMFLY polynomials for $4_1$ are the reduced ones (i.e., divided by the Schur polynomial in the corresponding representation) and the answers for $5_1^2$ are divided by two Schur polynomials, corresponding to the two representations. We use here the same notation as in Appendix~\ref{app:3str}.

\begin{equation}
H^{4_1}_{22}=\frac{1}{q^{16}A^{8}}
\left(\begin{array}{rrrrrrrrrrrrrrrrr}
0 & 0 & 0 & 0 & 0 & 0 & 0 & 0 & 1 & 0 & 0 & 0 & 0 & 0 & 0 & 0 & 0 \\
&&&&&&&&&&&&&&&& \\
0 & 0 & 0 & 0 & 0 & -1 & -2 & 0 & 2 & 0 & -2 & -1 & 0 & 0 & 0 & 0 & 0 \\
&&&&&&&&&&&&&&&& \\
0 & 0 & 0 & 2 & 1 & -1 & -2 & 2 & 6 & 2 & -2 & -1 & 1 & 2 & 0 & 0 & 0 \\
&&&&&&&&&&&&&&&& \\
0 & -1 & -1 & 2 & 1 & -5 & -7 & 0 & 6 & 0 & -7 & -5 & 1 & 2 & -1 & -1 & 0 \\
&&&&&&&&&&&&&&&& \\
1 & -1 & -1 & 4 & 4 & -3 & -5 & 4 & 13 & 4 & -5 & -3 & 4 & 4 & -1 & -1 & 1 \\
&&&&&&&&&&&&&&&& \\
0 & -1 & -1 & 2 & 1 & -5 & -7 & 0 & 6 & 0 & -7 & -5 & 1 & 2 & -1 & -1 & 0 \\
&&&&&&&&&&&&&&&& \\
0 & 0 & 0 & 2 & 1 & -1 & -2 & 2 & 6 & 2 & -2 & -1 & 1 & 2 & 0 & 0 & 0 \\
&&&&&&&&&&&&&&&& \\
0 & 0 & 0 & 0 & 0 & -1 & -2 & 0 & 2 & 0 & -2 & -1 & 0 & 0 & 0 & 0 & 0 \\
&&&&&&&&&&&&&&&& \\
0 & 0 & 0 & 0 & 0 & 0 & 0 & 0 & 1 & 0 & 0 & 0 & 0 & 0 & 0 & 0 & 0
\end{array}\right)
\end{equation}

\

\begin{equation}
H^{4_1}_{31}=\frac{1}{q^{18}A^{8}}
\left(\begin{array}{rrrrrrrrrrrrrrrrrrr}
0 & 0 & 0 & 0 & 0 & 0 & 0 & 0 & 0 & 0 & 0 & 0 & 0 & 1 & 0 & 0 & 0 & 0 & 0 \\
&&&&&&&&&&&&&&&&&& \\
0 & 0 & 0 & 0 & 0 & 0 & 0 & 0 & -1 & 0 & 0 & -1 & 1 & -1 & -1 & 0 & -1 & 0 & 0 \\
&&&&&&&&&&&&&&&&&& \\
0 & 0 & 0 & 0 & 1 & 0 & -1 & 2 & 1 & -1 & 5 & -1 & -2 & 4 & -2 & 1 & 3 & -1 & 1 \\
&&&&&&&&&&&&&&&&&& \\
0 & -1 & 1 & 0 & -5 & 3 & -2 & -6 & 9 & -5 & -6 & 7 & -8 & -1 & 5 & -5 & 0 & 0 & -2 \\
&&&&&&&&&&&&&&&&&& \\
1 & 2 & -3 & 4 & 2 & -7 & 12 & -1 & -8 & 15 & -8 & -1 & 12 & -7 & 2 & 4 & -3 & 2 & 1 \\
&&&&&&&&&&&&&&&&&& \\
-2 & 0 & 0 & -5 & 5 & -1 & -8 & 7 & -6 & -5 & 9 & -6 & -2 & 3 & -5 & 0 & 1 & -1 & 0 \\
&&&&&&&&&&&&&&&&&& \\
1 & -1 & 3 & 1 & -2 & 4 & -2 & -1 & 5 & -1 & 1 & 2 & -1 & 0 & 1 & 0 & 0 & 0 & 0 \\
&&&&&&&&&&&&&&&&&& \\
0 & 0 & -1 & 0 & -1 & -1 & 1 & -1 & 0 & 0 & -1 & 0 & 0 & 0 & 0 & 0 & 0 & 0 & 0 \\
&&&&&&&&&&&&&&&&&& \\
0 & 0 & 0 & 0 & 0 & 1 & 0 & 0 & 0 & 0 & 0 & 0 & 0 & 0 & 0 & 0 & 0 & 0 & 0
\end{array}\right)
\end{equation}

\

\begin{equation}
H^{5^2_1}_{1\otimes 321}=\frac{1}{q^{12}A^3(A - A^{-1})}
\left(\begin{array}{rrrrrrrrrrrrr}
0 & 0 & 0 & 1 & -1 & 1 & -1 & 1 & -1 & 1 & 0 & 0 & 0 \\
&&&&&&&&&&&& \\
-1 & 2 & -3 & 4 & -5 & 6 & -7 & 6 & -5 & 4 & -3 & 2 & -1 \\
&&&&&&&&&&&& \\
0 & 0 & 0 & 1 & -1 & 1 & -2 & 1 & -1 & 1 & 0 & 0 & 0
\end{array}\right)
\end{equation}

\

\begin{equation}
\begin{array}{l}
H^{5^2_1}_{2\otimes 3}=\frac{1}{q^{17}A^6(A - A^{-1})(Aq-A^{-1}q^{-1})}
\\ \\
\left(\begin{array}{rrrrrrrrrrrrrrrrr}
0 & 0 & 0 & 0 & 0 & 0 & 0 & 1 & 0 & -1 & 0 & 1 & 1 & -1 & -1 & 1 & 0 \\
&&&&&&&&&&&&&&&& \\
0 & 0 & 0 & -1 & 0 & 2 & 0 & -3 & -2 & 3 & 3 & -3 & -3 & 1 & 2 & 0 & -1 \\
&&&&&&&&&&&&&&&& \\
1 & -1 & -2 & 3 & 3 & -3 & -6 & 2 & 9 & -1 & -7 & -1 & 4 & 2 & -2 & -1 & 1 \\
&&&&&&&&&&&&&&&& \\
-1 & 0 & 3 & 1 & -5 & -3 & 5 & 5 & -3 & -5 & 1 & 3 & 0 & -1 & 0 & 0 & 0 \\
&&&&&&&&&&&&&&&& \\
0 & 1 & -1 & -2 & 1 & 2 & 1 & -2 & -1 & 1 & 0 & 0 & 0 & 0 & 0 & 0 & 0
\end{array}\right)
\end{array}
\end{equation}

\

\setlength{\arraycolsep}{2pt}

\begin{equation}
\begin{array}{l}
H^{5^2_1}_{2\otimes 5}=\frac{1}{q^{25}A^6(A - A^{-1})(Aq-A^{-1}q^{-1})}
\\ \\
\left(\begin{array}{rrrrrrrrrrrrrrrrrrrrrrrrr}
0 & 0 & 0 & 0 & 0 & 0 & 0 & 0 & 0 & 0 & 0 & 1 & 0 & 0 & 0 & -1 & 0 & 1 & 0 & 0 & 1 & -1 & -1 & 1 & 0 \\
&&&&&&&&&&&&&&&&&&&&&&&& \\
0 & 0 & 0 & 0 & 0 & -1 & 0 & 1 & 0 & 1 & 0 & -3 & -1 & 1 & -1 & 2 & 3 & -3 & -2 & 1 & -1 & 0 & 2 & 0 & -1 \\
&&&&&&&&&&&&&&&&&&&&&&&& \\
1 & -1 & -1 & 1 & -1 & 2 & 3 & -3 & -2 & 1 & -4 & 1 & 8 & -2 & -2 & 2 & -4 & -2 & 4 & 1 & -1 & 1 & -1 & -1 & 1 \\
&&&&&&&&&&&&&&&&&&&&&&&& \\
-1 & 0 & 2 & 0 & 0 & 1 & -4 & -2 & 4 & 0 & 0 & 4 & -2 & -4 & 1 & 0 & 0 & 2 & 0 & -1 & 0 & 0 & 0 & 0 & 0 \\
&&&&&&&&&&&&&&&&&&&&&&&& \\
0 & 1 & -1 & -1 & 1 & -1 & 0 & 2 & 0 & -1 & 1 & -1 & -1 & 1 & 0 & 0 & 0 & 0 & 0 & 0 & 0 & 0 & 0 & 0 & 0
\end{array}\right)
\end{array}
\end{equation}

\setlength{\arraycolsep}{6pt}

\newpage

\section{The form of blocks in the level-three projectors\label{app:projbl}}
In this Appendix, we present the projectors onto  level-three ``colored'' basis defined in Sec.~\ref{sec:level3col}. We also provide the matrices that diagonalize the presented projectors; they are the transition matrices to the level-three ``colored'' basis, as is explained inSec.~\ref{sec:level3col}. All the matrices are written in the basis where both $\mathcal{R}_1$ and $\mathcal{R}_4$ are diagonal and equal to

\begin{equation}
\begin{array}{l}
\mathcal{R}_1=\left(\begin{array}{cccccccc}
q\mathbf{1}
\\
&q\mathbf{1}
\\
&&q\mathbf{1}
\\
&&&q\mathbf{1}
\\
&&&&-\frac{1}{q}\mathbf{1}
\\
&&&&&-\frac{1}{q}\mathbf{1}
\\
&&&&&& -\frac{1}{q}\mathbf{1}
\\
&&&&&&&-\frac{1}{q}\mathbf{1}
\end{array}\right),
\\ \\
\mathcal{R}_4=\left(\begin{array}{cccccccc}
q\mathbf{1}
\\
&-\frac{1}{q}\mathbf{1}
\\
&&q\mathbf{1}
\\
&&&-\frac{1}{q} \mathbf{1}
\\
&&&&q\mathbf{1}
\\
&&&&&-\frac{1}{q} \mathbf{1}
\\
&&&&&& q\mathbf{1}
\\
&&&&&&&-\frac{1}{q} \mathbf{1}
\end{array}\right),
\end{array}
\end{equation}
where $\mathbf{1}$ denotes a unit matrix of size of the block where it is placed.
In this basis, both level-two projectors $\ ^*P^{(1)}_2\equiv\cfrac{1+q\mathcal{R}_1}{1+q^2}$,  $\ ^*
P^{(2)}_2\equiv\cfrac{1+q\mathcal{R}_4}{1+q^2}$ and
the level-three projector acting on the first 3 strands,
\begin{equation}
P^{(2)}_{21}=\cfrac{\left(R_1-R_2\right)^2}{[3]_q},
\end{equation}
is already diagonal. The level-three projector acting on the next 3 strands,
\begin{equation}
P^{(2)}_{21}=\cfrac{\left(R_5-R_4\right)^2}{[3]_q},
\end{equation}
has the block structure:
\begin{equation}
P^{(2)}_{21}=\left(\begin{array}{cccccccc}\wp_{3\otimes   2\otimes   1}\\&{\wp}_{3\otimes   11\otimes   1}\\&&{\wp}_{21\otimes   2\otimes   1}
\\&&&{\wp}_{21\otimes   11\otimes   1}\\&&&&{\wp}_{21\otimes   2\otimes   1}\\&&&&&{\wp}_{21\otimes   11\otimes   1}
\\&&&&&&{\wp}_{111\otimes   2\otimes   1}\\&&&&&&&{\wp}_{111\otimes   11\otimes   1} \end{array}\right).
\end{equation}
The structure of the matrix that describes transition to basis~(\ref{basiscol3}) is similar to the structure of~$P^{(2)}_{21}$:
\begin{equation}
U
=\left({\small\begin{array}{cccccccc}{U}_{3\otimes   2\otimes   1}\\&{U}_{3\otimes   11\otimes   1}\\&&{U}_{21\otimes   2\otimes   1}
\\&&&{U}_{21\otimes   11\otimes   1}\\&&&&{U}_{21\otimes   2\otimes   1}\\&&&&&{U}_{21\otimes   11\otimes   1}
\\&&&&&&{U}_{111\otimes   2\otimes   1}\\&&&&&&&{U}_{111\otimes   11\otimes   1} \end{array}}\right).
\end{equation}
Note, that there is in fact two representations
\mbox{$\underline{21}$, $\overline{21}$} coming from the
decomposition
\begin{equation}
([1]\otimes
[1])\otimes[1]=[3]+\underline{[21]}+\overline{[21]}+[111].
\end{equation}
We denote
them both as $21$ here, since the corresponding blocks in $P$ and
$\mathbf{U}$ are totally identic.

\par\smallskip

The straightforward calculation gives the form of the
constituent blocks presented below. The symbol $\emptyset$ means that the corresponding block is not present in the $R$-matrices and hence in the projectors. The symbol $0$ means that the corresponding block is present in the projectors and equals zero.

\paragraph{Blocks for $\mathbf{Q=[6]}$}
\begin{equation}
{\wp}_{3\otimes 2\otimes 1|6}=1,
\ \ {\wp}_{21\otimes 2\otimes 1|6}={\wp}_{3\otimes 11\otimes 1|6}=
{\wp}_{21\otimes 11\otimes 1|6}
={\wp}_{111\otimes 2\otimes 1|6}={\wp}_{111\otimes 11\otimes 1|6}=\emptyset, \end{equation}
\begin{equation}
U_{3\otimes 2\otimes 1|51}=1.\end{equation}
\paragraph{Blocks for $\mathbf{Q=[51]}$}
\begin{equation}
{\wp}_{3\otimes 2\otimes 1|51}=\left(\begin{array}{c}-s_{51}\\\phantom{-}c_{51}\end{array}\right)\left(\begin{array}{cc}-s_{51}&c_{51}\end{array}\right),\ \ \mbox{where}\ \ c_{51}=\frac{1}{[5]_q},\ s_{51}=\frac{\sqrt{[4]_q[6]_q}}{[5]_q},
\end{equation} \begin{equation} {\wp}_{21\otimes 2\otimes 1|51}=0,\ \ {\wp}_{3\otimes 11\otimes 1|51}=1,\ \
{\wp}_{21\otimes 11\otimes 1|51}
={\wp}_{111\otimes 2\otimes 1|51}={\wp}_{111\otimes 11\otimes 1|51}=\emptyset,\end{equation}
\begin{equation}
U_{3\otimes 2\otimes 1|51}=\left(\begin{array}{cc}c_{51}&-s_{51}\\s_{51}&\phantom{-}c_{51}\end{array}\right).
\end{equation}
\paragraph{Blocks for $Q=\mathbf{[42]}$}
\begin{equation}
{\wp}_{3\otimes 2\otimes 1|42}=\left(\begin{array}{c}-s_{42}\\\phantom{-}c_{42}\end{array}\right)\left(\begin{array}{cc}-s_{42}&c_{42}\end{array}\right),
\ \ {\wp}_{21\otimes 2\otimes 1|42}
=\left(\begin{array}{c}-s^{\prime}_{42}\\\phantom{-}c^{\prime}_{42}\end{array}\right)\left(\begin{array}{cc}-s^{\prime}_{42}&c^{\prime}_{42}\end{array}\right),
\end{equation} \begin{equation} {\wp}_{21\otimes 2\otimes 1|42}=0,\ \ {\wp}_{3\otimes 11\otimes 1|42}=1, \ \
{\wp}_{111\otimes 2\otimes 1|42} ={\wp}_{111\otimes 11\otimes 1|42}=\emptyset, \end{equation} \begin{equation}
U_{3\otimes 2\otimes 1|42}=\left(\begin{array}{cc}c_{42}&-s_{42}\\s_{42}&\phantom{-}c_{42}\end{array}\right),
\ \  U_{21\otimes 2\otimes 1}=
\left(\begin{array}{cc}c^{\prime}_{42}&-s^{\prime}_{42}\\s^{\prime}_{42}&\phantom{-}c^{\prime}_{42}\end{array}\right),
\ \ \mbox{where}\ \ c_{42}=\frac{2}{[3]_q},\ \ c^{\prime}_{42}=\frac{1}{[3]_q}. \end{equation}

\paragraph{Blocks for $Q=\mathbf{[411]}$}
\begin{equation} {\wp}_{21\otimes 2\otimes 1|411}=
\left(\begin{array}{c}-s_{411}\\\phantom{-}c_{411}\end{array}\right)\left(\begin{array}{cc}-s_{411}&c_{411}\end{array}\right),\
\ \ \ {\wp}_{3\otimes 11\otimes 1|411}=
\left(\begin{array}{c}c_{411}\\s_{411}\end{array}\right)\left(\begin{array}{cc}c_{411}&s_{411}\end{array}\right),
\end{equation} \begin{equation}
{\wp}_{3\otimes 2\otimes 1|411}={\wp}_{21\otimes 11\otimes 1|411}={\wp}_{21\otimes 11\otimes 1|411}=1,
\ \ {\wp}_{111\otimes 11\otimes 1|411} =\emptyset, \end{equation} \begin{equation}
U_{21\otimes 2\otimes 1}= U_{3\otimes 11\otimes 1}
=\left(\begin{array}{cc}c_{411}&-s_{411}\\s_{411}&\phantom{-}c_{411}\end{array}\right),\ \ \mbox{where}\ \ c_{411}=\frac{1}{[5]_q},\ s_{411}=\frac{\sqrt{[4]_q[6]_q}}{[5]_q}.
\end{equation}

\paragraph{Blocks for $Q=\mathbf{[33]}$}
\begin{equation} {\wp}_{21\otimes 2\otimes 1|33} ={\wp}_{21\otimes 11\otimes 1|33}=1,\ \
{\wp}_{3\otimes 2\otimes 1|33}=0, \ \
{\wp}_{3\otimes 11\otimes 1|33}={\wp}_{111\otimes 2\otimes 1|33}={\wp}_{111\otimes 11\otimes 1|33}=\emptyset.
\end{equation}
\paragraph{Blocks for $Q=\mathbf{[321]}$}
\begin{equation}
{\wp}_{3\otimes 2\otimes 1|321}={\wp}_{3\otimes 11\otimes 1|321}
={\wp}_{111\otimes 11\otimes 1|321}={\wp}_{111\otimes 2\otimes 1|321}=1, \end{equation}
The next case is the only one where $3\times 3$ block appear:
\begin{equation} {\wp}_{21\otimes 2\otimes 1|321}= \sigma\cdot {\wp}_{21\otimes 11\otimes 1|321}\cdot
\sigma=1-\mathrm{ v}\mathrm{ v}^{\dagger}, \end{equation}

where
\begin{equation} \mathrm{
v}^{\dagger}={\footnotesize\left(\begin{array}{ccc}\cfrac{1}{\sqrt{[2]_q[4]_q}}&\cfrac{1}{[2]_q}&\sqrt{\cfrac{[5]_q}{[2]_q[4]_q}}\end{array}\right)},
\ \ \mathrm{ v}^{\dagger}\mathrm{ v}=1, \ \
\sigma=\left(\begin{array}{ccc}&&1\\&-1\\1\end{array}\right). \end{equation}

The corresponding block in the mixing matrix is
\begin{equation}  U_{21\otimes 2\otimes 1}=
\sigma\cdot U_{21\otimes 11\otimes 1}\cdot\sigma={\footnotesize\left(\begin{array}{ccc}\cfrac{1}{\sqrt{[2]_q[4]_q}}&-\sqrt{\cfrac{[5]_q}{[2]_q[3]_q[4]_q}}&\sqrt{\cfrac{[4]_q}{[2]_q[3]_q}}\\\\
\cfrac{1}{[2]_q}&-\cfrac{1}{[2]_q\sqrt{[3]_q}}&-\cfrac{1}{\sqrt{[3]_q}}\\\\\sqrt{\cfrac{[5]_q}{[2]_q[4]_q}}&\sqrt{\cfrac{[3]_q}{[2]_q[4]_q}}&0\end{array}\right).}
\end{equation}

The blocks for the rest representations are obtained with help of the level-rank duality
(described, e.g., in~\cite{MMM2},\cite{IMMM1}),
\begin{equation}
\wp_{T_1T_2T_3|Q}(q)=\wp_{\tilde{T}_1\tilde{T}_2\tilde{T}_3|\tilde{Q}}(-q^{-1}),\ \ U_{T_1T_2T_3|Q}(q)=U_{\tilde{T}_1\tilde{T}_2\tilde{T}_3|\tilde{Q}}(-q^{-1}),
\end{equation}
where $\tilde{T}$ stands for the representation with the transposed $T$ diagram.

\section{Blocks in the three-strand level-two $\mathcal{R}$-matrices \label{app:colbl2}}
In this Appendix, we provide the result of the direct computation of the constituent blocks in the three-strand level-two $\mathcal{R}$-matrices defined in Sec.~\ref{sec:lev2str3}.
All the $\mathcal{R}$-matrices are given in the vertical framing. The symbol $\emptyset$ means that the corresponding block is not present in the $R$-matrices and hence in the obtained colored $\mathcal{R}$-matrices. The symbol $0$ means that the corresponding block is present in the $\mathcal{R}$-matrices and equals zero. We write  $\pm$ before elements of the matrices $\mathcal{R}_{T_1T_2}$ with $T_1\ne T_2$ to emphasize that they are defined up to a sign.

\paragraph{Blocks for $\mathbf{Q=[6]}$}
\setlength{\arraycolsep}{-24pt}
\begin{equation}
\begin{array}{lcr}
& \mathcal{R}_{(2\otimes 2)\otimes 2 |6}=q^4, &
\\ && \\
\mathcal{R}_{(2\otimes 2)\otimes 11|6}=\mathcal{R}_{(2\otimes 11)\otimes 2 |6}=\mathcal{R}_{(2\otimes 11)\otimes 11|6}=\mathcal{R}_{(11\otimes 2)\otimes 2 |6}=&&
\\
&&=\mathcal{R}_{(11\otimes 2)\otimes 11|6}=\mathcal{R}_{(11\otimes 11)\otimes 2 |6}=\mathcal{R}_{(11\otimes 11)\otimes 11|6}=\emptyset,
\end{array}
\end{equation}

\

\setlength{\arraycolsep}{6pt}
\begin{equation}
\begin{array}{c}
U_{2\otimes 2\otimes 2|6}=1,\\
U_{2\otimes 2\otimes 11|6}=
U_{11\otimes 2\otimes 2|6}=
U_{2\otimes 11\otimes 2|6}=
U_{2\otimes 11\otimes 11|6}=U_{11\otimes 2\otimes 11|6}=\mathcal{
U}_{11\otimes 11\otimes 2|6}=\mathcal{
U}_{11\otimes 11\otimes 11|6}=\emptyset.
\end{array}
\end{equation}
\paragraph{Blocks for $\mathbf{Q=[51]}$}
\begin{equation}
\begin{array}{c}
\mathcal{R}_{(2\otimes 2)\otimes 2 |51}=\left(\begin{array}{cc}q^4\\&-1\end{array}\right),\ \ \mathcal{R}_{(2\otimes 2)\otimes 11|51}=q^4,
\\ \\
\mathcal{R}_{(2\otimes 11)\otimes 2 |51}=\mathcal{R}_{(11\otimes 2)\otimes 2 |51}=\pm q^2,
\\ \\
\mathcal{R}_{(2\otimes 11)\otimes 11|51}=\mathcal{R}_{(11\otimes 2)\otimes 11|51}=\mathcal{R}_{(11\otimes 11)\otimes 2 |51}=\mathcal{R}_{(11\otimes 11)\otimes 11|51}=\emptyset,
\end{array}
\end{equation}

\

\begin{equation}
\begin{array}{c}
U_{2\otimes 2\otimes 2|51}={\footnotesize\left(\begin{array}{cc}-c_{51} &s_{51}\\ \\-c_{51}&-s_{51} \end{array}\right),}
\\ \\
U_{2\otimes 2\otimes 11|51}=U_{11\otimes 2\otimes 2|51}=1,\ \ U_{2\otimes 11\otimes 2|51}=-1,
\\ \\
U_{2\otimes 11\otimes 11|51}=U_{11\otimes 2\otimes 11|51}=U_{11\otimes 11\otimes 2|51}=\mathcal{U}_{11\otimes 11\otimes 11|51}=\emptyset,
\end{array}
\end{equation}
where
\begin{equation}
c_{51}=\frac{[2]_q}{[4]_q},\ \ \ s_{51}=\frac{\sqrt{[2]_q[6]_q}}{[4]_q}.
\end{equation}
\paragraph{Blocks for $\mathbf{Q=[42]}$}
\begin{equation}
\begin{array}{c}
\mathcal{R}_{(2\otimes 2)\otimes 2 |42}=\left(\begin{array}{ccc}q^4\\&-1\\&&q^{-2}\end{array}\right), \ \ \mathcal{R}_{(2\otimes 2)\otimes 11|42}=-1,
\\ \\
\mathcal{R}_{(2\otimes 11)\otimes 2 |42}=\mathcal{R}_{(11\otimes 2)\otimes 2 |42}=\mathcal{R}_{(2\otimes 11)\otimes 11|42}=\mathcal{R}_{(11\otimes 2)\otimes 11|42}=\pm q^2,
\\ \\
\mathcal{R}_{(11\otimes 11)\otimes 2 |42}=q^2,\ \ \mathcal{R}_{(11\otimes 11)\otimes 11|42}=\emptyset,
\end{array}
\end{equation}

\

\begin{equation}
\begin{array}{c}
U_{2\otimes 2\otimes 11|42}= U_{2\otimes 11\otimes 2|42}=U_{2\otimes 11\otimes 11|42}= U_{11\otimes 2\otimes 11|42}=U_{11\otimes 11\otimes 2|42}=1,
\\ \\
U_{11\otimes 2\otimes 2|42}=-1,\ \ U_{11\otimes 11\otimes 11|42}=\emptyset,
\\ \\
U_{2\otimes 2\otimes 2|42}={\footnotesize\left(\begin{array}{ccccc}\cfrac{[2]_q}{[3]_q[4]_q}
&&-\cfrac{[2]_q}{[4]_q}\
\sqrt{\cfrac{[5]_q}{[3]_q}}&&\cfrac{\sqrt{[5]_q}}{[3]_q}\\\\\sqrt{\cfrac{[5]_q}{[3]_q}}&&-\cfrac{[6]_q}{[3]_q[4]_q}&&-\cfrac{1}{\sqrt{[3]_q}}\\\\
\cfrac{\sqrt{[5]_q}}{[3]_q}&&\cfrac{1}{\sqrt{[3]_q}}&&\cfrac{1}{[3]_q}\end{array}\right).}
\end{array}
\end{equation}

\paragraph{Blocks for $\mathbf{Q=[411]}$}
\begin{equation}
\begin{array}{c}
\mathcal{R}_{(2\otimes 2)\otimes 2 |411}=-1,\ \ \mathcal{R}_{(2\otimes 2)\otimes 11|411}=\left(\begin{array}{cc} q^4\\&-1\end{array}\right),
\\ \\
\mathcal{R}_{(2\otimes 11)\otimes 2 |411}=\mathcal{ R}_{(11\otimes 2)\otimes 2 |411}=\left(\begin{array}{cc}\pm q^2\\&\pm q^{-2}\end{array}\right),
\\ \\
\mathcal{R}_{(2\otimes 11)\otimes 11|411}=
\mathcal{R}_{(11\otimes 2)\otimes 11|411}=\pm q^2,\ \
\mathcal{R}_{(11\otimes 11)\otimes 2 |411}= -1,\ \
\mathcal{R}_{(11\otimes 11)\otimes 11|411}=\emptyset,
\end{array}
\end{equation}

\

\setlength{\arraycolsep}{-80pt}
\begin{equation}
\begin{array}{lcr}
&U_{2\otimes 2\otimes 2|411}=1,\ \ U_{2\otimes 11\otimes 11|411}=1,\ \ U_{11\otimes 2\otimes 11|411}=-1,\ \ U_{11\otimes 11\otimes 2|411}=-1,\ \ U_{11\otimes 11\otimes 11|411}=\emptyset,&
\\ \\
&\setlength{\arraycolsep}{6pt}
U_{2\otimes 11\otimes 2|411}=\left(\begin{array}{rr} c_{411}&-s_{411}\\-s_{411}&-c_{411}\end{array}\right),
\setlength{\arraycolsep}{-80pt}&
\\
\setlength{\arraycolsep}{6pt}
U_{2\otimes 2\otimes 11|411}=\left(\begin{array}{rr} c_{411}&-s_{411}\\s_{411}&c_{411}\end{array}\right),
\setlength{\arraycolsep}{-80pt}
&&
\setlength{\arraycolsep}{6pt}
U_{11\otimes 2\otimes 2|411}=\left(\begin{array}{rr} c_{411}&s_{411}\\s_{411}&-c_{411}\end{array}\right),
\setlength{\arraycolsep}{-80pt}
\end{array}
\end{equation}
\setlength{\arraycolsep}{6pt}
where
\begin{equation}
c_{411}=\ \frac{[2]_q}{[4]_q}\ ,\ \ s_{411}=\ \frac{\sqrt{[2]_q[6]_q}}{[4]_q}.
\end{equation}

\paragraph{Blocks for $\mathbf{Q=[33]}$}
\begin{equation}
\begin{array}{c}
\mathcal{R}_{(2\otimes 2)\otimes 2 |33}=-1,\ \ \mathcal{R}_{(2\otimes 2)\otimes 11|33}=q^{-2},\ \ \mathcal{R}_{(2\otimes 11)\otimes 2 |33}=\mathcal{R}_{(11\otimes 2)\otimes 2 |33}=\pm q^2,
\\ \\
\mathcal{R}_{(2\otimes 11)\otimes 11|33}=\mathcal{R}_{(11\otimes 2)\otimes 11|33}=\mathcal{R}_{(11\otimes 11)\otimes 2 |33}=\emptyset,\ \ \mathcal{R}_{(11\otimes 11)\otimes 11|33}=q^2,
\end{array}
\end{equation}

\

\begin{equation}
\begin{array}{c}
U_{2\otimes 2\otimes 2|33}=1,\ \ U_{2\otimes 2\otimes 11|33}= U_{2\otimes 11\otimes 2|33}=U_{2\otimes 2\otimes 11|33}=-1,
\\ \\
U_{2\otimes 11\otimes 11|33}=U_{11\otimes 2\otimes 11|33}=U_{11\otimes 11\otimes 2|33}=\emptyset,\ \ U_{11\otimes 11\otimes 11|33}=-1.
\end{array}
\end{equation}

\paragraph{$\mathbf{Q=[321]}$}
\begin{equation}
\begin{array}{c}
\mathcal{R}_{(2\otimes 2)\otimes 2 |321}=\mathcal{R}_{(2\otimes 2)\otimes 11|321}=\left(\begin{array}{cc} -1\\&q^{-2}\end{array}\right),
\\ \\
\mathcal{R}_{(11\otimes 11)\otimes 11|321}=\mathcal{R}_{(11\otimes 11)\otimes 2 |321}=\left(\begin{array}{cc}q^2\\&-1\end{array}\right),
\\ \\
\mathcal{R}_{(2\otimes 11)\otimes 2 |321}=\mathcal{R}_{(11\otimes 2)\otimes 2 |321}=\mathcal{R}_{(2\otimes 11)\otimes 11|321}=\mathcal{R}_{(11\otimes 2)\otimes 11|321}=\left(\begin{array}{cc} \pm q^2\\&\pm q^{-2}\end{array}\right),
\end{array}
\end{equation}

\

\begin{equation}
\begin{array}{ll}
U_{2\otimes 2\otimes 2|321}=\left(\begin{array}{rr} -c_{321}&-s_{321}\\s_{321}&-c_{321}\end{array}\right),
&
U_{11\otimes 11\otimes 11|321}=\left(\begin{array}{rr}c_{321}&s_{321}\\-s_{321}&c_{321}\end{array}\right),
\\ \\
U_{2\otimes 2\otimes 11|321}=\left(\begin{array}{rr}c^{\prime\prime}_{321}&s^{\prime\prime}_{321}\\-s^{\prime\prime}_{321}&c^{\prime\prime}_{321}\end{array}\right),
&
U_{2\otimes 11\otimes 2|321}=\left(\begin{array}{rr}-c^{\prime}_{321}&-s^{\prime}_{321}\\-s^{\prime}_{321}&c^{\prime}_{321}\end{array}\right),
\\ \\
U_{11\otimes 2\otimes 2|321}=\left(\begin{array}{rr}- c^{\prime\prime}_{321}&s^{\prime\prime}_{321}\\s^{\prime\prime}_{321}&c^{\prime\prime}_{321}\end{array}\right),
&
U_{2\otimes 11\otimes 11|321}=\left(\begin{array}{rr}-c^{\prime\prime}_{321}&s^{\prime\prime}_{321}\\s^{\prime\prime}_{321}&c^{\prime\prime}_{321}\end{array}\right),
\\ \\
U_{11\otimes 2\otimes 11|321}=\left(\begin{array}{rr}-c^{\prime}_{321}&-s^{\prime}_{321}\\-s^{\prime}_{321}&c^{\prime}_{321}\end{array}\right),
&
U_{11\otimes 11\otimes 2|321}=\left(\begin{array}{rr}- c^{\prime\prime}_{321}&-s^{\prime\prime}_{321}\\s^{\prime\prime}_{321}&-c^{\prime\prime}_{321}\end{array}\right),
\end{array}
\end{equation}
where
\begin{equation}
c_{321}=\ \cfrac{1}{[2]_q}\ ,\ c^{\prime}_{321}=\ \cfrac{1}{[4]_q}\ ,\ \
c^{\prime\prime}_{321}=\sqrt{\cfrac{[3]_q}{[2]_q[4]_q}}\ ,\
s_{321}=\ \cfrac{\sqrt{[3]_q}}{[2]_q}\ ,\ \ s^{\prime}_{321}=\
\cfrac{\sqrt{[2]_q[6]_q}}{[4]_q}\
,\ s^{\prime\prime}_{321}=\sqrt{\cfrac{[5]_q}{[2]_q[4]_q}}\ .
\end{equation}

The blocks for the rest representations are obtained with help of the level-rank duality
(described, e.g., in~\cite{MMM2},\cite{IMMM1})
\begin{equation}
\mathcal{R}_{T_1T_2T_3|Q}(q)=\mathcal{R}_{\tilde{T}_1\tilde{T}_2\tilde{T}_3|\tilde{Q}}(-q^{-1}),\ \ U_{T_1T_2T_3|Q}(q)=U_{\tilde{T}_1\tilde{T}_2\tilde{T}_3|\tilde{Q}}(-q^{-1}),
\end{equation}
where $\tilde{T}$ stands for the representation with the transposed $T$ diagram.

\section{Blocks in the two-strand level-three $\mathcal{R}$-matrices\label{app:colbl3}}
In this Appendix, we provide the result of the direct computation of the constituent blocks in the two-strand level-three $\mathcal{R}$-matrices defined in Sec.~\ref{sec:lev3str2}.
All the $\mathcal{R}$-matrices are given in the vertical framing. The symbol $\emptyset$ means that the corresponding block is not present in the $R$-matrices and hence in the obtained colored $\mathcal{R}$-matrices. The symbol $0$ means that the corresponding block is present in the $\mathcal{R}$-matrices and equals zero. We write  $\pm$ before the matrices $\mathcal{R}_{T_1T_2}$ with $T_1\ne T_2$ to emphasize that they are defined up to a sign. The notation $\underline{21}$ is used for the representation $21$ symmetric in the first pair of strands and $\overline{21}$ for the representation $21$ antisymmetric in the first pair of strands.

\paragraph{Blocks for $\mathbf{Q=[6]}$}

\begin{equation}
\begin{array}{c}
\mathcal{R}_{3\otimes 3|6}=q^9,
\\ \\
\mathcal{R}_{3\otimes\underline{21}|6}=\mathcal{R}_{\underline{21}\otimes 3|6}=\mathcal{R}_{3\otimes\overline{21}|6}=\mathcal{R}_{\overline{21}\otimes 3|6}=
\\
=\mathcal{R}_{3\otimes 111|6}=\mathcal{R}_{111\otimes 3|6}=\mathcal{R}_{\underline{21}\otimes\underline{21}|6}=\mathcal{R}_{\underline{21}\otimes\overline{21}|6}
=\mathcal{R}_{\overline{21}\otimes\underline{21}|6}=\mathcal{R}_{\overline{21}\otimes\overline{21}|6}=
\\
=\mathcal{R}_{\underline{21}\otimes 111|6}=
\mathcal{R}_{111\otimes\underline{21}|6}=\mathcal{R}_{\overline{21}\otimes 111|6}=\mathcal{R}_{111\otimes\overline{21}|6}=\mathcal{R}_{111\otimes 111|6}=\emptyset.
\end{array}
\end{equation}

\paragraph{Blocks for $\mathbf{Q=[51]}$}

\begin{equation}
\begin{array}{c}
\mathcal{R}_{3\otimes 3|51}=-q^3,\ \ \mathcal{R}_{3\otimes\underline{21}|51}=\mathcal{R}_{\underline{21}\otimes 3|51}=\pm q^6,\ \ \mathcal{R}_{3\otimes\overline{21}|51}=\mathcal{R}_{\overline{21}\otimes 3|51}=\pm q^6,
\\ \\
\mathcal{R}_{\underline{21}\otimes\underline{21}|51}=\mathcal{R}_{\underline{21}\otimes\overline{21}|51}= \mathcal{R}_{\overline{21}\otimes\underline{21}|51}=\mathcal{R}_{\overline{21}\otimes\overline{21}|51}=\mathcal{R}_{3\otimes 111|51}=\mathcal{R}_{111\otimes 3|51}=
\\ \\
=\mathcal{R}_{\underline{21}\otimes 111|51}=\mathcal{R}_{111\otimes\underline{21}|51}=\mathcal{R}_{\overline{21}\otimes 111|51}=\mathcal{R}_{111\otimes\overline{21}|51}=\mathcal{R}_{111\otimes 111|51}=\emptyset.
\end{array}
\end{equation}

\paragraph{Blocks for $\mathbf{Q=[42]}$}

\begin{equation}
\begin{array}{c}
\mathcal{R}_{3\otimes 3|42}=q^{-1},\ \ \mathcal{R}_{3\otimes\underline{21}|42}=\mathcal{R}_{\underline{21}\otimes 3|42}=\pm q^2,\ \ \mathcal{R}_{3\otimes\overline{21}|42}=\mathcal{R}_{\overline{21}\otimes 3|42}=\pm q^2,
\\ \\
\mathcal{R}_{\underline{21}\otimes\underline{21}|42}=\mathcal{R}_{\overline{21}\otimes\overline{21}|42}= q^5,\ \ \mathcal{R}_{\underline{21}\otimes\overline{21}|42}=\mathcal{R}_{\overline{21}\otimes\underline{21}|42}=\pm q^5,
\\ \\
\mathcal{R}_{3\otimes 111|42}=\mathcal{R}_{111\otimes 3|42}=\mathcal{R}_{\underline{21}\otimes 111|42}=\mathcal{R}_{111\otimes\underline{21}|42}=\mathcal{R}_{\overline{21}\otimes 111|42}=\mathcal{R}_{111\otimes\overline{21}|42}=\mathcal{R}_{111\otimes 111|42}=\emptyset.
\end{array}
\end{equation}

\paragraph{Blocks for $\mathbf{Q=[411]}$}

\begin{equation}
\begin{array}{c}
\mathcal{R}_{3\otimes 111|411}=\mathcal{R}_{111\otimes 3|411}=\pm q^3,\ \ \mathcal{R}_{3\otimes\underline{21}|411}=\mathcal{R}_{\underline{21}\otimes 3|411}=\mathcal{R}_{3\otimes\overline{21}|411}=\mathcal{ R}_{\overline{21}\otimes 3|411}=\pm 1,
\\ \\
\mathcal{R}_{\underline{21}\otimes\underline{21}|411}=\mathcal{R}_{\overline{21}\otimes\overline{21}|411}=-q^3,\ \ \mathcal{R}_{\underline{21}\otimes\overline{21}|411}=\mathcal{R}_{\overline{21}\otimes\underline{21}|411}=\pm q^3,
\\ \\
\mathcal{R}_{3\otimes 3|411}=\mathcal{R}_{\underline{21}\otimes 111|411}=
\mathcal{R}_{111\otimes\underline{21}|411}=\mathcal{R}_{\overline{21}\otimes 111|411}=\mathcal{R}_{111\otimes\overline{21}|411}=\mathcal{R}_{111\otimes 111|411}=\emptyset.
\end{array}
\end{equation}

\paragraph{Blocks for $\mathbf{Q=[33]}$}

\begin{equation}
\begin{array}{c}
\mathcal{R}_{3\otimes 3|33}=-q^{-3},\ \ \mathcal{R}_{3\otimes\underline{21}|33}=\mathcal{R}_{\underline{21}\otimes 3|33}=\pm q^6,\ \ \mathcal{R}_{3\otimes\overline{21}|33}=\mathcal{R}_{\overline{21}\otimes 3|33}=\pm q^6,
\\ \\
\mathcal{R}_{\underline{21}\otimes\underline{21}|33}=\mathcal{R}_{\overline{21}\otimes\overline{21}|33}=-q^3,\ \ \mathcal{R}_{\underline{21}\otimes\overline{21}|33}=
\mathcal{R}_{\overline{21}\otimes\underline{21}|33}=\pm q^3,
\\ \\
\mathcal{R}_{3\otimes 111|33}=\mathcal{R}_{111\otimes 3|33}=\mathcal{R}_{\underline{21}\otimes 111|33}=
\mathcal{R}_{111\otimes\underline{21}|33}=\mathcal{R}_{\overline{21}\otimes 111|33}=\mathcal{R}_{111\otimes\overline{21}|33}=\mathcal{R}_{111\otimes 111|33}=\emptyset
\end{array}
\end{equation}

\paragraph{Blocks for $\mathbf{Q=[321]}$}

\begin{equation}
\begin{array}{l}
\mathcal{R}_{3\otimes\underline{21}|321}=\mathcal{R}_{\underline{21}\otimes 3|321}=\mathcal{R}_{3\otimes\overline{21}|321}=\mathcal{ R}_{\overline{21}\otimes 3|321}=\pm q^{-3},
\\ \\
\mathcal{R}_{111\otimes\underline{21}|321}=\mathcal{R}_{\underline{21}\otimes 111|321}=\mathcal{ R}_{111\otimes\overline{21}|321}=\mathcal{R}_{\overline{21}\otimes 111|321}=\pm q^3,
\\ \\
\mathcal{R}_{3\otimes 3|321}=\mathcal{R}_{3\otimes 111|321}=\mathcal{R}_{111\otimes 3|321}=\mathcal{R}_{111\otimes 111|321}=\emptyset.
\end{array}
\end{equation}

In the next case, we encounter a new phenomenon: representation
$[321]$ appears with the multiplicity $2$  in the decomposition of the tensor product
\begin{equation}
[21]\otimes [21]=[42]+[411]+\mathbf{2[321]}+[3111]+[2211].
\end{equation}
The corresponding $2\times 2$ block $\mathcal{R}_{21\otimes 21|321}$ turns to be
non-diagonal and even different for the crossings of types $\underline{21}\otimes
\underline{21}$, $\overline{21}\otimes \overline{21}$, and
$\underline{21}\otimes \overline{21}$:
\begin{equation}
\begin{array}{l}
\mathcal{R}_{\underline{21}\otimes\underline{21}|321}=\left(\begin{array}{cc}-c_{321}&s_{321}\\\phantom{-}s_{321}&c_{321}\end{array}\right)=
u\left(\begin{array}{cc}1\\&-1\end{array}\right)u^{\dagger},
\\ \\
\mathcal{R}_{\overline{21}\otimes\overline{21}|321}=\left(\begin{array}{cc}\phantom{-}c^{\prime}_{321}&-s^{\prime}_{321}\\-s^{\prime}_{321}&-c^{\prime}_{321}\end{array}\right)=
\tilde{u}\left(\begin{array}{cc}1\\&-1\end{array}\right)\tilde{u}^{\dagger},
\\ \\
\mathcal{R}_{\underline{21}\otimes\overline{21}|321}=\mathcal{R}_{\overline{21}\otimes\underline{21}|321}= \left(\begin{array}{cc}-c^{\prime\prime}_{321}&-s^{\prime\prime}_{321}\\-s^{\prime\prime}_{321}&\phantom{-}c^{\prime\prime}_{321}
\end{array}\right)=\tilde{\tilde{u}}\left(\begin{array}{cc}1\\&-1\end{array}\right)\tilde{\tilde{u}}^{\dagger},
\end{array}
\end{equation}
where
\begin{equation}
\begin{array}{lcr}
c_{321}=\cfrac{[3]_q^2}{[2]_q^2[4]_q},
&
c^{\prime}_{321}=\cfrac{[5]_q}{[2]_q^2[4]_q},
&
c^{\prime\prime}_{321}=\cfrac{1}{[2]_q^7[4]_q}+\cfrac{[3]_q^5[5]_q}{[2]_q^8[4]_q^2},
\end{array}
\end{equation}
and $u$, $\tilde{u}$ and $\tilde{\tilde{u}}$ are some orthogonal matrices. All the squared blocks equal to the $2\times 2$ unit matrices.

The blocks for the rest representations are obtained with help of the level-rank duality
(described, e.g., in~\cite{MMM2},\cite{IMMM1})
\begin{equation}
\mathcal{R}_{T_1T_2|Q}(q)=\mathcal{R}_{\tilde{T}_1\tilde{T}_2|\tilde{Q}}(-q^{-1}),
\end{equation}
where $\tilde{T}$ stands for the representation with the transposed $T$ diagram.

\end{document}